\documentclass{article}

\usepackage[]{graphicx}

\newcommand{\be}{\begin{equation}}
\newcommand{\ee}{\end{equation}}

\begin{document}


\title{Redshift of photons penetrating a hot plasma}         
\author{Ari Brynjolfsson \footnote{Corresponding author: aribrynjolfsson@comcast.net}}


\date{\centering{Applied Radiation Industries, 7 Bridle Path, Wayland, MA 01778, USA}}


\maketitle


\begin{abstract}  A new interaction, plasma redshift, is derived, which is important only when photons penetrate a hot, sparse electron plasma.  The derivation of plasma redshift is based entirely on conventional axioms of physics, without any new assumptions.  The calculations are only more exact than those usually found in the literature.   When photons penetrate a cold and dense electron plasma, they lose energy through ionization and excitation, through Compton scattering on the individual electrons, and through Raman scattering on the plasma frequency.  But when the plasma is very hot and has low density, such as in the solar corona, the photons lose energy also in plasma redshift, which is an interaction with the electron plasma.  The energy loss of a photon per electron in the plasma redshift is about equal to the product of the photon's energy and one half of the Compton cross-section per electron.  This energy loss (plasma redshift of the photons) consists of very small quanta, which are absorbed by the plasma and cause a significant heating.  In quiescent solar corona, this heating starts in the transition zone to the solar corona and is a major fraction of the coronal heating.  Plasma redshift contributes also to the heating of the interstellar plasma, the galactic corona, and the intergalactic plasma.  Plasma redshift explains the solar redshifts, the redshifts of the galactic corona, the cosmological redshifts, the cosmic microwave background, and the X-ray background.  The plasma redshift explains the observed magnitude-redshift relation for supernovae SNe Ia without the big bang, dark matter, or dark energy.  There is no cosmic time dilation.  The universe is not expanding.  The plasma redshift, when compared with experiments, shows that the photons' classical gravitational redshifts are reversed as the photons move from the Sun to the Earth.  This is a quantum mechanical effect.  As seen from the Earth, a repulsion force acts on the photons.  This means that there is no need for Einstein's Lambda term.  The universe is quasi-static, infinite, everlasting and can renew itself forever.  All these findings thus lead to fundamental changes in the theory of general relativity and in our cosmological perspective.
\end{abstract}




\tableofcontents

\makeatletter	   
\renewcommand{\ps@plain}{
     \renewcommand{\@oddhead}{\textit{Ari Brynjolfsson: Redshift of photons penetrating a hot plasma}\hfil\textrm{\thepage}}%
     \renewcommand{\@evenhead}{\@oddhead}
     \renewcommand{\@oddfoot}{}
     \renewcommand{\@evenfoot}{\@oddfoot}}
\makeatother     

\pagestyle{plain}


\section{Introduction}

Compton scattering consists of scattering of one incident photon on one electron, and it results in only one out-going photon.  The cross section is about $\Phi_C= 6.65\cdot 10^{-25}\:\rm{cm}^2$.  Only a very small amount of recoil energy is transferred to the electron.  The `double Compton scattering' consists of one incident photon scattered on one electron and results in two out-going photons.  The cross section is very small, or about $\Phi_C/137$.  The `multiple Compton scattering' is also very small and consists of one incident photon scattered on one electron and results in multiple (three or more) out-going photons.  Both double and multiple Compton scattering cross sections are quantum mechanical effects and could not be deduced in classical physics.  Coherent scattering on the electrons of atoms is usually called Rayleigh scattering when the initial and final states of the electrons are the same.  But if the initial and final electronic states differ, the corresponding incoherent scattering is often called either Raman scattering or Stokes scattering.  All these processes are well known, and not a subject of this article.  When the photons scatter on the plasma electrons in thermal equilibrium, the redshifts produced by these processes are small and usually insignificant.  If the scattering electron moves relative to the observer, we get a Doppler shift, but that does not change the nature of the interactions.

\indent The plasma-redshift theory, that is deduced in this article distinguishes itself from all the processes mentioned above.  It is about the interaction of one incident photon with a great many electrons in the plasma.  The theory for this scattering has never been dealt with before.  The plasma redshift is related to `double Compton scattering' and `multiple Compton scattering', but it distinguishes itself from these processes, because it is a new multiple scattering process on a great many electrons (not only one electron, as in double and multiple Compton scattering).  Although incoherent, it is not 
related to Raman scattering, or incoherent scattering on the plasma frequency.  The plasma redshift can usually be deduced using classical physics methods, but it requires quantum mechanics to derive the relevant damping.  If classical physics damping were used, the cross section would be zero.

\indent In Compton scattering, an incident photon with wavelength of 500 nm transfers energy of about $1.6\cdot 10^{-30}h\nu$ to the plasma per electron.  The corresponding energy transferred to the plasma in the plasma redshift is 
about 200,000 times larger, or $3.3\cdot 10^{-25}h\nu$ per electron. Compared with heating by Compton scattering, the heating by plasma redshift is very large and important for explaining the heating of the solar corona, the corona of galaxies and the intergalactic plasma.  This interaction is very important, although it has been overlooked in the past.

\indent Heitler (see, in particular, sections 23 and 33 of [1]) estimated that when one of the photons emitted in `double Compton scattering' (or `multiple Compton scattering') is far in the infrared, the cross section becomes large and approaches infinity as infrared photon energy approaches zero.  When higher order effects are taken into account, Heitler found the integrated cross section to be finite.  Gould [2] has made refined calculations with essentially the same result.  The results of both Heitler and Gould are based on the incorrect assumption that the photon interacts with only one electron.  However, when one of the outgoing infrared photons is far in the infrared, the interaction in a hot, sparse plasma involves always many electrons.  Collective effects are then very important and make this cross section much larger and significant in hot, sparse plasmas, such as those in the coronas of stars, while it is usually insignificant in cold, dense laboratory plasmas and in the denser and colder chromospheres of stars.

\indent  We should realize that the solution of the infrared problem by Heitler and Gould in case of a sparse hot plasma is incorrect, because they assumed that in the infrared limit the double and multiple Compton scattering involves only one electron.  That never happens in a hot sparse plasma, when the emitted and absorbed photons are very soft.  In this case the interaction always involves great many electrons, even in the sparsest plasma of intergalactic space.  Their assumed cross section must then be replaced by plasma-redshift cross section, which is deduced in this paper.

\indent In the hot, sparse plasmas of stars' coronas, the electrons keep each other at distances, which are very long compared with their de Broglie wavelengths.  The exchange effects, which play a role only over distances shorter or comparable to the de Broglie wavelengths, are therefore of little or no importance in the hot, sparse plasmas that we are dealing with in the following discussion.  Within reasonable boundaries all the electrons have different energy levels.  The quantum numbers of angular moments in the interactions between the electrons are large.  We may, therefore, treat these sparse and hot plasmas either quantum mechanically or semiclassically.  The quantum-mechanical equation for polarization of the plasmas by light is for $\hbar\omega\ll m_e c^2$ identical with the semiclassical equation.

\indent In section 2, we deduce the cross section for the redshift in a plasma free of magnetic fields.  Some of the details of the theory are shown in Appendix A.  The cross section for the plasma redshift depends on the photon width and the damping in the plasma.  In section 3, we elaborate on how the damping in the electron plasma varies with plasma temperature and also how the damping and the density affect both the coherence effects and the cross section for the plasma redshift of photons.  It is shown how the plasma redshift varies with the wavelength, electron temperature and density.  Only when the wavelength is less than a certain cut-off wavelength, which depends on the electron temperature and density, is the plasma redshift significant.  In section 4, we give examples of how the magnetic field affects the plasma redshift and the cut-off wavelength for the redshift.  This is especially important for explaining some of the phenomena in the Sun, such as the flares, loops and arches.  Also important for explaining the phenomena is the theory for transforming magnetic field energy to heat, which is developed in Appendix B.  The transformation is often initiated and accelerated by the plasma-redshift heating.  In sections 5.1 to 5.12, we compare the plasma-redshift theory with observations in the Sun, the Milky Way, and the intergalactic space.  These comparisons lead to fundamental changes in the theory of general relativity, and in our cosmological perspective. The changes in the theory of general relativity include a reversal of the gravitational redshift of photons, which causes a significant modification of the equivalence principle.  The solar redshift experiments show clearly this reversal.  The reversals of photons' redshifts are discussed in sections 5.6.2 and 6.  The changes in the cosmological perspective include replacing the big-bang model with a seemingly static model of the universe, because the plasma redshift leads to a hot intergalactic plasma, which can explain the entire cosmological redshift and the microwave background.  Furthermore, the reversal of photons' redshifts makes it possible that the universe is seemingly static without Einstein's Lambda term.  In section 7, we suggest additional experiments for confirming the findings.  In section 8, we give a summary and conclusions.


\section{Energy loss of photons as they penetrate a plasma}
For a photon's field moving along the x-axis, we can at $x = 0$ normalize the Poynting vector, $S$, to the energy flux of one photon, $\hbar \omega_0= h\nu_0$, per second and per square cm in vacuum, where $h$ is the Planck constant.  Even in a vacuum, the photon is never infinitely sharp but consists of a distribution of frequency components as indicated by
\be
S= \hbar\omega_0= \hbar\omega_0\frac{\gamma}{2\pi}\int\limits_{-\infty}^\infty
{\frac{{d\omega}}{{\left({\left({\omega - \omega_0}\right)^2} + {{\gamma ^2}/4}
\right)}}}\rm{,}
\ee
where $\gamma$ is the photon width [1].  For the dielectric constant $\varepsilon = (n - i\kappa)^2 = 1$ and therefore the refraction index $n = 1$ and absorption cooefficient, the imaginary part $\kappa = 0 {\rm{,}}$ this form of the field in the photon also follows from Eq.\,(3) below, which follows from Eq.\,(A29) in the Appendix A.  When the photon penetrates a plasma, the photon's virtual field will be modified by the dielectric constant $\varepsilon= (n - i \kappa )^2 {\rm{.}}$  From the solution, Eq.\,(A15), to the dynamical Eq.\,(A12) of the Appendix A, we get that the polarization is given by Eq.\,(A18).  From the polarization, we derive that the dielectric constant is given by Eqs.\,(A19) and (A20) of the Appendix A.  If the binding-energy frequency $\omega_q = 0$ and the collision damping $\alpha = 0{\rm{}}$ (because the collision damping, $\alpha {\rm{,}}$ is included in $\beta \omega^2{\rm{),}}$ we derive from Eq.\,(A20) that the dielectric constant is
\be
\varepsilon= \left(1-\frac{\omega_p^2}{\omega^2+\beta^2\omega^4}\right)\,-\; i\frac
{\beta\omega\omega_p^2}{\omega^2+\beta^2\omega^4}\rm{,}
\ee
where $\omega_p= \sqrt{4\pi e^2N_e/m_e}$ is the plasma frequency, and where
$\beta\omega^2$ is the radiation damping in the hot sparse plasma.

\indent We set the magnetic permeability equal to 1.  As shown in Eq.\,(A29) of Appendix A, we get from Eq.\,(1) and Eq.\,(2) at distance $x$ that
\be
S= \hbar\omega_0\frac{\gamma}{2\pi}\int\limits_{-\infty}^\infty{\frac{n}
{\varepsilon\bar\varepsilon}\frac{\left[\exp\left(-2\kappa\omega x /
c\right)\right]d\omega}{\left(\gamma^2/4+\left(\omega-\omega_0
\right)^2\right)}\rm{,}}
\ee
where $\bar\varepsilon$ is the complex conjugate of $\varepsilon$.

\indent We differentiate this expression with respect to $x$ and get that the photon's energy loss per cm is given by
\[ \frac{dS}{dx}= -\frac{\hbar\omega_0\gamma}{2\pi c}\int\limits_{-\infty}^\infty {\frac{2\kappa\omega n}{\varepsilon\bar\varepsilon}\frac{\left[\exp
\left(-2\kappa\omega x/c\right)\right]d\omega}{\left(\gamma^2/4
+\left(\omega-\omega_0\right)^2\right)}}. \]
For $x$ equal to 0, the energy loss per cm is then
\be
\frac{dS}{dx}= -\frac{\hbar\omega_0\gamma}{2\pi c}\int\limits_{-\infty}^\infty  {\frac{2\kappa\omega n}{\varepsilon\bar\varepsilon}\frac{d\omega}{\left(
\gamma ^2/4+\left(\omega-\omega_0\right)^2\right)}}.
\ee
From Eq.\,(2), see also Eq.\,(A22), we derive that
\[ \frac{2n\kappa\omega}{\varepsilon\bar\varepsilon} = \frac{\beta\omega^4\omega_p^2} {\left(\omega_p^2- \omega^2\right)^2 + \beta^2\omega^6}\rm{,}\]
and when we insert this expression into Eq.\,(4), we get that the photon's energy loss per cm is
\be
\frac{dS}{dx}= -\frac{\hbar\omega_0\gamma}{2\pi c}\int\limits_{-\infty}^\infty{
\frac{\beta\omega_p^2\omega^4}{\left[\left(\omega_p^2-\omega^2\right)^2+\beta^2
\omega^6\right]}\frac{d\omega}{\left[\gamma^2/4+\left(\omega-\omega_0
\right)^2\right]}}.
\ee
The right side of Eq.\,(5) can be integrated in the complex plane along the x-axis from $-\infty$ to $+\infty$ and then counterclockwise along the semicircle in the upper half plane.  The integral along the semicircle is zero.  The integral in Eq.\,(5) is then equal to $2\pi i$ times the sum of the residue of the poles in the upper half-plane, where $i$ is the notation for 
the imaginary component.  For \[
\omega_q= 0, \; \; \alpha= 0, \; \; \beta\gg\beta_0, \; \; \omega_0\gg\omega_p,
 \; \; \beta\omega_p \ll 1, \; \; \gamma \ll \omega_0, \]
the four poles in the upper plane are given by (see Eq.\,(A33) of Appendix A)
\be
\omega = \left\{ {\begin{array}{llll}   \rm{a} & = & + ~ \omega_p & \displaystyle +\; \; i\, \frac{\beta\omega_p^2}{2} \\
   \rm{b} & = & - ~ \omega_p & \displaystyle +\; \; i\,\frac{\beta \omega_p^2}{2} \\    \rm{c} & = &   ~        & \displaystyle +\; \; i\,\frac{1}{\beta}\left[1+\left(\beta\omega_p\right)^2\right]  \\
   \rm{d} & = & + ~\omega_0 & \displaystyle + \; \; i\,\frac{\gamma}{2}
\end{array}} \right\}{\rm{.}}
\ee

\indent From Eqs.\,(5) and (6), we get that
\be
\frac{dS}{dx}= -\hbar\omega_0\frac{8\pi}{3}r_e^2N_e\left[\frac{\gamma}{4\gamma_0}
+\frac{\gamma}{4\gamma_0}+\frac{\gamma}{2\gamma_0}\cdot\frac{\left(1-1/
\left(\beta\omega_0\right)^2\right)}{\left(1+1/\left(\beta\omega_0
\right)^2\right)^2}+\frac{1}{\left(1+\left(\beta_0\omega_0\right)^2\right)}\right]{\rm{,}}
\ee
where $\beta_0\omega_p^2/c= \left(8\pi /3\right)r_e^2N_e= 6.65 \cdot 10^{-25}N_e$ is the Compton cross section per cm of photon path when $N_e$ is the electron density per cubic cm, and where $\beta_0= 2e^2/\left(3m_e c^3\right)= 6.266 \cdot 10^{-24}$ is the classical damping constant and $\gamma_0= \beta_0\omega_0^2$
the classical damping for the incident photon.

\indent \textbf{The last term inside the brackets}, corresponding to the pole d in Eq.\,(6), is identical to the quantum mechanical Compton scattering cross section for soft photons, as deduced by Heitler [1].  In the Compton scattering, we set damping constant equal to the classical damping constant and the dielectric constant equal to 1, as Heitler did, because in the sparse plasmas of our interest the incident photon interacts with only one electron.  If the electron were bound in an atom with other electrons, we would get Rayleigh scattering.

\indent \textbf{The two first terms inside the brackets}, correspond to the poles a and b in Eq.\,(6).  These poles correspond to Raman scattering on the plasma frequency, $\omega_p {\rm{.}}$  In the treatment above, the oscillator strengths are positive as we assumed that they were in the ground state.  In the actual hot plasma of our interest the plasma oscillators are usually in thermodynamic equilibrium, and we have then about equal number of positive and negative oscillator strengths.  In thermodynamic equilibrium the emission and absorption cancel each other.  Nevertheless, these interactions cause small angular scatterings, which are insignificant in practically all experiments of our interest, because the densities of the plasmas of our interest are usually very low.  However, they can affect the observations of most distant supernovas as Eq.\,(52) shows.

\indent \textbf{The third term inside the brackets, the plasma-redshift term}, corresponds to pole c, the pole on the imaginary axis.  The plasma-redshift term is a new cross section term and the focus of our interest in this article.  It is due to loss (emission and absorption) of very low frequency components in the
photon field. This cross section has been overlooked in the past, most likely because when the damping factor $\beta$ in the radiation damping, $\beta\omega_0^2 {\rm{,}}$ is close to the classical value of $\beta_0 {\rm{,}}$ as it is in an ordinary laboratory plasma, this third term is practically zero and the cross section insignificant.  However, in a hot, sparse plasma both the damping factor, $\beta {\rm{,}}$ and the collective effects are very large; and this plasma-redshift term becomes significant.  As mentioned in the introduction, this term is also related to the emission of very soft quanta in double and multiple Compton scattering.  Those familiar with the deduction of Cherenkov radiation, which is emitted when fast charged particles penetrate dielectrics, may also find some resemblance between plasma redshift and Cherenkov radiation.  The classical mechanics cannot treat properly the radiation damping terms. We must therefore use quantum mechanics to determine the damping.  It then can be seen that $\beta$ can be very large in a hot, sparse plasma.  In the third term inside the brackets of Eq.\,(7), the value of $1/\left(\beta\omega _0\right)$ is then very small.  The plasma-redshift cross section becomes then equal to $3.326 \cdot 10^{-25}\left(\gamma/\gamma_0\right)N_e$ per cm of the photon path.  In the following section 3, we will see how $\beta$ changes with temperature and density of the plasma, and with the wavelength of the incident light.


\section{Damping factor $\beta$ and the cut-off wavelength for plasma redshift}

\subsection{Semi-classical treatment of the collision damping}

The collision damping $\alpha$ in Eq.\,(2) is often equated with $2/\tau {\rm{,}}$ where $\tau$ is the time between collisions; see {\textbf{Comment A4}} in Appendix A.   From the stopping theory for charged particles we know that also the frequencies of the Fourier field of the incident particles determine the energy transfer in the collisions.  The collision factors $\alpha = 2 / \tau {\rm{,}}$ and $\alpha + \beta_p \omega^2=\beta \omega^2$ depend on the frequency $\omega $ of the fields' Fourier components of the colliding electrons (see discussion below Eq.\,(A12) in the Appendix A).  For being effective in disturbing the oscillation of an electron in the Fourier field of the incident photon, the Fourier harmonic of the colliding electrons must have about the same or higher frequency as the photon. 

\indent  For example, the root $d$ in Eq.\,(6) corresponds to the center (principal) frequency, $\omega_0 {\rm{,}}$ of the incident photon field.  For disturbing the oscillation of the electron at the frequency $\omega_0 {\rm{,}}$ the frequency of the collision field must be about equal to or greater than $\omega_0 .~$  From the stopping theory of charged particles, we know that the energy absorbed per colliding electron in a small increment $dp$ of the impact parameter $p$ is proportional to $\{[x K_1(x)]^2 + [(x/\gamma) K_0(x))]^2\}(dp/p),$  where $K_0(x)$ and $K_1(x)$ are the modified Bessel functions of zero and first order, $ x = p \omega_0 / \gamma v {\rm{,}}$ and where $\gamma$ in this case is the relativistic factor $\gamma = 1/\sqrt{1-v^2/c^2} {\rm{,}}$ which is not to be confused with the same notation for the radiation damping.  Niels Bohr deduced these relations in 1913 and 1915 [3].  In the following, we will assume that this relativistic factor, $\gamma {\rm{,}}$ is about equal to one, which eliminates any confusion about the notations.  The quantity within the braces is about equal to one for $x = p \omega_0 / \gamma v < 1 {\rm{,}}$ and for $x \geq 1 {\rm{,}}$ it falls off exponentially.  When we integrate over all values of $p{\rm{,}}$ we get that the energy loss per second from the colliding incident electrons is for relativistic factor $\gamma \approx 1$
\[
\frac{{dE}}{{dt}}= \frac{{4\pi e^4 N_e}}{{m_e v}}\cdot\left[{\rm{ln}}\frac{{m_e v^2}}{{\delta \cdot h \nu}} - \frac{{\delta_1}}{{2}} \frac{{v^2}}{{c^2}}  \right]= 7.3425\cdot 10^{-10} \frac{{ N_e}}{{ v}}\cdot\left[{\rm{ln}}\frac{{m_e v^2}}{{1.147 \cdot h \nu}} - \frac{{0.815}}{{2}} \frac{{v^2}}{{c^2}}~{\rm{erg\,s}^{-1}} \right]{\rm{,}}
\]
where $\delta = 1.147 $ and $\delta_1 = 0815 {\rm{.}}~$ In case of a proton collision, we must add the term ${\rm{ln}}\,2 $ within the brackets, and replace the electron density, $N_e {\rm{,}}$ with the proton density, $N_p {\rm{.}}~$

\indent  In the solar corona, we may have that $N_e \approx 5.5\cdot 10^7~{\rm{cm}^{-3}} {\rm{,}}$ and $T\approx 2\cdot 10^6~{\rm{K,}}$ which corresponds to $v = 9.536\cdot 10^8~{\rm{cm\,s}^{-1}} {\rm{,}} $ and $m_e v^2 = 517.044~{\rm{eV}} {\rm{.}}~$  We get for $h\nu \approx 4 ~{\rm{eV}} $ that $dE/dt \approx 2 \cdot 10^{-10}~{\rm{erg\,s}^{-1}} {\rm{,}}$  which is small compared with $\gamma_0 \, h\nu \approx 2.3142\cdot 10^8 \cdot 6.4087\cdot 10^{-12} = 1.4831\cdot 10^{-3}~{\rm{erg\,s}^{-1}} {\rm{,}}$ where $\gamma_0 = 6.266\cdot 10^{-24} \cdot \omega ^2 $ is the classical damping.  Usually, we have that the quantum mechanical damping, $\gamma $ is in the range of $ 0.1 \gamma_0 \leq \gamma \leq 10 \gamma_0 {\rm{.}}~$  The Compton scattering in the corona is therefore not affected by the collision damping.  Similar estimates show that the collision broadening does not affect the Compton scattering low in the transition zone, and much less in interstellar and intergalactic space where the densities are much lower.

\indent When the frequency, $\nu_0 {\rm{,}}$ of the incident photon is very low, the value of $dE/dt$ becomes larger, and the value of $\gamma_0 = 6.266\cdot 10^{-24} \cdot \omega ^2 $ smaller.  For example, for $h\nu \leq 4\cdot 10^{-4} ~{\rm{eV}}  {\rm{,}}$ the collision broadening will affect the Compton scattering depending on the density.

\indent  The incident photon consists of a broad spectrum of frequencies as shown in the integrand of Eq.\,(1).  The low frequency components of the photon may act on several electrons coherently.  The corresponding classical damping, $\gamma_0 = 6.266\cdot 10^{-24} \cdot \omega ^2 {\rm{,}}$ is then emitted coherently from several electrons, which therefore magnifies the corresponding damping contribution.  These low energy components cannot keep up with the photon's higher frequencies, and are ripped off the photon and absorbed in the plasma.  This is similar to the Cherenkov radiation from charged particles penetrating dielectrics.  We usually think of the Cherenkov radiation as being all emitted, but a part of the Cherenkov radiation, the part that is close to the resonance, is actually absorbed immediately, just like the damping in the pole c of Eq.\,(6), which results in the third term within the brackets of Eq.\,(7).  The forces within the photon recreate the removed low-frequency components, just like the Fourier harmonics of the field of the charged particles are regenerated after being removed as Cherenkov radiation.  We will in the following establish the exact condition for this damping to become significant.  We will have to treat the phenomena in accordance with quantum mechanics in the following subsection.


\subsection{The plasma electron as a harmonic oscillator}

It can be shown in many ways that when the plasma is disturbed, the forces within the plasma will result in characteristic oscillations with eigenfrequency $\omega_p .~$  For example, we see this frequency in the dielectric constant given by Eq.\,(A20) and therefore in displacement given by Eqs.\,(A15), and in the polarization given by Eq.\,(A18), and therefore in Eq.\,(2),\,(5),\,and\,(6).  For $\omega_q = 0$, the plasma frequency, $\omega_p ,$ is the principal frequency for absorption in Eq.\,(5).  Each electron will oscillate as a classical oscillator with a restoring force proportional to the displacement r.  The electron will oscillate as a classical oscillator due to the polarization with the restoring force $m \ddot{r} = - k r $ and the frequency $\omega = \sqrt{k / m} .~$  For each plasma electron, we have that $k = 4 \pi N_e e^2 .~$  The force $ m \ddot{r} = - k r = - 4 \pi N_e e^2 r $ corresponds to the polarization given by Eq.\,(A18).  When we solve the classical equation $m \ddot{r} = - k r = - 4\pi N_e e^2 r,$ the solution is that of a classical harmonic oscillator with the frequency $\omega_p =  \sqrt{k / m} = \sqrt{4\pi N_e e^2/m} .~$  The plasma frequency $\omega_p $ is defined by Eq.\,(A21).  For $\omega_q = 0$ it is the characteristic eigenfrequency for each plasma electron, as Eq.\,(5) shows  

\indent   The forced oscillations of the electron will result in the usual radiation damping.  The positive ions will also act like harmonic oscillators, but their radiation damping is much smaller, because of their larger mass.

\indent  We can then treat the electron plasma quantum mechanically.  The plasma consists of great many oscillators.  For the electrons in a hot plasma the Hamiltonian for each oscillator is given by
\be
H_0= H_0^0+\frac{1}{2}m_e\omega_q^2\bar rr= - \frac{\hbar^2}{2m_e}\nabla^2+\frac{1}
{2}m_e\omega_q^2\bar rr,
\ee
where $\omega_q$ is principally any frequency in the plasma and $r$ is given by Eq.\,(A15).  The plasma frequency $\omega_p$ is the dominant frequency, however, and we will usually replace $\omega_q$ by $\omega_p$.  This non-relativistic Hamiltonian does not take into account the effects of magnetic fields, which will be treated subsequently. 

\indent The solutions corresponding to the Hamiltonian given by Eq.\,(8) are well known.  The energy levels when we include the finite lifetime of the states are
\be
E_\Lambda= E_{n,l}= \left(\Lambda+\frac{3}{2}\right)\hbar\left(\omega_q-i\frac
{\beta\omega_q^2}{2}\right),
\ee
where $\Lambda = 2n+l$ and where $n$ and $l$ can be any of the whole numbers: 0, 1, 2, ., ., ..  The imaginary part of the frequency is included to indicate the finite lifetime of the states and the magnitude of the damping.  In the case of magnetic fields, we must, in addition to the radial quantum number $n$ and the angular quantum number $l$, take into account the third quantum number $m$.

\indent When the magnetic field is zero, the states are degenerate; that is, several states can have the same energy for $\Lambda > 1$.  For example, for the states $\Lambda= 4$, we have $(n,l)= (2,0),(1,2),$ or $(0,4)$.  These three states in turn have the multiplicity of 1, 5, and 9, respectively, for a 
total of 15 states.  More generally, for each quantum number $\Lambda$, the total number of states is $\left(\Lambda+1\right)\left(\Lambda+2\right)\mathord{/}2$, all with the same energy.

\indent The radiation-damping factor $\beta$ may deviate significantly from the 
classical value $\beta_0$.  In the transitions from a state $\Lambda$ to all final states, the energy emitted is given by
\be
\begin{array}{l} \displaystyle 
I\left(\omega_q\right)= \hbar\omega_q A_{n,l}= \frac{4e^2\omega_q^4}{c^3}
\frac{\hbar}{2m_0\omega_q}\left[\frac{n_x+n_y+n_z}{3}\right] \\
\displaystyle = \hbar\omega _q \frac{{2e^2 \omega _q^2 }}{{3m_0 c^3 }}\Lambda  = \hbar \omega _q \Lambda \beta _0 \omega _q^2  = \hbar \omega _q \beta \omega _q^2 {\rm{ ,}} 
 \end{array}
\ee
where $\beta= \Lambda\beta_0$ can be very large.  In hot plasmas, practically all the oscillators are highly excited.  Their average $\Lambda$ values are about $\Lambda = 3kT/\left(\hbar\omega_q\right)$,
and Einstein's $A$ coefficients, $A_{\Lambda\to\Lambda-1}= \beta\omega_q^2= \Lambda\beta _0\omega_q^2$,
 are therefore large.  The radiation loss given by the redshift term in Eq.\,(7) then becomes relatively large and significant.  In good plasmas $\omega_p$ is the dominant frequency, and we can therefore replace $\omega_q$
 by $\omega_p$.

\indent For evaluating the value of this redshift term, we need to average it over all states in the hot plasma.  The number of possible states in a hot plasma with quantum number $\Lambda$ is
\be
g_\Lambda  =  \frac{{\left( {\Lambda  + 1} \right)\left( {\Lambda  + 2}
 \right)}}{2} \approx \frac{{\Lambda ^2 }}{2}{\rm{ }}{\rm{.}}
\ee
The statistical energy distribution of the oscillators in thermal equilibrium is given by
\be
n_\Lambda= \frac{g_\Lambda}{\exp\left(\alpha+\Lambda\hbar\omega_p/
\left(kT\right)\right)-1}\approx\frac{g_\Lambda\exp\left(-\alpha\right)}
{\exp\left(\Lambda\hbar\omega_p/\left(kT\right)\right)}\rm{.}
\ee
The last approximation is valid because $\rm{exp}\left(\alpha\right)$
 is very large in hot and sparse plasmas, which in turn means that the Boltzmann, Fermi-Dirac, and Bose-Einstein statistics all render the same 
result.  The normalized distribution function for the oscillator strengths is then given by
\be
P\left(n_\Lambda\right)d\Lambda= \left(\frac{\hbar\omega_p}{kT}\right)^3\frac
{\Lambda^2}{2}\exp\left(-\frac{\Lambda\hbar\omega_p}{kT}\right)d\Lambda\rm{.}
\ee

\indent The cut-off frequency for the plasma redshift can be determined by weighing the third term inside the brackets of Eq.\,(7) by the normalized distribution 
function, Eq.\,(13).  For $\beta\omega_0= \Lambda\beta_0\omega_0= x$,
 and $a = \hbar\omega_p/\left(\beta _0\omega_0kT\right)= 3.65\cdot 10^5 \lambda_0\sqrt{N_e}/T$,
 we get
\be
F_1 \left( a \right) = a^3 \int\limits_0^\infty  {\frac{{\left( {x^2  -
 1} \right)}}{{\left( {x^2  + 1} \right)^2 }}\frac{{x^4 }}{2}} \exp \left
( { - ax} \right)dx{\rm{.}}
\ee


\begin{table}
\centering
{\bf{Table 1}} \, \, $F_1(a)$ as a function of $a$; see Eq.\,(14).

\vspace{2mm}

\begin{tabular}{llllllcl}
	\hline
$a$ & $F_1(a)$ & $a$ & $F_1(a)$ & $a$ & $F_1(a)$ & $a$ & $F_1(a)$ \\
	 \hline \hline
0.0 & 1.000 & 1.0 & 0.571 & 2.0 & 0.228 & 6.0 & -0.070 \\
0.1 & 0.990 & 1.1 & 0.527 & 2.2 & 0.183 & 7.0 & -0.073 \\
0.2 & 0.962 & \textbf{1.163} & \textbf{0.500} & 2.4 & 0.144 & 8.0 & -0.071 \\
0.3 & 0.921 & 1.2 & 0.485 & 2.6 & 0.111 & 9.0 & -0.067 \\
\textbf{0.344} & \textbf{0.900} & 1.3 & 0.445 & \textbf{2.671} & \textbf{0.100} & 10.0 & -0.061 \\
0.4 & 0.872 & 1.4 & 0.407 & 2.8 & 0.082 & 20.0 & -0.024 \\
0.5 & 0.821 & 1.5 & 0.372 & 3.0 & 0.057 & 40.0 & -0.0071 \\
0.6 & 0.769 & 1.6 & 0.339 & 3.5 & 0.010 & 50.0 & -0.0047 \\
0.7 & 0.717 & 1.7 & 0.309 & 3.633 & 0.000 & 100.0 & -0.0012 \\
0.8 & 0.667 & 1.8 & 0.280 & 4.0 & -0.022 & 200.0 & -0.0008 \\
0.9 & 0.618 & 1.9 & 0.253 & 5.0 & -0.057 & $\infty$ & -0.0000 \\
\hline
\end{tabular}
\end{table}

\noindent  When we integrate Eq.\,(14) over the parameter $ x=\beta\omega_0 = \Lambda\beta_0\omega_0 ,$ we get
\[
 F_1( a ) = a^3 \,\left[ \frac{{1}}{{a^3}} - \frac{{1.5}}{{a}} +2 \,f(a)-\frac{{a}}{{2}}\, g(a) \right]
 = \left[ 1 + 2\, a^2 \left( a \, f(a) - 0.75 - \frac{{a^2}}{{4}}\, g(a) \right) \right] ,
\]
where the functions  $f(a)$ and $g(a)$ are given by
\[
 f(a) = \int\limits_0^\infty \frac{{ {\rm{exp}}( - a x)}}{{x^2+1}} \,dx\quad {\rm{and}} \quad g(a) = \int\limits_0^\infty \frac{{ x \,{\rm{exp}}( - a x)}}{{x^2+1}} \,dx .
\]

\noindent  For numerical values of $f(a)~{\rm{and}}~g(a)$, see reference [4].
The numerical values for the oscillator strength function, $F_1(a) {\rm{,}}$ are shown in Table 1. 

\indent In Eq,\,(7), we can then replace the product of $N_e$ and the third term within the brackets by $  N_e\cdot F_1\left(a\right)$, which is a measure of the oscillator strength in the plasma redshift term, the third term inside the brackets in Eq.\,(7).  Table 1 shows that $F_1\left(a\right)$ is close to 1 for small values of $a$, or for high frequencies $\omega_0$, or for small wavelengths $\lambda _0 $ in cm of the incident radiation.  From the definition of $a = \hbar \omega_p/\left(\beta _0\omega_0 kT\right)= 3.651\cdot 10^5 \lambda_0\sqrt{N_e}/T {\rm{,}}$ we have that the photon's cut-off wavelength is 
\be
\lambda _0  = \frac{2 \pi\, c}{{\omega_0 }} = 2.739 \cdot 10^{ - 6}\, a \, \frac{T}{{\sqrt {N_e } }}{\rm{ }}{\rm{.}} 
\ee
From Table 1, we can find the oscillator strength for a given value of $a {\rm{.}}~$ For example, for $a \leq 1.163 {\rm{,}}$ we have that $F_1(a) \geq 0.5 {\rm{,}}$ or that the oscillator strength is $ \geq 50 \, \%  {\rm{,}}$ where the 50\% cut-off wavelength, $\lambda _{0.5} {\rm{,}}$ for the redshift is determined by inserting 1.163 for $a$ into Eq.\,(15).  We get
\be
\lambda _{0.5} = 2.739 \cdot 10^{ - 6}  \cdot 1.163\frac{T}{{\sqrt {N_e} }} = 3.185 \cdot 10^{ - 6} \frac{T}{{\sqrt {N_e } }} ~\, {\rm{cm }}~ = 318.5  \frac{T}{{\sqrt {N_e } }} ~\, {\rm{\AA }}  {\rm{.}}
\ee
The 90\,\% and 10\,\% oscillator strengths are obtained for $a$ equal to 0.344 and 2.671, respectively.  The corresponding 90\,\% and 10\,\% cut-off wavelengths are obtained by inserting the corresponding values for $a$ into Eq.\,(15).

\indent When we in Eq.\,(7) replace $dS$ by $d\hbar \omega $ and $\omega_0$ by $\omega  {\rm{,}}$ and when we consider only the redshift term, the third term within the brackets, while disregarding the first, second, and fourth term within the brackets, we get

\be
 - \frac{{d\hbar \omega }}{{\hbar \omega }} = \frac{{4\pi }}{3}r_e^2 N_e \frac{\gamma }
{{\gamma _0 }}\,F_1(a)\, dx {\rm{.}}
\ee
When we then integrate each side and set $\lambda-\lambda_0= \Delta\lambda$, we get
\be
 - \int\limits_{\omega _0 }^\omega  {\frac{{d\omega }}{\omega }}  = {\rm{ln}}\left( \frac{{\omega _0 }}{{\omega }} \right) = {\rm{ln}}\left( 1 + \frac{{\Delta \lambda }}{{\lambda _0 }} \right)= {\rm{ln}}\, ( 1 + z ) = 3.326 \cdot 10^{ - 25} \int\limits_0^R {F_1 \left( a \right)\frac{\gamma }{{\gamma _0 }}N_e dx}  {\rm{,}} 
\ee
where $\Delta\lambda/\lambda = z  {\rm{.}}$

\indent Once the redshift is initiated in the transition zone to the solar corona, the redshift heating (due to absorption of the far infrared Fourier components of each photon) causes relatively rapid temperature increase and density decrease.  Below 50\,\% cut-off, given by Eq.\,(16) for $a=1.163 {\rm{,}}$ the oscillator strength function, given by Table 1, is less than 50\,\%, and above the 50\,\% cut-off it is more.  By averaging, we can often for each wavelength set the oscillator strength function equal to 1 above the 50\,\% cut-off and equal to zero below the 50\,\% cut-off.

\indent In the middle of the transition zone to the solar corona, we have (see the discussion in section 5.1) that $T \approx 500,000~{\rm{K}} {\rm{,}} $ and $N_e \approx 5\cdot 10^9 ~{\rm{cm}^{-3}} {\rm{.}}~$ These values correspond to $T\,N_e \approx 5 \cdot 10^{14} ~{\rm{K\,cm}^{-3}} {\rm{.}}~$ From Eq.\,(16) we get for these values that the 50\% cut-off wavelength is 500 nm; that is, photons with wavelength shorter than 500 nm will be redshifted more than 50\% of the maximum redshift.  Above the cut-off limit the temperature usually increases sharply and the density continues to decrease until the entire solar spectrum is redshifted.

\indent  Detailed analysis shows that for a quiescent corona, the redshift heating exceeds the X-ray and recombination cooling in the transition zone to the corona.  This causes the temperature to increase to about two million degrees (see sections 5.1 and 5.2).  Below this maximum temperature, a significant fraction of the heating in the upper transition zone leaks by conduction into the lower transition zone and helps compensate the greater recombination cooling.  The unevenness in the heat conduction and the effects from the magnetic field cause some turbulence in the transition zone.  For shorter frequencies, the cut-off penetrates deeper into the transition zone.  For example, for about the same pressure, we get from Eq.\,(16) that when $T\approx 200,000~{\rm{K}} {\rm{,}}$ the 50\% cut-off wavelength is
$\lambda_{0.5} \approx 121.5~{\rm{nm}} =1215~{\rm{\AA}} {\rm{.}}~$


\subsection{Photon width}

We see from Eq.\,(18) that the redshift is proportional to the photon width, $\gamma {\rm{.}}~$ Different broadening effects broaden the photon width.  For example, at the center of the solar disk, where the pressure in the line forming elements is greatest, the measured photon-width, $\gamma= \gamma_i$, of the Na I 589.592 nm resonance-line is about 17 times the classical width, $\gamma_0 = \beta_0\omega^2= 6.266\cdot 10^{-24}\omega^2 {\rm{,}}$ which in this case is about equal to the quantum mechanical width of the photons from the undisturbed sodium atom.  However, we have also that after the emission, when the photon penetrates and interacts with the electron plasma, the photon's initial width should approach the photon width, $\gamma_0 {\rm{,}}$ which is the natural quantum mechanical and classical width of photons interacting with an electron plasma.  We do not know exactly how fast the redistribution of the frequencies within the photon takes place, or how fast the photon width approaches the classical width, $\gamma_0 {\rm{;}}$ but we assume that the small incremental change in the width on the stretch $dx$ is proportional to the difference in the actual width and the final classical width and proportional to the plasma redshift.  We set
\be
d\gamma  =  - \xi \frac{{\left( {\gamma  - \gamma _0 } \right)\omega }}
{{\gamma _0 }}\frac{{4\pi r_e^2 }}{3}N_e dx{\rm{ ,}}
\ee
where $\gamma_0 $ is the classical width as well as the quantum mechanical width of photons penetrating and interacting with the electron plasma.  In Eq.\,(19), $\xi$ is an adjustment factor, and its value is to be determined experimentally.  (When we have a better theory for the forces within the photon, we may be able to determine this factor theoretically, but at this stage we suggest that it be determined experimentally.  A rough estimate for the resonance line of Na-I in the Sun indicates that  $\xi$ is about 0.25).  From Eq.\,(19) we determine $\gamma$ as a function of $x$, and insert that value into Eq.\,(18).  For oscillator strength function $F_1\left(a\right)$ equal to 1, Eq.\,(18) takes the form
\be
{\rm{ln}}\, ( 1 + z ) =  3.326 \cdot 10^{ - 25} \int\limits_0^R {N_e dx} \; + \; \frac{{\gamma _i  - \gamma _0 }}{{\xi \omega }} =  3.326 \cdot 10^{ - 25} \int\limits_0^R {N_e dx} \; + \; \frac{{\delta \lambda_i  - \delta \lambda_0 }}{{\xi \lambda}}{\rm{,}}
\ee
where $\gamma_i~{\rm{s}^{-1}}{\rm{,}}$ or $\delta \lambda_i~{\rm{cm}}{\rm{,}}$ are the initial photon widths Stark broadened, or broadened by collisions and pressure, while $\gamma_0 = \beta_0 \omega^2=6.266\cdot 10^{-24}\cdot \omega^2~{\rm{s}}^{-1} {\rm{,}}$ or $\delta \lambda_0 = 0.118~ {\rm{m\AA ,}} $ are the final photon widths, which are equal to the classical photon widths and independent of the pressure.  The second term on the right side of Eq.\,(20) is often a small correction to the first term.  For small redshifts, such as those in the corona of the Sun and in stars, it is usually significant and varies from line to line depending on the initial line strength and on the collision broadening and Stark broadening.

\indent  In the Sun, the collision broadening is small for some of the lines.  The redshift for these lines increases strongly from the center to limb, because the integration path in the corona is longer as we approach the limb and because the second term is small.  For these lines the center-to-limb variation is large, as we will see in section 5.6.  For other lines the collision broadening is large.  The second term on the right side of Eq.\,(20) is then large at the center of the solar disk, but it decreases usually as we get closer to the limb due to the lower pressure in the line forming elements.  The second term of these lines is then large at the center than at the limb.  This may partially cancel the center to limb variation caused by the first term.  The center to limb variations give us therefore good opportunity to check the theory against observations, as we will see in section 5.6.  

\indent  The second term of Eq.\,(20) is also very important in collapsars, such as the White Dwarfs, because of the large pressure broadenings, which causes the second term to be a large fraction of the total redshift.  When we perform the integration that results in Eq.\,(20), we can see that column density that results in the second term is relatively small, or $\int N_e\,dx \approx 10^{18} {\rm{.}}~$  We cannot discern the center-to-limb effect in the collapsars, but the redshift will vary from line to line depending on the photon width, which is caused mainly by pressure broadening (including Stark broadening).

\indent  Another interesting characteristic, is that even when the column density in the corona, the first term on the right side of Eq.\,(20), is relatively small, the second term is important as it requires only small column density, or only about $\int N_e \,dx \approx 10^{18}~{\rm{cm}^{-2}}{\rm{.}}~$ The electron density integral in interstellar space is large enough.  The collapsar would therefore show plasma redshift, even if they had little or no corona, such as a relatively cold collapsar.  This gives us another method to compare the theoretical prediction with observations; see section 5.6.4.


\section{Effect of magnetic fields}
In sections 2 and 3, we disregarded magnetic fields mainly because exact 
calculations that include their effects lead to significant complications.  Had we included the magnetic fields from the start, we might have lost sight of the simplicity and basic nature of the plasma redshift.  When the photon's polarization is in the direction of the magnetic field, the dielectric constant is largely unchanged; however, when it is in a plane perpendicular to the magnetic field, the dielectric constant is affected significantly.  Polarization produced by an external force in one direction may then cause a force on the charge in other directions.  The isotropic dielectric constant can be replaced by an anisotropic dielectric tensor.  This tensor complicates the mathematical treatment.  We must then solve Maxwell's equations together with the constitutive relations for current and magnetization in three dimensions.

\indent We are mainly interested in phenomena involving exchange of small energy
quanta or low frequencies.  From Maxwell equations and plane wave equations for the fields, we derive the homogeneous wave equation $\mathbf{k}\times\left(\mathbf{k}\times\mathbf{e}\right)+k_0^2\varepsilon \, \mathbf{e}= 0$, where \textbf{k} is the wave vector, \textbf{e} the 
polarization vector of the electromagnetic wave and $\varepsilon$ is the 
dielectric tensor (see, for example, Sturrock 1994 [5]).  The dispersion 
relation for waves propagating parallel to the magnetic field in the z direction is
\be
\left|{\begin{array}{ccc} 
   \displaystyle {\varepsilon_{xx}-k^2/k_0^2} & {\varepsilon_{xy}} & {\varepsilon_{xz}}  \\
   \displaystyle {\varepsilon_{yx}} & {\varepsilon _{yy}-k^2/k_0^2} & {\varepsilon _{yz}}  \\
   \displaystyle {\varepsilon_{zx}} & {\varepsilon_{zy}} & {\varepsilon_{zz}} 
\end{array}} \right| = 0.
\ee
Due to the rotational symmetry and definition of the axes, and because an
electric field when in the direction of the z-axis produces no coupling to the other axes, we have that: $\varepsilon_{xx}= \varepsilon_{yy}\rm{,} \; \varepsilon_{xy}= -\varepsilon_{yx}\rm{,} \; \varepsilon_{xz}= \varepsilon_{zx}= 0, \; \rm{and} \; \varepsilon_{yz}= -\varepsilon_{zy}= 0$.  In the very long wavelength limit, we make the approximation that the wave vector $k$ is independent of the dielectric constant.  We can then write the dielectric tensor in the form
\be
\left\{\begin{array}{ccc}
   \displaystyle {1-\frac{\omega_p^2\tilde\omega}{\omega\left({\tilde \omega ^2  - \omega _c^2 } \right)},} & \displaystyle {i \cdot \frac{{\omega _p^2 \omega _c
}}{{\omega \left( {\tilde \omega ^2  - \omega _c^2 } \right)}},} & 0  \\
    \displaystyle { - i \cdot \frac{{\omega _p^2 \omega _c }}{{\omega \left( {\tilde \omega ^2  - \omega _c^2 } \right)}},} & \displaystyle {1 - \frac{{\omega _p^2 \tilde \omega }}{{\omega \left( {\tilde \omega ^2  - \omega _c^2 } \right)}},} & 0  \\
    {0,} & {0,} & \displaystyle {1 - \frac{{\omega _p^2 }}{{\omega \tilde \omega }}} 
\end{array} \right\}{\rm{ ,}}
\ee
where $\omega_c= eB/\left(m_ec\right)$
is the cyclotron frequency, and where $\tilde\omega= \omega-i\beta\omega^2$.  From Eq.\,(21) we have that
\be
\left[\left(\varepsilon_{xx}-k^2/k_0^2\right)^2-\varepsilon_{{\rm{xy}}}
 \varepsilon_{{\rm{yx}}} \right]\varepsilon _{{\rm{zz}}}  = 0.
\ee
\indent The root $\varepsilon_{zz}= 1-\omega_p^2/\left(\omega\tilde\omega\right)= 
 0$, or $\omega_p^2= \omega^2-i\beta\omega^3$ corresponds to the plasmon oscillation along the z-axes.  For the other roots we have
\be
\left( {\frac{k}{{k_{\rm{0}} }}} \right)^2  \approx \varepsilon _{xx} \pm {\sqrt {\varepsilon _{xy} \varepsilon _{yx} }}  \approx 1 - \frac{{\omega
 _p^2 }}{{\omega \left( {\tilde \omega  \pm \omega _c } \right)}}{\rm{.}}
\ee
In the case when the minus sign in the denominator is valid, we have for $0<\omega\ll\omega _c$ and for $\omega\omega_c\ll\omega_p^2$,
that $v_{ph}/c= k_0/k\approx\sqrt{\omega\omega_c/\omega_p^2}\approx\sqrt{\omega} \sqrt{B/\left(4\pi ecN_e\right)}$, where
$v_{ph}$ is the phase velocity of the helicons traveling along the magnetic field lines or of the whistler waves in the ionosphere.  Our main interest,
 however, is the photon's attenuation factor.  We get:
\be
\frac{{2n\kappa \omega }}{{\varepsilon \bar \varepsilon }}  = \frac{{\beta \omega ^4 \omega _p^2 }}{{\left( {\omega _p^2  - \omega ^2  \pm \omega
 _c \omega } \right)^2  + \beta ^2 \omega ^6 }}{\rm{ }}.
\ee
The main result is that the magnetic field splits each of the poles for the dielectric constant in the complex plane into two poles if the photon's polarization is perpendicular to the field.  In the upper plane, seven poles would then replace the four poles given by Eq.\,(6).  But the sum of the oscillator strengths and the sum of the residues are the same as in a plasma free of magnetic fields.

\indent  At lower temperatures it often can be assumed that the dielectric constant is constant and equal to one.  We can set the magnetic field {\textbf{H}} equal to
 a rotation of a vector potential, {\textbf{H}} = curl {\textbf{A}}.  To 
the Hamiltonian operator (8), we must add two terms given by
\be
\frac{{ie\hbar}}{{mc}}\mathbf{A} \cdot \mathbf{grad} + \frac{{e^2 }}{{2mc^2 }}
\mathbf{A}^{2}  =  - \frac{e}{{2mc}}\mathbf{H} \cdot \mathbf{L} + \frac{{e^2 }}{{8mc^2 }}{\rm{H}}^{\rm{2}} {\rm{r}}^{\rm{2}} \,{\rm{sin}}^{\rm{2}} \theta \;{\rm{,}}
\ee
where $\mathbf{A= }\left(1/2\right)\mathbf{H}\times\mathbf{r}$ and $\mathbf{L}= \mathbf{r}\times\mathbf{p}$ are the vector potential and the angular momentum of a centrally bound electron, respectively, and $\theta$ is the angle between \textbf{r} and \textbf{H}.  The charge $e$ of the electron is a negative number.  We could in the usual way also add the electron spin term $-e\mathbf{H} \cdot \mathbf{S}/\left(mc\right)$.  At very low field strengths and low temperatures, the usual quantum mechanical calculations for bound electrons show that the angular momentum term and the spin dominate the second term.  The energy levels given by Eq.\,(9) will then split up into many very close states defined by the quantum number $m$, and the states become nondegenerate.

\indent For strong fields in plasmas at high temperatures (large \textbf{r}), the second term on the right side of Eq.\,(26) dominates.  In hot plasmas the first term and the spin term can then usually be disregarded.  The displacement and the line widths are then proportional to $\rm{B}^{\rm{2}}$.  For large \textbf{B} and \textbf{r} in hot plasmas, the problems can also be treated semiclassically. The electrons lose energy as they encircle the magnetic field lines.  This energy loss corresponds to an increase in the transition rates in the usual way.  The corresponding increase in the damping factor $\beta$ can be taken into account by multiplying Eq.\,(16) by a factor given by
\be
F_B  = \left( {1 + b\frac{{\omega _c^2 }}{{\omega _p^2 }}} \right) \approx \left( {1 + 1.3 \cdot 10^5 \frac{{B^2 }}{{N_e }}} \right){\rm{ ,}} 
\ee
where in the nonrelativistic approximation, we have set $b= 4/3$, and where $\omega_c$ and $\omega_p$ are the cyclotron and plasma frequencies, respectively, $B$ the field in gauss units, and $N_e$ is the electron density in $\rm{cm}^{-3}$.  When applying this factor to Eq.\,(16) we get for the 50\% cut-off wavelength that
\be
\lambda _{0.5}  = 3.185 \cdot 10^{ - 6} \left( {1 + 1.3 \cdot 10^5 \frac{{B^2 }}
{{N_e }}} \right)\frac{T}{{\sqrt {N_e } }}{\rm{.}}
\ee
At extremely high temperature, we must include a factor that takes into account relativistic effects.  However, for the redshift in the solar corona and in most astrophysical plasmas, this factor, which is about $\left.\left[1+\left(1-v^2/c^2\right)^{-1/2}\right]\right/2$, is not important.


\section{Comparing plasma-redshift theory with experiments}
Comparison of the plasma-redshift theory with an experimental finding often requires a thorough review of the experimental design and extensive calculations.  In sections 5.1 to 5.12, we will discuss briefly applications of the plasma-redshift theory, and compare its predictions with the observations.


\subsection{Transition zone to solar corona and the region of spicules}
The plasma redshift is significant for light with a wavelength less than a certain wavelength $\lambda_{0.5}$, which depends on the temperature, density and the magnetic field in accordance with Eq.\,(28).  According to Vernazza et al.~[6], the product of density and temperature in the transition zone to the solar corona is $N_eT\approx 5\cdot 10^{14} \: \rm{cm}^{-3}$ K.  When the temperature is 500,000 K, the electron density $N_e\approx 10^9 \: \rm{cm}^{-3}$, and the magnetic field B equal to zero, we get from Eq.\,(28) that photospheric light with wavelength less than cut-off wavelength $\lambda_{0.5}= 500$ nm is plasma redshifted.  If the magnetic field B is 20 gauss, the cut-off wavelength increases by 5\%; and if $B$ is 100 gauss the cut-off wavelength increases by 130\%.  That is, the cut-off wavelength increases as the magnetic field increases.  We have also that the cut-off wavelength $\lambda_{0.5}= 500$ nm reaches deeper into the transition zone, as the magnetic field increases.

\indent The notation $\lambda_{0.5}$ means that the redshift is 50\% of its maximum value.  Shorter wavelengths will be redshifted more than 50\% and longer wavelengths less than 50\%.  A more detailed evaluation is obtained by using Eq.\,(18) and by determining for each wavelength the value of the function $F_1\left(a\right)$, which is given by Eq.\,(14) and Table 1.  The plasma redshift of a photon means that the photon loses energy.  This energy loss consists of low energy quanta, which are immediately absorbed (evanescent) in the plasma and cause a corresponding increase in the plasma temperature.  According to Eq.\,(28), shorter wavelengths can be redshifted at lower temperature and or at higher densities.  For the magnetic field \textbf{B} about equal to zero, the cut-off for the shorter wavelengths is deep in the transition zone, and the cut-off for the longer wavelengths is high in the transition zone.  For this 
reason, we have that short-wavelength light is plasma redshifted slightly more than the long-wavelength light.  However, the transition zone is short so usually this is a small effect on the measured redshift, except for the very short-wavelength light.  Thus, the cut-off zone for the photospheric light is not sharp.  Besides the external light from the photosphere, we must sometimes take into account also the internal short wavelength light in the plasma.  This internal light in the plasma, mostly short wavelength light, below the $\rm{L}_\alpha$-limit, often has high intensity and high optical density.

\indent The region of spicules covers the upper chromosphere, where $2\cdot 10^4 \le T \le 2 \cdot 10^5$ K, and the transition zone, where $2\cdot 10^5  \le T \le 7 \cdot 10^5$ K.  The height of the spicules region is from about 2,150 km to about 15,000 km above the photosphere.  As described by Feldman et al.~[7], Friedman [8], Hollweg [9], and Goodman [10], this region is broader than the corresponding region in the models by Vernazza et al.~[6].  The models by Vernazza et al.~assume that the isothermal surfaces are stratified horizontally, while in fact they may sometimes be nearly vertical and roughly parallel to the surfaces of the spicules between huge plasma-redshift heated ``bubbles'', as described in the following paragraphs.

\indent The upper chromosphere is highly ionized and contains internally rather high intensity short-wavelength radiation emitted from highly excited states, including those of hydrogen and helium.  The pressures, temperatures, and densities in the plasma fluctuate.  In a small hot spot, a ``bubble'', the temperature may be about 100,000 K and the electron density $N_e\approx 4 \cdot 10^9 \: {\rm{cm}}^{-3}$, while the surrounding regions, the ``walls'' of the ``bubble'', contain slightly denser and colder plasma.  The denser ``walls'' may emit more of the internal light, which may bounce back and forth across the hot region.  The plasma-redshift heating is a first-order process in density, while the cooling processes are second order in density.  The plasma redshift causes, therefore, the short-wavelength photons to deposit some of their energy in the less dense hot spots, the bubbles.  The cooling due to recombination emission in the denser and colder ``walls'' is compensated to a lesser extent by the redshift heating.  The plasma redshift enhances, therefore, temperature inhomogeneity and makes the hot low-density region, the ``bubble'', hotter, while the denser surrounding regions, the ``walls'', become colder.  According to Eqs.\,(16) and (28), the 50\% cut-off wavelength for the above-mentioned density and temperature and for low magnetic fields is initially in the hot, low-density region about 50.3 nm; and the analogous 10\% cut-off wavelength, corresponding to $a$ = 2.671 in Table 1, is about 116 nm.  A magnetic field will increase the cut-off wavelength.  In the bubbles, the very short wavelength internal light in the plasma may then initiate significant plasma-redshift heating.

\indent The magnetic field enhances therefore the temperature inhomogeneity.  In addition, the conversion of field energy to heat is strongest in the hot regions (see Appendix B).  As the kinetic energy of a particle increases, its diamagnetic moment increases.  This fact in turn reduces the H-field inside the hot low-density region and partially transforms the field energy to heat, while the fields from the diamagnetic moments inside the bubble combine to strengthen the H-field in the colder high-density region, the ``walls''.

\indent As the temperature in the hot region increases, some of the light from the photosphere will also be plasma redshifted, first the short-wavelength light, and then the longer-wavelength light.  The plasma redshift deposits then a fraction of the photospheric photon energy in the ``bubbles'' in the transition zone, and accelerates conversion of magnetic field energy to heat.

\indent A rough estimate of the plasma-redshift heating from the photosphere is obtained from the integrated intensity $I(R)$ over all directions of solar light penetrating each location in the transition zone and the corona.  The light intensity from the photosphere decreases with $R$ as
\be
\begin{array}{l}
 \displaystyle I\left( R \right) \approx I_0 2\pi \left( {1 - \sqrt {1 - R_0^2 /R^2 } } \right) \\
  \displaystyle \; \; \;  = 2\sigma T^4 \left( {1 - \sqrt {1 - R_0^2 /R^2 } } \right) = 1.29 \cdot 10^{11} \left( {1 - \sqrt {1 - R_0^2 /R^2 } } \right)\; \; {\rm{ erg}}\,{
\rm{cm}}^{ - 2} \,{\rm{s}}^{ - 1} {\rm{,}}
\end{array}
\ee
where $R_0$ is the solar radius.  Low in the transition zone, we have that the parenthetical factor is close to 1, but farther away it approaches $0.5(R_0/R)^2$.  In the transition zone, we can multiply Eq.\,(29) by the redshift given by Eq.\,(18) per cm of integration, and get for the photospheric light that
\be
\begin{array}{l}
 \displaystyle \frac{{dQ_{heat} }}{{dt}} = I\left( R \right)\frac{{\delta \lambda }}{\lambda } = 1.29 \cdot 10^{11} \left( {1 - \sqrt {1 - R_0^2 /R^2 } } \right)  3.326 \cdot 10^{ - 25} F_1 \left(a \right)\frac{\gamma }{{\gamma _0 }}N_e \\
 \displaystyle \quad \quad \; = 4.29 \cdot 10^{ - 14} \left( {1 - \sqrt {1 - R_0^2 /R^2 } } \right) F_1 \left( a \right)\frac{\gamma }{{\gamma _0 }}N_e \; \; {\rm{  erg}}\,{\rm{cm}}^{ - 3} {\rm{s}}^{ - 1} {\rm{,}} 
 \end{array}
\ee
where $\delta\lambda/\lambda$ is the redshift per cm of the plasma.  This redshift heating density can be
compared with net cooling density from recombination emission and X-ray emission, which according to Sutherland and Dopita [11] has a maximum cooling density of about $4.5 \cdot 10^{-22} N_t N_e ~ {\rm{erg}}\, {\rm{cm}^{-3}}\, {\rm{s}^{-1}}$ at a temperature of about 180,000 K, where $N_t = 0.917 N_e$ is the number density of positive ions. At lower and higher temperatures the cooling rate is lower.  Below the temperature of about 30,000 K, and above about 800,000 K, the rate of cooling is less than $10^{-22} N_t N_e$.  We will see later that the excess redshift heating in the corona leaks into the transition zone and that the conversion of magnetic heating is significant and may double the redshift heating in the transition zone.  However, even when we disregard this additional heating, we see that the redshift heating given by Eq.\,(30) about balances the recombination emission and X-ray cooling, when the temperature is about one million degrees and the electron density $4.7 \cdot 10^8 ~ \rm{cm}^{-3}$ and $\left(\gamma / \gamma_0 \right) F_1 \left(a \right) \approx 1$.

\indent The initial photon width $\gamma_i$ is often large due to broadening by collisions with neutral atoms, Fourier field components of charged particles, the Stark effect, and stimulated emissions.  Holweger [12] found that the initial photon widths of the sodium resonance lines 588.995 and 589.592 nm, which are formed high in the photosphere, are about 0.202 pm.  This width is about 17 times the classical photon width of 0.0118 pm.  We do not know the average photon width, but for many lines it is between 1 and 20 times the classical width.  The second electron in the important $\rm{H}^-$-species is loosely bound and sensitive to collision broadening by neutral atoms, by the Fourier field harmonics of the fast moving electrons, and also by stimulated emissions and absorptions.  Many of the Ca-II and Mg-II lines have very broad photon widths, while some weak lines have small photon widths.  In the balance of heating and cooling, we should consider also other processes, and different modes of heat transport.  However, other authors, such as Vernazza et al.~[6], have done so.  The present focus is to evaluate only the 
photon-width portion directly related to the plasma redshift, which has not been considered by others.

\indent The increase in redshifts due to broadenings of photon widths is significant in the transition zone, but less in the corona, because in the corona the photon width is close to the classical width.  This estimate is based on comparison of observed redshifts of Fraunhofer lines with Eq.\,(20).   Presently, this estimate is imprecise and will have to be improved as the photon widths become better determined.  The heating by conversion of the magnetic field is most important in the spicules region, as discussed in Appendix B, because the field is stronger and because the mutual induction between the diamagnetic moments of the charged particles and the currents creating the magnetic field is larger low in the transition region than high in the corona.

\indent The redshift experiments indicate that the second term on the right side of Eq.\,(20) is on the order of $10^{-6}$.  The integrated heating derived from this second term is then on the order of $1.29 \cdot 10^{11} \cdot 10^{-6} = 1.3 \cdot 10^5 \; \rm{ erg} \, \rm{cm}^{-2} \rm{s}^{-1}$, which is deposited mainly in the transition zone.  In addition, we have the heating from the first term, which can be obtained by integrating the first term on the right side of Eq.\,(20) along the line of sight from the observer to the different points on the solar disk.  Much of the heating by this term, about $1.1 \cdot 10^5 \; \rm{ erg} \, \rm{cm}^{-2} \rm{s}^{-1}$, leaks by conduction into the transition zone, as we show in section 5.2.  The total plasma-redshift heating in the spicules region and in the corona is then about $2.4 \cdot 10^5 \; \rm{erg} \, \rm{cm}^{-2} \rm{s}^{-1}$. 

\indent In addition to this direct plasma-redshift heating, we have heating by conversion of the magnetic field to heat, which is often initiated by the plasma redshift, as described in Appendix B.  The magnetic heating consists of two major contributions:

\begin{enumerate}
\item The conversion of magnetic field energy to heat.  This conversion is often induced by the increase in the diamagnetic moments caused by the redshift heating.  The increase in the diamagnetic moments induces electromotive forces that oppose the magnetic field and the currents that generate it.  This effect is most prominent in the hot bubbles between the spicules in the transition zone.
\item The repulsion of the diamagnetic moments by the outward decreasing magnetic field.  This repulsion is important for accelerating the solar wind, as shown in section 5.3.
\end{enumerate}

\indent The plasma redshift has a tendency to create ``bubble'' like structures in the upper chromosphere, because the plasma redshift is a first-order process in the electron density, while the cooling processes are second (or higher) order processes in the density.  The plasma redshift therefore enhances the inhomogeneity in temperature.  Initially, the ``internal'' light in the plasma in the upper chromosphere contributes to the plasma redshift and the unevenness in temperature.  This ``internal'' light is rich in high-energy, short-wavelength photons from the Lyman series in hydrogen and helium transition lines, as hydrogen and helium are highly ionized in the upper chromosphere.  Subsequently, as the temperature in the hot ``bubbles'' increases, also the light from the photosphere is redshifted in the ``bubbles''.  If for a moment, we disregard the recombination cooling, and the heating by the magnetic field, the plasma-redshift heating in the transition zone according to Eq.\,(30) is for $\left(\gamma / \gamma _0 \right) F_1 \left(a \right) \approx 10$ about equal to $dQ_{heat} / dt \approx 4.3 \cdot 10^{-13} N_e \; \rm{erg} \, \rm{cm}^{-3} \rm{s}^{-1}$.  The rate of heating is then about 
\be
\frac{{dT}}{{dt}} \approx \frac{{4.3 \cdot 10^{ - 13} }}{{1.917 \cdot \frac{3}{2}k}} \approx 1000 \; \; {\rm{  K}}\,{\rm{s}}^{ - {\rm{1}}} {\rm{,}}
\ee 
\noindent  where $k$ is the Boltzmann constant.  The ``bubble'' temperature would then reach 500,000 K in about 500 seconds, which is on the order of the time for the formation of the spicules (see p. 124 of reference [8]).  It takes the spicules about 270 seconds to fall freely 10,000 km.  During the latter part of the ``bubble'' formation, the conversion of magnetic field energy to heat is relatively large, and the rate of heating of the ``bubbles'' increases exponentially as the ``bubbles'' explode to the surface and into the corona.  A spicule consists of the colder plasma in the walls of the bubbles.  This colder plasma is squeezed 5 to 15 km out into the transition zone and the lower corona by the expanding ``bubbles'' [8].  The spicule then falls down, which causes significant pressure variations and turbulence in this region.  The ``bubbles'' are filled with a hot, fully ionized plasma as they open up into the corona.

\indent  The heating produced by the conversion of the magnetic field varies with strength of the field.  If a field of 10 gauss is embedded in the plasma with a density of $N_e = 10^{10} \; \rm{cm}^{-3}$, it could heat the plasma to about million degrees.  If such a field in a bubble with a height of 10,000 km is destroyed every 1000 seconds the heating would be about $4 \cdot 10^6 \; \rm{erg} \, \rm{cm}^{-2} \, \rm{s}^{-1}$.  The magnetic field in the transition zone is estimated to be usually in the range of 1 to 10 gauss.  Similarly, if a field of 2 gauss in a bubble with a height of 10,000 km is embedded in the plasma density of $N_e = 10^{10} \; \rm{cm}^{-3}$, and converted to heat every 1000 seconds, it would correspond to an energy flux of $1.6 \cdot 10^5 \; \rm{erg} \, \rm{cm}^{-2} \, \rm{s}^{-1}$.  The conversion of magnetic energy is likely to fluctuate, and to correspond to an energy flux, which is usually between $1.6 \cdot 10^5 $ and $4 \cdot 10^6 \; \rm{erg} \, \rm{cm}^{-2} \, \rm{s}^{-1}$.
 
\indent The repulsion of a diamagnetic moment in an outward decreasing magnetic field is pronounced throughout the transition zone and the corona.  At high altitudes, the magnetic field is primarily responsible for the outward acceleration of the solar wind.  If the solar wind at the distance of the Earth is $S_{215R_0} = N_p v_p = 3 \cdot 10^8 \; \rm{cm}^{-2} \, \rm{s}^{-1}$, the energy flux at the solar surface required for overcoming the gravitational potential is about $0.53 \cdot 10^5 \; \rm{erg} \, \rm{cm}^{-2} \, \rm{s}^{-1}$.  This assumes that helium and trace elements account for 20\% of the proton energy requirements.  If the solar wind has helium and trace elements equal to their concentration in the Sun, the required energy flux would be 17\% higher.  Usually, the kinetic energy of the particles in the solar wind increases.  When the proton velocity has increased to about 600 km per second, the kinetic energy flux corresponds to about $0.53 \cdot 10^5 \; {\rm{ erg}}\,{\rm{cm}}^{ - 2} \, {\rm{s}}^{ - 1}$.  The total of potential and kinetic energy in the solar wind is then about $1.1 \cdot 10^5 \; {\rm{ erg}}\,{\rm{cm}}^{ - 2} \, {\rm{s}}^{ - 1} $.  In the quiescent corona, the total energy flux from the redshift heating, the magnetic heating (corresponding to conversion of 2 gauss of field energy to heat), and from the input into the solar wind is about $\left( {2.4 + 1.6 + 1.1} \right) \cdot 10^5 = 5.1 \cdot 10^5  \; {\rm{ erg}}\,{\rm{cm}}^{ - 2} \, {\rm{s}}^{ - 1}$ .  During flares the conversion of magnetic field energy to heat can be much greater, as described in section 5.5.

\indent According to McWhirter et al.~[13] and Withbroe [14, 15] the estimated energy input for the quiescent corona, including the energy needed to drive the solar wind, is about $4 \cdot 10^5$ to $6 \cdot 10^5 \; {\rm{ erg}}\,{\rm{cm}}^{ - 2} \, {\rm{s}}^{ - 1} $.


\subsection{Solar corona}
The energy transport from the lower and cooler solar atmosphere to the hotter transition zone and corona has remained a puzzle, as the plasma-redshift theory has not been available.  Several papers have dealt with the subject; see [13-27].  For explaining the observations, the different authors have tried to assume that some mechanisms deposited energy.  They have suggested that sound waves or acoustic flux, Alfv\'{e}n waves, electrical currents, magneto-hydrodynamic turbulence, etc.~transferred energy to the corona.  These explanations may be partially correct and do not necessarily conflict with the present explanation.  For example, the turbulence created by the redshift-heated bubbles produces chocks and acoustic fluxes that transplant some of the energy.  However, these acoustic fluxes should not be considered the primary cause, but ancillary to the primary cause, which is the plasma-redshift heating followed by conversion of magnetic field energy to heat.  Recently, reconnection of the field lines has also been mentioned.  Also this reconnection is more a phenomenological description than a causal explanation.  The reconnection is due to the redshift heating of the plasma followed by conversion of the magnetic field energy to heat as described in Appendix B.


\begin{figure}[t]
\begin{centering}
\includegraphics[scale=0.45]{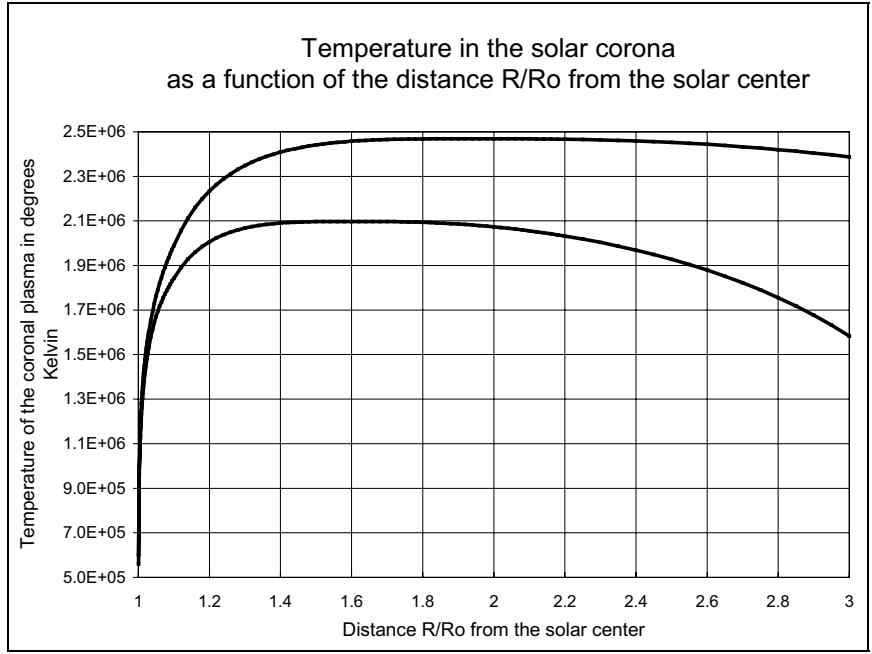}
\caption{The lower and upper curves correspond to different pressure in the transition zone, $N_e  T = 4 \cdot 10^{14}$ and $5 \cdot 10^{14} \; {\rm{ cm}}^{ - 3} {\kern 1pt}  {\rm{ K,}}$ respectively.  The solar wind flux at the distance of the Earth is assumed to be $N_e v_p  = 3.2 \cdot 10^8 \; {\rm{ cm}}^{ - 2} {\kern 1pt} {\rm{s}}^{ - 1}$.  The magnetic field in the transition zone is assumed to be 5.5 and 6.3 gauss, and to decrease nearly inversely with the square of the radius to solar center.}
\end{centering}
\end{figure}

\indent We will for a moment assume that in the transition layer and lower corona the densities are diffusion controlled and determined mainly by the gravitational potential and the temperature.  We have then that
\be
\begin{array}{l} 
\displaystyle N_e  = \frac{{N_0 T_0 R_0 }}{{TR}}\exp \left[ {\frac{{GM_S \mu m_H }}{{kTR_0 }}\left( {\frac{{R_0 }}{R} - 1} \right)} \right] \\
 \displaystyle \quad \; \; = \frac{{N_0 T_0 R_0 }}{{TR}}\exp \left[ {\frac{{1.41 \cdot 10^7 }}{T}\left( {\frac{{R_0 }}{R} - 1} \right)} \right]{\rm{ ,}}
 \end{array}
\ee

\noindent where $N_e$ in ${\rm{cm}}^{ - 3}$, $T$ in K, and $R$ in cm are the electron density, temperature, and radius to the solar center.  At the bottom of the corona, the same quantities are indicated with subscript zero.  $G$ is the gravitational constant, $M_S$ is the solar mass, and $\mu  = 0.61$ the average molecular weight relative to the proton mass.  The exponential factor accounts for the effect of the gravitational potential.  The dependence of $R$ in the factor before the exponential factor is derived from the diffusion equation with proper boundaries.  Other authors usually disregard this $R$ dependence, but it is significant.  (When we discuss the solar wind in section 5.3, we will see that we must modify this equation and take into account the outward force on the diamagnetic moments.  But for the moment we use this equation to help us analyze the phenomena.)  For deriving the electron distribution from Eq.\,(32), we must know the temperature as a function of $R$.  We can compare the temperature distribution as measured by Sturrock, Wheatland and Acton [22, 23] in the most important region from 1.01 to 2 solar radii with that derived from the plasma-redshift heating and the magnetic heating. 

\indent Fig.\,1 shows how the electron temperature varies below 3 solar radii when the pressure in the transition zone corresponds either to $N_e T = 4 \cdot 10^{14}$, or to ${\rm{5}} \cdot {\rm{10}}^{{\rm{14}}} \; {\rm{ cm}}^{ - 3} {\kern 1pt} {\rm{K}}$.  Most of the measured temperatures by Sturrock, Wheatland and Acton are between the two curves.


\begin{table}
\centering

{\bf{Table 2}} \, \, Temperatures and electron densities in the quiescent solar corona as a function of distance.

\vspace{2mm}

\begin{tabular}{llllll}
	\hline
Distance $r/R_0$ & Temperature & Electron & Distance $R/R_0$ & Temperature & Electron \\
in units of & $T$ in degrees & densities $N_e$ & in units of & $T$ in degrees & densities $N_e$ \\
solar radius $R_0$ & Kelvin & in $\rm{cm}^{-3}$ & solar radius $R_0$ & Kelvin & in $\rm{cm}^{-3}$ \\
\hline \hline
1.001 & 5.70 $\times 10^{5}$ & 7.45 $\times 10^{8}$ & 2.6 & 2.02 $\times 10^{6}$ & 3.35 $\times 10^{6}$ \\
1.002 & 8.81 $\times 10^{5}$ & 4.73 $\times 10^{8}$ & 2.8 & 1.92 $\times 10^{6}$ & 2.89 $\times 10^{6}$ \\
1.005 & 1.10 $\times 10^{6}$ & 3.65 $\times 10^{8}$ & 3.0 & 1.78 $\times 10^{6}$ & 2.54 $\times 10^{6}$ \\
1.01 & 1.27 $\times 10^{6}$ & 2.99 $\times 10^{8}$ & 3.2 & 1.60 $\times 10^{6}$ & 2.12 $\times 10^{6}$ \\
1.02 & 1.45 $\times 10^{6}$ & 2.38 $\times 10^{8}$ & 3.4 & 1.29 $\times 10^{6}$ & 1.78 $\times 10^{6}$ \\
1.05 & 1.69 $\times 10^{6}$ & 1.60 $\times 10^{8}$ & 3.6 & & 1.55 $\times 10^{6}$ \\
1.1 & 1.88 $\times 10^{6}$ & 1.03 $\times 10^{8}$ & 5.0 & & 7.36 $\times 10^{5}$ \\
1.2 & 2.06 $\times 10^{6}$ & 5.49 $\times 10^{7}$ & 6.0 & & 1.85 $\times 10^{5}$ \\
1.3 & 2.14 $\times 10^{6}$ & 3.40 $\times 10^{7}$ & 7.0 & & 8.12 $\times 10^{4}$ \\
1.4 & 2.17 $\times 10^{6}$ & 2.29 $\times 10^{7}$ & 8.0 & & 4.45 $\times 10^{4}$ \\
1.6 & 2.19 $\times 10^{6}$ & 1.23 $\times 10^{7}$ & 10.0 & & 1.81 $\times 10^{4}$ \\
1.8 & 2.19 $\times 10^{6}$ & 7.67 $\times 10^{6}$ & 30.0 & & 6.16 $\times 10^{2}$ \\
2.0 & 2.17 $\times 10^{6}$ & 5.85 $\times 10^{6}$ & 60.0 & & 1.23 $\times 10^{2}$ \\
2.2 & 2.14 $\times 10^{6}$ & 4.72 $\times 10^{6}$ & 100.0 & & 3.77 $\times 10^{1}$ \\
2.4 & 2.09 $\times 10^{6}$ & 3.92 $\times 10^{6}$ & 215 & & 6.38 \\
\hline
\end{tabular}
\end{table}

\indent In models A to F of the solar atmosphere by Vernazza et al.~[6], the product of temperature and electron density varies from $2.4 \cdot 10^{14}~{\rm{K\,cm}}^{-3}$ to $1.1\cdot 10^{15}~{\rm{K\,cm}}^{-3} $ at $T = 447,000~{\rm{K.}}~$  In the model C the product is $5.4 \cdot 10^{14}$ at the same temperature.  These values, especially the ones for model C, are consistent with measurements of temperatures and densities low in the corona as determined by Sturrock et al.~[22] and Wheatland et al.~[23], who for two points low in the quiescent corona found the product to be $4.79 \cdot 10^{14}$ at a temperature of $1.33 \cdot 10^6$ and $4.26 \cdot 10^{14}$ at a temperature of $1.64 \cdot 10^6$ K.


\begin{figure}[t]
\centering
\includegraphics[scale= .4]{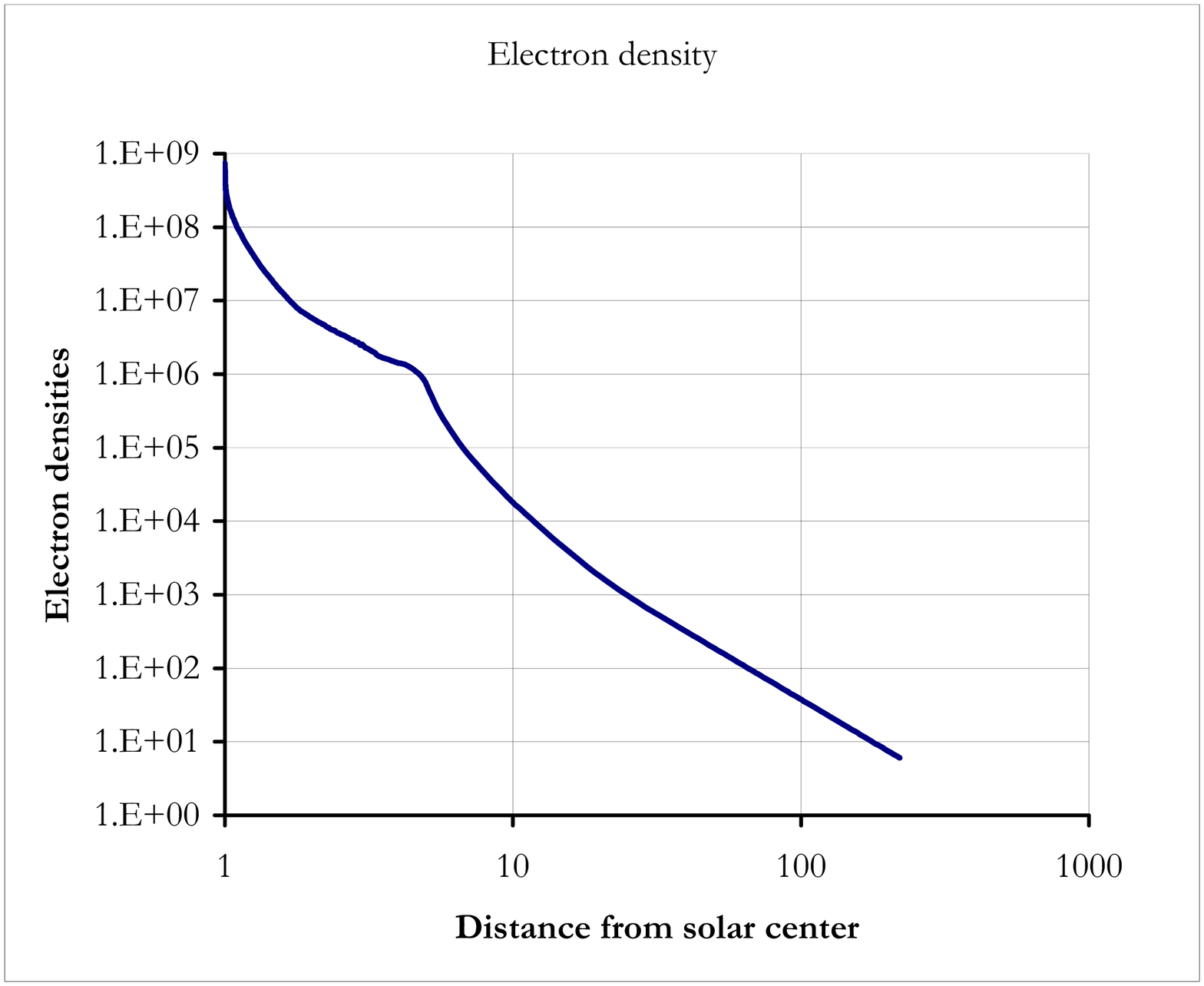}
\caption{The curve corresponds to $N_e T = 4.75 \cdot 10^{14} \; {\rm{ cm}}^{ - 3} {\kern 1pt} {\rm{K}}$ at the cut-off in the transition zone.  The solar wind flux at the distance of the Earth is assumed to be $N_e v_p  = 3.2 \cdot 10^8 \; {\rm{ cm}}^{ - 2} {\kern 1pt} {\rm{s}}^{-1}{\rm{.}}~$  The magnetic field in the transition zone is assumed to be 6 gauss, and to decrease nearly inversely with the square of the radius to solar center.  The hump around 4 to 5 solar radii is not well defined and is very sensitive to pulsations in the transition zone and the strength of the magnetic field.  It is in this example enhanced to remind us that this is an unstable region, where the proton temperature can occasionally drop significantly and cause relatively high densities.  Beyond this distance, the solar wind accelerates and causes the density to fall off more steeply.}
\end{figure}

\indent In Table 2, we list the electron temperature and density values for the entire corona.  We have assumed that in the middle of the transition zone $T = 475,000$ K, and $N_e  = 10^9 \; {\rm{ cm}}^{ - 3} $.  We also assumed that the magnetic field $B$ is so low that it does not affect the cut-off wavelength as given by Eq.\,(28), but it contributes to the heating.  In these examples, Spitzer's conductivity [28] is assumed to be $\kappa  = 10^{ - 6} T^{{5 \mathord{\left/
 {\vphantom {5 2}} \right.
 \kern-\nulldelimiterspace} 2}} $.  The electron densities in Table 2 are also shown in Fig.~2.   The calculations can be refined.  The values listed in Table 2 are approximate values for the temperatures and electron densities as functions of the radii to solar center.  For this discussion, the values should not be considered theoretically deduced values, but rather some values in fair agreement with experiments.  The values serve only as a reference for the discussion and for helping us understand what is going on.  However, the temperatures and pressures are consistent with the values observed by Sturrock, Wheatland and Acton [22, 23], and with the pressure in the transition zone as estimated by Vernazza et al.~[6].  The values reported in the literature [13-23] depend on time of observation, but during quiescent corona they are in reasonable agreement with those in Table 2. 

\indent According to Eq.\,(28), the 50\% cut-off wavelength is about 478.5 nm when the temperature is 475,000 K, electron density $10^9 \; {\rm{ cm}}^{ - 3}$, and the magnetic field only a few gauss.  The cut-off wavelength increases as the temperature increases and the density decreases.  At 1.005 solar radii the temperature is about 1,100,000 K, the electron density about $3.65 \cdot 10^8 \; {\rm{ cm}}^{ - 3}$, and the cut-off wavelength about 1,833.8 nm.  For this estimate the solar radius, $R_0 $ is assumed to extend to 475,000 K.

\indent The plasma-redshift heating in the upper layers covers the entire solar spectrum and is in excess of the recombination cooling and X-ray cooling.  Most of the excess heat transferred to the plasma by redshift of photons will leak by heat conduction into the transition zone.

\indent When we multiply the first term on the right side of Eq.\,(20) by the right side of Eq.\,(29) and disregard the second term on the right side of Eq.\,(20), we have that the plasma-redshift heating is
\be
\begin{array}{l} 
\displaystyle \frac{{dQ_{heat} }}{{dt}} = 1.29 \cdot 10^{11} \left( {1 - \sqrt {1 - R_0^2 /R^2 } } \right) \cdot 3.326 \cdot 10^{ - 25} N_e  \\
 \displaystyle \,\,\,\, = 4.3 \cdot 10^{ - 14} N_e \left( {1 - \sqrt {1 - R_0^2 /R^2 } } \right)\; \; {\rm{  erg}}\,{\rm{cm}}^{ - 3} {\kern 1pt} {\rm{s}}^{ - 1} {\rm{,}} 
 \end{array}
\ee
where the electron density, $N_e $, is in ${\rm{cm}}^{ - 3} $.  The second term of Eq.\,(20) and the transformation of the magnetic field energy to heat are most important in the transition zone, as discussed in section 5.1.

\indent The plasma-redshift heating obtained by Eq.\,(33) may be compared with the dominant cooling processes, which are due to recombination emissions and X-ray emissions.  Sutherland and Dopita [11] have estimated the cooling rate.  They find for the temperature range $8 \cdot 10^5  \le T \le 1.8 \cdot 10^6 $ that the cooling rate is given by ${{dQ_{heat} }} / {{dt}}  \approx 10^{ - 22} N_t N_e \; \; {\rm{  erg}}\,{\rm{cm}}^{ - 3} {\,} {\rm{s}}^{ - 1}$, where for solar abundance of helium and trace elements the number density of positive ions is $N_t  \approx 0.917N_e $.

\indent From Table 2, we get for the temperature and the electron density at 1.005 solar radii that the cooling rate is $3.35 \cdot 10^{ - 14} N_e\; \; {\rm{  erg}}\,{\rm{cm}}^{ - 3} {\, } {\rm{s}}^{ - 1} $.  According to Eq.\,(33), the redshift heating at the same location is $3.87 \cdot 10^{ - 14} N_e\; \; {\rm{  erg}}\,{\rm{cm}}^{ - 3} {\, } {\rm{s}}^{ - 1} $.  Thus, the plasma-redshift heating, disregarding the second term in Eq.\,(20) and the conversion of the magnetic field to heat, is slightly greater than the cooling.

\indent From Table 2, we get for the temperature and the electron density at 1.1 solar radii that the cooling rate is $0.945 \cdot 10^{ - 14} N_e\; \; {\rm{  erg}}\,{\rm{cm}}^{ - 3} {\, } {\rm{s}}^{ - 1} $.  According to Eq.\,(33), the redshift heating at the same location is $2.5 \cdot 10^{ - 14} N_e\; \; {\rm{  erg}}\,{\rm{cm}}^{ - 3} {\, } {\rm{s}}^{ - 1} $.  The plasma-redshift heating at this location is thus about 2.7 times greater than the cooling.  The excessive heating leaks by conduction into the transition zone.  These heating estimates disregard the second term in Eq.\,(20) and the conversion of the magnetic field to heat.  The actual heating is, therefore, greater and more heat leaks into the transition zone.  This surplus heating over cooling continues to more than about 1.5 solar radii, and the excessive heat leaks by conduction into the transition zone.

\indent At higher temperatures the recombination cooling decreases significantly, according to Sutherland and Dopita [11].  The plasma-redshift heating then far exceeds the cooling by recombination emissions and X rays.   However, because of the lower densities high in the corona (beyond 1.5 solar radii), the cooling from solar wind becomes significant.  The lifting of the solar wind in the gravitational field requires energy.  Lower in the corona and in the transition zone the gravitational field is stronger, but the high densities dominate the solar wind density, and the energy transferred to the solar wind is an insignificant fraction of the total cooling.  High in the corona, the cooling by the solar wind flux is important.  This cooling is reduced significantly by the repulsion of diamagnetic moments in the outward decreasing magnetic field, as discussed in section 5.3.  The gravitational cooling by the solar wind, nevertheless, results in a maximum temperature, which is usually between 1.9 and 2.6 million degrees at distances of about 1.8 to 2.0 solar radii.  In Table 2 the maximum temperature, about $2.2 \cdot 10^6$ K, corresponds to a solar wind flux of $N_p v_p  = 3.2 \cdot 10^8 \; {\rm{ cm}}^{ - 2} {\kern 1pt} {\rm{s}}^{ 
- 1}$ at the distance of the Earth.  The maximum temperature in the quiescent corona varies with the redshift heating, solar wind flux, and the heat conductivity coefficient, which varies with the strength and direction of the magnetic field.  Spitzer's conductivity coefficient [28] was used for estimating the values in Table 2.

\indent The temperature profile is obtained by an iteration method.  The electron temperatures and the variations with height, as shown in Table 2 and Fig.\,\,1, are similar to the average temperatures measured by Wheatland, Sturrock and Acton [22, 23], who examined the temperatures in the inner corona using long-exposure Yohkoh images of two regions of quiet corona.  Interestingly, for explaining their data they assumed ``ad hoc'' deposition of energy in the higher layers of the corona.  The excess energy then leaked into the transition zone.  Their ``ad hoc'' method for explaining the temperature variations between 1.1 to 1.8 solar radii mimics to some extent the deposition of energy by the plasma-redshift heating.  They, of course, could not explain how this heating came about. But their experimentally determined relation is consistent with the predictions of the plasma-redshift theory.


\subsection{Solar wind}
It has been difficult to explain many of the phenomena associated with the solar wind in the quiescent corona.  For example, it has been difficult to explain why the solar wind accelerates outwards.  Sheeley Jr.~et al.~[24] observed ``white light images'', which are believed to be associated with the solar wind from quiescent corona, accelerate outwards from a position of about 5 solar radii to about 30 solar radii and beyond.  Wang et al.~[25] observed similar phenomena closer to the Sun.  The outward acceleration of the solar wind continues beyond the position of the Earth, according to Withbroe, the solar wind velocities are higher in the coronal holes [14, 15].  Withbroe estimated the solar wind flux to be usually between about $N_p v_p  = 2.7 \cdot 10^8$ and about $3.2 \cdot 10^8 \; {\rm{ cm}}^{ - 2} {\kern 1pt} {\rm{s}}^{ - 1} $.  He also found that the mass flux in the solar wind varies much less than the particle density; that is, when the particle density increases, the velocity decreases.

\indent Some have tried to explain the acceleration as caused by Alfv\'{e}n waves.  However, the velocity $v = {B \mathord{\left/ {\vphantom {B {\sqrt {4\pi \rho } }}} \right.
 \kern-\nulldelimiterspace} {\sqrt {4\pi \rho } }}$ of the Alfv\'{e}n waves at about the distance of the Earth is much lower than that of the observed solar wind.  Close to the Earth the magnetic field B is usually less than 0.0001 gauss and the density usually about 5 protons per cubic cm.   The mass density including solar abundance of helium is then $1.17 \cdot 10^{ - 23}$.  The corresponding velocity of the Alfv\'{e}n waves is $8.2 \cdot 10^6 \; {\rm{ cm}}{\kern 1pt} {\kern 1pt} {\rm{s}}^{ - 1} $, while the measured solar wind velocity, according to Gosling [16] usually exceeds $4 \cdot 10^7 \; {\rm{ cm}}\,{\rm{s}}^{ - 1} $.  As Parker [20] and Spangler and Mancuso [29] have shown, the Alfv\'{e}n waves are unlikely to be significant for heating of the solar corona.

\indent In the following discussion, we will show that both the plasma-redshift heating and the repulsion of the diamagnetic moments by the magnetic field (see Eq.\,(B10) of appendix B) are important for explaining the observations.

\indent If the solar wind flux at the distance of the Earth is $S_{215R_0 } = N_p v_p  = 3 \cdot 10^8 ~ {\rm{ cm}}^{ - 2} {\kern 1pt}{\rm{s}}^{ - 1}{\rm{,}}$ then the corresponding flux, $S_{R \,}{\rm{,}}$ at $R$ is equal to the flux $S_{215R_0} $ multiplied by the factor $\left( {{{215R_0 } \mathord{\left/ {\vphantom {{215R_0 } R}} \right. \kern-\nulldelimiterspace} R}} \right)^2 {\rm{.}}~$  The increase, $dq_g {\rm{,}}$ in gravitational potential energy, $q_g {\rm{,}}$ of the flux per height increment $dR $ in cm is then 
\be
\frac{{dq_g }}{{dR}} = \frac{{GM_S \mu m_p }}{{R^2 }}\left( {\frac{{215R_0 }}{R}} \right)^2 S_{215R_0 }  \approx 2.54 \cdot 10^{ - 15} \left( {\frac{{R_0 }}{R}} \right)^4 S_{215R_0 }\; ~{\rm{ erg\,cm}}^{-3}\, {\rm{s}}^{-1} {\rm{,}}
\ee
where we have assumed 5\% abundance of helium in the solar wind, and $\mu  = 1.2$ instead of the frequently used factor of 1.4, which corresponds to 10\% helium.

\indent At 1.1 solar radii, the gravitational cooling by the solar wind is, according to Eq.\,(34), about $5.2 \cdot 10^{ - 7}~{\rm{ erg\,cm}}^{-3}\, {\rm{s}}^{-1} {\rm{,}}$ which is about 20\% of the plasma-redshift heating alone. The plasma-redshift heating at this location, according to Eq.\,(33) and Table 2, is 5 times greater or about $2.58 \cdot 10^{ - 6} {\rm{.}}~$  This estimate disregards the conversion of magnetic field energy to heat, which low in the corona is usually significant.  Thus, at this location and below it, the cooling by solar wind is only a small fraction of the plasma-redshift heating, and even smaller fraction of the entire heating.  The X-ray emission cooling at 1.1 solar radii, according to Sutherland and Dopita [11], is $9.7 \cdot 10^{-7} {\rm{.}}~$  The excess heating leaks by conduction in to the transition zone.

\indent At about 1.6 solar radii, the gravitational cooling by the solar wind is according to Eq.\,(34) about $1.16 \cdot 10^{ - 7} {\rm{.}}~$  The plasma-redshift heating at this location is according to Eq.\,(31) and Table 2 also about $1.16 \cdot 10^{-7} {\rm{.}}~$  This estimate disregards the conversion of magnetic field energy to heat, which often may be significant.  The radiation cooling at the same location is according to Sutherland and Dopita about $1.4 \cdot 10^{ - 8} {\rm{.}}~$

\indent Beyond about 1.6 solar radii, the gravitational cooling by the solar wind dominates the plasma-redshift heating.  Nevertheless, the temperature continues to increase and reaches maximum of about 2.2 million degrees at about 1.8 solar radii.  Beyond about 5 solar radii the solar wind accelerates although the gravitational cooling exceeds the plasma-redshift heating.  For explaining these apparent contradictions, we need to take into account not only direct conversion of magnetic field to heat, but also two important forces.

\begin{enumerate}
\item We must take into account that plasma redshift transfers the energy to the electrons, and that the energy-transfer from the electrons to the protons is a very slow process.  The light and hot electrons therefore diffuse outwards ahead of the protons and build up an electrical field, which drags the protons outwards or upwards in the gravitational field of the Sun.
\item We must take into account the repulsion of diamagnetic moments by the magnetic field, which is described quantitatively in Eq.\,(B10) of Appendix B.  Low in the corona, this magnetic repulsion force decreases the gravitational cooling significantly.  High in the corona, beyond about 5 solar radii, the outward diamagnetic repulsion force exceeds the gravitational attraction force and causes outward acceleration of the solar wind.
\end{enumerate}


\subsubsection{The stopping power}
The stopping power is about $9.3 \cdot 10^{ - 11} \; {\rm{ erg}}\,{\rm{s}}^{ - 1}$ for a 1000 eV incident electrons penetrating a hot electron plasma with temperature of about one million degrees and an electron density of $N_e  = 10^7 \; {\rm{ cm}}^{ - 3} $.  According to the conventional theoretical estimates, the stopping power is only about $1.2 \cdot 10^{ - 11} \; {\rm{ erg}}\,{\rm{s}}^{ - 1} $.  The conventional theory includes energy transfer to the plasma frequency, but not to the very low frequencies corresponding to the root c in Eq.\,(6).  This root is important only in very hot, sparse, non-degenerate plasma.  The rate of energy loss of 1000 eV electrons (even according to the conventional theory) is much greater than the rate of redshift heating per electron, as given by Eq.\,(33), anywhere in the corona and transition zone.  The electrons in the solar corona will be thermalized, therefore, and have a Maxwellian energy distribution.

\indent The rate of energy transfer from 1000 eV electrons to protons is much smaller.  The shielding effect by the electrons in the plasma prevents all but the highest frequency Fourier harmonics of the incident electrons' field from penetrating the shielding of the protons.  M{\o}ller's (as well as Mott's) hard collision cross-section for stopping of electrons is a very small fraction of the total stopping power.  The hard collision term in electron's interaction with protons is even much smaller due to the protons 1836 times bigger mass.  The main energy transfer from the electrons to the protons is therefore caused by the electrons' diffusion outwards ahead of the protons.  This outward diffusion of the electrons ahead of protons produces an electrical field, which drags the protons outwards.  The electron plasma shields the positively charged particles.  Their effective charge depends on the density and relative velocity of the electrons to the protons.


\subsubsection{Temperature of a plasma}
In this context, we should realize that the temperature measurements are usually based on the X-ray production and the ion excitation levels.  Mainly the electron temperature, and not the heavy ion temperature, determines both the X-ray production and ion excitations.  The temperatures measured by Sturrock et al.~[22] and Wheatland et al.~[23] are principally the electron temperatures and not necessarily the proton temperatures.  The outward forces on the diamagnetic moments and the outward electrical forces on the protons counteract the gravitational attractions.  This reduction in the gravitational attraction corresponds to reduction in the value of $G$ in Eq.\,(32).  The temperature, $T$, in the exponential term in Eq.\,(32) is the average temperature and is therefore usually lower than the electron temperature.  When the magnetic and electrical forces reduce the gravitational attraction, the temperature, $T$, inside the brackets in the exponential term, must be reduced for maintaining the value of the exponential term.


\subsubsection{Magnetic force on diamagnetic moments in the corona}
Let us first consider the outward force on each diamagnetic moment created by the outward decreasing magnetic field, as quantified in {Eq.}~(B10) in Appendix B.  At a point P in the corona this force is $F = \left( {{1 \mathord{\left/
 {\vphantom {1 2}} \right.
 \kern-\nulldelimiterspace} 2}} \right)mv_ \bot ^2 \left( {{n \mathord{\left/
 {\vphantom {n R}} \right.
 \kern-\nulldelimiterspace} R}} \right) $, where $m$ is the mass of the particle, $v_ \bot$ its velocity perpendicular to the field, $R$ the distance from solar center, and $n$ is the exponent in the function $B_P \left( {{{R_P } \mathord{\left/
 {\vphantom {{R_P } R}} \right.
 \kern-\nulldelimiterspace} R}} \right)^n $, which gives the radial decrease of the magnetic field $B$ at the location P.
 
\indent In the transition zone, the electron and the proton temperatures are usually less than or about equal to 500,000 K.  The average value of $\left( {{1 \mathord{\left/
 {\vphantom {1 2}} \right.
 \kern-\nulldelimiterspace} 2}} \right)mv_ \bot ^2  = kT \le 6.9 \cdot 10^{ - 11} \; {\rm{ erg}} $. Let us assume $n = 3$.  We get then that the force on the proton-electron pair is usually less than ${{6kT} \mathord{\left/
 {\vphantom {{6kT} R}} \right.
 \kern-\nulldelimiterspace} R} = {{4.1 \cdot 10^{ - 10} } \mathord{\left/
 {\vphantom {{4.1 \cdot 10^{ - 10} } R}} \right.
 \kern-\nulldelimiterspace} R}\;{\rm{ dyne}} $.  The corresponding gravitational force on the proton is about ${{ - 3.2 \cdot 10^{ - 9} } \mathord{\left/
 {\vphantom {{ - 3.2 \cdot 10^{ - 9} } {R{\; \rm{ dyne}}}}} \right.
 \kern-\nulldelimiterspace} {R{\; \rm{ dyne}}}}$.
The repulsion force acting on the diamagnetic moments of a proton-electron is then less than 13\% of the gravitational attraction.  (However, the magnetic dipoles creating the magnetic field are usually high in the atmosphere.  The exponent $n$ could then occasionally be very large.  If $n$ were equal to 25, the diamagnetic forces on the proton-electron pair would be ${{3.5 \cdot 10^{ - 9} } \mathord{\left/{\vphantom {{3.5 \cdot 10^{ - 9} } R}} \right. \kern-\nulldelimiterspace} R}\;{\rm{ dyne}}$; that is, greater than the gravitational attraction.  More generally, we get that when at this location the value of $n \ge 23 $, and when the kinetic energy component of the proton electron pair at the right angle to the magnetic field exceeds $2kT =1.39 \cdot 10^{-10} ,$ the magnetic repulsion will exceed the gravitational attraction.)   

\indent At a distance of 2 solar radii, the electrons' temperature according to Table 2 is about 2.17 million K, and therefore $kT \approx 3 \cdot 10^{ - 10} \; {\rm{ erg}} $.  The proton temperature may be 25\,\% of the electron temperature.  In this region, the magnetic field is likely to decrease with $n$ about equal to 2.1.  The diamagnetic repulsion force on the proton-electron pair is then about 49\% of the gravitational attraction.  For this reason, the solar wind cooling is only 51\,\% of the cooling given by Eq.\,(34), or $2.59 \cdot 10^{ - 8} \;{\rm{ erg}}\,{\rm{s}}^{ - 1} $.  The emission cooling, according to Sutherland and Dopita [11] is about 13\% of this value.  The total cooling, $2.91 \cdot 10^{ - 8} \; {\rm{ erg}}\,{\rm{s}}^{ - 1}$, can be compared with the plasma-redshift heating, which is about $3 \cdot 10^{ - 8} \; {\rm{ erg}}\,{\rm{s}}^{ - 1} $.  The small excess heating leaks by conduction outwards.

\indent The solar wind cooling between about 3 and 5 solar radii is significant and causes instabilities, which are most likely responsible for the formation of the white light images observed by Sheeley Jr. et al.~[24].  From about 5 solar radii, they observed the white light images accelerated outwards beyond 30 solar radii.  At about 5 solar radii, the velocities of the white light images were about $2 \cdot 10^6 \; {\rm{ cm}}\,{\rm{s}}^{ - 1} $.  If this is typical of the proton velocity, the corresponding proton temperature could be very low at 5 solar radii.  The electron temperature at this distance could be much higher, as the hot electrons diffuse outward from the hotter layers below.  The rate of plasma-redshift heating transferred from the electrons to the protons by means of collisions is small.  The electrical field created by the electrons as they diffuse outwards must therefore drag the protons outwards.  At 5 solar radii, we set the electrons' temperature at about 1.2 million degrees K.  For the electrons, we have then that $kT$ is about $1.66 \cdot 10^{ - 10}$ erg.  The proton's kinetic energy perpendicular to the magnetic field is likely to 
be only a small fraction of this.  The magnetic field is likely to decrease with $n > 2$, for example,  $n  \approx 2.1 .~$  The diamagnetic force acting on the proton-electron pair is then about ${{3.5 \cdot 10^{ - 10} } \mathord{\left/ {\vphantom {{3.5 \cdot 10^{ - 10} } R}} \right.
 \kern-\nulldelimiterspace} R}$ dyne.  The gravitational force acting on the proton-electron pair at 5 solar radii is ${{6.4 \cdot 10^{ - 10} } \mathord{\left/ {\vphantom {{6.4 \cdot 10^{ - 10} } R}} \right.
 \kern-\nulldelimiterspace} R}$ dyne.  The diamagnetic force counteracts the gravitational force and reduces the gravitational cooling to 45\,\%.  If the solar wind flux at the distance of the Earth is $S_{215R_0 }  = N_p v_p  = 3.2 \cdot 10^8 \; {\rm{ cm}}^{ - 2} \,{\rm{s}}^{ - 1}$, the 45 \,\% of cooling given by Eq.\,(34) is at 5 solar radii equal to $5.9 \cdot 10^{ - 10}\; {\rm{ erg}}\,{\rm{cm}}^{ - 3} {\kern 1pt} {\rm{s}}^{ - 1} $.  According to Sutherland and Dopita [11], the emission cooling is about 10\,\% of this value.  The total cooling is then about $6.5 \cdot 10^{ - 10}\;  {\rm{ erg}}\,{\rm{cm}}^{ - 3} {\kern 1pt} {\rm{s}}^{- 1} $,  which is about equal to the plasma-redshift heating given by Eq.\,(33) and Table 2.

\indent At about 10 solar radii, Sheeley Jr. et al.~[24] observed that the ``white light images'' had outward velocities of $v \approx 2 \cdot 10^7  \;{\rm{ cm}}\,{\rm{s}}^{ - 1} $, which is large compared with $2 \cdot 10^6 \; {\rm{ cm}}\,{\rm{s}}^{ - 1} $ at 5 solar radii.  When the protons are accelerated outwards between 5 and 10 solar radii, the collision frequency is not adequate to make their velocity distribution isotropic.  The diamagnetic moments of the protons at 10 solar radii would then be relatively small.  The thermal velocity of the electrons would be more isotropic and their diamagnetic moments large.  The electrons would then pull the protons outwards.  If the electrons have a thermal velocity of $kT$ equal to about $5 \cdot 10^{ - 10}\; {\rm{erg}}$, and the protons and helium about 10\,\% of that, the outward diamagnetic force on a proton for $n$ equal to 2.1 would be about ${{1.16 \cdot 10^{ - 9} } \mathord{\left/{\vphantom {{1.16 \cdot 10^{ - 9} } R}} \right. \kern-\nulldelimiterspace} R}$ dyne.  The gravitational attraction force on the protons at this location is about ${{3.2 \cdot 10^{ - 10} } \mathord{\left/ {\vphantom {{3.2 \cdot 10^{ - 10} } R}} \right. \kern-\nulldelimiterspace} R}$ dyne.  The magnetic repulsion thus exceeds the gravitational attraction by ${{8.4 \cdot 10^{ - 10} } \mathord{\left/{\vphantom {{8.4 \cdot 10^{ - 10} } R}} \right. \kern-\nulldelimiterspace} R}$ dyne.  This excess repulsion force on the proton corresponds to outward acceleration of 722 ${\rm{cm}}\,{\rm{s}}^{ - 2} $, which is in the range of 290 to 830 ${\rm{cm}}\,{\rm{s}}^{ - 2} $, observed by Sheeley Jr. et al.~[24].


\subsubsection{The significance of angular scattering}
 Eq.\,(B11) of Appendix B indicates that the angular scattering of the ${\rm{He}}^{{\rm{ +  + }}}$ ions is much greater than that of the protons.  The velocities of the helium-ions are therefore much more isotropic than that of the protons.  The average force on the diamagnetic moments produced by the helium-ions may then be much larger than the corresponding force on the protons.  Even the average outward velocity of the helium ions may then become larger than the corresponding velocity of the protons.  This possibility may explain the remarkable observation by Steinberg et al.~that helium ions often have velocities that are equal to or even exceed that of the protons [30].


\subsubsection{Synopsis}
  In the transition layer and low in the corona the transformation of magnetic field to kinetic energy and heat is important, as indicated in Appendix B.  High in the corona this transformation is less important, because of smaller mutual coupling to the field-generating currents.  High in the corona, and especially beyond 5 solar radii, the magnetic repulsion of the diamagnetic moments is important and transfers energy to the plasma.  The radial magnetic field must therefore decrease with $n$ greater than 2.  When the field decreases and approaches equilibrium with the thermal kinetic energy of the plasma, it bends around due to solar rotation and decreases until the field becomes amorphous.

\indent These examples serve mainly as illustrations of how to use the equations and the theory.  The temperature of the protons and the exact variation in the strength of the magnetic field are not known well enough to make exact predictions.  However, it appears that the experiments are in rough agreement with both the plasma-redshift theory and the magnetic repulsion theory (see Eq.\,(B10)), which appear to explain in a simple way phenomena that previously could not be explained.


\subsection{Far-reaching solar streamers}
During total solar eclipse, far-reaching streamers are seen radiating almost isotropically within the first 3 to 5 solar radii from the Sun;  see Fig.~37, p. 124 of reference [8].  It has been difficult to explain these isotropic streamers.  We are inclined to believe that the field lines from dipoles creating the magnetic fields would curve around, and that the field would decrease outwards approximately with $1/(R-R_1)^3$, where $R_1$ is the distance of the pertinent magnetic dipole from the solar center.

\indent The plasma redshift facilitates explanation of these phenomena.  The plasma redshift results in ``bubbles'' containing hot electron plasmas.  The ``walls'' of the ``bubbles'' contain slightly colder plasma, which may not be as fully ionized as the inside of the ``bubbles'', especially in the transition zone.  The electron pressure from the inside on the ``walls'' of each ``bubble'' is nearly isotropic.  The gravitational field affects the pressure of the protons much more than that of the electrons; the heavier protons inside the bubble and in the walls produce greater pressure at the bottom.  The electrical field reduces the vertical pressure difference, but the hotter interior of the bubble will nevertheless create more uniform pressure than the colder plasma in the walls.  Consequently, the bubble will be elongated in the vertical direction even at places where the magnetic field initially was horizontal.  This bubble structure is rather stable, even when the plasma is fully ionized and reaches far into the corona.  The reason is that the plasma redshift is a first-order process while the cooling is second order in density.  High in the corona the reduced mutual induction between the currents of charged particles encircling the 
field lines and the currents creating the magnetic field reduces the conversion of magnetic field to heat.  The outward decreasing magnetic field pushes the diamagnetic moments outward and pushes simultaneously the top of the bubble outward, as given by Eq.\,(B10) in Appendix B.  We find, therefore, that the magnetic field radiates nearly isotropically from the Sun and reduces nearly proportional to ${1 \mathord{\left/
 {\vphantom {1 {R^2 {\rm{,}}}}} \right.
 \kern-\nulldelimiterspace} {R^2 {\rm{,}}}}$ rather than $1/(R-R_1)^3$.  The remarkable observations of streamers in solar corona are consistent with predictions of plasma-redshift theory and Eq.\,(B10) in Appendix B.


\subsection{Solar flares}
Solar flares come in many sizes and have many forms.  At the end of a large solar flare, when a large ``bubble'' breaks the surface, a hot plasma gushes into the corona and the hot plasma may reach the Earth, increase auroral activity, influence the magnetic field, and disturb power lines.  The plasma-redshift theory facilitates the explanations of the flare phenomena.

\indent In section 5.1, we saw that plasma redshift leads to formation of hot ``bubbles'' in the upper chromosphere.  The spicules consist of colder plasma squeezed between the expanding ``bubbles'' in the transition zone.  The flare phenomena are related to the ``bubbles'' and the spicules in the transition zone and have similar explanation, but the flares are formed below the spicules region.  Usually, the condition for a plasma redshift in deeper layers of the chromosphere and the reversing layers is not present.  The temperatures are too low, the densities too high, and the magnetic fields not strong enough for initiating plasma redshift.  However, hot gas containing relatively strong magnetic fields can occasionally move up from layers below the photosphere into the chromosphere.  As described in Appendix B, the partially ionized plasma cools down when it moves through the photosphere.  The B-field then approaches the H-field, which only slowly can transform into heat.  The rate of transformation of the magnetic field energy to kinetic energy of the charged particles increases as the plasma temperature increases.  At a certain location within this plasma close to the chromosphere, the temperature may be about 20,700 K and electron density about $1.211 \cdot 10^{10} \; {\rm{ cm}}^{ - 3} {\rm{.}}$  (These values are close to the temperature and densities at height of 2154 km in model A by Vernazza et al.~[6].)  As shown by De La Beaujardire et al.~[31], the B-field in a large flare may be in the range of 1000 to 1,600 gauss.  Eq.\,(28) indicates that the 50\% cut-off wavelength is then about 70 to 170 nm.  This plasma is highly ionized and short wavelength radiation intensity internal in the plasma is high. Because the plasma redshift is a first-order process while the cooling is a second-order process in density, the hot areas have a tendency to get hotter and cold areas colder.  A slight imbalance in temperature will therefore concentrate the plasma-redshift heating in a ``bubble'', which then becomes much hotter than its surrounding region.  Initially, this imbalance is caused mainly by short wavelength light internal to the hot plasma.  As the temperature increases the light from the photosphere starts to contribute to the heating in the hot bubble.  The heating causes transformation of magnetic field energy to heat, as explained in Appendix B.  At higher temperatures, the rate of transformation of the magnetic field to heat increases nearly exponentially.  The increase in the diamagnetic moments with the heating causes the B-field to decrease inside the ``bubble'' volume.   This causes an electromotive force that seeks to reduce the currents creating the magnetic field.  The corresponding decrease of the magnetic field in the walls of the ``bubble'' is reduced or even reversed by the return field from the diamagnetic moments inside the ``bubble''.  The field in the walls has a tendency to intrude into the ``bubble'' and increase the heating of the intruding plasma particles.

\indent An initial magnetic field of about 1,600 gauss could create a pressure that could balance the thermal pressure in the plasma corresponding to $N_e T \approx 3.8 \cdot 10^{20} \; {\rm{ cm}}^{ - 3} \,{\rm{K}}{\rm{.}}$

\indent It may be difficult to detect the initial formation of the bubble through the layers above, as the short wavelength radiation inside the bubble has usually high optical density.  The bubble will expand, preferably along the field lines.  When the field is horizontal the bubble may span 100,000 km.  The temperature can increase to or exceed 100 million degrees K, and sometimes it could possibly even reach much higher temperatures before the bubble breaks through the surface and the hot plasma erupts into the corona.  However, the heat conductivity coefficient becomes very large and will usually prevent these extreme temperatures unless the magnetic field is very strong.  The hottest temperatures can nevertheless produce significant amounts of X rays, and even neutrons [31, 32].  At these high temperatures, some fusion may take place.

\indent The magnetic heating will magnify the redshift heating in the bubble.  The high temperatures, large magnetic fields, and high pressure and the magnetic field lines will also broaden the width of the photons and increase the rate of plasma-redshift heating in the bubbles to about $8.6 \cdot 10^{ - 13}\; {\rm{ erg}}\,{\rm{s}}^{ - 1}$ per electron.  Disregarding the cooling, the rate of temperature increase of the plasma would then be similar to that in Eq.\,(31) or 
\be
\frac{{dT}}{{dt}} \approx \frac{{8.6 \cdot 10^{ - 13} }}{{1.9 \cdot \frac{3}{2}k}} = 2000 \; \;{\rm{K}}\,{\rm{s}}^{ - 1} {\rm{,}}
\ee
where $k$ is the Boltzmann constant.  The bubble temperature would then reach 100 million degrees K in about 50,000 seconds or about 14 hours.  When the bubble becomes hot the rate of transformation of the magnetic field to heat increases, as shown in Appendix B.  The last 7 to 10 hours may then be shortened significantly.  In the initial stages, much of the recombination cooling does not escape, but is absorbed or reflected in the walls of the bubble.  The rough estimate in Eq.\,(35) serves only to indicate the order of magnitude.

\indent For nearly vertical field lines, we should expect smaller flares because it is easier for the bubbles to expand along the field lines and open up to the corona.  The bubbles can also be initiated at smaller field strengths and higher in the atmosphere, which makes the corresponding flares smaller.

\indent  Loops and arches are sometimes seen reaching far into the solar corona.  These loops and arches have been difficult to explain; but the plasma-redshift heating helps explain these phenomena.

\indent When a large ``bubble'' with horizontal field lines approaches the surface, a dome consisting of the ``wall'' of the bubble may be seen.  As the dome rises the wall is likely to split into arches along the field lines, because any weakness caused by ripples or fluctuation will enhance the splitting of the dome along the field lines.  When the hot plasma penetrates between the field lines, the redshift heating will push the arches formed in this way apart.  The arches are relatively stable structures.  The redshift heating is proportional to the density, while the cooling is a second order in density.  Inside an arch the cooling dominates and the density inside the arch is higher than that outside the arch.  But the pressure inside the arch may be lower than the pressure in the hot sparse plasma outside the arch.  The redshift heating outside the arch pushes the sparse plasma on the outside into the arch, while the cold emission-cooled plasma leaks down in both ends of the arch.  This structure is relatively stable and can sometimes last for days and stretch far into the corona.  These phenomena have been difficult to explain, but plasma redshift gives a simple explanation of the observations.


\subsection{Plasma redshift of spectral lines and gravitational redshift}
Plasma redshifts of solar Fraunhofer lines is significant and requires revision of both Einstein's gravitational redshift and the standard interpretation of solar redshifts.


\subsubsection{Plasma redshift of solar Fraunhofer lines}
The redshift of solar lines is small, and the second term on the right side of Eq.\,(20) is usually significant.  Most photon widths are broadened due to the effects of collisions with: 1) the neutral atoms, 2) the Fourier field of charged particles (Stark effect), and 3) the photon field, which result in stimulated absorptions and emissions.  The photon widths can also be broadened by magnetic fields.  The Doppler effect caused by movements in the line-forming elements affects the line width, but not the photon width.  The Doppler shifts can be very large in the spicules region; but in the reversing layer and in the photosphere, the Doppler redshifts are nearly equal to Doppler blue shifts.  The average Doppler shifts of photospheric lines are usually smaller than those often assumed for explaining the large deviation of the observed shifts from the expected gravitational redshifts.

\indent The broadenings of the widths of the photons vary with temperature and density and therefore with the depth of the line formation in the photosphere and the reversing layers.  Usually, a photon's width varies across the solar disk.  It is often significantly smaller close to the limb because of lower density.  This then reduces the limb effect, as in the case of the sodium and potassium resonance lines.  It is complicated to estimate the broadening effects, as the theories for the different broadenings are often inadequate.  It is sometimes necessary, therefore, to rely on experimentally determined values for the Sun.  For the relative variations of the collision broadenings of the lower and upper levels of each transition, it is often useful to use as a rough guide the collision excitation functions:
\be
Y_L  = NT^{{1 \mathord{\left/
 {\vphantom {1 2}} \right.
 \kern-\nulldelimiterspace} 2}} {\rm{exp}}\left( {{{ - E_L } \mathord{\left/
 {\vphantom {{ - E_L } {kT}}} \right.
 \kern-\nulldelimiterspace} {kT}}} \right), \; \; \; {\rm{and}} \; \; \; Y_U  = N T^{{1 \mathord{\left/
 {\vphantom {1 2}} \right.
 \kern-\nulldelimiterspace} 2}} {\rm{exp}}\left( {{{ - E_U } \mathord{\left/
 {\vphantom {{ - E_U } {kT}}} \right.
 \kern-\nulldelimiterspace} {kT}}} \right){\rm{,}}
\ee
where the $E_L$ and $E_U$ are the energies of the lower and upper levels, and where $N$ and $T$ are densities of the neutral and charged particles and $T$ their temperatures from model C of the solar atmosphere developed by Vernazza et al.~[6].  The variations in the functions with height are illustrated in Fig.~3.  Besides the excitation functions $E_L$ and $E_U$, the broadenings depend on electronic configuration, the angular momentum for each level, and transition probabilities for each transition.

\indent These excitation functions show that deep in the photosphere the broadening for each line decreases steeply with increasing height.  Analysis of the redshift measurements shows that when $E_{L} \geq 3.5 ~ \rm{eV}$ (and the upper level about 5.8 eV), the collision broadening is usually small, and the photon width is then close to the natural quantum mechanical width, except deep in the photosphere where the temperatures and pressures are high.  When $E_{L} \leq 3 ~ \rm{eV} ,$ the collision broadening is often significant even above the temperature minimum in the solar atmosphere.


\begin{figure}[t]
\centering
\includegraphics[scale=.4]{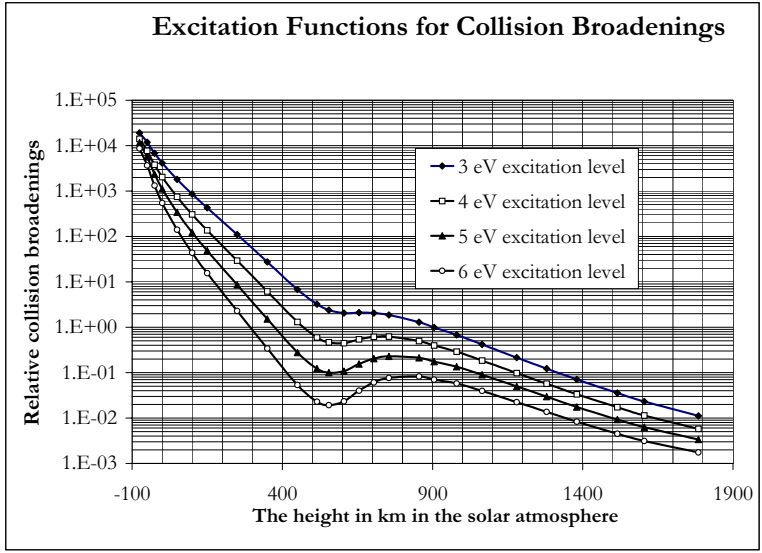}
\caption{The abscissa gives the height in km above the optical density $\tau _{500}  = 1$ in the solar photosphere.  The ordinate gives in arbitrary units the excitation function $Y=NT^{1/2} \, {\rm{exp}} \left(-E/kT \right)$ for the collision broadening.  The values of the number density $N$ of hydrogen and the temperature $T$ are from model C of Vernazza et al.~[6].  The value of the excitation potential $E$ of the energy level in the atom is equal to 3, 4, 5, and 6 eV for the four curves, respectively.  The curves indicate how the collision broadening varies with the height of formation.  The collision broadening varies also with the transition rate and the angular momentum change for each transition.}
\end{figure}

\indent Weak lines are formed at great depths.  When their widths are collision-broadened, the photons have maximum widths at the center of the solar disk.  This broadening of weak lines causes increased redshift at the center.  Because the lines are weak and formed relatively deep in the photosphere, the broadening decreases sharply with height, as Fig.~3 indicates.  We observe therefore a decrease in the redshift as the line of sight moves away from the center of the solar disk and up in the photosphere.  For these lines, the observations show that the redshift often reaches a minimum at distances between about 0.4 to about 0.7 solar radii from the center of the solar disk.  Beyond this distance the redshift of these weak lines increases towards the limb due to increased path length low in the corona in accordance with the first term, the integral term, of Eq.\,(20).

\indent Slightly stronger lines are formed higher in the photosphere.  These lines often have low excitation potentials and are therefore more easily collision broadened.  When they are formed around the temperature minimum in the solar atmosphere, the collision broadening varies less with height.  These lines show then a plasma redshift, which increases towards the limb in accordance with the first term of Eq.\,(20), similar to that shown for the lines listed in Table 3.  When at the center the line is formed at the temperature minimum and at the limb just above the temperature minimum, the second term will enhance the limb effect.

\indent In case of stronger lines, which at the center are formed close to and slightly above the temperature minimum, the temperature increase may dominate the decrease in density.  As the line of sight moves away from the center, the broadening effects may then first increase slowly and reach a low maximum before they decrease again because of reduced density.  The redshift of collision-broadened lines formed at this height varies therefore less with the distance towards the limb, as the increase in redshift caused by the first term of Eq.\,(20) is sometimes compensated by the decrease in the width and the redshift caused by the second term.  The lines may then have relatively small center-limb variation.  The lines with small center-to-limb effect are then also characterized by having relatively small strength at the limb relative to the line's strength at the center of the solar disk, such as the sodium and potassium resonance lines.


\begin{table}
\centering

{\bf{Table 3}} \, \, Solar redshift experiments by Adams and Higgs and plasma redshift theory.

\vspace{2mm}

\begin{tabular}{lllll}
       \hline
Distance $r/R_0$ from & Measurements & Measurements & Averages of & Estimates$^1$ of \\
center of solar disk in & by Higgs of & by Adam of & $(\Delta\lambda/\lambda)\cdot 10^6$ & Plasma redshift \\
units of solar radius $R_0$ & $(\Delta\lambda/\lambda)\cdot 10^6$ & $(\Delta\lambda/\lambda)\cdot 10^6$ & & $(\Delta\lambda/\lambda)\cdot 10^6$ \\
\hline \hline
0.999 & & & & 3.12 \\
0.998 & 3.03 & 2.67 & 2.85 & (3.03) \\
0.995 & 2.83 & 2.56 & 2.69 & (2.86) \\
0.990 & 2.62 & 2.44 & 2.53 & (2.66) \\
0.985 & 2.47 & 2.35 & 2.41 & (2.53) \\
0.980 & 2.36 & 2.35 & 2.35 & (2.43) \\
0.975 & 2.27 & 2.21 & 2.24 & (2.36) \\
0.974 & & & & 2.35 \\
0.970 & 2.19 & 2.15 & 2.17 & (2.27) \\
0.950 & 1.96 & 1.96 & 1.96 & (2.08) \\
0.949 & & & & 2.04 \\
0.925 & 1.80 & 1.77 & 1.78 & (1.84) \\
0.900 & 1.70 & 1.61 & 1.66 & 1.72 \\
0.875 & 1.64 & 1.48 & 1.56 & (1.60) \\
0.800 & & & & 1.40 \\
0.700 & & & & 1.24\\
0.600 & & & & 1.14\\
0.500 & & & & 1.08 \\
0.000 & 1.08 & 1.08 & 1.08 & 1.00 \\
\hline 
\end{tabular} \\
\raggedright{\ \ \ \ \ \ \ {$^1$ The values in the parentheses are obtained by interpolation.}}
\end{table}

\indent For estimating the redshift, we used the coronal electron densities shown in Table 2.  For the stretch from 3 to 5 solar radii, we used slightly smaller hump, because the average height of this hump is slightly smaller than indicated by Table 2.  The redshift given by Eq.\,(20) is integrated from the cut-off region in the Sun to the observer on the Earth.  The integral is rather insensitive to the electron densities variations beyond about 3 solar radii because of the low electron densities.  In case of large electron densities in the transition zone and lower corona, the cut-off region moves up; and for low electron densities, the cut-off region moves down.  The plasma-redshift integrals just above the cut-off region vary therefore less than if the cut-off zone did not move up with increase in plasma density.  This region is the densest and contributes most to the integral in Eq.\,(20).  The plasma-redshift integral, which uses the average densities, relative to the cut-off, is therefore a relatively good estimate.  As mentioned in the last paragraph of section 5.2, the temperatures and therefore the densities compare well with the average temperatures measured by Wheatland, Sturrock and Acton [22,\,23].

\indent The measurements of the redshift are usually made when the solar atmosphere is quiescent, and the average values are usually obtained by integrating over long time periods.  The electron densities listed in Table 2 are close to average values for a quiescent corona, because the temperature is consistent with the average estimates of Sturrock et al.~[22] and Wheatland et al.~[23].  For comparison, see the estimates by Newkirk [21], McWhirter et al.~[13], and Withbroe [14, 15].


\begin{figure}[t]
\centering
\includegraphics[scale= .5]{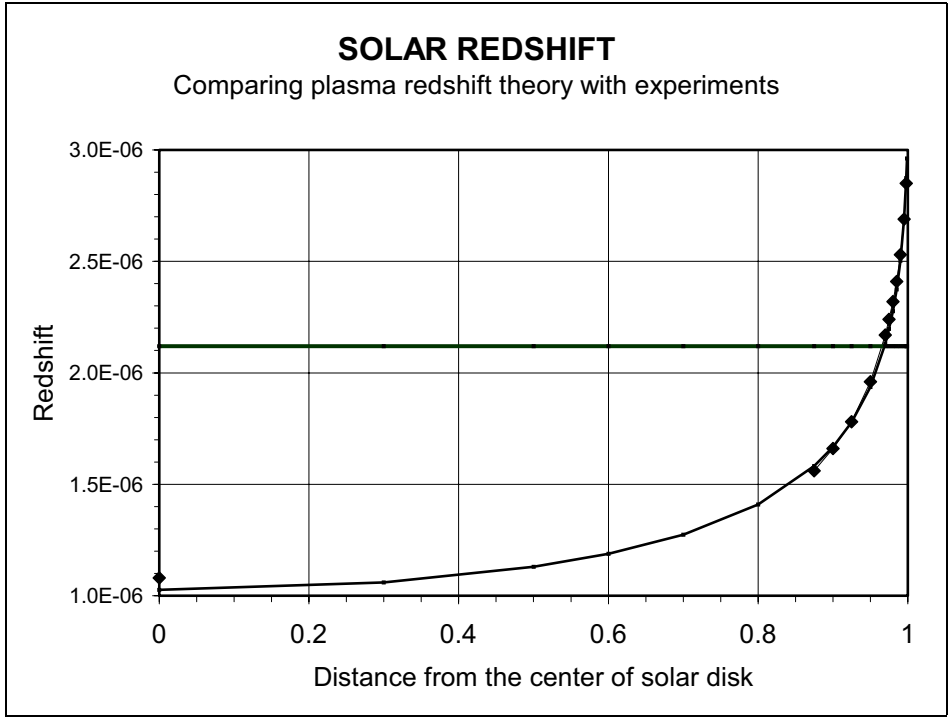}
\caption{The diamonds show the average redshift measured by Adam [33] and Higgs [34] (see Table 3), while the curve shows the redshift predicted by the plasma-redshift theory when using the coronal electron densities listed in Table 2 and shown in Fig.~2.  The horizontal line at $2.12 \cdot 10^{ - 6}$ shows the redshift predicted by Einstein's classical gravitational theory.}
\end{figure}

\indent Solar redshift measurements by Adam [33] and Higgs [34] of the 629.78, 630.15, and 630.25 nm lines from Fe-I are listed in Table 3.  The ionization potentials of the upper levels are: 4.191, 5.620, and 5.653 eV, and of the lower levels: 2.223, 3.654, and 3.686 eV, respectively, for the three lines.  At the center, the formation of these lines is below the temperature minimum; but close to the limb, the formation is above the temperature minimum.  The excitation functions mentioned above indicate that the collision broadening of the average of the three lines does not vary much with height for these lines.  Estimates of the plasma redshift in the last column are obtained by using the integral term in Eq.\,(20).  This is equivalent to assuming that the average photon widths of these three lines are about equal to the classical width of the photons, and that the collision broadening does not vary much across the solar disk.

\indent In these experiments, the authors made extreme efforts to obtain accurate average redshift for each line at each position all the way to the extreme limit of the solar limb.  The steep variation close to the limb requires that the opening angle is small and the integration time long.  As can be seen from Table 3, the results of Adam and Higgs differ slightly.  Such differences are understandable when we think of the variation in the electron densities with the flares and the sunspot cycle.  As Table 3 and Fig.~4 indicate, the agreement between the plasma-redshift theory and the experiments is very good.  The redshift varies from center to limb with the electron density integral.

\indent   The center-to limb variations may vary slightly from one time to another.  The size and frequency of coronal holes vary with the sunspot cycle.  Coronal holes are more frequently over the polar region than over the equatorial region.  In coronal holes, the electron densities are usually low.  The limb effect in the north-south direction will therefore sometimes be smaller than that along the equator.  This is consistent with the measurement of the center-to-limb variation of the Fe line 557.6 nm by Brandt and Schr\"{o}ter [35], who found significant difference between center-to-limb redshifts in south-north and east-west directions.  Their measurements were made in April 1978 and a few in September 1979.  In March the Sun's south pole is tilted about 7 degrees towards the Earth and in September away from the Earth.  During the times of observations, there were sunspots at high latitude but very few at low latitude.  Brandt and Schr\"{o}ter considered the conventional explanation of possible Doppler shifts.  Meridional flows towards the equator could possibly cause the decrease in the redshifts close to the polar region.  They rejected that explanation.  They found that the required hypothetical currents of about 250 m $\rm{s}^{-1}$ were too large and contradicted other observations.

\indent Usually, the redshift increases steeply close to the limb.  It is then important that the entrance slit have a small width, preferably less than about 1 arcsec.  In the limb region, the line of sight penetrates the spicules region, which is highly turbulent.  It is then necessary to use long exposure time to get good averages.  Some researchers have not taken this effect adequately into account and find then smaller limb effect.  Some of the conventional extrapolations of the redshifts from points far away from the limb to points close to the limb are invalid, because the redshifts often increase more steeply towards the limb than that assumed.  Frequently, the extrapolations sought to approach Einstein's gravitational redshift at the limb.

\indent   In magnetically active regions, the redshifts are usually found to be greater than the redshifts far away from them; see Cavallini et al.~[36].  These larger redshifts are often interpreted as due to reduced upward Doppler shifts.  The magnetic fields are assumed to reduce the upward movements of photospheric currents.  Other researchers explain this increased redshift as due to general down draft in the active regions.

\indent According to Eq.\,(28), we expect the cut-off region for the plasma redshift to reach deeper into the transition zone over a magnetically active region.  The magnetic fields over the active region will, therefore, often increase the redshifts.  Concurrently, the magnetic field may reduce the movements in the photospheric layers.  However, the Doppler blue shifts should be expected to nearly balance the Doppler redshifts.  General down draft is also a possibility, but independent evidence of that is lacking.  The magnetic field may reduce the temperature and shift the minimum temperature in the atmosphere, which in turn will affect the distribution in collision broadenings.  The shift of the minimum temperature to deeper layers will steepen the increase in collision broadening with depth, which is consistent with the changing tilt of the bisector as observed by Cavallini et al.~[36].  The bisector segments close to the continuum are redshifted the most, while the bisector segments closer to the bottom of the line are redshifted less and sometimes blue shifted relative to the regions outside the active regions (most likely 
due to lower minimum temperature in the active region).  According to the plasma redshift theory, this 
effect and the lowering of the cut-off zone by the magnetic field are the likely causes of the increased redshift in the active regions.


\subsubsection{Gravitational redshift}
The greatest surprise is that the plasma redshift appears to explain the solar redshift without Einstein's gravitational redshift, as Table 3 and Fig.~4 show.  Einstein's gravitational redshift has been proven beyond reasonable doubt in a great many experiments.  It follows from reasonable extension of the special theory of relativity (TR) to theory of general relativity (TGR).  The fact that the plasma-redshift theory explains the observed solar redshift of Fraunhofer lines contradicts the generally accepted view that the solar lines are gravitationally redshifted.

\indent  However, this apparent contradiction has a simple explanation.  The experiments (on the Earth and in space) that appeared to prove the gravitational redshift are not able to detect the actual blue shift of photons, which would reverse the gravitational redshift.  The length of the wave packet defining the photons in these experiments is much larger than the height difference between emitter and absorber.  Einstein's theory of general relativity (TGR) and all the experiments (apart from solar redshift experiments) are entirely in the domain of classical physics.  The design, execution and conventional interpretation of the experiments disregard quantum mechanical effects.

\indent  The photons are gravitationally redshifted when emitted in the Sun, because of the time dilation; but during their travel from the Sun to the Earth, they lose their gravitational redshift, and are not gravitationally redshifted when they arrive on the Earth.  The change in frequency, as the photons travel from the Sun to the Earth, is due to increased gravitational potential as the photons travel from the Sun to the Earth.  In the quantum mechanical treatment, the photons' coordinate frequencies adjust to the gravitational potential as if the photons were particles.  Thorough analysis consistent with quantum mechanics shows that the photons are repelled, as seen by a distant observer at rest relative to the Sun but at a large distance free of gravitational field.  This observer will see reversal of the gravitational redshift of photons, as the photons move from Sun to Earth.  An observer at rest at the location of the photon close to the gravitating body (Sun) will see the photon as weightless.  (The bending of light in the gravitational field is independent of the frequency or the weight or the weightlessness of the photon.  This bending depends on the velocity of light and warping of space, which are unaffected by photon's frequency.)  We may be surprised, because this fact contradicts conventional thinking and what we have been taught.  This fact contradicts Einstein's field equations and the equivalence principle, which surmise that all energy forms are gravitationally attracted proportionally to their inertial mass.  In the special theory of relativity, it is proven that $E = m_i \, c^2 {\rm{,}}$ where $m_i$ is the inertial mass.  Einstein surmised that inertial mass, $m_i ,$  including the inertial mass of the photon, was always equivalent to the gravitational mass, $m_g $ of the same particle.  Conventional thinking has surmised without adequate proof that the photons, like particles, were always attracted, as seen by an observer at rest at the location of the photon.  As we will see, correct interpretation of the many relevant experiments shows that the gravitational field does not attract the photons.  The photons are being pushed outwards, as seen by a distant observer in a coordinate system free of gravitational fields.  This repulsion leads to reversal of photons' gravitational redshifts and simplifies explanation of many phenomena.  The repulsion of photons counterbalances the gravitational attraction in a coordinate system at rest at the position of the photon.  This weightlessness of photons in the standard system of reference makes Einstein's lambda term unnecessary, and brings stability and harmony to the cosmological perspective, as we will see.

\indent  Einstein's thought experiment with an observer in a closed elevator box is interesting and almost correct, but not completely.  The gravitational field lines are almost parallel, but not completely.  The gravitational field has always a gradient, while the inertial field does not.  It can be shown that mathematically, a gradient towards a point makes a difference.  For the present discussion, we need only recognize that there is a difference between an inertial field and gravitational field.

\indent Let us consider an atom in the Sun.  All frequencies of the atom, all the energy levels, and all the frequencies corresponding to the energy differences between two levels in the nucleus and in the atomic electron configuration are gravitationally redshifted, according to Einstein's classical TGR, as seen by a distant observer in a coordinate system free of gravitational fields.  As we bring that atom from the Sun to the Earth, the gravitational redshifts of the frequencies disappear.  During the travel of the atom from the Sun to the Earth the levels and the frequencies are blue shifted, and that blue shift cancels the gravitational redshift (or more exactly, the difference in gravitational redshift in the potentials of the Sun and the Earth).  The gravitational redshift depends on the potential energy.  In case of the atom, we can say that the energy that we transfer to the atom by lifting it out of the solar gravitational field is the cause of the blue shift of all the frequencies of the atom.  The frequencies of the photon particle behave similarly.  The photons' frequencies increase as the photon moves out of the gravitational field of the Sun.  According to Einstein, on the other hand, the photons do not change frequency as they move from the Sun to the Earth, as seen in a coordinate system free of gravitational fields.  In some articles and textbooks on this subject, a different description is often given of the gravitational redshift.  Different reference systems are used at the point of emission and absorption (detection).  Close to the Sun, where the photons are emitted, the rate of clocks in the local reference system is slower than the rate of clocks in the reference system at rest on the Earth.  It is then stated that the photons lose energy as they move up the gravitational potential from the Sun to the Earth, because the frequencies of the photons are slower than the frequencies of corresponding transition in atoms on the Earth.  Actually, however, according to Einstein, the photons' frequencies stayed constant during their travel from the Sun to the Earth.  The lower frequencies of the photons when they arrived on the Earth are due to the slower rate of clocks in the coordinate system close to the Sun, where the photons were emitted.  In fact, a maser clock in the Sun would be seen to move at a slower rate than the clocks on the Earth.  In the Sun, the frequencies of the atoms and the photons are slower, like the photon frequencies that would control the maser clocks in the Sun.  According to Einstein, the photon's ``coordinate frequency'' is constant and the photon energy is constant, as the photon moves from the Sun to the Earth.  $\hbar \, \omega = H$ is a constant of motion, where $H = E + V$ is the total energy of the photon, $E$ its kinetic energy, and $V$ its potential energy.  In his arguments, Einstein considered that light consists of classical electromagnetic waves; that is, that the wave packet for the photons reached from the Sun to the Earth and beyond.  Accordingly, he argued, equally many waves have to arrive on the Earth as leave the Sun, and the frequency has to be constant.  Einstein's logic is correct, and his assumptions are reasonable.  But although the assumptions (premises) are reasonable, they appear to conflict with nature, as solar redshift experiments indicate.  According to Galileo, nature must tell us what is right, not the assumptions made in a thought experiment.

\indent  In the quantum mechanically modified TGR, we have that, for lifting the atom out of the solar gravitational field, we have to use energy to overcome the gravitational attraction.  However, although the photons frequencies increase in the same way as the frequencies of the atoms, we do not have to apply energy to bring the photon out of the gravitational field.  In a reference system free of gravitational fields a repulsive coordinate force acts on the photon; but in a standard reference system at rest at the location of the photon, the photon is weightless.  We could then say that in the standard reference system a repulsion force counterbalances exactly the gravitational attraction.

\indent  Einstein's argument that equally many waves must arrive on the Earth as leave the Sun is a classical physics argument applied to an infinite train of electromagnetic waves, which is impermissible in quantum mechanics.  If the photons reached from the Sun to the Earth, as in classical physics, Einstein would be right.  The photons of some masers used in many experiments have this characteristic.  They may reach from the Sun to the Earth and well beyond.  However, the photons in the visible light are very short, on the order of a few meters.  As a measure of their length we can take ${{\lambda ^2 } \mathord{\left/
 {\vphantom {{\lambda ^2 } {\left( {\Delta \lambda } \right){\rm{,}}}}} \right.
 \kern-\nulldelimiterspace} {\left( {\Delta \lambda } \right){\rm{,}}}}$
where $\lambda$ and $\Delta \lambda$ are the wavelength and the photon width.  In an undisturbed sodium atom the resonance line has wavelength 589 nm, and a quantum mechanical photon width, which is 0.0116 pm, about the same as the classical photon width.  The length of the photon is then about 30 m.  In the Sun the photon width at the center of solar disk is about 17 times broader, and the length of the photon therefore only about 1.8 m.

\indent  The solar light photons during their travel to the Earth have plenty of time to change their frequency and adjust to the gravitational potential field.  In contrast, the 14.4 keV photons used in the experiments by Pound and Repka, Jr. [37, 38], and Pound and Snider [39] did not have enough time (only 75 ns) to change their frequency, as they traveled the distance of only 22.5 m between emitter and absorber in the weak gravitational potential of the Earth.  As we will see, these photons need more than $19~{\mu}{\rm{s}} ~$ for changing the frequency.  The difference in gravitational potentials of the 14.4 keV photon at the position of the emitter and the absorber in the experiments [37-39] is:

\be
\Delta E = \left( {{{h\nu } \mathord{\left/{\vphantom {{h\nu } {c^2 }}} \right.
 \kern-\nulldelimiterspace} {c^2 }}} \right)981 \cdot 2250 = 5.67 \cdot
 10^{ - 23}\; \; {\rm{ erg,}}
\ee

\noindent   where the factor 981 is the gravitational acceleration in the laboratory, and the factor 2250 cm is the height difference.  From both the uncertainty relation and the transition theory in quantum mechanics, the minimum transition time between the two states of the photon is:

\be
\Delta t \approx {h \mathord{\left/{\vphantom {h {\left( {2\pi \Delta E} \right)}}} \right.
 \kern-\nulldelimiterspace} {\left( {2\pi \Delta E} \right)}} = 1.9 \cdot 10^{ - 5} \;\; {\rm{ s,}}
\ee

\noindent  where $h$ is Planck's constant.  The wave packet needed to define that small energy difference has a length of about $c\,\Delta t \approx 5.6 ~ km.~$  The length of the 14.41 keV photon is about 
$L\approx 2\pi c\tau = 270~ m,$ where $\tau =143~ns$ is the lifetime of the 14.41 keV transition in the nucleus.  During the travel from the emitter to absorber, the photons would usually experience smaller potential difference.  Therefore, the minimum time for transition exceeds $19~\mu \rm{s} .~$  It is impossible for the photons to adjust to the new potential or to change frequency during their short travel time of 75 ns from the emitter to the absorber in these experiments.  On the other hand, the atoms and the nuclei in the emitter and absorber had plenty of time to adjust to the gravitational potential difference.  Every transition in quantum mechanics takes time.  Photons require time to change from one state to another, even if the states overlap and are continuous.  In the experiments [37-39] the photons had no chance to adjust to the gravitational potential, while they moved from the emitter to the absorber.  These experiments are inconclusive, because the experimental design did not make it possible to detect the change in frequency as the photons traveled from emitter to absorber.

\indent  In the rocket experiments by Vessot et al.~[40], the maser photons are too long.  Importantly, the potential difference is too small.  In the space experiments by Krisher et al.~[41], the laser photons (signals) used in the experiment are extremely long.  They reach from the Sun to beyond Saturn.  None of the experiments ``proving'' the gravitational redshift makes it possible to detect the relevant reversal of the gravitational redshift of photons coordinate frequency.

\indent  Quantum mechanics, in accordance with Bohr's correspondence principle, approaches and becomes identical with the classical mechanics in the limit of macroscopic experiments, and in the limit of slow variation in the gravitational field.  A correct quantum mechanically modified TGR does not contradict the classical physics experiments, like the experiments by Pound and Rebka [37, 38], Pound and Snider [39], rocket experiments by Vessot et al.~[40], and the space experiments by Krisher et al.~[41], because the experiments are inconclusive and do not make it possible to detect the frequency change of the photon.  In the gravitational deflection experiments by Riveros and Vucetich [42], and in the experiments on the time delay of radar echoes by Shapiro et al.~[43], it is important to realize that 50\,\% of the measured effect is due to variation in the speed of light in the gravitational field, and 50\,\% is due to warping of space.  The changes in the speed of light and the warping of space are caused by the mass of the Sun (the star or the galaxy).  The speed of light is independent of the photon's frequency.  In the modified TGR both the speed of light and the warping of space are the same as in the classical TGR.  According to Fermat's principle, the path of the photon depends on the speed of light and the warping of space.  In these cases, the photons' velocities had to adjust to the gravitational potential during their time of fight.  Quantum mechanical analysis of the experiments shows that this is possible, although a small but finite lag may occur, especially in case of long wavelengths at large distances from Sun.


\subsubsection{Results of solar redshift experiments compared with theory}
There is thus a fundamental difference between solar redshift experiments and the classical physics experiments that have been assumed to prove Einstein's classical TGR.  The solar redshift experiments reveal the reversal of the gravitational redshift.  The other experiments do not.

\indent It will be argued that the solar redshift experiments have been proven to confirm the gravitational redshift.  For example, it will be argued that the elaborate works of Dravins et al.~[44] and Dravins [45] have proven beyond reasonable doubt that the difference between the observed redshift and the predicted redshift can be explained as due to Doppler shift in the line-forming elements.  The analysis by Dravins et al.~is supported by many observations.  The explanation of shifts of lines as we move from the granules to the intergranular lanes appeared convincing.

\indent Essential premises for the conventional theory by Dravins et al.~are:

\begin{enumerate}
\item Einstein's gravitational redshift of solar lines when observed on Earth applies.
\item Most of the deviations from the gravitational redshift are due to Doppler shifts.
\end{enumerate}

\noindent There is no place for a plasma redshift, which follows from basic laws of physics.  For obtaining a match between the assumed velocity distribution and observed C-form of the line median, and for explaining the large blue shifts of most lines at the center of the solar disk relative to the gravitational redshift, Dravins et al.~suggested that these observations derive from up-streaming currents of about 1200 ${\rm{m}}\,{\rm{s}}^{ - 1}$ over 75\% of the area in the granules and down-streaming currents of about 3600 ${\rm{m}}\,{\rm{s}}^{ - 1}$ over 25\% of the area in the intergranular lanes.  The actual division between these regions is more gradual and variable.  The assumed large asymmetry in the area and velocities between up and down streaming currents was necessary to explain the observations.  These assumptions were supported by observation of apparent line shifts in the granules and intergranular lanes.  (We will see later when we report on the observations by Miller et al. that the observed shifts in the granules and intergranular lanes are in the wings and not the core of the lines.)  It is well known that the observed widths of the Fraunhofer lines exceed the Doppler broadening caused by thermal motion.  Vernazza et al.~[46] explain the additional broadening as caused by turbulence with velocities on the order of 1 to 1.4 ${\rm{km}}\,{\rm{s}}^{ - 1}$.  But the assumed asymmetry in the flow is not supported by independent evidence.

\indent The predictions of the Doppler shift theory by Dravins et al.~are very often the same as the predictions of the plasma-redshift theory.  The Doppler shift theory by Dravins et al.~explains that weak lines, which are formed deep in the atmosphere, have small redshifts, because the velocities of the up-streaming currents appeared to increase with depth in the atmosphere.  (Apart from the observed shifts of the lines, there is no independent verification of increasing velocities with depth.  The increasing density with depth would be consistent with decreasing velocities with depth.)   As Eqs.\,(18) and (20) show, the plasma redshift increases with photon width.  The plasma redshift predicts that weak lines, which are formed deep in the photosphere, have small redshifts because their photon widths are usually small.  The photon width is usually associated with the strength of a line, which is associated with the height of the line's formation in the atmosphere. However, there are important exceptions to this rule.  Exceptionally, some weak lines have large photon widths and therefore large plasma redshifts.   The Doppler shift theory, on the other hand, predicts small redshifts for the weak lines independent of the 
photon width.

\indent Pierce and LoPresto [47] found for the Ni\,I lines a redshift-strength relation similar to that for the Fe I lines.  However, for similar equivalent widths the redshift of the Ni\,I lines was significantly greater than that of the Fe\,I lines.  This is contrary to what is to be expected if the redshift is a function of depth of the line formation, as suggested in the theory of Dravins et al.~[44].  The observation is in agreement with the prediction of the plasma-redshift theory.  For same equivalent widths, the nickel lines have broader photon widths than the iron lines, because nickel concentration is only about 4\% of the iron concentration in the solar atmosphere.

\indent For Ti\,I lines, Pierce and LoPresto [47] found a similar relation, but the discrepancy between the measured redshifts and those expected from the equivalent widths was greater.  In the Sun, the concentration of titanium is about 0.4\% of that of iron.  These observations contradict the conventional theory by Dravins et al., but are consistent with the plasma-redshift theory.

\indent The 3d-4p lines 849.8 and 854.2 nm lines of Ca\,II have exactly the same upper level (same voltage of 3.15 eV and same angular momentum) as the 393.3 nm resonance line.  The relative rate constants and strengths of these three lines are roughly proportional to 1, 9 and 136, respectively.  However, the photon widths of the two weaker lines are on the order of or greater than the photon width of the strong 393.3 nm resonance line.  The lower levels of the 849.8 and 854.2 nm are easily broadened by Stark effect, as shown by Griem [48], and can be broader than the lower level of the resonance line.  St.~John [49] measured the center redshift of the strong 393.3 nm resonance-line to be about 1296 ${\rm{m}}\,{\rm{s}}^{ - 1}$, while Herzberg [50] measured the redshift of the much weaker 849.8 and 854.2 nm lines from the same upper level to be about 1533 and 1933 ${\rm{m}}\,{\rm{s}}^{ - 1}$, respectively.  This is consistent with the plasma-redshift theory, but contradicts the Doppler shift theory by Dravins et al.~[44]

\indent Miller et al.~[51], who used high spatial resolution technique, about 0.28 square arcsec resolution, measured the redshift and the form of the bisector in the granules and separately in the intergranular lanes.  These measurements can be expected to find huge differences in the redshifts.  The down streaming currents in the intergranular lanes could result in large redshifts far in excess of the gravitational redshift and the up-streaming currents in the granules could result in large blue shifts.  However, the measurements showed that in the core of the line, including the bottom of the line, the redshift and the C-form of the bisector are almost the same (within about 20) ${\rm{m}}\,{\rm{s}}^{ - 1}$ in the granules and the intergranular lanes.  This finding was the case for all three lines investigated (Fe\,I lines 406.5, 543.4, and 523.3 nm).  The wings (not the core) of the lines were redshifted 75 to 200 ${\rm{m}}\,{\rm{s}}^{ - 1}$ in the intergranular lanes as compared with that in the granules.  In redshift experiments, the core (and not the wings) of the line defines the redshift.

\indent These observations contradict the assumption by Dravins et al.~[44], but are consistent with the plasma-redshift theory.  They do not contradict the expectation that the turbulence is significant for broadening of the lines.   The turbulence does not appear to affect the redshift of the core of the line, which means that the Doppler redshifts about compensate the Doppler blue shifts. This is consistent with the expectations of many early (before 1960) researchers in this field.  Importantly, the core and the vertex of the line determine the observed redshift as usually measured.  The redshifts in the wings, 75 to 200 ${\rm{m}}\,{\rm{s}}^{ - 1}$, may possibly be due to Doppler shifts.  However, even if they are, they appear to be much smaller than those expected.

\indent The theory by Dravins et al.~[44] explains that the C-form of a line's bisector is due to variation in the upward and downward currents.  The assumed larger areas of the upward currents in the photosphere cause the prevailing blue shift relative to the expected gravitational redshift at the center of the solar disk.  The plasma-redshift theory explains the C-form of the bisector as caused mainly by the spectrum of photon widths for each line.  The collision broadening causes a distribution in the widths of the photons; even without the collision broadening, the widths have a distribution.  This distribution in collision broadenings varies with temperatures and densities in the atmosphere, as well as the states of the colliding atoms.  The conventional approach of averaging the collision broadenings and the collision shifts is not permissible in the plasma-redshift theory, if our focus is the line form.  The plasma redshift, which depends on the photon's width, is independent of the direction of the currents in the line-forming elements.  The plasma redshift has a distribution for each line independent of the concurrent Doppler shift.  The photons with the largest width are redshifted the most.  The fact that Miller et al.~[51] observed that the core of the bisector in the granules is about the same as that in the intergranular lanes is consistent with the plasma-redshift theory; however, these observations contradict the Doppler shift theory by Dravins et al.~[44].

\indent In the plasma-redshift theory, we often can disregard the collision shift of the lines, because it is usually much smaller than the shift caused by the collision broadening.  The collision shift may be on the order of 10\% of the collision broadening.  The redshift of the wavelength caused by the collision broadening is for the sodium resonance lines about 4 times the broadening, or about ${{\delta \lambda } \mathord{\left/
 {\vphantom {{\delta \lambda } \xi }} \right.
 \kern-\nulldelimiterspace} \xi } = 4\delta \lambda$ for the 588.995 and 589.592 nm lines of Na\,I.  The collision shift would then be only 2.5\% of the redshift caused by the collision broadening.  However, if the collision shift is significant, it must be included.


\subsubsection{Redshifts of stars}
\textbf{O and B stars.}  Similar analyses for other stars show that like the Sun they must have coronas and intrinsic redshifts, especially the hot stars.  The redshift heating per electron of the corona increases with the fourth power of the star's temperature, as Eqs.\,(29) and (30) show.  The redshift heating can then compensate cooling from higher density plasma.  The increase in the redshift is reduced slightly, because the increased densities in the transition zone lead to increased heights of the cut-off zone for the plasma redshift.  As Eq.\,(32) indicates, a large radius of a star will usually extend the high-density region of the corona.  The intrinsic redshift of a star should therefore increase with the temperature and the radius of the star.  As we do the analysis of the redshift observations, we should take into account the strength and the formation depth of the line.  The pressure broadening will usually be different from that in the Sun, because the temperatures will often be higher and the densities smaller in the line forming elements.  In the hot large stars, many of the strong lines are formed high in the in the chromosphere and close to the spicules region.  The movements in the line forming elements affect mainly those lines that are formed close to the spicules region.  The limb effect and the spread in the redshift values can be expected to be large and vary from line to line.  Like the Sun, these hot stars will have a turbulent spicules region.  Crude estimates for Spica lead to an intrinsic redshift of 5 to 10 ${\rm{km}}\,{\rm{s}}^{ - 1} {\rm{.}}~$  The plasma redshift theory predicts thus large redshifts especially in the large and hot O and B stars.  These predictions are consistent with the observed K-effect, which was discovered long ago, but could not be explained; see Arp [52].  

\textbf{Collapsars.}  The observed redshifts of dwarf stars, such as the white dwarf (WD) Sirius B, can be explained as plasma redshifts without the conventionally assumed gravitational redshifts, which will have been cancelled by corresponding blue shifts, as the photons move out of the intense gravitational fields; see section 5.6.2, and an analysis of the theory and the relevant experiments by Brynjolfsson [53].  

\indent  The plasma redshifts in collapsars are usually mainly due to the second term in Eq.\,(20),
\[
{\rm{ln}}\, ( 1 + z ) =  3.326 \cdot 10^{ - 25} \int\limits_0^R {N_e dx} \; + \; \frac{{\gamma _i  - \gamma _0 }}{{\xi \omega }} =  3.326 \cdot 10^{ - 25} \int\limits_0^R {N_e dx} \; + \; \frac{{\delta \lambda_i  - \delta \lambda_0 }}{{\xi \lambda}}{\rm{,}} \quad \quad \quad {\rm{(20)}}
\]
\noindent  where the photon widths, $\gamma_i {\rm{}}$ and $\delta \lambda_i {\rm{,}}$ for a constant temperature are Stark broadened increasingly with the pressure.  The classical photon widths, which are equal to the quantum mechanical widths of photons penetrating a plasma are given by $\gamma_0 = \beta_0 \omega^2=6.266\cdot 10^{-24}\cdot \omega^2~{\rm{s}}^{-1} {\rm{,}}$ and $\delta \lambda_0 = 0.118~ {\rm{m\AA ,}} $ These classical photon widths are independent of the pressure.  The factor $\xi $ for the heavily broadened lines is likely to be between 0.25 to 0.36.  (In the heavily broadened lines, it is likely to approach $e^{-1}= 0.36 {\rm{.)}}~$  The Stark broadening of the photons is often about 10 times the Stark shift.  The plasma redshift in collapsars is on the order of 27 to 40 times the Stark shift, which often corresponds to about 0.2 to 4 ${\rm{km\,s}}^{-1}{\rm{.}}~ $  The second term in Eq.\,(20), which increases with the pressure and the gravitational potential, corresponds then often to a redshift of about 10 to 70 $ {\rm{km\,s}}^{-1}$ in dwarf stars.  Therefore, it is often difficult to distinguish plasma redshift from the conventionally assumed gravitational redshift.

\indent  The second term on the right side of Eq.\,(20) requires only a small column density of the electrons, $N_e \approx 10^{18}~{\rm{cm}^{-2}} {\rm{.}}~$  Therefore, even when the collapsar is relatively cold and has a corona with a column density, $N_e < 10^{18}~{\rm{cm}^{-2}} {\rm{,}}$ the second term is valid, because the column density in interstellar space would usually be adequate to produce the second term in Eq.\,(20).  The photon widths and the pressure broadenings of the photons vary from line to line.  Therefore the plasma redshift varies from line to line, while the conventionally expected gravitational redshift should be relatively constant independent of the line.  In collapsars, the variation of the redshift from line to line helps distinguish the plasma redshift from the conventionally expected gravitational redshift, which should be constant independent of the line observed.  In the first approximation, we do not need to know the star's velocity.  The question is only: ``Does the redshift vary from line to line?''  Also, the wings of the lines are formed at slightly higher densities in the collapsar, such as the white dwarfs. The plasma redshift may then cause the wings to be redshifted slightly more than the center portion of the line.  This is only a small effect.

\indent  It has usually been difficult to determine the redshifts of collapsars.  In 1925, Adams [54] determined the gravitational redshift of Sirius B to be about $21 ~{\rm{km \, s}^{-1} } {\rm{.}}~$  He observed significant variation in the redshift from line to line, but took an average.  In 1926, Str\"{o}mberg [55] corrected the redshift to be about $19 ~{\rm {km \, s}^{-1} } {\rm{.}}~$  This value matched the previous predictions by Eddington [56].  For 46 years, these estimates were considered a proof of Einstein's gravitational redshift.  In 1971, Greenstein et al.~[57]  estimated the redshift to be about $89\pm 16~{\rm{km \, s}^{-1} } {\rm{.}}~$  Again it was found to match the theoretical expectations.  Greenstein et al. measured the redshift of the ${\rm{H_{\alpha}}} $-line and found a weighted average of $94.6~{\rm{km \, s}^{-1} } {\rm{,}}$ and of the ${\rm{H_{\beta}}} $-line to be $73.5~{\rm{km \, s}^{-1} } {\rm{.}}~$  In accordance with the conventional gravitational theory, Greenstein et al. took a weighted average to derive the value $89\pm 16~{\rm{km \, s}^{-1} }{\rm{-value.}}~$  The difference of $21.1 = 94.6 - 73.5 ~{\rm{km \, s}^{-1} } {\rm{}}$ between the redshift of the ${\rm{H_{\alpha}}}$ and the ${\rm{H_{\beta}}} $-line does not prove that the difference is real, but it could be.  The photon width and Stark broadening of the ${\rm{H_{\alpha}}} $-line is also greater than of that of the ${\rm{H_{\beta}}} $-line.  The observed difference is consistent with the plasma redshift theory.  In 2005, Barstow et al.~[58] measured only the ${\rm{H_{\alpha}}} $-line and found therefore a small spread and a redshift of about $80.42\pm 4.83~{\rm{km \, s}^{-1} } {\rm{,}}$ and the temperature about $25,193\pm 37~{\rm{K} } {\rm{.}}$  We don't know what it would be had he instead observed the ${\rm{H_{\beta}}} $-line. 

\indent  The intrinsic redshift of a star depends slightly on the direction of the star's magnetic dipole axis.  The observed redshift could then vary slightly with time, provided the dipole axis wobbles, or if the magnetic field varies, confer the 11 years cycle in the Sun.

\indent  Rather than proving the gravitational redshifts, the explanations and the analyses of the observations, usually assume the validity of the gravitational redshift.  This is often stated explicitly in the analyses of the observations.  When the gravitational redshift does not fit the expectation, some thing else is assumed wrong.  In case of Sirius B, the interference from Sirius A is a real problem.  If the gravitational redshift was a sure thing, the conventional analysis of the data is justified.  But if we question its validity, we must analyze the subject more carefully.

\textbf{Quasars.}  It is reasonable to assume that quasars and other active galactic nuclei have large intrinsic redshifts caused by the plasma redshift (see Brynjolfsson [59]).  In quasars the measured redshift is useful for characterizing their corona.  For the quasar 3C 273 with redshift of 0.158, the electron density integral is $\, \left( \ln {1.158} \right)/\left( 3.326 \cdot 10^{ - 25} \right) = 4.41 \cdot 10^{23} \; {\rm{ cm}}^{ - 2} .\,$  This redshift integral, when combined with the other features, such as the bolometric temperature(s), including X rays and infrared spectrum, can help us to obtain a rough estimates of their absolute magnitude, distance, and even mass.  The estimate of a rather simple quasar such as 3C 273 is rather complicated, as it will be necessary to take into account different parts of the spectrum including the X rays, the ultraviolet component, the visual spectrum, the infrared, spectrum, the production and cooling by microwaves, production and cooling of dust emitting infrared radiation, the optical density of the different radiations, the magnetic field, etc.  In addition to the usual line broadenings, we may also have that the center-limb effect broadens the observed lines significantly.  The plasma redshift is a first order process in the density, while the cooling effects are usually second order in density.  This has a tendency to produce cloud formations in the corona, as in the case of the galactic corona.  These clouds may result in absorptions lines with smaller redshifts than the emission lines.  A crude estimate, based on the assumption that the magnetic field in the corona is small, places the quasar 3C 273 at a distance on the order of 8 Mpc, while the conventional theory derives a distance of about 680 Mpc, if the Hubble constant is about $H_0 = 65~{\rm{km\,s}}^{-1}{\rm{Mpc}}^{-1}{\rm{.}}~$  The blobs, which move away from the quasar at angular speed of 0.65 mas per year, would not move at a superluminal speed, but at about $3\cdot 10^4 \; {\rm{km}}\,{\rm{s}}^{ - 1}.~$  This rough estimate indicates that the absolute magnitude is about $-17.6.$


\subsection{Galactic corona}
Lyman Spitzer in the 1950's [60] made it plausible that the Milky Way galaxy has a corona.  Since then many observations have confirmed this fact.  The dispersions in the frequencies of microwaves along the lines of sight from pulsars indicate that we not only have a hot corona around the Milky Way, but that we also have hot interstellar H\,II regions within the Milky Way.  These H\,II regions stretch far beyond the Str\"{o}mgren radii of nearby stars.  It has been difficult to find sources for the required heating of the corona, and the heating and the large dimensions of the H\,II regions [60-67].  The plasma-redshift heating helps explain the observations.  X rays and ultraviolet radiation can heat some regions to intermediate temperatures, but plasma redshift can increase the temperature to more than a million degrees.  These high temperatures reduce the density and the rate of cooling.  The fully ionized plasma regions can therefore reach far away from the hot stars.  The H\,II regions around stars can retain their high temperature for a long time because of the continuous plasma-redshift heating.

\indent These hot H\,II regions have often been explained 'ad hoc' as due to past supernovas even when no remnant of a supernova was found, because it was thought that only supernovas could explain the high temperatures.  These supernova bubbles were assumed to move out into the corona and supply the needed heating of the corona of our Galaxy [60, 66, 67].  The transition zones to the H\,II regions and the galactic corona have absorption and emission lines similar to those in the transition to the solar corona.  Beyond the transition zone, the hot plasma is usually more difficult to detect; but some lines indicating a million degree temperatures are sometimes observed.


\subsubsection{Heating of the corona}

We will see that X rays from intergalactic plasma combined with plasma redshift heating by light sources within Milky Way are the main sources for heating of the corona.  The intergalactic plasma is heated mainly by plasma redshift of intergalactic light from the galaxies; see section 5.8. 

\indent  Analyzing the spectrum from the supernova 1987A in the Large Magellanic Cloud (LMC), Pettini et al.~[68] found that the $6374.51~ {\rm{\AA}}$-line in Fe\,X caused a pronounced trough in the spectrum.  This absorption line is produced preferentially when the electron temperature is about 1.25 million K.  If formed at one location, the trough would be relatively narrow.  The full width at half maximum (FWHM) would be about $32 \; {\rm{km} }\, {\rm{s} }^{ - 1}$, provided the turbulence is small.  The trough stretched from about 210 to about $365~ {\rm{km} }\,{\rm{s} }^{ - 1}$ for a total width of about ${155~ \rm{km} }\, {\rm{s} }^{ - 1}$ [68] (see section IIIa and Figs. 1 and 2 of that source).

\indent The distance of supernova 1987A in the LMC is about 51 kpc, the galactic longitude about 280 and the galactic latitude about $-32$ degrees.  Its movement along the line of sight is uncertain, as its redshift is a combination of a plasma redshift and a Doppler shift.  In the past, we interpreted all shifts as Doppler shifts; and Pettini et al.~interpreted the width as due to Doppler shifts in the intervening plasma [68].  In light of the plasma-redshift theory, we realize that most shifts are combinations of the two.  We will assume for a moment that the displacement of the beginning of the Fe X absorption, at about $210~{\rm{km\,s}^{-1}} ,$ is mostly due to a Doppler shift caused by the velocity of the solar system relative to the point where the line of sight to the LMC cuts the surface of our Milky Way.  Some of the redshifts within our Galaxy will be due to the plasma redshifts in the H\,II regions.  Based on measurements within our Galaxy, the average electron density is about 0.016 [61] to about 0.025 ${\rm{cm}}^{ - 3}$ [63].  For distances of about 2 to 3 kpc, the integrated plasma redshift within our Galaxy is likely to be about 10 to $23~ {\rm{km} }\, {\rm{s}}^{ - 1}.~$  The electron temperature of the 
galactic corona may deviate from the optimum temperature of 1.25 million K for the Fe\,X line, and would then contribute less to the absorption trough.

\indent We will assume for a moment that the width of $155~ {\rm{km} }\, {\rm{s} }^{ - 1}$ (from 210 to 365) of the trough as observed by Pettini et al.~[68] is a rough measure of the plasma redshift along the line of sight in the corona, where the absorption by the $6374.51~ {\rm{\AA} }$-line in Fe\,X is significant.  A white light from a supernova could produce such a trough.  The light absorbed close to the supernova would have the largest plasma redshift.  The light absorbed in the transition zone to the corona would have the smaller redshift, and the lines formed in the H\,II regions would have redshifts smaller than any part of the trough.

\indent When using Eq.\,(20) and assuming that the second term on the right side is insignificant, we get that the average electron density along the coronal line segment $D$ is given by
\be
z = \frac{{155}}{{3 \cdot 10^5 }} = 3.326 \cdot 10^{ - 25} \left( {N_e } \right)_{av} D\, 3.086 \cdot 10^{21} , \quad {\rm{  or  }} \quad \left( {N_e } \right)_{av} =  \frac{{0.50}}{D}{\rm{.}}
\ee
If the distance $D$ of the line segment within the corona to the supernova 1987A in LMC is $(51-3)= 48$ kpc, we get that the average electron density is 
$\left( {N_e } \right)_{av}\approx 0.01~ {\rm{cm}}^{ - 3},$ and that the electron column density is 
$\left( {N_e } \right)_{av} D\, 3.086 \cdot 10^{21}= 1.6 \cdot 10^{21} \; {\rm{ cm}}^{ - 2} {\rm{.}}~$  From Figs.~1 and 2 in reference [68], it appears possible that the absorption trough reaches to about 400 ${\rm{km}}\,{\rm{s}}^{ - 1}.~$  The electron's density and the column density would then be about $0.0128 \; {\rm{ cm}}^{ - 3}$ and  $2 \cdot 10^{21} \; {\rm{ cm}}^{ - 2} ,$ respectively.

\indent  Pettini et al.~[68] believed, however, that absorption trough for the $\lambda = 6374.51~\rm{\AA}$-line stretched only over the range $v_H \approx 215-270~\rm{km\,s}^{-1} ,$  and that the remaining parts of the absorption trough were due to contamination by telluric lines and diffuse interstellar bands.  Pettini et al.~did not know the plasma redshift.  Their assumptions are therefore reasonable.  As explained in the following, the high concentration of Fe\,X is most likely due to ionization by X rays from about $3\cdot 10^6~\rm{K}$ intergalactic plasma, with an average X-ray energy of about 260 eV.  These X rays will deposit their energy in the colder plasma low in the corona and in the transition zone to the corona.  The ionization potential of Fe\,IX is 235~eV, and of Fe\,X it is 2.62 eV.  The X rays from the intergalactic plasma therefore have about the right energy to create a relatively high concentration of the Fe\,X-ions in the colder plasma, even in plasma with temperatures well below $1.25\cdot 10^6~\rm{K}.$   

\indent  In support for the contamination of the line, Pettini et al.~point out the difference in the form of the absorption trough in the spectra taken April 9 1987 and July 20 1987 (see Fig.~2 of that source).  However, it is more likely that during the 102 days, a colder plasma cloud formed in (or moved into) the line of sight to the supernova thereby enhancing the absorption trough closer to the supernova, as observations on July 20 indicate.  Plasma redshift, because it is a first order process, has the tendency to create hot bubbles separated by walls of colder plasma.  These walls of colder plasma form and disappear in the line of sight.  The walls could also move into or out of the line of sight.  Other lines formed in these walls would simultaneously show an increased and decreased strength.  Plasma redshift therefore gives a reasonable explanation of the observations.  

\indent Pettini et al.~[68] assumed the iron concentration to be similar to that in the Sun.  They assumed collisional ionization based on thermal equilibrium in the plasma.  From the depth and area of the absorption trough, they determined the column density and found a value that corresponds to about $N_{\rm{H}}\approx 3.2\cdot 10^{21}~{\rm{cm}^{-2}}.{~}$  Pettini et al.~ considered several possibilities for explaining this unexpected high value, but concluded that the above-mentioned column density was most likely a reasonable estimate for the corona between our Milky Way and the LMC [68].  However, their assumptions may not be right.  The iron concentration may deviate from that in the Sun.  The oscillator strength assumed for the line may be inaccurate.  For example, the line may be broadened by Stark effect in the hot plasma.  A high concentration of Fe\,X may be produced by the high-energy X rays from about $3\cdot 10^6~\rm{K}$ hot plasma in intergalactic space (see section 5.11).  As mentioned above, the high-energy X rays from this hot plasma can penetrate, without much attenuation, the fully ionized hydrogen component of the plasma.  They are very efficient in ionizing the iron (and other trace elements) in the coronal plasma.  At temperatures well below $1.25\cdot 10^6~\rm{K} ,$ this X-ray ionization will make the Fe\,X concentration appear much higher than if the excitations and ionization were made primarily by the $1.25\cdot 10^6~\rm{K} $ electron plasma.

\indent  If the line is formed within a narrow region with a width of about $55~ \rm{km\,s}^{-1} ,$ as assumed by Pettini et al., and if the column density is about $N_{\rm{H}}\approx 3.2\cdot 10^{21}~{\rm{cm}^{-2}}$, as they derived, then the pressure would be unreasonably high.  For example, if this column density were the integral of the density over 10 kpc or less, the hydrogen density would be 
$N_H \geq 0.104~{\rm{cm}^{-3}},$ which when combined with a temperature of $1.25\cdot 10^6~\rm{K} $ corresponds to an unreasonably high pressure at a distance greater than 8 kpc from the galactic center.  Even if the hydrogen density is integrated over the entire distance of 50 kpc to the supernova, the average density of about $N_H =0.0207~{\rm{cm}^{-3}},$ would for $1.25\cdot 10^6~\rm{K} $ correspond to $N_e T = 1.2 N_H T= 3.1\cdot 10^4~{\rm{cm}^{-3}}\,{\rm{K}},$ which is also too high.

\indent  The ionization and heating in the transition zone to the corona would be by the X rays from intergalactic space, by the heat conduction from the corona, and by the plasma redshift of the galactic light.  Similarly, the heating of the corona would be by the plasma redshift of the galactic light, by the X rays from intergalactic space, and at the outer edges by the heat conduction from intergalactic plasma. 

\indent Peebles [69] (see that source, pages 45-49), using the data of van Albada and Sancisi [70] makes it plausible that the corona of our Galaxy contains significant mass.  His evaluation indicates that the coronas of some galaxies contain a halo with a ``dark mass'' density, which beyond about 5 kpc decreases approximately proportionally to $R^{ - 2} {\rm{,}}$ where $R$ is the distance to the center of the galaxy.  For explaining the Tully-Fisher relation for large velocities of distant hydrogen clouds Milgrom [71] suggested modifying Newton's gravitational laws.  The evaluation by Bottema et al.~[72] was consistent with this suggestion.

\indent According to Eq.\,(39), the theory of plasma redshift leads to an average plasma density along the line of sight to supernova 1987A, of $\left( {N_e } \right)_{av}\approx 0.01~ {\rm{cm}}^{ - 3}$ over a distance of about 48 kpc in the corona of the Milky Way.  Although this density is about 20 times higher than that usually assumed, it is most likely about right.  This average includes the relatively dense plasma in the transition zone.  In addition to the relatively high plasma densities made possible by the X-ray ionization and the plasma-redshift heating, we may have significant ``cloud'' formations caused by the fact that the plasma-redshift heating is a first order in density while the cooling is of higher order in density.  The velocities of the ions and electrons close to surface of a galaxy, where the density is relatively large, are likely to be isotropic relative to the local rotational velocity of the plasma.  When the plasma diffuses outwards, its velocity perpendicular to the radius to the center of the Milky Way will remain constant.  If that plasma condenses and forms hydrogen clouds at any distance from the galaxy, the velocity of the hydrogen atoms perpendicular to the radius of the galaxy will remain constant and will be roughly equal to the corresponding velocity of the plasma close to the galaxy.  This can explain, without the ``dark matter'', the observed constancy of the velocity of the atomic hydrogen perpendicular to the radius as measured by the 21 cm line.  The mass of the ``clouds'' and stars in the corona can be significant.  There may therefore not be any need for ``exotic dark matter particles'' [69] (see p.48 of that source) or for changing the gravitational laws.  The plasma redshifts produced by the relatively dense plasma in clusters of galaxies together with intrinsic plasma redshifts of many objects will, when interpreted as Doppler shifts, result in unreasonably large velocities for some of the members of the cluster.  Also this misinterpretation of the observations has lead to the ``dark matter'' hypothesis.  The plasma-redshift theory can explain the velocity distribution in clusters without the ``dark matter'' hypothesis.

\indent  We will assume that the light intensity at the surface of the Galaxy corresponds to about $n_G 10^{11}$ solar like stars at its center.  The factor $n_G$ is most likely between about 1 and 3.  The mass of the Milky Way is often estimated to be about $2\cdot 10^{11}$ solar masses, but the light intensity may be less than that of $2\cdot 10^{11}$ suns.  For a Hubble constant $\rm{H_0} = 0.7,$ Peebles [69] (see Eq.\,(3.40) of that source) estimates that inside the distance of the Sun from the 
galactic center the mass is $1.6\cdot 10^{11}$ solar masses.   The light intensity varies with the direction to the center.  For example, it may be greater over the poles (the bulge) of the Milky Way, and it may vary along the surface; for example, it may be larger over the arms.  The heating-rate caused by the plasma redshift of the galactic light (GL) is similar to Eq.\,(33)
\be
\frac{{dQ_{\rm{GL}}}}{{dt}} = 2.15 \cdot 10^{ - 14} n_G 10^{11} N_e \frac{{R_0^2 }}{{R^2 }} = n_G 1.09 \cdot 10^{ - 24} \frac{{N_e }}{{R^2 }}\; \;\, {\rm{  erg}}\,{\rm{cm}}^{ - 3} {\kern 2pt} {\rm{s}}^{ - 1} {\rm{,}}
\ee
where $N_e$ is the electron density in ${\rm{cm}}^{ - 3}$, $R$ the radius in kpc to the galactic center, and  $R_{0}$  the solar radius in kpc.

\indent  We can compare the plasma-redshift heating of the Galactic light given by Eq.\,(40) with the free-free emission cooling.  The cooling by trace elements may be smaller, partly because of reduced concentration of the trace elements and partly because reabsorption of some of the recombination emission stretching over many kpc in the corona.  The energy emitted in the free-free emission cooling is
\be
\frac{{dQ_{ff} }}{{dt}} = 1.426 \cdot 10^{ - 27} \sum {Z_i^2 N_t N_e T^{{1 \mathord{\left/{\vphantom {1 2}} \right.
 \kern-\nulldelimiterspace} 2}} } \left\langle {g_{ff} } \right\rangle ~~ {\rm{ erg}}\,{\rm{cm}}^{ - 3} \,{\rm{s}}^{ - 1} . 
\ee
For solar abundance of trace elements the Gaunt factor is  $\left\langle {g_{ff} } \right\rangle = 1.29$ for a temperature of 1.1 million K (but for insignificant concentration of trace elements the Gaunt factor is closer to 1.4).  This continuum X-ray cooling is then about
\be
\frac{{dQ_{ff} }}{{dt}} = 2.57\cdot 10^{ - 24} N_e^2 ~~ {\rm{ erg}}\,{\rm{cm}}^{ - 3} \,{\rm{s}}^{ - 1} . 
\ee
\noindent  where in Eq.\,(41) the value of $\Sigma Z_i^2N_i \approx 1.33 N_e {\rm{.}}~$  (This cooling does not include the recombination cooling, which is especially important at low temperatures.)  When equating the cooling given by Eq.\,(42) with the heating given by Eqs.\,(40), we get
\be
N_e \approx \frac {{n_G \cdot 0.425}}{{R^2}}\quad ~~ {\rm{cm}^{-3}}. 
\ee
\noindent  For $n_G = 1$ and for R equal to 8, 16, and 40 kpc from the galactic center, we get that the values of $N_e$ are equal to 0.0066, 0.0017 and $0.0003~{\rm{cm}}^{ - 3}$, respectively.  For $n_G = 2$ the values of $N_e $ are 0.0133, 0.0033, and $ 0.0005~{\rm{cm}}^{ - 3},$ respectively.  Some of the free-free emission is reabsorbed, which increases the $N_e$-values slightly.  These densities of $N_e$ are too low for explaining the observations by Pettini et al. [68].  We conclude therefore that plasma redshift is not adequate for heating the relatively dense corona of the Milky Way.

\indent   The density, $ 0.0005~{\rm{cm}}^{ - 3},$  at about 1.1 million K at distance of 40 kpc would result in significant flow of plasma into intergalactic space.  It appears thus that the density in intergalactic space could be significant as the cosmological redshift in sections 5.8 and 5.9 shows.  The CMB and the X rays in section 5.10 and 5.11 confirm those densities in intergalactic space.   We must then also conclude that the light escaping from the galaxies will be redshifted and result in heating of the intergalactic plasma, which then produces X rays that return to the energy emitted from the galaxies back to the coronas of galaxies.    The X rays from intergalactic space will ionize the trace elements, such as iron.  The colder regions, especially in the transition region to the corona will increase the average density significantly. 

\indent   In intergalactic space, the average heating rate caused by the plasma redshift of the cosmic microwave background (CMB) radiation will balance the corresponding cooling of the intergalactic plasma caused by emitting the CMB at the original frequency instead of the redshifted frequency.  However, because of the higher plasma density (the plasma-redshift heating is proportional to $N_e$) in the corona of the Milky Way galaxy, the heating will be greater than in intergalactic space.  This heating is partially offset by greater cooling because of higher temperature of the emitted microwaves from the coronal plasma, because the emitted energy is proportional to $3NkT_e {\rm{,}}$ where $N$ is the total number of particles and $T_e$ is the average particle temperature; see Eq.\,(C20) in Appendix C.  In spite of the lower temperature in the corona than in intergalactic space, we have a net heating.   As we show in section 5.10 (see the numerator in Eq.\,(62)), the energy density in the CMB is about $u_{\rm{CMB}}= a\cdot T_{CMB}^4 = 4.19\cdot 10^{-13}{~\;\rm{erg\,cm}^{-3}} .~$  The average energy flux at each location integrated over all directions is then about $c\cdot u_{\rm{CMB}} = 1.26\cdot 10^{-2}{~\;\rm{erg\,cm}^{-2}\,s^{-1}} .~$  For this rough estimate, we can assume that CMB radiations is approximately constant and isotropic.  From Eq.\,(28), we find that at the high temperatures (above about 300,000 K) and low densities in the corona all the frequencies are plasma redshifted.  We get then that 
\be
\frac{{dQ_{\rm{CMB}}}}{{dt}} =c\cdot u_{\rm{CMB}}\cdot 3.326\cdot 10^{-25}\, N_e = 4.18\cdot 10^{-27}\, N_e ~ \; {\rm{  erg}}\,{\rm{cm}}^{ - 3} {\kern 2pt} {\rm{s}}^{ - 1} {\rm{.}} 
\ee
\noindent  The actual CMB heating of the corona is about 33\,\% less because of the microwave emission-cooling, or 
\be
\frac{{dQ_{\rm{CMB}}}}{{dt}} =c\cdot u_{\rm{CMB}}\cdot 3.326\cdot 10^{-25}\,\frac{{T_{int}-T_{gal}}}{{T_{int}}} N_e = 2.8 \cdot 10^{-27}\, N_e ~ \; {\rm{  erg}}\,{\rm{cm}}^{ - 3} {\kern 2pt} {\rm{s}}^{ - 1} {\rm{.}} 
\ee
\noindent  From Eq.\,(40) and (45), we see that the heating by the CMB equals the plasma-redshift heating from the galactic light when $R\approx 20 {\sqrt{n_G}}~\;{\rm{kpc}.}~$  For smaller distances the light from the Milky Way galaxy dominates heating by CMB, and for larger distances the CMB dominates the galactic heating until it becomes about equal to CMB cooling in intergalactic space.  We must then Sum the heating given by Eq.\,(40) and Eq.\,(45) and equate this sum with the free emission cooling given by Eq.\,(42).  We get
\be
N_e \approx \frac {{n_G \cdot 0.425}}{{R^2}}\quad + \quad 1.1\cdot 10^{-3} ~~ {\rm{cm}^{-3}}. 
\ee
\noindent  For $n_G = 1$ and for R equal to 8, 16, and 40 kpc from the galactic center, we get that the values of $N_e$ are equal to 0.0077, 0.0027 and $0.0013~{\rm{cm}}^{ - 3}$, respectively.  For $n_G = 2$ the values of $N_e$ are 0.0144, 0.0044, and $ 0.0016~{\rm{cm}}^{ - 3},$ respectively.  Some of the free-free emission is reabsorbed, which increases the $N_e$-values slightly. 

\indent  More important is, however, that the cooling given by Eq.\,(41) is an underestimate.  It does not include the recombination cooling.  According to Sutherland and Dopita [11] the local continuum X-ray cooling in a plasma with solar abundance at $T_e = 1.1 \cdot 10^6 ~ \rm{K}$ is only about 3\,\% of the total local X-ray cooling in plasma with solar abundance of trace elements. 

\indent   We see thus that when we take the heating by the CMB and the cooling by trace elements into account, the densities of $N_e$ and the corresponding trace elements, including iron, are too low for explaining the observations by Pettini et al. [68].  We conclude therefore that plasma-redshift heating of the corona is inadequate for producing the required Fe\,X concentration.  {\it{The X rays from intergalactic space appear essential for explaining the high Fe\,X concentration observed by Pettini et al. in the corona.}} See also section 5.11 and in particular the penultimate row in Table C1 of Appendix C.


\subsubsection{Effect of magnetic field}
We have disregarded the repulsion of the diamagnetic moments in a divergent magnetic field; see section B2 in Appendix B.  The repulsion requires that the energy density, of the magnetic field be greater than the kinetic energy density of the particles, which corresponds to 
$H\approx 10^{ -5 }~{\rm{gauss}}.~$  According to Clarke et al.~[73], the magnetic field in a few extended regions (10 kpc) of the intracluster plasma is about $5 \cdot 10^{ - 6}$ to $10^{ - 5}$ gauss.  The corresponding energy density in these few clouds with detectable magnetic fields is then comparable, as is to be expected (see section B1 of Appendix B), to the kinetic energy density.  It is therefore not clear if the diamagnetic repulsion is significant.

\indent At places where the magnetic field is relatively strong, most likely close to the surface of the galaxy, the field would facilitate the outward streaming of the plasma.  It could then deposit significant energy into the plasma, and the densities would increase beyond the above estimates.  The magnetic repulsion could possibly facilitate outflow of hydrogen from the Galaxy, which could contribute to a high density in the galactic corona, which in turn may lead to subsequent precipitation of hydrogen into clouds.  However, clear evidence for such effects from magnetic fields is lacking.


\subsubsection{Cloud formation in the galactic corona}
The plasma-redshift heating is first order in density while the cooling is second order in density.  This disparity produces instabilities.  The hot low-density regions become hotter and the cold high-density regions colder.  This fact has some similarities to stability of loops and arches in the solar corona mentioned at the end of section 5.5.  The phenomena are also related to the formation of the spicules in the transition to solar corona and to the formation of flares.  The imbalances caused by the plasma redshift and the cooling could also explain the relative stability of hydrogen clouds; that is, the fact that the plasma redshift in the hot regions has a tendency to push the hot plasma into the colder regions, such as arches in the solar corona or clouds in the galactic corona.

\indent Within the corona of the Milky Way Galaxy, we have many clouds, including the clouds in the Leading Arm of the LMC, in the bridge between the Large and Small Magellanic Clouds, and in the Magellanic Stream, which trails the Clouds in our Milky Way corona [74, 75].  The pressure in the hotter and sparser plasma outside a cloud could be higher than the pressure inside it.  The pressure in the hotter plasma outside a cloud could then push the hot plasma into the cloud and cause it to grow, or the outside pressure could more or less balance the pressure in the cloud and cause the cloud to be relatively stable.

\indent When the temperature in the cloud's surface decreases, the emission cooling increases and the redshift heating decreases.  This instability may cause the plasma to cool to a very low temperature.  The rate of cooling has several relative minima, which depend on the concentration of the trace elements.  The cooling rate usually has a relative minimum around 20,000 K.  A relatively large fraction of the cloud's surface may then have a temperature around 20,000 K.  The density is then about 50 times greater than that at 1 million K.  Other minima are lower, and the center of the cloud could be much colder and density higher, while the surface of the cloud will be warmer due to collisions and X rays.  The mass contained in a cloud may be significant.  Putman [75] estimates that the Magellanic Bridge between the two Magellanic Clouds may contain about 100 million solar masses; and the Leading Arm may contain 10 million solar masses [74].

\indent If the plasma contains magnetic field, the field lines will concentrate in the cloud as the hot 
plasma outside the cloud pushes the plasma into the cloud, similar to that in the relatively stable arches in the Sun's corona.  The cloud may then often have a tendency to form structures or filaments that are elongated like the clouds in the Leading Arm and in the Magellanic Stream [74, 75].  Kaz\`{e}s et al.~[76] searched for the Zeeman effect in clouds and detected only one with about $11.4 \cdot 10^{ - 6}$ gauss, which indicates that the magnetic field energy density is usually small or only about equal to the kinetic energy density in the surrounding hot plasma.

\indent Some clouds could form far away from the Galaxy, even in intergalactic space.  It has been difficult to explain the cloud formation and the structures observed, but the plasma redshift with its tendency to create instabilities, hot ``bubbles'' between colder regions, gives a natural explanation of the observed phenomena.


\subsubsection{Positrons}
Positrons will increase the redshift heating without contributing much to the X-ray cooling.  Positrons are formed by gamma rays with energy in excess of 1.02 MeV.  Many processes can form positrons, such as decay of nuclei and of pions.  Aharonian and Atoyan [77] make a case for the importance of high-energy inelastic proton interaction for explaining the gamma-ray emission in the range of $10^8$ to $10^{11}$ eV in the galactic disk.  These interactions must also apply to intergalactic space.  The high-energy protons may result in formation of pions and electron-positron pairs in intergalactic space.  Such processes will contribute to the heating of intergalactic plasma.  In the intergalactic sparse plasma, the hot positrons would have a long lifetime [78]; and according to Gould [79], the X-ray intensity produced by a 5 million K electron-positron plasma is reduced by factor of $v^2 /c^2  \approx 10^{ - 3} $ when compared to the X-ray intensity in an electron-proton plasma at the same temperature.

\indent The large intensity of positron annihilation radiation in the galactic center was clearly demonstrated by Purcell et al.~[80] and Kinzer et al.~[81].  Dermer and Skibo [82] have found that there appears to be a ``fountain'' of positrons streaming into the galactic bulge of the Milky Way.  They suggest that the positrons are injected within about 100-200 pc of the galactic center and are carried up by the hot gas to annihilate in the polar regions of the galaxy at heights mostly within a few kpc.  Other galaxies and quasars are likely to be rich sources of electron-positron pairs.  The background X-ray spectrum of Trombka et al.~[83] shows clearly a hump corresponding to the annihilation photons in the X-ray background.  However, it appears that most of the annihilation radiation is due to point sources, and is not from a dispersed intergalactic plasma.

\indent It appears possible that the positrons with the electrons will diffuse into intergalactic space far away from the galaxies, due to their low gravitational potential relative to the protons.  The positron-electron plasma might then fill most of the intergalactic space with the proton-electron plasma confined mainly to the corona of galaxies.  The positrons would produce a plasma redshift similar to that of the electrons.  The X-ray cooling would be much smaller, but the annihilation radiation should be pronounced.  If we are to be able to observe the annihilation radiation, the radiation must be able to penetrate the relatively dense plasma around the galaxies.  It should be detectable although the intensity is weakened by the absorption and scattering.  Annihilation radiation from intergalactic plasma has not been firmly established, as most of the observed annihilation radiation has been traced to point sources.  It is unlikely, therefore, that even far away from the galaxies a positron-electron plasma replaces the proton-electron plasma.  Future research should, nevertheless, consider this remote possibility.


\subsubsection{Observations of X rays from galactic corona}
The Draco cloud is located at $\left( {{\rm{l,}}\,{\rm{b}}} \right) \approx \left( {90^\circ ,\; + 39^\circ } \right)$ and at a distance in excess of 300 pc.  Borrows and Mendenhall [84] observed that the Draco cloud decreased the X-ray intensity from the corona behind it.  The shadow covered X rays in the range below about 0.4 keV.  {\it{This shadow appears to be a footprint of the 1 to 4 million degree hot coronal and intergalactic plasma.}}  Both Herbstmeier et al.~[85], and Wakker and Van Woerden [86] give examples where the intermediate and high-velocity clouds throw a shadow in the 250 eV region.  The shadows indicate that also beyond the high-velocity clouds the plasma temperature may reach about 1 to 4 million degrees.  This is just another indication of the intergalactic plasma.  The intergalactic X-ray intensity will be discussed in section 5.11 and Appendix C.


\subsubsection{The redshifts within the clusters of galaxies}
Clusters of galaxies are often shown as having an elongated distribution in the radial direction (z-direction); for example, Virgo cluster, and Coma cluster (``Finger of God'' pointing at us) are elongated significantly in the radial direction (see Fig.~3.6 in reference [69]).  Now we are inclined to interpret this elongation in the radial direction as caused by plasma redshift.  As the line of sight penetrates the relatively dense plasma between galaxies in the cluster, the redshift increases and causes the objects at the back of the cluster to have relatively large plasma redshifts.  This explanation applies also to the anomalous redshift of the Centaurus cluster (see Fig.~5.4 of reference [69]).  The observed redshifts are significantly larger than the redshifts expected from the Tully-Fisher relation for determining the distances.  Sometimes, the distances can be determined independent of the redshift. The excessive redshifts are usually assumed to be due to Doppler shifts.  This misconception may then lead to excessive intracluster velocities, which then lead to an assumption of a ``dark matter'' for explaining the dynamics.  The plasma redshift may thus resolve: 1) the ``dark matter'' 
problem, 2) the problem of preferential elongation of galaxy clusters along the line of sight, and 3) the anomalous redshift of some members in the Centaurus cluster.

\indent In the transition zone to the corona, and at the surface of the clouds formed in the corona, the temperatures may be too low and densities too high for the cut-off condition given by Eq.\,(28) to be fulfilled for the 21 cm H I line.  The redshift of the 21 cm line may then be slightly smaller (and the blue shift larger) than the redshift (blue shift) of lines in the visual spectrum.  This tendency, although a small effect, is seen in Table 3 of reference [86].


\subsubsection{Synopsis}
The simplified examples above serve only as a point of reference and for illustration.  The actual facts may deviate significantly from the assumed values.  For example, the light and X-ray intensities from the galactic center may not be isotropic.  Close to the axes and in the ``bulge'' of the Milky Way Galaxy, the light and X-ray intensities as well as the magnetic field may be relatively larger than at lower latitudes.  Even in case of such a modification, the plasma-redshift heating, together with X-ray heating and conduction heating from intergalactic plasma would predict a galactic corona with much greater densities than those usually assumed.

\indent The light intensity equal to $n_G 10^{11}$ Suns in Eq.\,(40) is arbitrary and serves only as a point of reference.  It appears, however, that the heating by plasma-redshift of galactic light and CMB, and the intense ionization by the X rays from the intergalactic plasma are needed for explaining the high densities deduced from the observations by Pettini et al.~[68].  The intergalactic plasma is heated by plasma redshift of intergalactic light.  This explains the cosmological redshift, the microwave background, and the X-ray background, as shown in sections 5.8 to 5.11.  

\indent The conventional explanation, which assumes that supernovae supply the heating to both the H II regions within the Milky Way and to its corona, is inadequate for explaining the observations.  The past estimates have usually assumed a value of $N_e T \approx 500~ \rm{cm}^{-3} \rm{K}$ instead of the above estimated value of $N_e T \approx 10,000~ \rm{cm}^{-3} \rm{K} $ at a distance of about 8 to 10 kpc from the galactic center.  The required heating by supernovae [60] was therefore only about 1/400 of the present estimates, which are based on the plasma-redshift heating.  The higher densities predicted by the plasma-redshift theory are essential for explaining the observations.


\subsection{Cosmological redshift}
For the cosmological redshifts, the first term on the right side in Eq.\,(20) is usually large compared with the second term on the right side.  In Eq.\,(18), therefore, we can usually in the extended plasmas of space set $F_1 \left( a \right) = 1 $, because the intergalactic plasma is usually very hot and the electron density very low.  We can also set $\gamma  = \gamma _0 $, because the redshifts are relatively large compared with the initial photon width.   From Eq.\,(18), we get then for large redshifts that
\be
\ln \left( {1 + z} \right) \approx 3.3262 \cdot 10^{ - 25} \int\limits_0^R
 {N_e dx}.
\ee
\noindent If the electron density is nearly constant, the right side is proportional to $R$, and we get that
\be
R = \frac{{\ln \left( {1 + z} \right)}}{{3.3262 \cdot 10^{ - 25} N_e }} = \frac{{3.0064 \cdot 10^{24} }}{{N_e }}\ln \left( {1 + z} \right)~~{\rm{cm }}{\rm{.}}
\ee
\noindent From Eq.\,(48), we derive the average electron density (or the sum of electron and positron density) in intergalactic space provided the plasma redshift explains the entire cosmological redshift.   When the integration distance $R = 3.085 \cdot 10^{24}\; {\rm{ cm}}{\rm{,}}$ or one Mpc, the redshift is $z = {{H_0 } \mathord{\left/{\vphantom {{H_0 } c}} \right.
 \kern-\nulldelimiterspace} c}{\rm{,}}$ where $H_0$ is the Hubble constant.  Press [87] has evaluated statistically the different estimates of $H_0$ and finds that it is most likely between 72 and 77 ${\rm{km}}\,{\rm{s}}^{ - 1}\,{\rm{Mpc}}^{-1}$.  Initially, we will therefore set the Hubble constant equal to the average 74.5 ${\rm{km}}\,{\rm{s}}^{ - 1}\,{\rm{Mpc}}^{-1}$.  We get then from Eq.\,(47) and (48) that
\[
\ln \left( {1 + z} \right) = \ln \left( {1 + \frac{{H_0 }}{{c }}} \right) \approx \frac{{H_0 }}{{c}} = 3.326 \cdot 10^{ - 25} \left( {N_e } \right)_{av} 3.085 \cdot 10^{24} {\rm{.}} 
\]
\noindent  The relation between plasma density and the Hubble constant is then
\be
\left( {N_e } \right)_{av}  =
\frac{{H_0}}{{3.076 \cdot 10^5 }} \approx \frac{{74.5}}{{3.076 \cdot 10^5 }} = 2.42 \cdot 10^{ - 4} \left( \frac{{H_0}}{{74.5}} \right)\,~ {\rm{ cm}}^{ - 3} {\rm{,}} 
\ee
\noindent  which shows that the average density of electrons is equal to $\left( {N_e } \right)_{av}  = 2.42 \cdot 10^{ - 4} \left( {{{H_0 } \mathord{\left/{\vphantom {{H_0 } {74.5}}} \right. \kern-\nulldelimiterspace} {74.5}}} \right)~{\rm{ cm}}^{ - 3} {\rm{.}}~$

\indent  From Eqs.\,(48) and (49), we derive that the distance is given by
\be
 R = \frac{{c }}{{H_0}}\, \ln \left( {1 + z} \right) ~~{\rm{Mpc ,}}
\ee
\noindent  where $c$ is the velocity of light in ${\rm{km\,s}}^{-1}{\rm{,}} $ and $ H_0 ~ {\rm{km \, s}}^{-1} {\rm{Mpc}}^{-1} {\rm{}}$ is the Hubble constant.  The ratio $c /H_0 $ is usually refered to as the Hubble length.

\indent  The relation between distance and redshift in big-bang cosmology is more complicated, because of the time dilation factor and the adjustment parameters for ``dark mass'' and ``dark energy''; see Brynjolfsson [59] (in particular, sections 3.1 and 3.2 and Fig.\,1 of that source).  The big-bang cosmologists usually assume that $\Omega_{\Lambda} = 1 - \Omega_M {\rm{.}}~$  They then adjust $\Omega_{\Lambda}$ to give the best agreement with their observations.  The optimum value of $\Omega_{\Lambda}$ may then vary with the redshift z.  In plasma redshift, there is only one distance-redshift relation, the one given by Eq.\,(50).

\indent The Compton scattering reduces the number of photons that reach the observer from a distant supernova.   The reduction in light intensity, $I$, is given by
\[
I = \frac{{I_0  \cdot \exp \left( { - a\left( R \right) - b N_e R - 2b N_e R} \right)}}{{R^2 }}{\rm{ ,}}
\]
where the factor $\exp \left( { - a\left( R \right)} \right)$ accounts for mostly Galactic and host galactic extinctions.  The factor $\exp \left( { - bN_e R} \right)$ accounts for reduction in bolometric light intensity caused by the plasma redshift.  The factor $\exp \left( { - 2bN_e R} \right)$ is due to Compton scattering.  The cross section $2b$ for the Compton scattering is twice the cross-section $b$ for the plasma redshift.  From Eqs.\,(47), we get that $b N_e R =\ln (1+z){\rm{.}}~$ We get then that
\be
I = \frac{{I_0  \cdot \exp \left( { - a\left( R \right)} \right)}}{{\left( {1 + z} \right)^3 R^2 }}{\rm{.}}
\ee
\vspace{2mm}

{\textbf{Raman scattering.}}  We have so far disregarded the Raman scattering on the plasma frequency.  The electron-plasma is in thermodynamic equilibrium, and the photons energy loss in Raman scattering should therefore usually equal the energy gain.  However, the Raman scattering causes small angular scattering, which for very distant small objects, like the distant supernovas, could be significant.  The effect of this small angular scattering depends on how the observed light intensity is integrated.

\indent For a single Raman scattering the deflection angle is about equal to $\alpha  = {{\omega _p } \mathord{\left/{\vphantom {{\omega _p } {\omega ,}}} \right.
 \kern-\nulldelimiterspace} {\omega ,}}$ where $\omega _p  = 5.64 \cdot 10^4 N_e^{{1 \mathord{\left/{\vphantom {1 2}} \right.
 \kern-\nulldelimiterspace} 2}}  \approx 878$ Hz is the plasma frequency in intergalactic space, and $\omega  \approx 2\pi  \cdot 10^{15}$ Hz for the observed 300 nm photons.  For the large redshifts of space, the photon width is equal to the classical photon width.  We get then from Eq.\,(7) that the cross section for Raman scattering on the plasma frequency is $\left( {{\omega  \mathord{\left/{\vphantom {\omega  {\omega _p }}} \right.
 \kern-\nulldelimiterspace} {\omega _p }}} \right)3.326 \cdot 10^{ - 25} \;\rm{cm}^2$.  The number of interactions during photons travel over a distance $R$ is then about $\left( {{\omega  \mathord{\left/{\vphantom {\omega  {\omega _p }}} \right.
 \kern-\nulldelimiterspace} {\omega _p }}} \right)3.326 \cdot 10^{ - 25} N_e R, $ where product of electron density and the distance can be obtained from Eq.\,(48).  For a supernova with redshift $z = 0.97$, the number of interactions is then about $\left( {{\omega  \mathord{\left/{\vphantom {\omega  {\omega _p }}} \right.
 \kern-\nulldelimiterspace} {\omega _p }}} \right)\ln \left( {1 + z} \right) = \left( {{\omega  \mathord{\left/{\vphantom {\omega  {\omega _p }}} \right.
 \kern-\nulldelimiterspace} {\omega _p }}} \right)0.678.~$  The scattering angle after many interactions is a gaussian distribution with an average of 
\be
\begin{array}{l}\displaystyle
 \theta  = \sqrt {\frac{{\omega _p }}{\omega }\ln \left( {1 + z} \right)} \; {\rm{; \;\; and }}\; \; \; R\theta  \approx \frac{{3.0064 \cdot 10^{24} }}{{N_e }}\left( {\frac{{\omega _p }}{\omega }} \right)^{1/2} \left( {\ln \left( {1 + z} \right)} \right)^{3/2} {\rm{, \; \; \; or}} \\
 \displaystyle \quad \quad \quad \quad \quad \quad \quad R\theta  \approx \frac{{7.14 \cdot 10^{26} }}{{\sqrt {N_e \, \omega } }}\left( {\ln \left( {1 + z} \right)} \right)^{3/2} {\rm{.}} 
 \end{array}
\ee
\indent For $z = 1.7 $, the angle is $\theta  \approx 3.73 \cdot 10^{ - 7}$ and the distance $R\theta  \approx 5.73 \cdot 10^{20}$ cm.  At large distances, the angular spread may exceed the size of the supernova.  The method of light intensity measurements may then not include all the scattered light, because the peak of the light intensity at the center of the image will be reduced and scattered out to larger angles. It may sometimes be difficult to distinguish the supernova light from the background, which may include scattering from other stars in the host galaxy.   The short wavelengths are scattered slightly less than the long wavelengths.  This difference could affect the evaluation.  If any of the scattered light is not included, the star will appear dimmer.  Density fluctuations will increase the angular scattering, and other effects may also play a role.  We will in the following assume that the measurements include correctly the scattered light.  (The supernovae researchers, for example, measure the light-intensity background from the galactic region containing the supernova before or well after the supernova explosion.)
\vspace{2mm}

{\textbf{Magnitude-redshift relation.}}  Increase of an object's observed magnitude by $\delta m$ is defined, as a decrease of the light intensity by $10^{ - 0.4\delta m} ,$ and the absolute magnitude $M$ of an object is its magnitude at 10 parsec.  We have then from Eq.\,(51) that the object's observed magnitude, $m {\rm{,}}$ is given by
\be
m = 5\log R + 1.086a + 7.5\log \left( {1 + z} \right) - 5\log \left( {10} \right) + M {\rm{,}}
\ee
where the distance $R$ to the star is in the unit of parsec, and $a = a(R)$ is the absorptions coefficient in Eq.\,(51).  In this equation, we insert $R$ from Eq.\,(50).  We get then that the relation between magnitude and plasma redshift, $z{\rm{,}}$ is
\[
  m - 1.086a = 5\log \left( \frac{{c \cdot 10^{6}\, \ln \left({1 + z} \right)}}{{ H_0 }} \right) + 7.5\log \left( {1 + z} \right) - 5 + M{\rm{,}} 
\]
or
\be
m - 1.086a = 5\log \left( {\ln \left( {1 + z} \right)} \right) + 7.5\log \left( {1 + z} \right) + 5\log \left( \frac{{c \cdot 10^{6} }}{{H_0 }} \right) - 5 + M {\rm{.}} 
\ee
This magnitude-redshift relation in the plasma-redshift cosmology may be compared with the conventional equation (see Eq.\,(23) of Sandage [88]), which is based on the big bang and expansion of the universe, but disregards both possible acceleration and deceleration of this expansion
\be
m - 1.086a = 5\log \left( z \right) + 5\log \left( {1 + z} \right) + 5\log \left( {\frac{{c \cdot 10^{6} }}{{ H_0 }}} \right) - 5 + M{\rm{.}}
\ee
The third term on the right side of Eq.\,(54) is equal to the third term on the right side of Eq.\,(55).  In these equations, $c$ is in ${\rm{ km}\,\rm{s^{-1}}}$, $H_0$ in ${\rm{ km}\,\rm{s^{-1}}\,\rm{Mpc^{-1}}}{\rm{.}}~$

\indent We may subtract Eq.\,(55) from Eq.\,(54) and get that difference in expected magnitude is
\be
\Delta m_d  = 5\log \left( {\frac{{\ln \left( {1 + z} \right)}}{z}} \right) + 2.5\log \left( {1 + z} \right){\rm{.}}
\ee
For $z$ equal to: 0.5, 1.0, and 2.0, we find $\Delta m_d$ is: $- 0.0149,\; - 0.0433,\; - 0.1081$, respectively; see Table 4.  If a redshift less or equal to $z = 0.5$ is used to determine the magnitude-redshift relation, the deviations at larger $z$-values would be reduced significantly and make it still more difficult to distinguish between two radically different theories.


\begin{table}[h]
\centering
{\bf{Table 4}} \, \, The variation in $-\Delta m_d$ with the redshift $z$ as defined in Eq.\,(56).

\vspace{2mm}

\begin{tabular}{llllllll}
\hline
$z$ & $-\Delta m_d$ & $z$ & $-\Delta m_d$ & $z$ & $-\Delta m_d$ & $z$ & $-\Delta m_d$ \\
\hline \hline
0.1 & 0.0008 & 0.6 & 0.0200 & 1.1 & 0.0496 & 1.6 & 0.0820 \\
0.2 & 0.0030 & 0.7 & 0.0254 & 1.2 & 0.0560 & 1.7 & 0.0885 \\
0.3 & 0.0062 & 0.8 & 0.0312 & 1.3 & 0.0624 & 1.8 & 0.0951 \\
0.4 & 0.0102 & 0.9 & 0.0371 & 1.4 & 0.0689 & 1.9 & 0.1016 \\
0.5 & 0.0149 & 1.0 & 0.0433 & 1.5 & 0.0754 & 2.0 & 0.1081 \\
\hline
\end{tabular}
\end{table}


\begin{figure}[t]
\centering
\includegraphics[scale=.5]{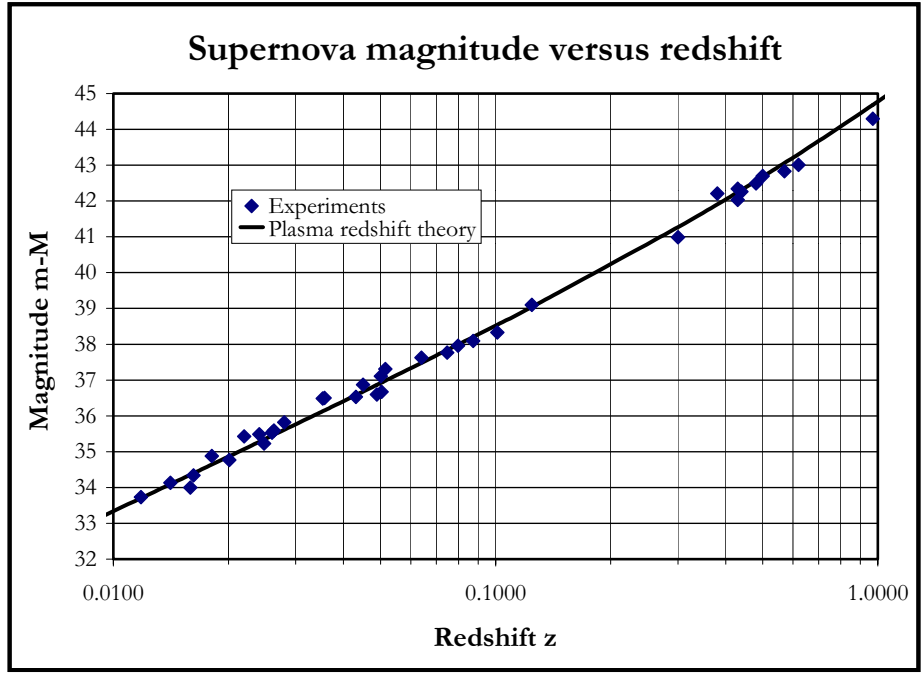}
\caption{The magnitude m-M of 37 supernovas versus their redshifts z.  The experimental points, indicated by the diamonds, are from Riess et al.~[91] (see in particular Tables 6 to 10 of that source).  The curve shows the theoretical magnitude-redshift relation given by Eq.\,(54), which is based on the plasma-redshift theory.  The derived Hubble constant that best fits the data is $H_0 = 65.23~\rm{km\,s}^{-1} ~ \rm{Mpc}^{-1}.~$ }
\end{figure}

\indent The observations of supernovas by Perlmutter et al.~[89, 90] and Riess et al.~[91, 92] are consistent with the magnitude-redshift relation, Eq.\,(54), valid for the plasma-redshift cosmology.  The Hubble constant in the third term of Eq.\,(54) is closely related to the average electron density, as shown in Eq.\,(49).  The best fit to all the data in Fig.\,5 is obtained for $H_0 = 65.23~\rm{km\,s}^{-1} ~ \rm{Mpc}^{-1},$ which corresponds to $N_e = 2.12\cdot 10^{-4}~{\rm{cm}}^{-3}{\rm{.}}~$  

\indent  As Fig.\,5 shows, the observations are consistent with the plasma-redshift theory.  Most supernovas are within a galaxy, which has a corona.  This corona increases the redshift of the supernova inside the corona.  Therefore, nearby supernovas tend to be above the line, while distant supernovas will more often be below the line.  The observations are also in agreement with Eq.\,(55).  However, in Eq.\,(55), we have assumed that the expansion neither decelerated nor accelerated.  In the big bang scenario, this is not a reasonable assumption.  The masses, including ``dark masses'', in the universe are expected to slow the expansion or decelerate the expansion.  It turned out that the expected deceleration was too large.  It was necessary to introduce a counterbalancing expansion force the ``dark energy''.  This ``dark energy'' results in expansion force, which counteracts the deceleration and may even accelerate the expansion.  The deceleration of the expansion, due to an attraction of the masses, would reduce the expansion with increasing time.  In the long past the stars would then be moving apart faster than in the recent past.  For a given redshift, a distant star would be closer and therefore brighter, and the magnitude $m$ smaller than if there were no deceleration.  Perlmutter et al.~[89, 90] and Riess et al.~[91, 92] have considered these different possibilities.  When they assumed reasonable deceleration, the distant supernovas were observed dimmer than expected.  Only if the acceleration approximately compensated the deceleration, as in Eq.\,(55), was there a reasonable fit.  We see thus that while the experiments confirm the plasma-redshift theory, they can only fit the big bang hypothesis if an expansion force approximately compensates (or even partially reversed) the attraction of the masses in the universe.

\indent The plasma redshift follows from the conventional laws of physics.  It explains the cosmological redshift without these artificial parameters of big bang, ``dark matter'' and ``dark energy''.  There is no need and no place for big-bang hypothesis or cosmological-expansion hypothesis.  In his static model of the universe, Einstein introduced $\Lambda$ to counter the gravitational attraction.  Its meaning of Einstein's $\Lambda$ has been modified in the scenario of big bang and expanding universe and several lambdas have been introduced to explain the contradictions with experiments.  In plasma redshift cosmology there is no need for $\Lambda$ or any expansion.  The conventional laws of physics explain all the observation, provided calculations are made more exact, and provided conventional quantum mechanical explanations are used to explain the observations.

\indent The weightlessness of photons (which was deduced from the solar redshift experiments) counteracts the gravitational attraction of matter and leads to continuous renewal of matter; see sections 5.6.2 and 6, and reference [53].  The more matter concentrates, the more is converted to photons.  The photons gain energy as they move away from a gravitating nucleus.  If the photons are energetic enough, they can reform matter in a continuous renewal process lasting forever.  The plasma-redshift theory leads, thus, to a self-regulating static universe, without the need for Einstein's $\Lambda ,$ or ``dark energy'', without the need for a big bang, without need for ``dark matter'', and without the need for ``black holes'', as we will see in section 6.


\subsection{Time dilation and large cosmological redshifts}

Subsequent to the publication of the data that formed the basis for Fig.\,5, Riess et al.~[93] published an updated set of data, which extended the maximum redshift values from $z = 0.97$ to about $z = 1.755{\rm{.}}~$  This new set contained a total of 186 SNe\,Ia shown in Fig.\,6 while the previous data containing only 37 SNe\,Ia are shown in Fig.\,\,5.  With the new supernovae, it was difficult to fit the high-$z$ values to the theoretical curve without back-correcting the absolute magnitude $M_{exp}$-values for $M$ as reported by Riess et al.~[93].  {\it{The comparison between the experimental data and the plasma redshift theory in Fig.\,6 show that there is no cosmic time dilation.}}


\begin{figure}[h]
\centering
\includegraphics[scale=.5]{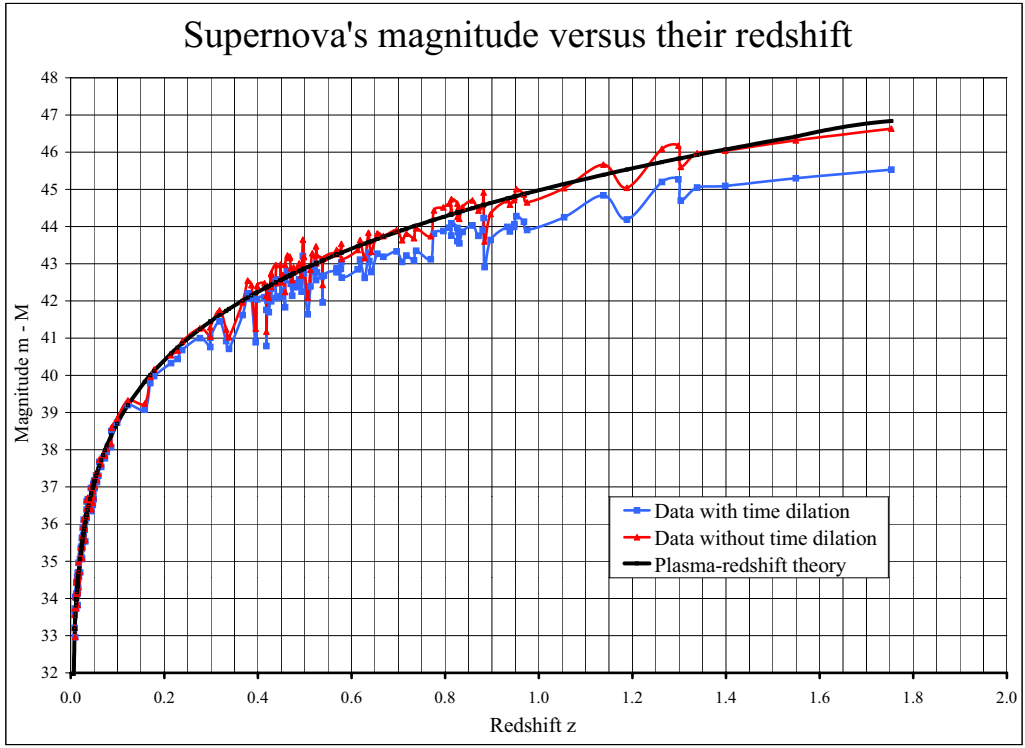}
\caption{The magnitudes, m-M, of supernovas on the ordinate versus their redshifts, $z$ from 0.0 to 2, on the abscissa.  The data include all 186 supernovas reported by Riess et al.~[93] (see the expanded Tables 5 of that source).  The lower data points noted with small rectangles and a blue curve show the absolute magnitudes,  $M_{exp}{\rm{,}}$ as reported by Riess et al.~[93].  The data points noted with small triangles and a red curve show the same absolute magnitudes corrected for the time dilation, $M = M_{exp} - 2.5 \ln (1+z){\rm{.}}~ $   The black curve shows the theoretical predictions of the plasma-redshift theory given by Eq.\,(54).  The Hubble constant that best fits the data is  ${\rm{H}}_0 = 59.44 ~{\rm{km\,s}}^{-1}\,{\rm{Mpc}}^{-1}{\rm{.}}~$}
\vspace{2mm}
\end{figure}

\indent  Eqs.\,(54) and (55) make it clear that for confirming the magnitude-redshift relation, we must know the absolute magnitude $M {\rm{ }}$ of the objects, which in this case are the magnitudes of SNe\,Ia.  If all the SNe\,Ia were a standard candle, we would need to know the value of $M$ for only one of them.  But the supernovae SNe\,Ia are not standard candles.  Those that are close to us, $z < 0.1{\rm{,}}$  vary slightly.  For these supernovae, the researchers have found that the light curves of the brighter supernovae decay more slowly, and their width $w\approx s{\rm{, }}$ where $s$ is called the stretch factor, increases nearly proportional to the brightness or intensity.  For $z < 0.1$ the assumed time dilation $(1+z)$ is small and does not affect the width or the stretch factor $s$ significantly.  For the more distant supernovae, the supernovae researchers assume a time dilation and write the observed width, $w{\rm{, }}$ of the light curve as: $w = (1+z)\, s {\rm{, }}$ where $s$ is the assumed width of the light curve if the supernovae was nearby.  The factor $(1+z)$ is the cosmic time dilation factor.  The supernovae researchers estimate the stretch factor $s$ by dividing the observed width $w$ by $(1+z)$ and assume that the so derived $s$ is the lightcurve width, if this supernovae was close by.  From this reduced width $s$ they derive the absolute magnitude, $M_{exp}{\rm{, }}$ of the distant supernova. 

\indent  Goldhaber et al.~[94] investigated 42 high-redshift supernovae and found that the width, $w =(1+z) \, s {\rm{,}}$ was experimentally proportional to $(1+z)$ with a large noise in $s$ around zero.  They considered this as an indication, or even a proof that the width is proportional to $(1+z){\rm{.}}~$ However, it should be clear to everyone that we could also find that $w$ was proportional to $s$ and that there was a noise in the $(1+z)$ factor.  Their finding does not, therefore, prove anything about the time dilation.  We might or might not have time dilation.  However, if we assume for the moment that there is a cosmic time dilation, then the distant supernovae do not show any Malmquist bias, because, according to Goldhaber, the value of $s$ averages out to about zero.  We are at the limit of observations and the number of supernovae should increase significantly with increasing distance.  We are therefore likely to see more of the brightest supernovae, and the Malmquist bias in our observations should be significant.  The data used by Goldhaber et al. thus contradict an expected reasonable Malmquist bias, and indicate thereby that the cosmic time dilation is false (see Brynjolfsson [95], and in particular Eqs.\,(3) and (4) of that source).

\indent  Recently, Foley et al.~[96] claim to have ``A Definitive Measurement of Time Dilation in the Spectral Evolution of the Moderate-redshift Type Ia Supernova 1997ex''.  The authors selected one supernova out of several hundred.  Actually, it had been selected about 7 years earlier, as indicating ``time dilation'' in accordance with preconceived ideas; see Filippenko and Riess [97] and in particular page 7 of that source.  {\it{In proper statistics we cannot pick and choose those elements (supernovae) that we like or think fit our ideas.}}  The supernova 1997ex was not among the supernovae forming the bases for Figs.\,\,5 and 6.  The width (brightness) of 1997ex after reduction by the time dilation is even greater than the average width of the reference samples.  Yet Foley et al. claim that the null hypothesis, that is, "no time dilation", is excluded at significance level of 99\,\%.  {\it{For proper statistical evaluation of the null hypothesis (no time dilation), we need a statistically valid samples of the spectral features for supernovae with the same width at $z = 0{\rm{,}}$ as the observed width of the supernova 1997ex at $z = 0.361{\rm{.}}~$}}  No such comparison was made.  I find, therefore, that both their method of selecting the supernova and their application of statistics invalidates their conclusion.  The conclusion that I draw is that the data by Goldhaber et al.~[94], who investigated 42 high-redshift supernovae, are most likely the best data we have. Their data show an absence of a reasonable Malmquist bias if the time dilation is assumed.  Their data indicate, therefore, that the time dilation is false.   The supernovae data used in Fig.\,\,6, indicate strongly that the time dilation is false.

\indent   In section 5.6.4, we mentioned that in plasma-redshift cosmology we should expect quasars to have large intrinsic redshifts.  This is also supported by the lensing data (see Brynjolfsson [59]).  But let us for a moment assume that the quasars are at cosmological distances in accordance with the usual consensus in the astronomical community.  We have then as Hawkins showed [98] that the time scale of quasar variation does not increase with redshift as required by the time dilation.

\indent  The expansion hypothesis affects strongly the brightness-redshift relation.  Peebles [69]  (see in particular Eqs.\,(6.41) to (6.44) of that source) shows that the big-bang cosmology leads to $ i = (1+z)^{-3}\, i_0 {\rm{,}}$ where $ i $ is the surface brightness or the energy flux per unit area, per solid angle, and per frequency interval in the expanding reference system, while $i_0$ is the corresponding energy flux in a nonexpanding system.  When we integrate over the frequencies, we get further reduction with the increasing redshift, because of the photons redshift.  We get then $ I = (1+z)^{-4}\,I_0 {\rm{.}}$   Recently, Eric Lerner [99] has compared these relations with great many observations and finds that they contradict observation.  Thus, also his analysis shows that the cosmic time dilation is false.  

\indent  The plasma redshift explains the observed redshift without any expansion, or cosmological time dilation; see Fig.\,6.  It explains also the CMB and the observed X-ray intensity, see in sections 5.10 and 5.11.  Plasma-redshift cosmology needs neither ``dark matter'' nor ``dark energy''.  The plasma redshift is based only on conventional physics, as we know it from the laboratory experiments.  We have only made more exact calculations than those usually found in the literature (see ``Comment A2'' to ``Comment A5'' in Appendix A).  Plasma redshift is not invented for explaining something.  Instead, the conventional physics, without any new assumptions, leads to such phenomena as the cosmological redshift and the CMB.  The plasma redshift is an integral part of conventional physics.  There is no place for the big bang cosmology.

\indent  The magnitude-redshift relation, Eq.\,(54) is particularly significant, because it shows that the plasma redshift does not only predict $N_e = 1.95\cdot 10^{-4} \cdot (H_0/60) ~{\rm{cm}}^{-3}{\rm{,}}$ but predicts also correctly the corresponding dimming that is caused by Compton scattering.  In the big-bang cosmology the distances are often assumed to be proportional to $d_{bb}= (c/H_0) \cdot z {\rm{,}}$ which for large redshifts is much larger than $d = (c/H_0) \cdot \ln (1+z)$ given by Eq.\,(50) for plasma-redshift cosmology.  See also the first terms on the right sides of Eqs.\,(54) and (55).  For correcting some of the additional dimming caused by the large $d_{bb}$, the big-bang cosmologists were forced to assume also a variable ``dark energy'' in addition to ``dark matter''.  Neither of these fudge factors are needed in the plasma-redshift explanations of the many observed phenomena (see Fig.\,1 for the distance-redshift relation in an article by Brynjolfsson [59]).

\indent  The second term on the right side in Eq.\,(54) consists of two contributions: $2.5 \,\log \left( {1 + z} \right)$ for the energy loss to plasma redshift when the photons penetrate the intergalactic plasma, and $5.0 \,\log \left( {1 + z} \right)$ for the loss of photons (in a narrow beam geometry) by Compton scattering on the plasma electrons.  In the big-bang cosmology the time dilation is an integral part.  In Eq.\,(55) the second term consists of two terms: $2.5 \,\log \left( {1 + z} \right)$ for the energy loss in the redshift and $2.5 \,\log \left( {1 + z} \right)$ for the energy loss caused by cosmic time dilation.  Riess et al [93] include this term, $5 \,\log \left( {1 + z} \right){\rm{,}}$ by defining the ``luminosity distance'' as, $R_L= R \cdot (1+z) {\rm{.}}~$

\indent  When I preapared Fig.\,5, I used the data for $M$ as reported by Riess et al.~[91-92].  I could get a reasonable fit to the theoretical curve, because the highest redshift was about $z = 0.97{\rm{.}}~$  It can be seen, however, that the three points with the highest redshifts are slightly below the theoretical curve, and for $z < 0.1$ the average of the points is slightly above the curve.  This is because I had not corrected the reported values for $M$ for the false time dilation.  When I later wanted to apply the expanded data by Riess et al.~[93] this trend in the data was exasperated.  The data did not fit very well to the theoretical curve, the high-z value for the supernovae were consistently too far below the curve.  Riess et al.~[91-93] had when evaluating the absolute magnitude divided the intensity by the false time dilation factor $(1+z)$.  This reduces their estimate of the light intensity of the supernova, and increases their estimate of the absolute magnitude to a value $M_{exp}{\rm{.}}~$  We must therefore back-correct the reported data and replace $M_{exp}{\rm{}}$ by $M{\rm{,}}$ which is then given by
\[
M = M_{exp} - 2.5 \, \ln (1+z) {\rm{.}}
\]
When we in this way back correct the data as reported by Riess et al.~[93], we get the upper curve in Fig.\,6, which compares well with the theoretical predictions of the plasma redshift theory.   The lower curve is for the uncorrected data.  In this case, the redshift on the abscissa is linear (for the purpose of spreading out the high-z values) and not logarithmic as in Fig.\,5.

\indent  The plasma redshift predicts that a part of the observed redshift is due to the corona of the Milky Way Galaxy and the corona of the host galaxy.  For this reason, we have reduced all the redshifts, $z{\rm{,}}$ by an amount $\Delta z = 0.00185.~$  This is nearly an insignificant correction, but in principle a correction on this order of magnitude should be applied when using the plasma-redshift theory.  This corresponds to reducing the average redshift by $\Delta z = 0.000925~$ for each galaxy.  This correction for the intrinsic redshift of the galaxies does not affect the form of the curve, but it affects slightly the value of the Hubble constant.   The corresponding Hubble constant is ${\rm{H}}_0 = 59.4 ~{\rm{km\,s}}^{-1}\,{\rm{Mpc}}^{-1}.~$  The main reduction of the Hubble constant is due to the elimination of the false time dilation.

\indent  In Fig.\,(6), we used all of the 186 supernovae reported.  The distribution in the $M$-values around the theoretical curve is nearly gaussian with a standard deviation for an individual sample of about $\sigma _{\rm{M}}= 0.30$ of a magnitude; see Brynjolfsson [95].  This variance in the data is equal to that obtained by Riess et al.~[93] when using in addition to the big-bang expansion hypothesis both the ``dark energy'' and the ``dark matter'' parameters for minimizing the variance.  The plasma-redshift cosmology has no need for such adjustable parameters.  As Fig.\,6 shows, the data support with high accuracy the plasma redshift theory, which has no time dilation, no dark matter, and no dark energy.


\subsection{Cosmic microwave background radiation}
The cosmic microwave background (CMB) radiation has a spectrum and intensity corresponding to a radiation from a thermal blackbody cavity with temperature of $T_{CMB} = 2.728 \pm 0.002 ~ {\rm{K\,,}} $ as estimated by Fixen et al.~[100] (Peebles [69] used an estimate of $2.736 \pm 0.017$ K; see his Eq.\,6.1.).  This isotropic radiation is often been mentioned as a strong proof for the big bang hypothesis.  It has been difficult to find any other reasonable explanation for it.  In spite of the frequently quoted ``proofs'' and the contention that only the big bang hypothesis can explain the CMB, we will show that the plasma-redshift cosmology gives a rather simple explanation of the microwave background.

\indent We saw in sections 5.7, 5.8 and 5.9 that the plasma redshift, which follows from basic axioms of physics, leads to relatively high densities and high temperatures in intergalactic space.  These same densities and temperatures explain not only the cosmological redshift, but also the CMB, and the X-ray background, as we will see.

\indent  From the Hubble constant, derived from the experiments in Fig.\,6 and Eq.\,(49), we determine the electron density to be
\be
\left( {N_e } \right)_{av} = 1.95 \cdot 10^{- 4}\left( {\frac{{H_0 }}{{60}}} \right)\,~ {\rm{ cm}}^{ - 3} {\rm{. }} 
\ee
\noindent  The value of the Hubble constant, $H_0 = 74.5$, estimated by Press [87] and used in Eq.\,(49), is likely to be too large, because the researchers did not take adequately into account the intrinsic redshifts of galaxies.  The value is also affected by the false time dilation, which was assumed to be valid.  We have in this case used a Hubble constant of $H_0 \approx 60~\rm{km\,s}^{-1} ~ \rm{Mpc}^{-1}{\rm{}}$ for the intergalactic space, as indicated by the analysis of the SNe\,Ia in Fig.\,6, which depends less on the intrinsic redshifts of galaxies, and is not affected by a false time dilation.   

\indent  The intergalactic plasma will gradually absorb the photon energy and create a blackbody cavity with a radius, which is about equal to the plasma-redshift distance, $R_{pl}=\kappa_{pl}^{-1} {\rm{,}}$ where $\kappa_{pl} $ is the absorption coefficient in the plasma redshift.  We get
\be
R_{pl}  = \frac{1}{{3.326 \cdot 10^{ - 25} \left( {N_e } \right)_{av} }} = 1.542 \cdot 10^{28} \left( {\frac{{60}}{{H_0 }}} \right)~{\rm{ cm}}{\rm{.}}
\ee 
\noindent  The plasma redshift absorption length, $R_{pl}{\rm{,}}$ is equal to the Hubble length, $c/H_0{\rm{.}}~$ The Compton scattering length, $R_C {\rm{,}}$ which is one half of $R_{pl} {\rm{,}}$ is given by
\[
R_C  = \frac{1}{{6.652 \cdot 10^{ - 25}  \cdot \left( {N_e } \right)_{av} }} = 7.71 \cdot 10^{27} \left( {\frac{{60}}{{H_0 }}} \right){\rm{ cm}}{\rm{.}}
\]
\noindent  The CMB radiation, which is emitted by the plasma, can be scattered many times and will be uniform and isotropic.

\indent  The plasma-redshift absorption is $\kappa_{pl} = 1/R_{pl} = 3.3262\cdot 10^{-25}\cdot (N_e)_{av}= 6.486\cdot 10^{-29}~{\rm{cm}}^{-1}{\rm{,}}$ where $R_{pl}~{\rm{and}}~(N_e)_{av}$ are given by Eqs.\,(58) and (57).  As shown in section C1.7 of Appendix C, this leads to  
\be
 R_{pl}\, j_{\nu} = I_{\nu} = B_{\nu}(T_{CMB}) =   \frac{{2 h {\nu}^3}}{{c^2}} \, \frac{{1}}{{\,\,e^{h\nu/kT_{CMB}}\,-\,1 \, }} ~~{\rm{erg \, cm}}^{-2} \, {\rm{s}}^{-1}\,{\rm{sr}}^{-1}\,{\rm{Hz}}^{-1} \, {\rm{.}}
\ee
\noindent This equation shows that the photon intensity, $I_{\nu} {\rm{,}}$ emitted by the intergalactic plasma has a perfect blackbody spectrum.  The plasma redshift absorption, which is independent of frequency, is the dominant absorption (usually by several orders of magnitudes), as shown in sections C1.2 to C1.5 of Appendix C.  The plasma acts as a blackbody cavity.  The ``free'' electrons are mainly responsible for the emission.  The walls of this blackbody cavity consist of the particle density, $N_e (x) {\rm{,}}$ along the line of the radius, $R_{pl} {\rm{,}}$ from the observer.  The column density, $\int N_e (x)\,dx = (N_e)_{av}\, R_{pl} {\rm{,}}$ forms the wall.  The radiation pressure, $p{\rm{,}}$ inside this cavity is given by $p = u/3 {\rm{,}}$ where $u$ is the photon's energy density.  We have then that
\be
 p = \frac{{u}}{{3}}=  \frac{{4\pi}}{{3\,c}} \int\limits_0^\infty  B_{\nu}(T_{CMB})\,d\nu =  \frac{{4\pi}}{{3\,c}} \int\limits_0^\infty  \frac{{2 h {\nu}^3}}{{c^2}} \, \frac{{d\nu}}{{\,\,e^{h\nu/k T_{CMB} }\,-\,1 \, }} = \frac{{4\,\sigma}}{{3\,c}}\, T_{CMB}^4 ~~{\rm{erg \, cm}}^{-3} \, {\rm{.}}
\ee
\noindent    This photon pressure, $p{\rm{,}}$ in the usual way must be equal to particle pressure, $N k T_e {\rm{,}}$ in the ``walls'' of the cavity, because of the second law of thermodynamics.  We get
\be
 \frac{{4\,\sigma }}{{c}}\,T_{CMB}^4 = a\,T_{CMB}^4 = 3 N k T_e ~~{\rm{dyne \, cm}}^{-2}  {\rm{,}}
\ee
where in Eq.\,(60) and (61), the partial pressure of CMB is $p = u/3 = (a/3)\, T_{CMB}^4 ~ {\rm{dyne \, cm}}^{-2} {\rm{,}}$ and where Stefan-Boltzmann constant for energy density is $a = 7.566\cdot 10^{-15}~ {\rm{dyne \, cm}}^{-2}\, {\rm{K}}^{-4}  {\rm{.}}~$  For the particle density, we use the approximation: $N \approx N_p + N_{He} + N_e = (2.3/1.2)\, N_e \approx 1.917\, N_e~ {\rm{cm}}^{-3}{\rm{.}}~ $  The electron density predicted by the plasma-redshift cosmology in section 5.9 is $N_e = 1.95\cdot 10^{-4}\cdot (H_0/60)~ {\rm{cm}}^{-3}{\rm{.}}~$  If the photons' energy density is mainly due to the CMB radiation with temperature $T_{CMB} = 2.728 \pm 0.002 {\rm{}}$ [100], we get that $T_e$ is
\be
T_e = \frac{{a\,T_{CMB}^4}}{{3 N k}} = \frac{{4.1902 \cdot 10^{-13} }}{{3\cdot 1.917\cdot (N_e)_{av}\, 1.3807\cdot 10^{-16} }} = 2.706\cdot 10^6 \cdot\frac{{60}}{{H_0}}~~{\rm{K}}\,{\rm{,}} 
\ee
\noindent  where $(N_e)_{av} = 1.95\cdot 10^{-4}\cdot (H_0/60)~ {\rm{cm}}^{-3}{\rm{,}}$ and $k = 1.3807\cdot 10^{-16}$ is the Boltzmann constant.

\indent  The blackbody radiation is well defined through $N_e$ and $T_e{\rm{,}}$ which are averaged over the distance $R_{pl}\approx 5000~{\rm{Mpc}}{\rm{.}}~$ {\it{We see that the intensity, $4\pi \int I_{\nu}\, d\nu = c \, a \, T_{CMB}^4 = 3 \, c\, p {\rm{,}}$ of the CMB, according to Eqs.\,(59) to (61), is proportional to the pressure $p{\rm{.}}~$}} In spite of the temperature variations caused by the ``bubble'' formations in the plasma, the CMB temperature, $T_{CMB}{\rm{,}}$ is about the same in the cold and hot regions of space, provided the pressure is the same.  The pressure should vary much less than the temperatures and the densities.  The large dimensions $R_{pl}\approx 5000~{\rm{Mpc}}{\rm{,}}$ the constancy of the pressure, and the fact that the plasma redshift absorption and emission dominate by several orders of magnitude the other absorption and emission processes from $\nu = 10^9~{\rm{Hz}} $ to $\nu = 5\cdot 10^{11}~{\rm{Hz,}} $  (see sections C1.2 to C1.5 of Appendix C) helps us understand why $T_{CMB}{\rm{}}$ is so well defined and isotropic, as observations indicate.  Close to galaxy clusters, where the pressure is higher, the CMB temperature should be slightly higher as measurements indicate.  This is, especially, clear within our Milky Way.
\vspace{2mm}

{\textbf{Intensities at frequencies below the CMB are consistent with the plasma redshift theory.}}  Below  about $\nu \approx 10^9~{\rm{Hz}}{\rm{,}}~$ the plasma-redshift cut-off gradually sets in.  The cut-off begins in the coldest filaments of space and then in the colder regions.  Just below the cut-off, the intensity accumulates and increases therefore as the frequency decreases.

\indent   According to Eqs.\,(16), (57) and (63), we have for average density and temperature that the 50\,\% cut-off for plasma redshift is at $\lambda \approx 684~{\rm{cm,}}$ or at the frequency of $\nu = 44~{\rm{MHz}}{\rm{.}}~$  In the colder regions of space the density and temperature may be about 10 times lower and 10 times higher, respectively, than the average.  The 50\,\%cut-off wavelength is then $\lambda \approx 216 ~{\rm{cm,}}$ or at the frequency of $\nu = 139~{\rm{MHz}}{\rm{.}}~$ The hotter regions may have about 10 times lower density than the average and 10 times higher temperatures.  The 50\,\%cut-off is then at the wavelength of $\lambda \approx 2164~{\rm{cm,}}$ or at the frequency of $\nu = 14~{\rm{MHz}}{\rm{.}}~$   

\indent  Keshet et al.~[101] have evaluated the observed intensities in this region; see in particular their Fig.\,6.  They find that the intensity increases with decreasing frequency approximately as $I_{\nu}\propto \nu^{-1.3}$, which is about what we should expect.  The cut-off frequency, $\nu = 44~{\rm{MHz}}{\rm{,}}$ for the average density and temperature is in the middle of the intensity increase, as we should expect.  Like the cut-off in the middle of the transition zone to the solar corona, this is another beautiful and independent confirmation of the plasma-redshift theory.

\indent  The experimental data contain a significant contribution from the Milky Way Galaxy, especially, at low galactic latitude towards the Galactic center; see Fig.\,(8) of Keshet at al.~[101].  This is consistent with plasma-redshift theory.  As Eq.\,(61) illustrates, the emitted microwave energy in ${\rm{erg \, cm}}^{-3} \, {\rm{}}$ increases proportional to the pressure, $p$.  The pressure in the Milky Ways corona may be about 20 times higher than in the intergalactic space (see section 5.7), but at high Galactic latitudes the column density is much lower.  At high galactic latitudes, the intergalactic space then determines the intensity in the $3 {\rm{~to~}}300$ MHz frequency-range.  Even in the direction of the anticenter, the intensity is close to that of the Galactic poles.  As we reduce the latitude towards the Galactic center, the density increases significantly, and the emission from the Galactic plasma becomes more pronounced.  Due to the increased densities, the cut-off frequency for the plasma redshift increases.  We see therefore increased intensity of the background in the 10 to 300 MHz frequency-range.  The observed variation in Fig.\,8 of Keshet at al.~[101] is consistent with the plasma redshift theory.  It reinforces the above conclusion that the observations of the intensity below about $10^9$ Hz confirm the predictions of the plasma-redshift cosmology.
\vspace{2mm}

{\textbf{Intensities at frequencies above the CMB are consistent with the plasma redshift theory.}}  We observe the continuum of redshifted photons from the high-frequency photons emitted by stars, active galactic nuclei (AGN), supernovae, and the high-pressure coronal plasma surrounding these objects.  These objects are mostly characterized by point sources.  These photons are gradually redshifted in the plasma surrounding these sources, in the coronas of the stars, the galaxies, the galaxy clusters, and in the intergalactic plasma.  Some of these radiations are absorbed in condensations of colder plasma and in cloud formations usually in the coronas of galaxies.  We should observe, therefore, a continuum intensities of photon frequencies above about $7\cdot 10^{11} ~ {\rm{Hz}} $ to about $10^{16} ~ {\rm{Hz}} $ and beyond.  The sources of the high-energy radiations are usually found in and close to the galaxies.  The energy content is sometimes estimated to be on the order of 10\,\% of the CMB radiation.  If these other radiations contribute about 10\,\% of the CMB radiation, the average temperature is about $ T_e \approx 3\cdot 10^6\cdot (60/H_0)~{\rm{K,}}$ with a high-energy tail skewing the thermal distribution.  These predictions of the plasma-redshift theory are consistent with observations.
\vspace{2mm}
   
\indent The average energy density of the X rays in intergalactic space may be about 22\,\% of the energy density in the CMB radiation.  Most of that X-ray intensity is from the plasma, and the X-ray heating will then balance the X-ray cooling.  Peebles [69] (see in particular Eq.\,(5.143) of that source) estimated for a Hubble constant of $H_0 = 60 ~ {\rm{km\,s}^{-1}}\,{\rm{Mpc}^{-1}}$ that the average luminosity density from the brightest galaxies is about
$7.8\cdot 10^{-33} ~ \rm{erg\,s}^{-1}\,\rm{cm}^{-3}.~$  When we multiply this value by $R_{pl}$ and divide by c, we get $4.01\cdot 10^{-15} ~ \rm{erg}\,\rm{cm}^{-3}$ for the energy density of the light in intergalactic space from the brightest galaxies.  This is only about 0.95\,\% of the CMB.  But Peebles [69] (see his remarks below his Eq.\,(5.168) of that source) also indicated that the average light luminosity in intergalactic space could be 10 times higher, corresponding to about 10\,\% of the energy density in the CMB.  If we add this to the very rough 22\,\% estimate, the average temperature per particle could be as high as about 1.32 times that derived in Eq.\,(62).  Due to the uncertainty in the X-ray and light intensities, we will usually assume that the average temperature in intergalactic space is between 10\,\% and 32\,\% higher than that in Eq.\,(62), or about
\be
 {T_{av} } \approx (3 {\rm{~to~}} 3.6) \cdot 10^{ 6} \left( \frac{{60 }}{{H_0}} \right)\,~{\rm{ K}} .
\ee
\indent When analyzing the phenomena in the transition zone to the solar corona in section 5.1, the solar flares and arches in section 5.5, and the galactic corona in section 5.7, we saw that the plasma redshift has a tendency to create large hot ``bubbles'' with relatively cold plasma in the ``walls'' of the ``bubbles''.  We expect to see similar phenomena in intergalactic space, where the ``bubble'' surfaces or the ``walls'' may have temperatures similar to that in the transition zone to the galactic corona, while the temperatures of the interiors of the ``bubbles'' could be high.  In large ``bubbles'' the temperatures per particle may exceed 10 million K.  The increase in Spitzer's thermal conductivity coefficient will limit the high temperatures, and the X-ray absorption will counteract the decrease in the temperature in the walls of the bubbles.  These bubble structures will also have a tendency to build walls or bridges between galaxies (see the discussion in section 5.7 about the bridge between the Milky Way and the LMC).  It is likely that the magnetic field will influence the structures, and that it will be aligned with the walls.  This ``bubble'' structure in intergalactic space will affect statistical variance in the measurements of the redshift versus distance and the determination of the Hubble constant.


\subsection{X-ray intensity from intergalactic plasma}
In sparse hot plasma of intergalactic space, the free-free absorption, $\kappa'_{\nu} {\rm{,}} $  for the spectrum of the emitted X rays is small when compared with the plasma-redshift absorption; see sections C1.3 to C1.5 of Appendix C.  Therefore, the corresponding free-free absorption length, $R'_{\nu}=1/\kappa'_\nu {\rm{,}} $ is many orders of magnitude larger than the plasma redshift distance $R_{pl}{\rm{.}}~ $   If $R'_{\nu}$ is used as an integration distance for the X-ray intensity from each cubic cm, the X-ray intensity would be very large.  Many physicists believe, therefore, incorrectly that the high densities required by the plasma redshift would lead to much too high X-ray intensities.  It is important for these physicists to realize that when disregarding the absorption from trace elements the absorption length to be used for the integration is $R_{pl} = 1/\kappa_{pl} {\rm{,}}$  and not $R'_{\nu} {\rm{;}}$ see section C1.4 and C1.5 in Appendix C.

\indent  When we include the absorption by trace elements, the actual absorption length, $R_{\nu} {\rm{,}}$ for the X rays in the intergalactic plasma and in the Milky Way's corona is even much shorter than the plasma-redshift distance, $R_{pl}= 1.54\cdot 10^{28}(60/H_0) {\rm{;}}$ see the last line in Table C1 in Appendix C.

\indent  When we then compare the X-ray intensity predicted from the plasma-redshift densities in intergalactic space, about $N_e = 1.95 \cdot 10^{-4}~ {\rm{cm}}^{-3} {\rm{,}}$ we get values that are about equal to that observed.  For example, as Eq.\,(C21) shows, the predicted X-ray intensity for $h\nu = 729~{\rm{eV}}$ is  
\[
I_{\nu} = 8.73  ~~{\rm{keV \, cm}}^{-2} \, {\rm{s}}^{-1}\,{\rm{sr}}^{-1}\,{\rm{keV}}^{-1}\,{\rm{,}}
\]
which matches the observations by by Kuntz et al.~[102], Kuntz and Snowden [103], De Luca and Molendi [104], 
 Vecchi et al.~[105],  and Barcons et al.~[106].  See further discussion of comparison of predictions with experiments in section C3 of Appendix C.  {\it{It can thus be seen that the X-ray intensity predicted by the plasma redshift cosmology matches that observed.  This is thus another beautiful example of how the predictions of the plasma-redshift cosmology match the observations.}}


\subsection{Mass density of the plasma in intergalactic space}

The average mass density of the plasma in intergalactic space is
\be
\rho  = \frac{{1.4}}{{1.2}}\left( {N_e } \right)_{av} m_p  = \frac{{1.4}}{{1.2}}\,\frac{{H_0 }}{{3.0764 \cdot 10^5 }}\,1.67 \cdot 10^{ - 24}  = 
 3.806 \cdot 10^{ - 28} \left( {\frac{{H_0 }}{{60}}} \right)\,~{\rm{ g}}\,{\rm{cm}}^{ - 3} {\rm{,}}
\ee
\noindent  where the factor 1.4 is the mass of hydrogen and trace elements, mainly helium, per proton; and the factor 1.2 is the approximate number of electrons per proton.  The average baryonic mass density derived in Eq.\,(64) is about 56 times larger than the conventionally assumed mass density for a closed universe and $ H_0 = 60 {\rm{,}}$ which is $ \rho_{crit} = 1.88 \cdot 10^{-33} {H_0}^2 \approx 6.77 \cdot 10^{-30}~{\rm{ g}}\,{\rm{cm}}^{ - 3}{\rm{.}}~$  For this critical density value, see Eq.\,(5.67) of [69].  The conventionally assumed baryonic mass density of about $N_B \approx (1.4\pm 0.3)\cdot 10^{-7}\,N_p~{\rm{cm}}^{-3}$ would result in $\rho =(1.4 \pm 0.3)\cdot 10^{-7}\cdot 1.67 \cdot 10^{-24}= (2.34\pm 0.5)\cdot10^{-31}~{\rm{g\,cm}}^{-3}{\rm{;}}$ see Eq.\,(6.27) of Peebles [69].

\indent  This baryonic density in the big-bang cosmology is about 0.0006 times the baryonic density given by Eq.\,(64).  (It is no wonder that the big-bang cosmologists needed to sprinkle some ``dark matter'' here and there.  Their denial of the intrinsic plasma redshifts, however, has been the main cause for the need of dark matter.)  Most of the baryonic matter in Eq.\,(64) is rather evenly spread in intergalactic space, and is often difficult to detect.  It affected the observations mostly through the plasma redshift, that is, the cosmological redshift.  The big-bang cosmologists explained this cosmological redshift through their big-bang hypothesis. 

\indent  To the density in Eq.\,(64), we should add the average of the additional mass density of the coronal plasma around the galaxies and galaxy clusters, and the average mass density of stars in any form, including quasars, neutron stars, dwarf stars, and planets, and of neutral gas and dust particles.  These additional masses are usually estimated to be a small fraction of the mass estimate in Eq.\,(64).  

\indent Positrons would produce plasma redshift analogous to that produced by the electrons.  If most of the intergalactic space were filled with electron-positron plasma, the density in intergalactic space would be small.  When we add the density of the electrons and positrons to that of conventionally assumed density (Peebles [69]; see Table 20.1 in that source), we get about
\be
\rho _0  = \rho _{ep}  + \rho _B  = 1.8 \cdot 10^{ - 31} \left( {\frac{{H_0 }}{{60}}} \right) + 2.7 \cdot 10^{ - 31} \left( {\frac{{H_0 }}{{70}}} \right)^2  \approx 4.5 \cdot 10^{ - 31} \;~ {\rm{ g}}\,{\rm{cm}}^{ - 3} .
\ee
\noindent  However, if the intergalactic plasma contained significant amount of positrons, the positron annihilation photons would be observed.  The fact that photons from positron annihilation in intergalactic space are not observed, rules out this possibility.  Significant positron annihilation is observed from active galactic nuclei and at the center of the Milky Way, as discussed later.

\indent The fact that the photons are weightless, as seen by an observer in a standard reference system at rest at the location of the photon (and repelled by the gravitational field, as seen by a distant observer) causes fundamental changes in Einstein's field equation.  It is conventionally assumed that the gravitational field attracts any form of mass and energy, because $m_i \, c^2 = E_{kin}$ in the special theory of relativity, and because the equivalence principle in GTR assumes equivalence of gravitational mass and inertial mass, $m_i = m_g .~$  The failure of this last assumption in case of photons causes us to suspect that other forms of energy may also be exempted.  However, as shown by Brynjolfsson [107], the experimental evidence indicates that only photons are weightless.  For example, the electro-magnetic field around charged particles is not weightless, as definitely confirmed in the experiments by Adelberger et al.~[108] and Su et al.~[109].  We assume, therefore, that the assumption that gravitational field attracts all forms of energy remains valid except for photons.  Estimates of the photons energy density indicate that photon's lack of a corresponding gravitational mass does not affect the density estimates given by Eqs.\,(64) and (65).  The modification of Einstein's classical physics field equations is small, except that the important $\Lambda$-term is not needed, because the weightlessness of photons eliminates that need for the $\Lambda$-term, as we will see in section 6.

\indent The classical field equations and Einstein's static model lead to
\be
\frac{1}{{R^2 }} =  \frac{{4\pi G\rho }}{{c^2 }}{\rm{. }}
\ee
\noindent  This is same as Eq.\,(14) by Einstein in [110]; see also Eq.\,12.121 in M{\o}ller's monograph [111].  As is usually done, we assume for a moment that Newtonian constant, $G{\rm{,}} $ of gravity is a constant, and that the velocity of light depends on the gravitational field in the conventional way.  We get then for the average mass density given by Eq.\,(64), that the curvature radius is about $1.68 \cdot 10^{27}{\rm{cm.}}~$  Had we used the conventional big-bang density of $ \rho_{crit} \approx 6.77 \cdot 10^{-30}~{\rm{ g}}\,{\rm{cm}}^{ - 3}{\rm{,}}$   the curvature radius would be $R \approx 1.26 \cdot 10^{28}{\rm{cm.}}~$  Had we used the density given by Eq.\,(65), the curvature radius would be $R \approx 4.88 \cdot 10^{28}{\rm{cm.}}~$  These values for the radius in the curvature are deduced here only for reference, because many physicists use them and object that the high densities derived in Eq.\,(64) are too high, because they lead to too small a curvature radius in accordance with Eq.\,(66).  In big bang cosmology, even without ``dark matter'', we should observe curvature.  However, no curvature is observed.  At least one of the assumptions leading to Eq.\,(66) must be wrong.  For solving the problem, the big-bang cosmologists have introduced a time variable ``dark energy'' in addition to ``dark matter''.  These adjustable parameters are not needed in the plasma-redshift cosmology.

\indent  Einstein accepted Newton's phenomenological equations and he extrapolated their use to infinity.  He was concerned because this extrapolation for a static infinite universe leads to infinite gravitational potential.  His original equations lead to a space curvature given by Eq.\,(66).  The curvature alleviated the problem of infinite gravitational potential.

\indent   Dark matter is often introduced "ad hoc" for explaining the observed velocity distribution.  This often comes about because all shifts are explained as Doppler shifts.  Taking into account the intrinsic redshifts can often solve these problems.  The frequent radial elongation of clusters, such as ``Finger of God'', (which put us in the center of the universe) come about when the line of sight to the objects in the back of the cluster penetrates the dense portion close to the center of the cluster gaining thereby significant increment of plasma redshift.   When this elongation is removed the actual density of the cluster is increased, which is often adequate for explaining the observed velocity distribution.  In these clusters there can be a problem with using the shift of the 21 cm line as an indicator.  Depending on the temperature and density of the plasma in the cluster, the parameter $a=3.65 \cdot \lambda_0 \sqrt{N_e} /T $ used in Table 1, can in some cases result in a blue shift, the negative values of $F_1(a)$, (due to the long wavelength and relatively high density plasma) and in other cases in a redshift (relatively low density plasma).  Use of shifts in the visible spectrum is then more reliable.

\indent  In the big bang cosmology, the velocity distribution determined from the 21 cm line from hydrogen in the corona of the galaxies is often a problem.  The dark matter is then invented to explain the observations.  We also have modification of Newtonian dynamics (MOND) [71] trying to explain the observations.  In big bang cosmology the material in the corona is assumed to move in elliptical orbits, or even slowly inwards.  Plasma redshift makes it reasonable that matter moves with a constant rotational velocity in the corona as the plasma drifts outwards from the galaxy into intergalactic space.  Ones in a while, high velocity clouds fall into the galaxy from intergalactic space.  Matter is constantly moving in and out of the galaxy.  This also explains why the trace element concentration in intergalactic space is similar to that in the galaxies.

\indent  When in plasma redshift we talk about a limit of gravitational attraction, we are not talking about inventing new laws or changing the nature.  We are showing that the laws already confirmed do not permit extrapolating the Newtonian laws to infinity, as Einstein did.  We find that the applications of the present laws, which include the plasma redshift, can explain the observed phenomena, such as the cosmological redshift, the CMB, and the phenomena in the solar corona.  My present rough estimates are that this limit on the Newtonian laws is well beyond the dimensions of a galaxy clusters.

\indent   {\textbf{The limit on Newtonian laws.}}  In the classical physics it is conventionally surmised, as Einstein did, that somehow the gravitational field is able to transfer to a particle both the strength and the direction of the gravitational field, even when the field at very large cosmological distances is extremely weak.  In quantum mechanics, this is not permissible assumption, as it disregards well-proven quantum effects.  The hot particles in stars and intergalactic plasmas are bombarded by other particles that change their directions and rotations often many times (even when we consider only gravitational effects in the collisions), while the extremely weak gravitational field is trying to manifest its action on the particles.  According to the uncertainty principle and transition theory in quantum mechanics, the direction and strength of the gravitational field may not have adequate time to act on the particles between the collisions.  In quantum mechanics, the transition theory requires always a certain time $\delta t{\rm{}}$ to manifest the transition to a new state.  This (or something similar to the uncertainty relation) is likely to apply also to gravitational fields.  It takes a certain time, $\delta t \geq \hbar/\delta E {\rm{,}}$ for the gravitational field to manifest itself.  A particle, for example a proton, embedded in the hot plasma is likely to exchange rotational, vibrational, and translational energy with the Fourier harmonics of the electromagnetic field (and of the gravitational field) of the surrounding particles before the extremely weak gravitational field can assert itself, both in respect to direction and potential energy.  It is analogous to the gravitational redshift experiments by Pound and Snider [39].  The photons, which were emitted in the basement, did not have time enough to realize that they had moved up to the gravitational potential on the top level, because the time-of-flight from the basement to the top level of Jefferson Laboratory, where the photons were absorbed, was only $7.5\cdot 10^{-8}~{\rm{sec}}{\rm{, }}$ when a minimum time required according to the uncertainty principle was $\delta t \geq 1.9 \cdot 10^{-5}~{\rm{sec}}{\rm{.}} $  The photons did not therefore have time to adjust to the new potential and change their frequency.  The solar photons when they move from the Sun to the Earth, on the other hand, have usually plenty of time to adjust to the potential and be blue shifted.  

\indent Rough estimates, based on the uncertainty principle, show that the gravitational potential, the collision frequency, collision cross section, and temperature do not permit the extrapolation of Newton's equation for gravitational attraction to very large cosmological distances.  Such an extrapolation of Newton's equation has no basis in physical theory or any observation.  For the present discussion (because we do not have a good gravitational theory), we use only that there is no experimental evidence supporting the extrapolation of Newton's phenomenological equations to extremely large cosmological distances.  {\it{Consistent with observations, we can assume that the universe is infinite without curvature on a large scale and is ever lasting.}}


\section{Weightlessness of photons}
{\textbf{Solar redshift experiments detected the reversal of the gravitational redshift.}}  In section 5.6, the redshifts of the solar lines are explained as due to the plasma redshift; see for example Fig.\,4.  The good fits between the predicted plasma redshifts and the observed shifts indicate that the Fraunhofer lines, when observed on the Earth, are not gravitationally redshifted.  We explained that the photons' frequencies are gravitationally redshifted in the Sun, but that the gravitational redshift is reversed as the photons move from the Sun to the Earth.  This blue shift cancels the gravitational redshift.

\indent   When an atom moves from the Sun to the Earth, the gravitationally redshifted energy levels and frequencies of the atom are blue shifted such as to cancel the gravitational redshift in the Sun.  In quantum mechanics, the frequencies of photons behave like the frequencies of particles; that is, the frequencies increase with the gravitational potential.  Brynjolfsson [107] shows that a simple and natural extension of the classical TGR to quantum mechanics leads to reversal of the gravitational redshift during photons' time of flight from the Sun to the Earth.  This change in frequency is closely related to the increase in velocity of light with the gravitational potential.  Importantly, this modified TGR is consistent with all observations.

\indent The transition from quantum mechanics to classical mechanics is in accordance with Bohr's correspondence principle.  In the classical limit, therefore, the quantum mechanically modified theory of general relativity becomes observationally identical to the classical theory, and can explain the outcome of the gravitational-redshift experiments by Pound and Rebka [37,38], Pound and Snider [39], Vessot et al.~[40], and Krisher et al.~[41].  The experiments on the bending of light analyzed by Riveros and Vucetich [42] and experiments by Shapiro et al.~[43] pertaining time delay of signals are independent of the frequency, and remain unaffected by this change in theory of relativity.

\indent  In the classical theory, the light cannot escape from the ``black hole limit''.  In the modified theory, however, the gravitational redshift is reversed.  As seen by a distant observer, the gravitational field repels the photon.  We know from collision experiments in the laboratory that under the extreme pressure in the collisions, the matter is transformed to photons.  It is reasonable that close to the ``black hole limit'', the matter is converted to photons.  When the highly gravitationally redshifted photons move away from the brink of the ``black hole limit'', the gravitational redshift is reversed.   According to well-known laws of physics, the most energetic of the photons could then reform matter at a distance from the ``black hole limit''.  Considering the pressure and the kinetic energy, the photons could have energy far exceeding the proton-antiproton pair production.  Thus, matter can renew itself eternally.

\indent  In the quantum-mechanically modified TGR, the total of mass and energy is conserved at all times.  For example, the potential energy that the particles lose when they fall towards the gravitating body (to be transformed into photons) is returned to the photons as they stream outwards (and are transformed into particles).  This all happens in accordance with conventional laws of physics, as we know them from the experiments.  In classical TGR, on the other hand, the particle and its energy gradually disappear into a ``black hole''.  In classical TGR, it is usually surmised that all matter and energy in the universe will gradually be sucked into the ``black hole''.  Usually, it is surmised further that eventually some mystical forces will release the particles and their energy from the bondage in the ``black hole''.

\indent  As the annihilation of matter to photons at the brink of the black hole limit and recreation of matter is only a possible extrapolation, we should consider if any observation supports this extrapolation.

\indent Narayan [112] finds that accreting black holes (or more correctly, objects which he believes are accreting black holes) almost always emit a substantial fraction of their luminosity in hard X rays and gamma rays.  The spectrum of the rays can be approximated by a power-law with a photon index $\alpha _N$ as
\be
N_E\, dE \propto E^{ - \alpha _N }\, dE{\rm{,}}
\ee
where the index is usually in the range of 1.5 to 3.  He also notes:

\begin{enumerate}
\item  Electron-positron annihilation features have been seen only in black hole candidates.
\item  Relativistic jets of active galactic nuclei and X-ray binary jets are all unambiguous signatures of black holes.
\end{enumerate}

It is more likely that such very high-energy jets are emitted from a brink of a black hole limit, if the photons are pushed away from the black hole, and if the photon's energy increases with the gravitational potential, as it does in the quantum mechanically modified theory, instead of retaining the redshifted energy, as the photon does in the classical theory.

\indent  At the center of a large neutron star-like object, the density may greatly exceed that of the atomic nucleus, as pointed out by Ruderman [113].  The transition from hadronic matter to quark-gluon matter may well occur at the center of an object somewhat similar to a neutron star, as pointed out by Olive [114].  We don't know at what density such a phase transition occurs, but it may occur before the ``black hole limit''.  We know from high-energy laboratory experiments that heavy pressure created in ion collisions lead to many very unstable particles, which may quickly decay into high-energy photons.  As the gravitational field repels these photons, they could create particle pairs.  It appears possible that close to ``black-hole limit'', we not only have the formation of electron-positron pairs, but at higher energies also proton-antiproton pairs and even higher-mass particles.  It appears to me far-fetched to think that only the accumulation of the entire mass in the universe could lead to such a condition, and recreation of matter.

\indent  At the brink of a black hole limit of an object close the center of a galaxy, or an active galactic nucleus, the kinetic energy of the accreting mass particles and mounting pressure could result in photons with energy in excess of the rest energy of most of the particles.  This rest energy is reduced by the gravitational redshift factor.  Laboratory experiments show that when the particle's kinetic energy exceeds the rest energy, the particles can transform into photons.  As seen by a distant observer, the gravitationally redshifted high-energy photons, which are repelled and gain energy as they move away from the ``brink of a black hole limit'', can recreate matter at any distance from the center.  The condition for the formation of proton-antiproton pairs would be there, and the condition for formation of the elements would have some similarities to that assumed in the big bang scenario.

\indent  This scenario possibly could explain not only the electron-positron formations, which, according to Narayan [112], are an unambiguous signature of what he thought was a black hole, {\it{but also why a large amount of hydrogen, according to Burton [115], streams outward from the center of our Milky Way galaxy.}}  This center is believed to contain a black hole.

\indent   Burton [115] points out that the redshifts of the 21 cm line from the region around the center of the Milky Way galaxy show {\it{that a tremendous amount of hydrogen moves at high speed away from the galactic center}}.  The flux of total gas flow is on the order of 1 to 2 solar masses per year.  The plasma redshift would modify slightly the flow diagrams used by Burton [115].  The observed asymmetrical flow, as reported by Burton, would become more symmetrical.  Thereby, the plasma redshift would simplify the interpretation of the observed flow.

\indent  For example, along the line towards the center of the Galaxy, the electron density causes a significant plasma redshift.  When we take the plasma redshift into account, the velocities of the hydrogen streaming towards us from the center would be greater than those reported by Burton.  Analogously, the velocities of the hydrogen streaming away from us on the other side of the center would be smaller than those reported by Burton.  In fact, reasonable numbers for the plasma redshift would make the observed asymmetrical flow, as reported by Burton, become more symmetrical.  An average electron density in the midplane of about 0.02 ${\rm{cm}}^{-3}$ (an average of 0.016 from Reynolds [61] and 0.025 from Cordes et al.~[63]) over 6.5 kpc from observer towards the hydrogen at 1.5 kpc from the center results in a plasma redshift of 40 ${\rm{km}}\,{\rm{s}}^{-1} {\rm{.}}~$  This value when added to the observed 53 ${\rm{km}}\,{\rm{s}}^{-1}$ for the arm close to the center gives an average of 93 ${\rm{km}}\,{\rm{s}}^{ - 1} {\rm{.}}~$   On the other side of the galactic center the plasma redshift is greater as the line of sight passes through the hot high-density area closer to the nucleus.  The average electron density in that region may be on the order of 0.08 ${\rm{cm}}^{ - 3}$ over 3 kpc.  The corresponding plasma redshift is about 74 ${\rm{km}}\,{\rm{s}}^{ - 1} {\rm{,}}$ which together with the 40 ${\rm{km}}\,{\rm{s}}^{ - 1}$ reduces the observed velocity of 210 ${\rm{km}}\,{\rm{s}}^{ - 1}$ away from the center to about $\left[ {210 - \left( {74 + 40} \right)} \right] = 96 \; {\rm{ km}}\,{\rm{s}}^{ - 1}  {\rm{.}}~$  The actual outward flow from the Galactic center would then be nearly symmetrical at about 95 ${\rm{km}}\,{\rm{s}}^{ - 1} {\rm{.}}~$  This outflow is lower than the average outflow $(53+210)/2 = 131$ $\rm{km}\,\rm{s}^{-1}$ without the plasma redshift.  Importantly, the plasma redshift indicates that the outflow of hydrogen from the center is approximately symmetrical and very large.

\indent  The plasma redshifts assumed in this example are uncertain, and better numbers can be obtained by analyzing the entire set of data.  However, such a modified analysis would not change the essence of the prior conclusion {\it{that a tremendous amount of hydrogen appears to be streaming away from the center of the Galaxy.}}

\indent  Burton mentions that Oort [116, 117] suggested a particular interpretation of the observed velocity flow around the nucleus, which was worked out by van der Kruit [118].  In this interpretation the galactic nucleus ejected gas at high velocity and at an angle of 25 to 30 degrees with respect to the galactic plane.  This material would then return to the plane at a few kpc from the center. 

\indent  In light of the reversal of the gravitational redshift when the photon moves away from gravitating center, we are inclined to modify slightly this scenario by suggesting that the high-energy photons from transformation (annihilation) of nuclear matter are ejected mainly along the vortex-axes of an object at the brink of the ``black hole'' limit.  In both ends of the vortex, the high-energy photons would gain energy as they move outwards, and if photon energy exceeds the rest mass energy, they could recreate matter.  (The ratio of photon energy and rest mass energy is independent of the gravitational potential in the modified TGR.)

\indent  Initially, when very little matter covers up the ends of the vortex, we observe two jets, one from each end, beaming far away from many objects believed to be black holes.  We will first see ``knots'' or ``lumps'' on the beams or jets, because, as is well known from laboratory experiments and theory for pair production, matter enhances the transformation rate of photons to particle pairs.  Occasionally, these ``lumps'' and ``knots'' may coalesce as they are being pushed away, and could possibly form quasars, about the way Halton Arp sees it in his monograph, Seeing Red, [52].  For example, the largest ``lump'' in M87 already now emits more X rays than the core of M87.  
 
\indent  In some cases the beams may be less focused, and as more and more hydrogen is formed and covers up the ends of the vortex (sometimes one end of the vortex before the other), the jets will shorten, and we see a bulge form over the vortex.  When the density in the bulge increases, the protons that are formed will diffuse from the axes of the vortex, and the recombination-emission cooling would result in neutralization of the hydrogen, which would then, due to gravity, leak down to the center plane at some distance, but close to the nucleus.  It is natural for systems affected by the plasma redshift to form structures of colder streams separated by redshift-heated sparse plasma in between.  As mentioned under the discussion of the spicules in the transition zone to the solar corona, and the discussions about the structures in the solar corona and the galactic corona, such structures are relatively stable, because the regions with sparse density will be hotter and the plasma-redshift heating per particle relatively greater than in the denser regions.  The hydrogen would then flow down from the bulge into the center plane, and then outwards into the relatively stable arms of the galaxy, the way Oort saw it.  Due to the tendency of the plasma redshift to create ``bubble'' structures, a relatively hot redshift-heated plasma will separate the arms.  The arms structures of the galaxies have been difficult to understand, but the plasma-redshift heating helps explain the arm structures.

\indent  Presently, this scenario should be considered only a working hypothesis.  However, both the plasma-redshift theory and the reversal of the gravitational redshift with increasing gravitational potential (or reversal of blue shift with decreasing potential) appear helpful in explaining in a simple way these remarkable and difficult to explain observations.


\section{Possible future experiments}
I have failed to conceive of a reliable and practical laboratory experiment for testing the pertinent theorems.   The plasma redshift is likely to play an important role in future designs of fusion experiments, because of its unique feature of transferring heat energy to a fully ionized plasma.  In such experiments, it is important to use the fact that the initial plasma-redshift, as given by Eq.\,(18), is proportional to the photon width $\gamma {\rm{.}}~ $ The source of the light used should therefore be designed to produce large photon widths, for example, by use of light sources at high pressures and high temperatures.  X-ray frequencies are usually needed for exceeding the plasma-redshift cut-off at relatively high densities.  In hydrogen fusion experiments, we could make use of plasma-redshift heating, but it would be difficult to test quantitatively the different relations.  The fusion equipment are usually too small and the plasma often not in thermodynamic equilibrium, which would make it difficult to measure conclusively small plasma redshifts.  I have therefore opted to use different astronomical observations for testing the predictions of the theory.


\subsection{Tests for confirming the plasma redshift}
We can confirm the plasma redshift by observing the shifts of spectral lines of stars as they graze the limb of the Sun during eclipse of the Sun by the Moon.  Each and every spectral line from a star will be redshifted slightly as the line of sight to the star grazes the limb of the Sun.  All the lines will have the natural classical photon widths, because they will all have penetrated an adequate column density of a plasma to obtain this width.  Therefore, all the lines passing at a certain distance from the solar limb will gain the same redshift increment, $\delta z {\rm{,}}$ as they penetrate the solar corona, provided the wavelength is shorter than the cut-off wavelength in the densest part of the corona they penetrate.

\indent   Table 5 shows for a quiescent corona the expected plasma-redshift increment, $\delta z {\rm{,}}$ as a function of the distance, $R/R_0 {\rm{,}}$ where $R_0$ is the solar radius to the plasma-redshift cut-off at $T \approx 5\cdot 10^5~{\rm{K}} $, and $R$ is the shortest distance from the solar center to the line of sight to the star.  


\begin{table}[h]
\centering
{\bf{Table 5}} \, \, Redshift $z$ of light from stars grazing the Sun

\vspace{2mm}

\begin{tabular}{cccccc}
\hline
$R/R_0$ & $z$ & \quad~~$ R/R_0$ & $z$ & \quad~~$ R/R_0 $  & $z$  \\
\hline \hline
1.05 & $3.70\cdot 10^{-6}$ & \quad~~1.30 & $1.30\cdot 10^{-6}$ & \quad~~1.8 & $0.52\cdot 10^{-6}$  \\
1.10 & $2.59\cdot 10^{-6}$ & \quad~~1.40 & $1.02\cdot 10^{-6}$ & \quad~~2.0 & $0.41\cdot 10^{-6}$  \\
1.20 & $1.75\cdot 10^{-6}$ & \quad~~1.60 & $0.70\cdot 10^{-6}$ & \quad~~2.5 & $0.24\cdot 10^{-6}$  \\

\hline
\end{tabular}
\end{table}

\indent  The fluctuations in the electron densities in the corona may require observations of many lines from many stars for the purpose of getting a good statistically valid value.  It is important to select a time period when and where the solar corona is quiescent and without major eruptions.  The effect from intense light scattering from the corona must be reduced as much as possible to allow observations close to the limb.  The wavelengths used must be significantly shorter than the cut-off wavelength, $\lambda_{0.5}{\rm{,}} $ given by Eq.\,(28).  Use of a sharp and narrow focus on the star will reduce the background of the scattered light.  A good experimental design should result in definite confirmation of the plasma redshift.  It is an advantage to use lines that are displaced from the corresponding lines in the Sun.  For example, the lines from the star may be displaced by the Doppler shift produced by the star's peculiar velocity.


\subsection{The gravitational redshift}
In Fig.\,4 the predicted plasma redshifts are compared with the observed solar redshifts.  This comparison shows that the solar Fraunhofer lines are not gravitationally redshifted when observed on the Earth.  The experiments indicate that the gravitationally redshifted photons in the Sun are blue shifted during their travel from the Sun to the Earth.  In these solar redshift experiments the photons had about 8.3 minutes to adjust to the gravitational potential.

\indent   In the laboratory experiments by Pound and Rebka [37-38] and Pound and Snider [39], the photon's travel time from the emitter to the absorber is only about $\delta t_1 = 7.5\cdot 10^{-8}~{\rm{s,}}$ which is thus the maximum time the photon had for adjusting their frequency to the gravitational potential.  This is much shorter than the minimum time, $\delta t_2 = 1.9\cdot 10^{-5}~{\rm{s,}}$ required, according to the uncertainty relation in quantum mechanics.  The length of the photon is about 270 m, which is about 12 times the distance, 22.5 m, between emitter and absorber (detector).  These experiments by Pound, Rebka and Snider do not therefore meet the basic requirements of quantum mechanics.  They do not, therefore, prove the gravitational redshift, because in these experiments the photons had no chance in responding to the gravitational potential difference between the emitter and the absorber.  The experiments only confirm the time dilation difference between the emitting and absorbing nuclei in the gravitational field.  They do not show if the photons are weightless, attracted, or repelled in the gravitational field.  Any such claim is not supported by a proper analysis of the experiments.

\indent   Only the solar redshift experiments are conclusive.  To a distant observer in a field free coordinate system, the solar redshift experiments indicate that the photon's frequency increases, as the photons move out of the gravitational field.  As seen by a distant observer, the gravitational field with a force that is numerically equal to but opposite to the usually assumed attraction repels a photon.  In a local system of reference the photons are weightless.  This is contrary to the commonly held opinion, which assumes that the photon has a weight and is attracted in the gravitational field, as if it had a mass $m_{ph}= h\nu/c^2 {\rm{.}}~$
\vspace{2mm}

 {\textbf{Einstein's two assumptions.}} Einstein showed that it is reasonable to assume that photon's frequency is lower when emitted close to the Sun, because of time retardation or the slowing of all clocks including atomic clocks with a decreasing gravitational potential.  We agree with this assumption and consider that it is well proven fact.  For example, in the experiments by Pound and Rebka, and Pound and Snider [37-39], the experiments measured the frequency shift (or the difference in the time retardation) at the emitter position and the absorber (detector) position.

\indent  In addition to time retardation, Einstein assumed that the emitted photon frequency stays constant as the photon moves from the Sun to the Earth.  This assumption may or may not appear reasonable in classical physics, but it is not reasonable in quantum mechanics.  We disagree with this assumption, because when a photon emitted in the Sun is observed on Earth its frequency is not gravitationally redshifted, as Fig.\,\,4 and the analyses of the many solar redshift experiments in sections 5.6.1\,\,to\,\,5.6.4 show.  This disagreement affects only the assumption that the photon's frequency stays constant as the photon moves through different gravitational potentials.  It affects gravitational redshift of photons with photon lengths much shorter than the distance traveled.  It does not affect the bending of light or the slowing of the speed of light with the gravitational potential, because these effects are independent of the photon's frequency.  The equivalence principle is valid except for photons; see Brynjolfsson [107]

\vspace{2mm}

\indent  The experiments by Vessot et al.~[40] and by Krisher et al.~[41], which have been assumed to prove the gravitational redshift, are invalid as a proof of Einstein's gravitational redshift.  The researchers used continuous transmission frequencies of about $2.2 \cdot 10^9~{\rm{Hz,}}$ and the experiments are therefore in the domain of classical physics.  Even if this continuous train of waves broke up into photons, the photon width would be about the classical width of  $\gamma = 6.266\cdot 10^{-24}\, \omega^2 = 1.2\cdot 10^{-3} ~{\rm{s}^{-1}} {\rm{,}}$ which corresponds to a lifetime of about $\tau = 1/\gamma = 835~{\rm{s.}}$  The value of $\gamma/\omega \approx 8.7 \cdot 10^{-14}  {\rm{}}$ is close to the estimated frequency stability.  The length of such a photon would be about $L \approx 2 \pi c \tau = 1.6 \cdot 10^{14}~{\rm{cm ,}}$ or about 10.5 AU.  These experiments, which used distances much smaller than 10 AU, were clearly in the domain of classical physics and had no chance of observing possible quantum mechanical effects, such as the frequency adjustment to the gravitational potential; see Brynjolfsson [107]
\vspace{2mm}

\indent  {\textbf{Experiments using the 14.41 keV line of $\mathbf{^{57} {\rm{\textbf{Fe.}}}}~$}}  For confirming the blue shift of photons during their time of flight, we could modify the laboratory experiment reported by Pound and Rebka [37, 38] and Pound and Snider [39] by increasing the height difference from 22.5 m to more than about 1000 m.  The 14.4 keV photons from $^{57} {\rm{Fe}}$ have a lifetime of $\tau = 1.43 \cdot 10^{-7}~{\rm{s,}}$ and a photon length of about $L = 2 \pi c  \tau = 270~ {\rm{m.}}~$   For less than 270 m height difference there is practically no reversal of the photon redshift and the photon behaves like a classical photon.  Beyond 270 m, the reversal is only gradual and partial.  When we increase the height difference, the absorption of the 14.4 keV photons becomes too large in any gas including helium over the height of 1000 m.  A large evacuated cylinder between emitter and absorber must then replace the helium filled plastic cylinder in the experiments by Pound et al.~[37-39].  It is also necessary to increase the source strength and the sensitivity of the detectors.  These modifications increase the cost of the experiment significantly.  Use of existing structures, such as, mines, boreholes, caves, or cliffs, for building the large (about 1 km) evacuated steel cylinder, will facilitate the design and reduce the cost.  But the experiments would be costly.
\vspace{1mm}

\indent  {\textbf{Experiments using the 77.3 keV line of $\mathbf{^{197} {\rm{\textbf{Au.}}}}~$}}  We could use higher energy photons and shorter photon lengths.  The 77.3 keV line in gold has a short lifetime, $\tau = 2.73\cdot 10^{-9}~{\rm{s,}}$ and therefore a relatively short photon length, $L = 2 \pi c \tau = 514 ~{\rm{cm.}}~$ However, the short photon length or short lifetime makes the resonance width of the line relatively large, $\Gamma = 2.412\cdot 10^{-7} {\rm{eV.}}~$  The value of $\Gamma/E = 2.412\cdot 10^{-7}/77300 = 3.12 \cdot 10^{-12} {\rm{.}}~ $  The gravitational redshift over a height difference of 300 m is $g h/c^2 = 9.81 \cdot 300/(3\cdot 10^8)^2 = 3.27 \cdot 10^{-14}{\rm{.}}~$  The classical gravitational redshift difference over the height of 300 m is therefore about 1\,\% of the half-width of the line, which is slightly greater than the 0.76\,\% in the experiments by Pound et al.~[37-39].  If the space between the emitter and absorber is filled with pure helium, the absorption in helium will be about 60\%.   It is therefore possible to use the helium filled plastic cylinder, instead of evacuated steel cylinder between the emitter and the absorber.  In the solar redshift experiments the photons had about 8 minutes to change with the gravitational potential.  In these experiments, the photon have only about $10^{-8}~{\rm{s,}}$ to adjust the frequency to the potential.  We don't know how long time it takes the photons to adjust to a new potential.  But we surmise that the blue shift of photons will lag only a few photon lengths behind the actual gravitational potential.  We surmise therefore (without any concrete evidence) that this short time, $10^{-8}~{\rm{s,}} $ in the experiment using gold is adequate to demonstrate a partial blue shift.  It may be adequate to use even a smaller height difference, maybe 30 m, which is reasonable to do as a first attempt.
\vspace{1mm}

\indent  {\textbf{Experiments using the other sources.}}  We could also consider other sources, such as, the 23.87, 27.72, and 73.0 keV gamma rays from $^{119} {\rm{Sn,}}$ $^{129}{\rm{I,}}$ and $^{193}{\rm{Ir,}}$ respectively.

\indent  We could modify the experiments Vessot et al.~[40] by increasing the height and length of the rocket path, and by increasing the frequency used by a factor of about 1,000.


\section{Summary and conclusions}
By using more exact calculations than those usually applied, we have deduced from conventional axioms of physics a new cross section for interaction of photons with a hot electron plasma.  This important cross section, plasma redshift, has been overlooked in the past; most likely, because it is usually insignificant, and because it cannot be detected in conventional laboratory experiments.  The plasma redshift is important only in a very hot and low-density electron plasma.  The cross section is given by Eq.\,(18).  In this equation, the oscillator strength function, $F_1 \left( a \right),$ is given by Eq.\,(14), and the numerical values are shown in Table 1.  The plasma redshift is small for large values of $a = \hbar \omega_p / \left( \beta _0 \omega_0kT \right) = 3.65 \cdot 10^5 \lambda_0\sqrt{N_e}/T {\rm{,}}$ but becomes significant as $a$ decreases.   We have that for $a$ equal to 0.344, 1.163, and 2.671 the oscillator strength function is 90\%, 50\%, and 10\%, respectively, of its full value.  For $a \geq 3.633 {\rm{,}}$  the oscillator strength function  $F_1 (a) $ has even small negative values, that is small blue shifts.  These small negative values (blue shifts) are usually not important.   Instead of the parameter $a$, we can use the cut-off wavelength defined by Eq.\,(15).  The 50\% cut-off wavelength is given by Eq.\,(16) and when magnetic fields are significant by Eq.\,(28), which gives the relation between the cut-off wavelength for the plasma redshift, the electron temperature, electron density, and the magnetic field.

\indent  The plasma redshift of photons results in transfer of very small quanta to the plasma.  This does not change the direction of the photons significantly.  The photon's energy loss by plasma redshift is in some respect analogous to a charged particle's energy loss by Cerenkov radiation.  The energy lost by the photons is immediately absorbed in the electron plasma and results in significant heating.  As shown in sections 5.1 to 5.5, the plasma-redshift heating together with the magnetic heating contributes significantly to the heating of the solar corona and to the heating responsible for many observable and interesting phenomena in the solar atmosphere.

\indent  In section 5.1, we use Eq.\,(28) to show that the cut-off wavelength for plasma redshift of photons predicts well the onset of the plasma-redshift heating in the transition zone to the solar corona.  The 50\,\% cut-off wavelength is about 500 nm when the temperature is about 500,000 K, the electron density about $N_e  = 10^9 \; {\rm{ cm}}^{ - 3} ,$ and magnetic field less than 10 gauss.  These values correspond to about the middle of the transition zone to solar corona.  For shorter wavelengths, the cut-off wavelength reaches deeper into the transition zone, while for the longer wavelengths, the cut-off reaches higher.  The cut-off zone reaches deeper into the transition zone as the magnetic field increases.  For example, the cut-off wavelength in the above example increases by 130\% when the magnetic field increases from 10 to 100 gauss.

\indent  In section 5.1, we also show how the plasma redshift together with transformation of magnetic field to heat, as described in Appendix B, facilitates explanation of the spicules in the transition zone to the corona.  Without the plasma-redshift theory, the formation of the spicules has been difficult to explain. 

\indent  In sections 5.2, 5.3, and 5.4, about the solar corona, the solar wind, and the solar streamers, we find that the plasma-redshift theory together with the theory for conversion of magnetic field to heat, as described in Appendix B, is consistent with many observations that have been difficult to explain.  In section 5.2, we show that above the cut-off zone, the plasma redshift exceeds the emission cooling and the excess heat leaks by conduction into the lower transition zone.  Gradually, the gravitational cooling by the solar wind increases relative to the plasma redshift heating; and it exceeds the plasma redshift heating and results in maximum temperature at about 2 solar radii.  The repulsion of the diamagnetic moments, as described by Eq.\,(B10), reduces this gravitational cooling by the solar wind.  At about 5 solar radii, the magnetic repulsion force exceeds the gravitational attraction. This results in an outward acceleration of the solar wind.  Due to Eq.\,(B11), this kind of acceleration of the heavier ions, such as helium ions, can sometimes be greater than that of the protons.  The plasma redshift transfers its energy to the electrons.  This heat energy causes the electron temperature often to exceed the proton temperature and increases the solar wind as shown in section 5.3.  The explanation of solar streamers, described in section 5.4, is related to the explanation of the spicules in section 5.1.  In addition to the plasma redshift these explanations make use of magnetic repulsion of the diamagnetic moments.

\indent  In section 5.5, we demonstrated that the plasma-redshift theory is helpful in explaining solar flares.  It is shown that for large magnetic fields, the plasma redshift can be initiated even deep in the chromosphere.  The plasma redshift heating can then also initiate conversion of the magnetic field energy to heat.  The heating that results from plasma redshift and the conversion of the magnetic field energy to heat can then initiate many hitherto unexplained phenomena, including the flares, loops and arches.

\indent  In sections 5.6.1, we compare the theory with the solar redshift experiments.  We find that the plasma redshift explains significant fraction of the observed redshifts of the solar Fraunhofer lines.  It leads thereby to the conclusion that the solar lines are not gravitationally redshifted when observed on the Earth.  For the evidence, see for example Table 3 and Fig.~4.

\indent  In section 5.6.2, we discuss the gravitational redshift and how the observations lead to a quantum theory for reversal of the gravitational redshift.  It is concluded that while the photons in the solar spectrum are gravitationally redshifted when in the Sun, as predicted by Einstein's classical TGR, the photons are usually not gravitationally redshifted when they arrive on the Earth.  The photons' frequencies are blue shifted during photons' time of flights from the Sun to the Earth, somewhat analogously to the frequencies of atoms when the atoms move from the Sun to the Earth.  This does not conflict with any of the experiments [37-41] that have been assumed to prove the gravitational redshift, because these experiments, due to quantum effects, were unable to detect if the photons were attracted or repulsed by the gravitational field.  The solar redshift experiments can discern if the photons are attracted or repulsed, and these experiments show clearly that the gravitational redshift is reversed as the photons move outwards from the Sun.

\indent  In section 5.6.3, we focus on the comparison of the present and conventional explanations of the solar redshift.  While the conventional theory often makes similar prediction to that of the present theory, there are a few crucial observations, which contradict the predictions of the conventional theory, but confirm those of the plasma-redshift theory.

\indent  In section 5.6.4, we mention that as in case of the solar corona, the plasma redshift appears to be able to explain in a reasonable way the relatively large redshifts of many bright stars, including the redshifts of collapsars and quasars.  The details of these explanations, however, were not perused in the present paper, because the extrapolations from solar corona appear plausible.  Detailed calculations are necessarily elaborate, and require often extrapolation from known facts.  Such extrapolations introduce uncertainties.  For example, we don't know much about the intensity and structure of the magnetic field in quasars.  We also don't know if the coronas of quasars are isotropic.

\indent  In section 5.7, we find that the plasma redshift can explain the observed corona of the Milky Way and the coronas of other galaxies.  It leads also to hot intergalactic plasma filling the intergalactic space, as the kinetic energies of the particles in the coronal plasmas exceed the gravitational potential energy of the particles.

\indent  In section 5.8, we show that the plasma redshift can explain the entire cosmological redshift, if the average electron density in intergalactic space is
\be
\left( {N_e } \right)_{av}  = \left( {N_{ep} } \right)_{av}  \approx 1.95 \cdot 10^{ - 4} \left(\frac{{H_0}}  {{60}} \right)\,~{\rm{ cm}}^{ - 3}.
\ee
The agreement between the theory and experiments is good; see Figs.\,(5) and (6).  The recently discovered dimming of distant supernovae is predicted well by the plasma-redshift theory.  No ``dark energy'' or ``dark matter'' are needed to explain the observations.

\indent  In section 5.9, we show that experimental data for the high-redshift supernovae indicate that there is no time dilation, and that the universe therefore is quasi-static.

\indent  In section 5.10 and Appendix C, we show that the plasma-redshift theory can also explain the cosmic microwave background (CMB).  It is found that when the intergalactic space is filled with electron-proton plasma with density equal to that required for explaining the cosmological redshift, the isotropic and well-defined CMB follows as a necessary consequence.  This hot plasma with a density $N_e = 1.95 \cdot 10^{ - 4}~{\rm{cm}}^{-3} $ and a thermal particle temperature of $T=2.706 \cdot 10^6~{\rm{K}} $ leads to the isotropic CMB with a well-defined blackbody temperature of $T_{CMB}=2.728~{\rm{K}}.~ $  The CMB is dominant between $10^9 ~{\rm{and}}~ 1.7\cdot 10^{11} ~{\rm{Hz .}}~ $   The cosmic spectrum below and above these limits is also consistent with the predictions of the plasma-redshift theory.

\indent  Radiations beyond CMB produce additional pressure on the wall of the blackbody ``cavity''.  Corresponding to these additional radiations, the average particle temperature per particle in intergalactic space exceeds the temperature $T=2.706 \cdot 10^6~{\rm{K}}{\rm{.}} ~$ The average particle temperature is then
\be
 {T}_{av} \approx (3.0\;\; \rm{to} \;\; 3.6) \cdot 10^6 \left(\frac {{60}}{{H_0 }} \right)\,~{\rm{ K}}.
\ee
The upper limit of $T_{av}$ includes contributions from the CMB, the X rays, and the intergalactic light.  

\indent The corresponding average density in intergalactic space is estimated to be about
\be
 {\rho}_{av} \approx 3.8 \cdot 10^{ - 28} \left(\frac{{H_0}}{{60}} \right)\,~{\rm{ g}}\,{\rm{cm}}^{ - 3}.
\ee
This average density in the universe is then about 56 times higher than the density usually assumed for a closed universe when $H_0 \approx 60 ~{\rm{km}}\,{\rm{s}}^{ - 1}\,{\rm{Mpc}}^{ - 1}$.  The plasma redshift leads thus to a much higher density in intergalactic space than that conventionally assumed.

\indent  In section 5.11 and in Appendix C, we show that the high density and temperature of the intergalactic plasma and in the Galactic corona explain well the observed X-ray background.  Some experts in the field will object that the high density and high temperature in space will result in an X-ray intensity that is too high.  Closer scrutiny shows, however, that the predicted X-ray intensity is consistent with observations.  This is because the redshift length (the inverse of plasma-redshift absorption) is many times shorter than the usually assumed absorption length (the inverse of the conventional X-ray absorption coefficient).  The integral of the X-ray luminosity degraded by the Compton scattering and the plasma redshift over the redshift distance results in modest X-ray intensity.  This soft X-ray intensity is nevertheless significant for maintaining uniform ionization in space, and it helps ionize the corona of galaxies and other objects.  This soft X-ray background from intergalactic plasma does not contradict the observations that find that most of the harder X rays from intergalactic space are due to point sources. 

\indent  The softer X-ray intensity is due mostly to relatively colder filaments between huge bubbles heated by the plasma redshift.  Like in the transition zones to the solar and galactic coronas, the bubble structure is due to the fact that the plasma redshift is first order process in density, while the cooling processes are usually second order in density.  Some of the slightly harder X rays are produced in the hot bubbles.  The harder X rays are also emitted from the hot dense plasma in large galaxy clusters and from other point sources.  We show that the X rays from intergalactic space are consistent with the observations.

\indent  In section 5.12, we discuss the high average density in the universe.  Some experts will also object that the high density in space will lead to too large a curvature of space.  Weightlessness of photons in a local standard reference system means that Einstein's field equations will have to be modified.  This modification is most likely small, although fundamentally very important.  However, Einstein's extrapolation of Newton's phenomenological equations for gravitation to very large distances and even to infinity has no experimental support and no support from conventional quantum mechanical physics.  It is most likely incorrect.  Distant collisions by hot particles with the proton particles (and other particles) will change the angular momentum and the gravitational energy transfer to the proton, before the gravitational potential change can assert itself, as the hot proton rotates and moves around.  Quantum mechanical theory requires a finite time for the proton to interact with the gravitational field.  Even with the high densities of Eq.\,(64), the space could be flat isotropic and infinite, which is consistent with observations.

\indent  In section 6, we show that the reversals of the gravitational redshifts of photons as they move out of solar gravitational field are supported by other important observations.  It was concluded in sections 5.6.2 and 5.6.3 that the plasma redshift could explain the observed redshift of solar Fraunhofer lines without the gravitational redshift, which was expected from the classical theory of general relativity (TGR).  It was concluded that the photons' frequencies are gravitationally redshifted in the Sun; but as the photons move from the Sun to the Earth the frequencies are blue shifted, which reverses their gravitational redshift.  The present quantum mechanically modified TGR leads to reversal of photons gravitational redshift.  This unexpected result of the plasma redshift contradicts the fundamental assumption in physics and in TGR that a photon's frequency is a constant of motion as the photon moves through gravitational fields, or that the frequency of a photon stays constant as the photon moves from one gravitational potential to another.  Conventional theory assumes that all other forms of mass and energy are redshifted, as the corresponding particles move to lower potential.  The equivalence of mass and kinetic energy in the special theory of relativity means that the inertial mass of the photon is $m_i = {h\nu }/c^2 .~$ According to equivalence principle, the gravitational mass $m_g = m_i$ and the photon should be attracted by the gravitational field.  Section 6 serves to show that, independent of the solar redshift experiments, the hydrogen streaming away from the galactic center supports the contention that the photons redshift is reversed, as the photons move out of the gravitational field.

\indent  It is argued that the reversal of the gravitational redshift removes the need for black holes, which are fictions created by extrapolating Newton's phenomenological equation far beyond their experimental foundation.  It is more likely that matter, instead of being sucked into a black hole, is transformed to photons at the brink of a black hole.  (This brink of a black hole may possibly consist of a vortex at the center of a large, fast spinning, and flattened neutron star-like object, which most likely has large magnetic fields, possibly sometimes on the order of or even exceeding $10^{15}$ gauss.)  The photons gravitational redshift is then reversed and the photons can reform matter, such as electron-positron pairs and proton-antiproton pairs and other particles at a distance in the usual way known from the laboratory experiments.  This extrapolation requires more experimental and theoretical support.

\indent  The experimental evidence for the correctness of this extrapolation is found in the intense positron annihilation spectrum detected close to the galactic centers, and in the large amount of hydrogen observed streaming away from the Milky Way center.  The reversal of photon's gravitational redshift and transformation of a matter at the centers of galaxies (and quasars) bring a self-regulating stability into our cosmological perspective.  Although it was not discussed in section 6 (because the evidence is not clear), the weightlessness of photons in a standard system of reference may also facilitate explanation of the large amount of energy released in supernova explosions.

\indent  The plasma redshift appears to eliminate five major deficiencies in Einstein's cosmological model for a static universe:

\begin{enumerate}
\item Plasma redshift can explain the cosmological redshift.
\item Plasma redshift can explain the cosmological microwave background.
\item Plasma redshift resolves the Olbers' paradox.  If starlight were not attenuated, as it traveled through intergalactic space, the sky would be bright as the stars in an infinite universe.  The attenuation of the light intensity by the plasma redshift of light by intergalactic electrons resolves this problem.
\item Einstein's cosmological model has significant instability, which is caused by the tendency of matter to concentrate due to gravitational attraction.  Plasma redshift, when compared with solar redshift, leads to reversal of photons gravitational redshifts and to the renewal of matter at the center of galaxies and quasars.  The eternal renewal of matter removes this gravitational instability.
\item In Einstein's static model of the universe, the stars will run out of energy and will have a finite lifetime.  Plasma theory leads to reversal of photons gravitational redshifts.  A reasonable extrapolation of that finding is that matter is eternally renewed at the centers of galaxies and quasars.  As shown in section 6, the observations support this extrapolation. 
\end{enumerate}

\indent  Plasma redshift, which is based on basic axioms of physics, leads thus to fundamental changes in our cosmological perspective and to changes in gravitational theory. 

\indent The problem of ever-increasing time and ever-increasing entropy is resolved when we realize that we are usually observing only one half of the material-photon cycle.  We usually focus on the physical changes from particle creation of material through its changes (which define the time) towards annihilation, while often disregarding the other half of the time cycle, the creation of photons and their transformation to matter in an everlasting renewal process at the centers of the galaxies, and most likely at the centers of quasars.
\vspace{2mm}


\noindent  \textbf{Acknowledgements}
I am indebted to my colleague Dr. Chia-P. Wang, Weston, MA for his comments when reviewing this paper.


\renewcommand{\theequation}{A\arabic{equation}}
\setcounter{equation}{0}
\section*{Appendix A}


\section*{A1\, \,Electromagnetic Waves in Dielectrics
}

We consider a homogeneous and isotropic medium with a dielectric constant, $\epsilon ,$ and a permeability, $\mu .$  Initially, these material constants do not vary with the coordinates nor with time.  When using Gaussian (cgs) system of units, we get from Maxwell's electrodynamic theory that 
\be
{\mathbf{\nabla}} \times {\mathbf{E}} = - \frac{{1}}{{c}} \, \frac{{\partial{\mathbf{B}} }}{{\partial{t} }} = - \frac{{\mu}}{{c}} \, \frac{{\partial{\mathbf{H}} }}{{\partial{t} }}
\ee
\be
{\mathbf{\nabla}} \cdot {\mathbf{D}} = {\mathbf{\nabla}} \cdot \epsilon \, {\mathbf{E}} = 0
\ee
\be
{\mathbf{\nabla}} \times {\mathbf{H}} = \frac{{1}}{{c}} \, \frac{{\partial{\mathbf{D}} }}{{\partial{t} }} = \frac{{\epsilon}}{{c}} \, \frac{{\partial{\mathbf{E}} }}{{\partial{t} }}
\ee
\be
{\mathbf{\nabla}} \cdot {\mathbf{B}} = {\mathbf{\nabla}} \cdot \mu \, {\mathbf{H}} = 0
\ee
\noindent  where the dielectric constant, $\epsilon ,$ and the permeability, $\mu ,$ depend only on the frequency and where c is the velocity of light. 
\vspace{2mm}

\parbox{5.5in}{ {\textbf{Comment A1.}} We can obtain the corresponding equations in the mks (rationalized) system of units by replacing $ \epsilon $ with $  {\epsilon} {\,} ( {\epsilon_0} {\,} c) , $ and $ {\mu} $  with $ {\mu} {\,} ({\mu_0} {\,} c) , $ where $ {\epsilon_0} $ and $ \mu_0 $ are the dielectric constant and permeability in vacuum, and where $ \epsilon_0 \, \mu_0 = 1 / c^2 . $    }

\vspace{2mm} 
  
We assume that the field varies sinusoidally as the real part of ${\rm{exp}}\, (i\, \omega \, t) .~$ For facilitating the calculations, we use complex notations for the different quantities; and in the usual manner, we use their modulus for comparison with experiments. 

\indent We write the general solutions to these equations on the form:

\be
E_y = \frac {{A(\omega)}}{{ \epsilon \, \sqrt{\mu} }}\, {\rm{exp}} [ i\,\omega (t - x \, \sqrt{\epsilon \, \mu}/c ) ] ,
\ee
\be
E_x = E_z = 0 
\ee
\be
H_z = \frac {{A(\omega)}}{{ \mu \, {\sqrt{\epsilon}}  }}\, {\rm{exp}} [ i\,\omega (t - x \, \sqrt{\epsilon \, \mu}/c ) ] , 
\ee
\be
H_x = H_y = 0
\ee
\vspace{2mm}

\parbox{5.5in}{ {\textbf{Comment A2.}} It is possible (as is often done) to set $B(\omega) = A(\omega) /\sqrt{\epsilon \, \mu}.~$   The coefficients in front of the exponential factors in the equations above would then be  ${B(\omega)}/{ \sqrt{\epsilon } } \quad {\rm{and}} \quad {B(\omega)}/{ \sqrt{\mu } }$, respectively.   Use of these coefficients leads to the conventionally used solutions and usually to a misleading assumption about the variations of the fields with $\epsilon $ and $\mu .~$  The forms of the coefficients $A(\omega) /(\epsilon \, \sqrt{\mu}) $ and $ A(\omega) /(\mu \, \sqrt{\epsilon }),$ in front of the exponential factors in Eqs.\,(A5) and (A6) are mathematically correct and physically simpler to interpret, because $A(\omega)$ is independent of $\epsilon $ and $\mu .~$  For example, for $\mu = 1 ,$ we have that ${\mathbf{\nabla}} \cdot {\mathbf{D}} = 4 \pi \rho ,$ where  $\epsilon \, {\mathbf{E}} = \mathbf {D} .$ Therefore, when we put a dielectric material around a charge, the vector $\mathbf D$ is unchanged; that is, the quantity $ A(\omega )$ is a constant when $\epsilon $ changes. }


\section*{A2\, \,Fourier Spectrum of Photons in Dielectrics}

We think of an atom free of external forces emitting a photon as it decays exponentially from an excited state with a lifetime of $\tau  = 1/\gamma {\rm{.}}$  The energy difference between the two states corresponds to a cyclic frequency $\omega _0  = 2\pi \nu _0 {\rm{.}}$  Using a gaussian (cgs) system of units, we have in a homogeneous electron plasma with the permeability $\mu  = 1{\rm{,}}$ and the dielectric constant $\varepsilon {\rm{,}}$ that electrical field $E_y$ and the magnetic field $H_z$ of a photon moving in the x-direction can be Fourier analyzed and we get at the point $(t, x) = (t, 0)$ that
\be
E_y \left( {t,\;\varepsilon } \right) = \int\limits_{ - \infty }^\infty{\frac{{E_y \left( {0,\;1} \right)}}{{\;2\pi \varepsilon \left( {\gamma/2 + i\left( {\omega  - \omega _0 } \right)} \right)}}\exp \left( {i\omega t} \right)d\omega },
\ee
and for the magnetic field, we have analogously that
\be
H_z \left( {t,\;\varepsilon } \right) = \int\limits_{ - \infty }^\infty{\frac{{E_y \left( {0,\;1} \right)}}{{\;2\pi \sqrt \varepsilon  \left( {\gamma /2 + i\left( {\omega  - \omega _0 } \right)} \right)}}\exp \left({i\omega t} \right)d\omega }.
\ee


\section*{A3\, \,Equation of Motion for Electrons}

In the following we will consider a simple case of an isotropic and uniform plasma without any constant magnetic field and with permeability $\mu  = 1.~$  (The effect of magnetic fields is considered in section 4 of the main paper.)  

\indent  The classical dynamical equation of motion in Gaussian (cgs) units is at the position of $x = 0$ usually approximated by (see Eq.\,(21-16) of Panofsky and Phillips [119] and Eq.\,(15.1) of Becker [120])
\be
m {\ddot r} - m {\beta}_{0} \dot {\ddot r} + m {\omega}_q^2 r = e {E} \exp \left( {i {\omega} t} \right),
\ee
\noindent  where the radiation damping constant is $\beta_0 = 2 e^2/(3 m c^3) = 6.266 \cdot 10^{-24}$ and $\dot {\ddot r} = - {\omega}^2 {\dot r} ,$ and where $\dot{r} ,$ $\ddot r ,$ and $\dot {\ddot r}$ are the first, second, and third time derivative of the complex radius $r$ in the oscillation of the electron with charge $e$ and mass $m$ in the electrical Fourier field harmonic with electrical field modulus ${E}$ and the frequency $\omega {\rm{.}}~$

\indent Our focus is the dynamical equation in plasma.  Instead of this conventional equation Eq.\,(A11), the equation of motion for a plasma electron acted upon by the electrical field's Fourier component, $\left( {{A \mathord{\left/{\vphantom {A \varepsilon }} \right.
 \kern-\nulldelimiterspace} \varepsilon }} \right){\rm{exp}}\left( {i\,\omega t} \right)$, may be approximated by
\be
m {\ddot r} +  m \alpha {\dot r} -  m \beta_p \dot{\ddot r} + m {\omega}_q^2 r = e \,\frac{{A}}{{\varepsilon }} \, \exp \left( {i {\omega} t} \right),
\ee
\noindent  where the first term on the left side Eqs.\,(A11) and (A12) is the acceleration of the electron with charge $e$ and mass $m$ in the field $E\, {\rm{\exp}} \left( {i {\omega} t} \right)= ( A/\epsilon )\, {\rm{\exp}} \left( {i {\omega} t} \right) ,$ on the right side.  The second term on the left in Eq.\,(A12) accounts for the collision damping and ${\alpha} = 2/{\tau},$ where $\tau$ is the time between collisions.  This collision damping is very large in a hot plasma, but it is often omitted in the conventional Eq.\,(A11).  The third term on the left in Eq.\,(A12) accounts for the emitted radiation by the electron when it is accelerated in the external field on the right side.  We call it the radiation damping term.  We note it by $ m \beta_p \dot {\ddot r} $ rather than $ m \beta_{0} \dot {\ddot r} $ to make it clear that $ \beta_p $ could deviate from $ \beta_{0} = 2 e^2 /(3mc^3 ) .~$  If a field with only the frequency $\omega_0$ acts on the electron, we have that $ m \beta_p \dot {\ddot r} = m \beta_{0} \omega_{0}^2 {\dot r} .~$  But if several frequencies of the field act on the electron simultaneously, then $ m \beta_p \dot {\ddot r} $ can deviate from $ m \beta_{0} \omega_0^2 {\dot r}.~ $  This is especially important in hot plasmas.  In the following, we will often combine the second and the third term and use the notation $\beta \omega^2 $ for $ (\alpha + \beta_p \omega^2) .~$ Later on we show that we can add the collision term to $ m \beta_p {\dot {\ddot r}},$ and replace it by $ m \beta \dot {\ddot r} = m \beta \omega^2 {\dot{r}} .~$  The collision field can be replaced with a the Fourier harmonics of the fast moving electrons in the hot plasma.  These Fourier fields will cause the plasma electrons to oscillate and lose the radiation energy in a similar way as the incident photon field on the right side of Eqs.\,(A11) and (A12).  The fourth term, $m {\omega}_q^2 r ,$ accounts for any ``elastic'' force that binds the electron to a certain equilibrium position.  This is an eigenstate and because it is assumed to be stable, it will not result in additional radiation damping unless acted on.  This binding of the electron is characterized by its ``eigenfrequency'', $\omega _q {\rm{.}}~$ 

\indent  The validity of Eqs.\,(A11) and (A12), specifically the radiation damping term, the third term of (A12), has often been questioned in the literature.  The problems raised can be traced to the fact that we do not have a reliable model of the electron structure [119 - 125].

\indent  As shown by Dirac [124] already in 1938, it is reasonable to assume that this form of the equation is valid, as it leads to correct quantum mechanical results.  We can expect deviations as shown by Hartemann and Kerman [125], when the radiation intensity becomes extremely high or when the wavelength approaches the classical electron radius.  These limits are well beyond the application of intensities and photon energies in focus of the present article.  

\indent In general, the force, $F\left( {\omega ,\,\varepsilon } \right){\rm{,}}$ acting on an electronic oscillator differs from the average field,  which is $ \left( {{A \mathord{\left/  {\vphantom {A \varepsilon }} \right.  \kern-\nulldelimiterspace} \varepsilon }} \right)\exp \left( {i\,\omega t}\right){\rm{.}}$  We have more generally that
\be
F\left( {\omega ,\;\varepsilon } \right) = \frac{A}{\varepsilon }\exp \left( {i\omega t} \right) + k4\pi P\left( \omega  \right),
\ee
where $P\left( \omega  \right) = N_e {\kern 1pt} e\,r\left( \omega  \right)$
is the polarization when $N_e$ is the electron density and $r\left( \omega  \right)$ the solution of Eq.\,(A11).  In an amorphous isotropic matter, we can usually set the shielding factor $k = 1/3$.  More generally this shielding factor, $k$, is a complex tensor.  In a fully ionized plasma without magnetic fields, we have that $\omega _q  = 0{\rm{,}}$ and $k = 0$.  The effects of magnetic fields complicate the calculations.  Their effects are treated in section 4 of the main paper. 

\indent When the electron is bound, we can set the collision damping factor $\alpha  = {2 \mathord{\left/{\vphantom {2 {\tau {\rm{,}}}}} \right.
 \kern-\nulldelimiterspace} {\tau {\rm{,}}}}$ where $\tau$ is the time between collisions.  The collisions with electrons and ions together with the collisions with neutral atoms are important for determining the width of the emissions and absorption lines.  In fully ionized plasmas, where $\omega _q  = 0{\rm{,}}$ the collision fields of the free electrons and ions can be Fourier analyzed and can be considered a part of the external fields affecting the third term in Eq.\,(A12).  We can then set $\alpha  = 0{\rm{,}}$ and replace the classical damping term, $\left( {\alpha  + \beta _p \omega ^2 } \right){\rm{,}}$
by the quantum mechanical damping term, $\beta \omega ^2 {\rm{,}}$ which in hot plasmas is much greater than the classical radiation damping, $\beta _p \omega ^2 {\rm{,}}$ as shown in section 3.

\indent In Eq.\,(A11), we have neglected the magnetic field force
\be
\frac{{\dot r}}{c}\,e \,\frac{A}{{\sqrt \varepsilon  }}\,\exp \left( {i\omega t} \right),
\ee
because it is very small for low-energy photons of light, and because in the first approximation it does not perform any work.  Its main component is at right angle to the velocity, $\dot r{\rm{,}}$ along the incident electrical photon field, and at right angle to the magnetic photon field $H_z .~$  Its main component is thus opposite to the photon's pressure on the electron.  This force is thus responsible for the loss and change of momentum of the photon, as it is deflected and its energy attenuated.

\indent  It is important to include $\epsilon $ in the denominator on the right side of Eq.\,(A12).  In the customary treatment, which uses Eq.\,(A11), the Fourier harmonics are usually obtained assuming that $\epsilon = 1 $ (see Eq.\,(21-16) in Panofsky and Phillips [119] or Eq.\,(15.1) in Becker  [120]).  Eq.\,(A11), without the $\epsilon$ in the denominator on the right side, leads to a Compton scattering cross section being equal to $\sigma_0 = 8\pi r_0^2 / 3 ,$ where $r_0$ is the electron radius (see Eqs.\,(21-3) to (21-27) in Panofsky and Phillips [119]) instead of $\sigma_0 = 8\pi r_0^2 / [3 (1+\beta_0^2 \omega_0^2)] .~$  We show later that in this case we usually can set $\beta_p = \beta_0 .~$  The classical term $\beta_0^2 \omega_0^2$ is usually small and insignificant.  It is of interest to note that the quantum mechanical treatment by Heitler [1] leads to the correct form $\sigma_0 = 8\pi r_0^2 / [3 (1+\beta_0^2 \omega_0^2)] .~$  It follows from the finite damping width of the photon.  This form corresponds to the last term within the brackets in Eq.\,(A34) in the following.  When the dielectric constant and the collision damping are very large, Eq.\,(A12) results in three additional terms within the brackets of Eq.\,(A34).
\vspace{2mm}

\parbox{5.5in}{ {\textbf{Comment A3.}}  {\it{The right side of Eq.\,(A12) differs from Eq.\,(A11) by the dielectric constant $\epsilon $ in the denominator}} on the right side, where $A = D$ and $D$ is the displacement field, and $A/\epsilon$ is the modulus of the Fourier harmonics of the photon field.  When we solve Maxwell's equation, we should select $A/(\epsilon \sqrt{\mu})$ as the solution and not the form $E ,$ and we should select $A( \mu \sqrt{\epsilon})$ and not the form $H .~$  This is for the purpose of indicating how the photon's Fourier field varies with the dielectric constant and the permeability.  In a plasma, we initially set $\mu$ equal to 1.  The $D$-field (for example, the displacement field from an electron) is for $\mu = 1$ independent of the dielectric constant so that the $E$-field varies like $ D/\epsilon = A/\epsilon = E .~$  This form, $A/\epsilon ,$ shows clearly how $E$ varies with the dielectric constant.  Unfortunately, these solutions to Maxwell equation are usually not among the forms selected in the standard literature.  That fact has often lead to difficulties and misleading equations in the conventional literature, even in high quality references such as Panofsky and Phillips [119] and Becker [120].  This correction is crucial for deriving the plasma redshift.  Without the factor $1/\epsilon$ on the right side of Eq.\,(A12) there is no plasma redshift.}
\vspace{2mm}

\noindent  The principal solution to Eq.\,(A12) is
\be
r = \frac{e}{m}\frac{{A/\varepsilon }}{{\left\{ {\omega _q^2  - \omega ^2  + i\left( {\alpha  + \beta_p \omega ^2 } \right)\omega } \right\}}}\exp\left( {i \omega t} \right).
\ee
\noindent  To this solution, we may add the solutions of the homogeneous differential equation, that is, the solutions
\be
\Delta r = \left( {C_1 /\alpha } \right) + \left[ {C_2 \exp \left( { - k_1 t} \right) + C_3 \exp \left( {k_1 t} \right)} \right]\exp \left( {t/2 \beta_p } \right),
\ee
where
\be
k_1  = \sqrt {\frac{\alpha }{\beta_p } + \frac{1}{{4\beta_p^2 }}} .
\ee
However, we can disregard these last mentioned solutions, which have been discussed by Dirac [124].

\noindent Eq.\,(A15) may be compared with Eq.\,(21-19) of Panofsky and Phillips [119], but take note of the subtle differences.
\vspace{2mm}

\parbox{5.5in}{ {\textbf{Comment A4.}} In their Eq.\,(21-19) of [119], which corresponds to Eq.\,(A15) above,  Panofsky and Phillips do not have any collision term and not the factor $1/\epsilon .~$  The same applies to Eq.\,(15.2a) by Becker [120].  But in his Eqs.\,(15.6) to (15.10) of [120], Becker considers the collision damping but disregards $1/\epsilon $ in the numerator.  When the conventional approximations of references [119] and [120] are applied to a hot sparse plasma, the approximations they used prevent us from discovering the plasma redshift.  We emphasize that Eq.\,(A12) and (A15) are the correct forms as they are more exact than the conventional equations in the literature.  These correct forms are essential for deduction of the plasma redshift.}
\vspace{2mm}

\noindent The polarization is given by
\be
P\left( \omega  \right) = N_e \,e\,r,
\ee
where $N_e$ is the number of plasma electrons per ${\rm{cm}}^{ 3 }$, $e$ the electronic charge, and where $r$, the displacement of each of the electrons, is given by Eq.\,(A15).  The dielectric constant is defined as
\be
\varepsilon \left( \omega  \right) = 1 + 4\pi \frac{{P\left( \omega 
\right)}}{{\left( {A/\varepsilon } \right)\exp \left( {i \omega t} \right)}}.
\ee
When in this expression for the dielectric constant, we insert Eqs.\,(A15) and (A18), we get
\be
\varepsilon  = 1 + \frac{{4\pi N_e e^2 /m}}{{\left\{ {\omega _q^2  - \omega ^2  + i \left( {\alpha  + \beta_p \omega ^2 } \right)\omega } \right\}}} = 1 + \frac{{\omega _p^2 }}{{\left\{ {\omega _q^2  - \omega ^2  + i\left( {\alpha  + \beta_p \omega ^2 } \right)\omega } \right\}}},
\ee
where 
\be
\omega _p  = 2\pi \nu _p  = \sqrt {\frac{{4\pi N_e e^2 }}{m}}  = 5.642 \cdot 10^4 \sqrt {N_e } {\rm{,}}
\ee
is the cyclic plasma frequency. 

\indent If we write the complex dielectric constant on the form $\varepsilon  = \left( {n - i \cdot \kappa } \right)^2 {\rm{,}}$ we get from Eq.\,(A20) that
\be
\frac{{2n\kappa \omega }}{{\varepsilon \bar \varepsilon }} = \frac{{\left( {\alpha  + \beta_p \omega ^2 } \right)\omega _p^2 \omega ^2 }}{{\left( {\omega _q^2  + \omega _p^2  - \omega ^2 } \right)^2  + \left( {\alpha  +\beta_p \omega ^2 } \right)^2 \omega ^2 }} = \frac{{ \omega _p^2 \,\beta \,\omega ^4 }}{{\left( {\omega _q^2  + \omega _p^2  - \omega ^2 } \right)^2  + \beta^2 \omega ^6 }}{\rm{.}}
\ee
\noindent where in the last form, we have replaced the damping constant $(\alpha + \beta_p \omega^2)$ by $\beta \omega^2 .~$  We have deduced this form for later use in calculating the attenuation of the photons in a plasma.    The value of $\beta $ can be very large in a hot plasma.  We show later that $\beta$ is on the order of  $\left( 3 k T / \hbar \omega_p \right) \beta_0 {\rm{.}}~$

\vspace{2mm}

\parbox{5.5in}{ {\textbf{Comment A5.}}  The form of the dielectric constant in Eq.\,(A20) above is similar to the form of the dielectric constant given by Becker [120] (see Eqs.\,(25.5) and (25.9) of that source), except that Becker uses the form $\gamma = 2 e^2 \omega^2 / (3mc^3)$ instead of $[\alpha + \beta_p \omega^2 ] \approx [\alpha + 2 e^2 \omega^2 / (3mc^3)]$ in the denominator of Eq.\,(A20).  (We draw attention to that Eq.\,(25.6) of Becker [120] applies to ponderable matter and does not apply to plasmas.)  Becker then proceeds to integrate over the Fourier harmonics as if $\gamma$ was small and a constant $\gamma = 2 e^2 \omega_q^2 / (3mc^3),$ (see his Eq.\,(26.8) of Becker [120]).  That is he disregards the solutions that could lead to the plasma redshift and the scattering on the plasma frequency.  Panofsky and Phillips do the same in reference [119] (see Eq.\,(21-26) of that source).  We, on the other hand, will integrate over the Fourier harmonics as if the damping factor, the imaginary part, of the denominator of Eq.\,(A20) could be large and that the damping constant varies with the frequency $\omega$ as $ \beta \omega^2 = (\alpha + \beta_0 \omega^2 ),$ because only by using the exact form for the damping of the electrons and the exact integration can we obtain the cross section for the plasma redshift.  In the conventional literature on plasmas it is even customary to disregard altogether the imaginary part in Eq.\,(A20) by assuming that $\epsilon = 1 - \omega_p^2 / \omega^2 .~$  Clearly, {\it{such approximations can never lead to plasma redshift.~}}  We emphasize that the forms of Eqs.\,(A20) and (A22) above are more accurate than those used in the conventional approximations }

\vspace{2mm}


\section*{A4\, \,Time Average of the Poynting Vector}

The Poynting vector is $\mathbf{S} = {{c\left( {\mathbf{E} \times \mathbf{H}} \right)} \mathord{\left/{\vphantom {{c\left( {\mathbf{E} \times \mathbf{H}} \right)} {4\pi }}} \right.
 \kern-\nulldelimiterspace} {4\pi }} = {{c\left( {E_y } \right)_\Re  \left( {H_z } \right)_\Re  } \mathord{\left/{\vphantom {{c\left( {E_y } \right)_\Re  \left( {H_z } \right)_\Re  } {4\pi {\rm{,}}}}} \right. \kern-\nulldelimiterspace} {4\pi {\rm{,}}}}$ where the last expression is in the direction of the x-axis, and where the subscript $\Re$ means the real value of the indexed quantity.  For obtaining the time average $\mathord{\buildrel{\lower3pt\hbox{$\scriptscriptstyle\smile$}}\over S}$ of the Poynting vector at $x = 0$ and at $t = 0$, we insert the Fourier transforms for $E_y$ and $H_z$ for the photon pulse with a decay time $\tau  = {1 \mathord{\left/{\vphantom {1 \gamma }} \right.
 \kern-\nulldelimiterspace} \gamma }$ and integrate.  We get
\be
\mathord{\buildrel{\lower3pt\hbox{$\scriptscriptstyle\smile$}}
\over S}  = \frac{{c\gamma }}{{4\pi }}\int\limits_{ - \infty }^\infty  {\frac{{A\left( {\omega ,\;0} \right)}}{\varepsilon }d\omega } \int\limits_{ - \infty }^\infty  {\frac{{A\left( {\omega ',\;0} \right)}}{{\sqrt \varepsilon  }}d\omega '} \int\limits_{ - \infty }^\infty  {\exp \left( {i \left( {\omega  + \omega '}\, \right)t}\, \right)dt}.
\ee
The last integral is the Dirac delta-function, $2\pi \delta \left( {\omega  + \omega '} \right){\rm{.}}$  We can then set $\omega ' =  - \omega {\rm{,}}$
and the corresponding function of $- \omega$ equal to the complex conjugate function of $\omega {\rm{.}}$  We add the complex conjugate of the average Poynting vector and divide by 2, and get\footnote{It is easily verified that the Poynting vector is also equal to the product of energy density and the light velocity in the medium.}
\be
\mathord{\buildrel{\lower3pt\hbox{$\scriptscriptstyle\smile$}}
\over S}  = \frac{{c\gamma }}{4}\int\limits_{ - \infty }^\infty  {\left[ {\frac{1}{{\sqrt {\bar \varepsilon } \varepsilon }} + \frac{1}{{\sqrt \varepsilon  \bar \varepsilon }}} \right]A\left( {\omega ,\;0} \right)\;\bar A\left( {\omega ,\;0} \right)} \;d\omega .
\ee
The bar over the quantity means the complex conjugate.  We have also that
\be
\frac{1}{{\sqrt {\bar \varepsilon } \varepsilon }} + \frac{1}{{\sqrt \varepsilon  \bar \varepsilon }} = \frac{{2n}}{{\varepsilon \bar \varepsilon }}\:{\rm{;}}
\ee
and when we for the Fourier transforms insert expressions from Eqs.\,(A9) and (A10), we get
\be
\mathord{\buildrel{\lower3pt\hbox{$\scriptscriptstyle\smile$}}
\over S}  = \frac{{c\gamma }}{4}\int\limits_{ - \infty }^\infty  {\frac{{2n}}{{\varepsilon \bar \varepsilon }}\frac{{E_y^2 \left( {x = 0,\;t = 0,\;\varepsilon  = 1} \right)}}{{4\pi ^2 \left( {\gamma ^2 /4 + \left({\omega  - \omega _0 } \right)^2 } \right)}}} \;d\omega {\rm{,}}
\ee
where $E_y$ is the amplitude of the electrical field at $x = 0$, $t = 0$, and $\varepsilon  = 1;$ and $\omega _0  = 2\pi \nu _0$ is the center frequency of the photon in vacuum\footnote{It can be shown that the form of these equations in the mks (rationalized) system of units is identical to their form in cgs system.}.   If we in the above deduction include the variation with $x$, we get
\be
\mathord{\buildrel{\lower3pt\hbox{$\scriptscriptstyle\smile$}}
\over S}  = \frac{{c\gamma }}{4}\int\limits_{ - \infty }^\infty  {\frac
{{2n}}{{\varepsilon \bar \varepsilon }}\frac{{E_y^2 \left( {x = 0,\;t =
 0,\;\varepsilon  = 1} \right)}}{{4\pi ^2 \left( {\gamma ^2 /4 + \left({\omega  - \omega _0 } \right)^2 } \right)}}} \left[ {\exp \left( { - 2\kappa \omega x/c} \right)} \right]d\omega.
\ee
We will normalize the Poynting vector $\mathord{\buildrel{\lower3pt\hbox{$\scriptscriptstyle\smile$}}\over S}$ at $x = 0$ to the average energy flux of one photon, $\hbar \omega _0  = h\nu _0 {\rm{,}}$ per second and per ${\rm{cm}}^{\rm{2}}$ in vacuum, where $h$ is the Planck constant; that is, we set
\be
\frac{{cE_y^2 \left( {x = 0,\,t = 0,\,\varepsilon  = 1} \right)}}{{8\pi ^2 }}\gamma  = \hbar \omega _0 \frac{\gamma }{{2\pi }}{\rm{.}}
\ee
We get then that in dielectric medium, the Poynting vector corresponding to a flux of one photon per second and per ${\rm{cm}}^{\rm{2}}$ at the distance $x$ from the source is
\be
\mathord{\buildrel{\lower3pt\hbox{$\scriptscriptstyle\smile$}}
\over S}  = \hbar \omega _0 \frac{\gamma }{{2\pi }}\int\limits_{ - \infty }^\infty  {\frac{n}{{\varepsilon \bar \varepsilon }}\frac{{\left[ {\exp \left( { - 2\kappa \omega x/c} \right)} \right]}}{{\left( {\gamma ^2 /4 + \left( {\omega  - \omega _0 } \right)^2 } \right)}}} d\omega {\rm{,}}
\ee
where $\omega _0$ is the center frequency of the photon in vacuum.


\section*{A5\, \,Photon's Energy Loss in a Plasma}

We will consider an electronic plasma, where the magnetic permeability is equal to one.  When we differentiate with respect to $x$ the average Poynting vector 
$\mathord{\buildrel{\lower3pt\hbox{$\scriptscriptstyle\smile$}}
\over S}$ as given by Eq.\,(A29), we get the decrease in photon energy flux per cm,
\be
\frac{{d\mathord{\buildrel{\lower3pt\hbox{$\scriptscriptstyle\smile$}}
\over S} }}{{dx}} = \frac{{d\hbar \omega _0 }}{{dx}} =  - \hbar \omega _0 \frac{\gamma }{{2\pi }}\frac{1}{c}\int\limits_{ - \infty }^\infty  {\frac{{2\kappa \omega n}}{{\varepsilon \bar \varepsilon }}\frac{{\left[ {\exp \left( { - 2\kappa \omega x/c} \right)} \right]}}{{\left( {\gamma ^2 /4 + \left( {\omega  - \omega _0 } \right)^2 } \right)}}} d\omega {\rm{.}}
\ee

\noindent  When we then insert Eq.\,(A22) into Eq.\,(A30) and set $x = 0$, we get

\be
 \frac{{d\hbar \omega _0 }}{{dx}} =  - \frac{{\hbar \omega _0 \gamma }}{{2\pi c}}\int\limits_{- \infty }^{\infty}  \frac{{ \omega _p^2  \beta \omega^4 }}{{\left[ \left( {\omega _q^2+ \omega _p^2  - \omega ^2 } \right)^2  +  \beta^2 \omega ^6  \right]}}\frac{{d\omega }}{{\left\{ {\gamma^2 /4 + \left( {\omega  - \omega _0 } \right)^2}\right\} }},
\ee

\noindent  which gives the photon's energy loss per cm at $x = 0$ due to the ``frictional'' forces resisting the forward movement of the photon.

\indent The expressions in the denominator have eight complex roots.  The four roots in the upper plane are:
\be
\omega  = \left\{ {\begin{array}{llll}
   \rm{a} & = & \displaystyle + \sqrt {\omega _{qp}^2  - \frac{{\left( {\alpha  + \beta \omega _{qp}^2 } \right)^2 }}{4}} & \displaystyle + ~\;{i \frac{{\left( {\alpha  + \beta \omega _{qp}^2 } \right)}}{2}}  \\
   {{\rm{b}}} & = & \displaystyle - \sqrt {\omega _{qp}^2  - \frac{{\left( {\alpha  + \beta \omega _{qp}^2 } \right)^2 }}{4}} & \displaystyle  + ~\;{i \frac{{\left( {\alpha  + \beta \omega _{qp}^2 } \right)}}{2}}  \\
   {{\rm{c}}} & = & & \displaystyle + ~\;{i \left( {\frac{1}{\beta } + \alpha  + \beta \omega _{qp}^2 } \right)}  \\
   {{\rm{d}}} & = & + \omega _0 & \displaystyle + ~\;{i \frac{\gamma }{2}} 
\end{array}} \right\}{\rm{.}}
\ee

\indent For integrating Eq.\,(A31), we use complex integration.  We select a path along the x-axis from $- \infty$ to $\infty$ and then along a semicircle in the upper half-plane from $+ \infty$ to $-\infty {\rm{.}}$  The integral along the semicircle is equal to zero.  The integral along the x-axis is therefore equal to $2\pi i$ times the sum of the residues in the poles in the upper half-plane.  Of the four poles in the upper half-plane, one pole, corresponding to the root d in Eq.\,(A32), is due to the roots of the expression inside the braces in the denominator of Eq.\,(A31), while the remaining three poles are due to the roots of the expression inside the brackets in the denominator. It is of interest to keep these contributions separate, because the different roots correspond to distinct interactions.

\indent The root d would exist even when the dielectric constant is equal to one.  This root corresponds to the conventional Compton scattering, which is due to interactions of the photon with individual electrons.  This Compton scattering is due to high frequency Fourier components around the center frequency $\omega _0
$ of the photon for which the dielectric shielding is usually unimportant.

\indent The three remaining roots in the upper plane are of different character.  They are due to the expression in the first pair of brackets, and to the dielectric constant not being equal to one.  Conventional calculations usually disregard these roots.  They correspond to collective interactions of the photon field with several plasma electrons.  As the photon with its associated virtual photon field penetrates and disturbs the plasma, it leaves behind a wake of collective oscillations, which carry away the energy given up by the photon to the plasma.  Two of the three roots, the roots a and b in Eq.\,(A32), correspond to Stokes scattering or Raman scattering or resonance scattering, where the frequency, $\omega _{qp}  = \sqrt {\omega _q^2  + \omega _p^2 } ,$ is the frequency that causes the Stokes scattering.  As can be seen, the absorption frequency $\omega _{qp}$ differs slightly from the eigenfrequency $\omega _q {\rm{.}}$  (In unionized matter the absorption frequency has a slightly different form.)

\indent The imaginary root c in Eq.\,(A32) is of a different nature.  This root, which is important only in a hot sparse plasma, has not been considered before.  It corresponds to energy loss in what we will call ``plasma redshift'' of the photon as it penetrates a hot electron plasma.  In a hot sparse plasma, the collision damping, $\alpha$ in Eq.\,(A31), is very important.  When $\omega _q  = 0{\rm{,}}$ we can equate the collision field with the Fourier harmonics of the fast moving electrons.  As shown in section 3 of the main paper, the quantum mechanical damping term $\beta \omega ^2$ in the plasma oscillations then replaces the classical damping term $\left( {\alpha  + \beta _0 \omega ^2 } \right){\rm{,}}$ where 
$\beta _0 \omega ^2$ is the classical radiation damping.

\indent In this article, the focus is on the plasma redshift, because the Stokes scattering and the Compton (Thomson) scattering are well known and have been estimated accurately by others, for example, by Heitler [1].  We can then in the following simplify the Stokes scattering term and set $\omega _q  = 0.$

When evaluating the integral of Eq.\,(A31), we will assume that the following six conditions are fulfilled:
\[
\begin{array}{l} \displaystyle
 1)\;\;\omega _q  = 0;\quad 2)\;\;\alpha  = 0;\quad 3)\;\;\beta  \gg \beta _0  = 6.266 \cdot 10^{ - 24} ;\;\;4)\;\;\beta \omega _p  \ll 1;\\
 {\rm{5)}}\;\;\omega _{\rm{0}}  \gg \omega _p ;\;\;{\rm{and}}\;\;6)\;\;\gamma  \ll \omega _0 . 
 \end{array}
\]
As we will see, these conditions are usually fulfilled for the plasmas of main interest in this article.

\indent The four roots in the upper plane of the denominator are then:
\be
\omega  = \left\{ {\begin{array}{llll}
\displaystyle {{\rm{a}}} & = & + \omega _p & \displaystyle + ~\; i\frac{{\beta \omega _p^2 }}{2} \\
  {{\rm{b}}} & = &  - \omega _p & \displaystyle + ~\; i\frac{{\beta \omega _p^2 }}{2}  \\
   {{\rm{c}}} & = & & \displaystyle + ~\; i\left( {\frac{1}{\beta } + \beta \omega _p^2 } \right)  \\
   {{\rm{d}}} & = & +\omega _0  & \displaystyle + ~\; i\frac{\gamma }{2}   
\end{array}} \right\}{\rm{.}}
\ee
The roots ``a'' and ``b'' are very close to the real axis, while the root ``c'' is purely imaginary.  In hot, sparse plasmas $\beta$ is small, although it is very large compared with $\beta _0 {\rm{.}}$

\indent The results of the integration on the right side of Eq.\,(A31) is then
\be
\begin{array}{l} 
\displaystyle \frac{{d\hbar \omega _0 }}{{dx}}  = - 2\pi i\frac{{\hbar \omega _0 \gamma }}{{2\pi c}}\left[ {{\mathop{\rm Re}\nolimits} s\left( {\rm{a}} \right) + {\mathop{\rm Re}\nolimits} s\left( {\rm{b}} \right) + {\mathop{\rm Re}\nolimits} s\left( {\rm{c}} \right) + {\mathop{\rm Re}\nolimits} s\left( {\rm{d}} \right)} \right] \\
 \displaystyle \quad  =  - \frac{{\hbar \omega _0 \gamma \omega _p^2 }}{{c\omega _0^2}}\left[ {\frac{1}{4}\quad  + \quad \frac{1}{4}\quad  + \quad \frac{1}{2}\frac{{\left( {1 - 1/\left( {\beta \omega _0 } \right)^2 } \right)}}{{\left( {1 + 1/\left( {\beta \omega _0 } \right)^2 } \right)^2 }}\quad  + \quad \frac{{\beta _0 \omega _0^2 }}{{\gamma \left( {1 + \left( {\beta _0 \omega _0 } \right)^2 } \right)}}} \right]{\rm{.   }} 
 \end{array}
\ee
We use the notation $\gamma$ for the actual quantum mechanical width of the incident photon, while the classical photon width is given by
\[
\gamma _0  = \beta _0 \omega _0^2  = \left( {\frac{2}{3}\frac{{e^2 }}{{mc^3 }}} \right)\omega _0^2  = 6.266 \cdot 10^{ - 24} \omega _0^2 {\rm{.}}
\] 
Eq.\,(A34) can then be written on the form
\be
\frac{{d\hbar \omega _0 }}{{dx}} =  - \hbar \omega _0 6.65 \cdot 10^{ -25} N_e \left[ {\frac{\gamma }{{2\gamma _0 }} + \frac{\gamma }{{2\gamma_0 }}\frac{{\left( {1 - 1/\left( {\beta \omega _0 } \right)^2 } \right)}}{{\left( {1 + 1/\left( {\beta \omega _0 } \right)^2 } \right)^2 }} + \frac{1}{{1 + \left( {\beta _0 \omega _0 } \right)^2 }}} \right]{\rm{.}} 
\ee
The first term inside the brackets corresponds to Stokes scattering (or Raman scattering).  It corresponds to the collective scattering on the plasma frequency.  In the quantum mechanical treatment of highly excited plasma, some of the oscillator strengths are negative while others are positive.  The incident photon can then absorb and emit a plasma frequency photon, and thereby increase or decrease its energy to $\hbar \left( {\omega _0  + \omega _p } \right)$ or $\hbar \left( {\omega _0  - \omega _p } \right)\,{\rm{.}}$  In thermodynamic equilibrium, the two processes will average out.  There will be an equal number of positive and negative oscillator strengths.  However, when one is observing very distant supernovae, the narrow beam geometry may scatter some of the photons enough to remove them from the observed intensity.

\indent The second term inside the brackets of Eq.\,(A35) corresponds to the plasma redshift.  This interaction has not been discovered before.  Usually, the collision damping $\alpha$ has been disregarded and the damping $\beta \omega ^2$ equated with $\beta _0 \omega ^2 {\rm{.}}$  This term was then insignificant as $\beta _0 \omega _0  \ll 1.~$  However, in section 3 of the main paper, we show that in a hot sparse plasma the damping in the plasma oscillations is $\beta \omega ^2 {\rm{,}}$ which then replaces the classical damping term $\left( {\alpha  + \beta _0 \omega ^2 } \right){\rm{,}}$ where $\beta _0 \omega ^2$ is the classical radiation damping.  We may then have that $\beta \omega _0  > 1{\rm{,}}$
provided the plasma is very hot and sparse.  The required conditions for experimental determination have not been present in the laboratory experiments.  There are thus good reasons why this interaction has neither been theoretically nor experimentally discovered previously.  The energy loss in the plasma-redshift scattering is all absorbed in the plasma.  It is about 50\% of the total Thomson (Compton) scattering.  For $\beta \omega _0  < 1{\rm{,}}$ the plasma-redshift term can be slightly negative, but the corresponding blue shift is usually insignificant.  This miniscule blue shift occurs mostly in relatively cold and dense plasmas.  However, in hot, sparse plasmas, the value of $\beta$ can be very large and $\beta \omega _0  > 1.~$  The second term inside the brackets is then important for explaining many astrophysical phenomena.  In section 3 of the main paper, we use quantum-mechanical treatment to estimate $\beta {\rm{.}}$  We can then see how $\beta$ varies with the temperature and the density of the plasma and the condition for 
$\beta \omega _0  > 1.~$  In section 4 of the main article, we show how the magnetic field affects $\beta {\rm{.}}$

\indent Most of the plasma-redshifted photons are not scattered out of the path in the narrow beam geometry, because the individual energy losses are so small.  The angular scattering of the photons is therefore so small in intergalactic space (smaller than the angular scattering on the plasma frequency) that it does not affect the observed intensity of the distant stars and supernovae.

\indent The fourth term, the Compton scattering term, is calculated 
assuming that $
\beta  \approx \beta _0$ for incident photons with relatively high frequencies $\omega _0  = 2\pi \nu _0 {\rm{.}}$  That is, we assumed that in accordance with the conventional Compton (Thomson) scattering the photons interact with individual electrons, and not coherently with several electrons in the plasma.  If the incident photon's frequency is very low, the plasma electrons may act collectively in the field of the incident photons (confer Rayleigh scattering on atoms).   Compton scattering is then affected.

\indent The recoil energy in the Compton scattering, which is absorbed in the plasma, is for $\hbar \omega _0  \ll mc^2$ only an insignificant fraction of the incident photon energy.  However, the scattered photons are usually removed from the narrow beam observation of a star.  The Compton scattering causes then dimming of the distant stars and galaxies, as shown in sections 5.8 and 5.9.


\renewcommand{\theequation}{B\arabic{equation}}
\setcounter{equation}{0}
\section*{Appendix B}


\section*{B1\, \,Conversion of Magnetic Field Energy to Heat in a Plasma}

The diamagnetic moments in closed shells of atoms are small.  But at high temperatures the atoms are ionized and form a plasma.  The electrons and the ions in a magnetic plasma encircle the magnetic field lines and produce thereby large diamagnetic moments with direction opposite the magnetic field.  These large diamagnetic moments in the plasma often cause the \textbf{B}-field to be significantly smaller than the \textbf{H}-field.  For simplifying the analysis, it is usually assumed in the following that the cyclotron frequency is much larger than the plasma frequency.

\indent An increasing magnetic field induces rotational electrical field, which accelerates the electrons and the ions as they encircle the field lines; that is, the increasing field increases the velocities of the electrons and ions, and therefore the plasma temperature.  From Maxwell's theory, we have, using gaussian (cgs) system of units, that
\be
\nabla  \times \mathbf{E} =  - \frac{1}{c}\frac{{\partial \mathbf{B}}}{{\partial t}} =  - \frac{\mu }{c}\frac{{\partial \mathbf{H}}}{{\partial t}} - \frac{\mathbf{H}}{c}\frac{{\partial \mu }}{{\partial t}}{\rm{,}}
\ee
where $\nabla  \times \mathbf{E}$ is the rotational electrical field created by the change in the magnetic induction $\mathbf{B} = \mu \mathbf{H}$ with time $t$.  In the plasma $\mu$ is less than 1, and when \textbf{H} increases $\mu$ could decrease or increase; but let us assume for a moment that $\mu$ is constant.  Then when \textbf{H} increases, the \textbf{B}-field increases proportional to the \textbf{H}-field, and the rotational electrical field transfers energy to the encircling charged particles producing diamagnetic moments.

\indent If the \textbf{H}-field is constant while we transfer heat to the plasma, the \textbf{B}-field will decrease; and if the plasma emits heat while the \textbf{H}-field is constant, the \textbf{B}-field will increase and approach the \textbf{H}-field.  If $\mu$ is constant, the work performed on a particle with charge $e$ encircling the field lines is for one revolution 

\be
\oint {e \mathbf{E}} \cdot {\mathbf{dr}} = e \int \int \left( \nabla \times \mathbf{E} \right)_n df =  - \frac{e}{c}\frac{{\delta \mathbf{B}}}{{\delta t}}\pi r_ \bot ^2  =  - \frac{e}{c}\frac{{\mu \delta \mathbf{H}}}{{\delta t}}\pi r_ \bot ^2 \mathbf{  ,}
\ee

\noindent  provided the rate of change in the field is so slow that it can be considered constant during one revolution of the particle.  The radius in the cyclotron movement at right angle to \textbf{B} is $r_ \bot   = {{v_ \bot  } \mathord{\left/
 {\vphantom {{v_ \bot  } {\omega  =  {{v_ \bot  } \mathord{\left/
 {\vphantom {{v_ \bot  } {\left( {2\pi \nu } \right)}}} \right.
 \kern-\nulldelimiterspace} {\left( {2\pi \nu } \right)}}}}} \right.
 \kern-\nulldelimiterspace} {\omega  = {{v_ \bot  } \mathord{\left/
 {\vphantom {{v_ \bot  } {\left( {2\pi \nu } \right)}}} \right.
 \kern-\nulldelimiterspace} {\left( {2\pi \nu } \right)}}}}{\rm{,}}$
where $v_ \bot$ is the velocity of the particle perpendicular to the \textbf{B}-field.  When ${{\partial \mathbf{B}} \mathord{\left/
 {\vphantom {{\partial \mathbf{B}} {\partial t}}} \right.
 \kern-\nulldelimiterspace} {\partial t}}$ and ${{\partial \mathbf{H}} \mathord{\left/{\vphantom {{\partial \mathbf{H}} {\partial t}}} \right.
 \kern-\nulldelimiterspace} {\partial t}}$ are positive, the work is positive and the particle's energy increases.

\indent However, $\mu$ is usually not a constant in a plasma.  For example, when the heating by the plasma redshift increases the temperature and the kinetic energy of the charged particles in the plasma, the diamagnetic moment may increase from $\mathbf{M}$ to $\mathbf{M} + \Delta \mathbf{M}{\rm{.}}$  For constant \textbf{H}-field the induction may then decrease from $\mathbf{B} = \mathbf{H} - 4\pi \mathbf{M}
$ to $\mathbf{B} - \Delta \mathbf{B} = \mathbf{H} - 4\pi \left( {\mathbf{M + }\Delta \mathbf{M}} \right){\rm{.}}$

\indent The mutual inductance between the diamagnetic moments and the currents that create the \textbf{H}-field may sometimes be small.  The plasma redshift results in a hot bubble surrounded by colder plasma.  When the plasma redshift increases, the temperature in the hot bubble increases, and the \textbf{B}-field in the hot bubble decreases.  However, that decrease in the \textbf{B}-field results often in a corresponding increase in the \textbf{B}-field in the colder surroundings, the walls of the hot bubble.  The coupling between the initial diamagnetic moment in the hot bubble and the more distant currents creating the \textbf{H}-field is then reduced significantly.

\indent Let us for a moment assume that these distant currents are constant and therefore the \textbf{H}-field is constant; and let us see if we can quantify some of these changes.  When we multiply Eq.\,(B2) by the number of turns per second $\nu  = {\omega  \mathord{\left/{\vphantom {\omega  {2\pi {\rm{,}}}}} \right. \kern-\nulldelimiterspace} {2\pi {\rm{,}}}}$ we get that the energy gained per second is
\be
\frac{{\partial A}}{{\partial t}} =  - \frac{{\partial {\rm{B}}}}{{\partial t}}\frac{e}{c}\nu \pi r_ \bot ^2 {,}
\ee
where for $e$ positive the number of turns is positive when the rotation vector is opposite to the direction of the \textbf{B}-field.  The force from a magnetic induction field \textbf{B} on the moving particle with charge $e$ and velocity $\mathbf{v}$ is $\mathbf{F} = e\left( {{\mathbf{v} \mathord{\left/{\vphantom {\mathbf{v} c}} \right.
 \kern-\nulldelimiterspace} c}} \right) \times \mathbf{B}{\rm{,}}$ which leads to $\vec \omega  =  - {{e\mathbf{B}} \mathord{\left/{\vphantom {{e\mathbf{B}} {mc{\rm{,}}}}} \right.\kern-\nulldelimiterspace} {mc{\rm{,}}}}$ as shown in Eq.\,(B4).

\indent The work $\partial A$ in Eq.\,(B3), which is transferred to the particle, increases its kinetic energy as the \textbf{B}-field decreases due to the plasma-redshift heating.  The increase is proportional to the rate of change in the field \textbf{B}.  The kinetic energy also increases with the diamagnetic moments.  In a fully ionized plasma, the radius $r_ \bot$ in a particle's orbit around the field-lines is not a constant during the increase or decrease in \textbf{B}.  We can also show that the particles usually make a great many turns before being affected significantly by collisions in sparse plasmas of our main interest.

\indent Our present interest is to focus on the particle's interactions with large magnetic fields.  We can equate the centripetal force with the centrifugal force and get
\be
\left| {\frac{{ev_ \bot  B}}{c}} \right| = \left| {\frac{{mv_ \bot ^2 }}{{r_ \bot  }}} \right|\quad  \to \quad \left| {r_ \bot  } \right| = \left| {\frac{{mv_ \bot  c}}{{eB}}} \right| = \left| {\frac{{mr_ \bot  \omega c}}{{eB}}} \right|\quad  \to \quad \vec \omega  =  - \frac{{e\mathbf{B}}}{{mc}}{\rm{.}}
\ee
The last expression is obtained by keeping track of the directions of the different vector quantities.
\indent From Eqs.\,(B3) and (B4), we get
\be
\partial A = mv_ \bot  dv_ \bot   =  - \partial B\frac{e}{c}\nu \pi r_ \bot ^2  =  - \partial B\frac{e}{c}\frac{{eB}}{{2\pi mc}}\pi \frac{{m^2 v_ \bot ^2 c^2 }}{{e^2 B^2 }}{\rm{; \; or \; }}2\frac{{dv_ \bot  }}{{v_ \bot  }} =  - \frac{{dB}}{B}{\rm{,}}
\ee
which when integrated gives
\be
v_2^2 B_2  = v_1^2 B_1 {\rm{ ,}}
\ee
where $B_1$ and $v_1$ are the lower and $B_2$ and $v_2$ are upper integration limits for the field \textbf{B} and for the particle's velocity $v_ \bot$ perpendicular to the induction field \textbf{B}.

We can simplify the analysis by assuming that the field \textbf{H} and the number density of particles are constant.  We multiply Eq.\,(B6) by $3m/2$ and by $
{H \mathord{\left/
 {\vphantom {H {\left( {8\pi } \right)}}} \right.
 \kern-\nulldelimiterspace} {\left( {8\pi } \right)}}{\rm{,}}$ and sum the different particles.  We get

\[
\quad \quad \displaystyle \frac{{HB_2 }}{{8\pi }}\sum \frac{{3mv_2^2 }}{{2}}  = \frac{{HB_1 }}{{8\pi }}\sum \frac{{3mv_1^2 }}{{2}}\, {\rm{,}}\quad \quad {\rm{or  }}
\]
\be
 \displaystyle \quad \frac{{\sum  \displaystyle \frac{{3(m v_2^2  - m v_1^2) }}{{2 }} }}{{\displaystyle \frac{{HB_1  - HB_2 }}{{8\pi }}}} = \displaystyle \frac{{\sum \displaystyle \frac{{3mv_1^2 }}{{2 }} }}{{\displaystyle \frac{{HB_2 }}{{8\pi }}}}\,{\rm{. }}\quad
\ee

\indent When we apply this equation to the reversing layers of the Sun, we find that the initial energy density ${{HB_1 } \mathord{\left/{\vphantom {{HB_1 } {\left( {8\pi } \right)}}} \right.
 \kern-\nulldelimiterspace} {\left( {8\pi } \right)}}$ of the magnetic field is usually much larger than the charged particles' kinetic energy density $ \sum 3{mv_1^2 }/2 {\rm{.}}~$ ($m v_1^2$ is the kinetic energy density perpendicular to the field, which is 2/3 of the total kinetic energy density).  At an optical density of $\tau _{500}  = 1$ at 500 nm in the solar photosphere, the electron density is about $N_e  = 6.4 \cdot 10^{13} \; {\rm{ cm}}^{ - 3}$ and the temperature about $T = 6420$ K.  The kinetic energy density in charged particles' movements is then about $\sum 3{mv_1^2 }/2  \approx 3 N k T /2  \approx 85 \;{\rm{ erg}}\,{\rm{cm}}^{ - 3}{\rm{.}}~$  In and above the photosphere the field is sometimes about 1000 gauss and the energy density in the field is then about 40,000 ${\rm{erg}}\,{\rm{cm}}^{ - 3}$.  Higher, at an optical density of $\tau _{500}  = 2.64 \cdot 10^{ - 7} {\rm{,}}$  the electron density is about $N_e  = 2.4 \cdot 10^{10} \; {\rm{ cm}}^{ - 3}$ and temperature about $T = 21,000$ K.  The kinetic energy density is then about $\sum 3{mv_1^2 }/2 \approx 0.104 \;{\rm{ erg}}\;{\rm{cm}}^{ - 3} {\rm {.}}$  We have for large fields that $B \approx H {\rm{.}}$  Even when the field is only 100 gauss, its energy density is about $400 \;{\rm{ erg}}\,{\rm{cm}}^{ - 3} {\rm{.}}$  Therefore, the kinetic energy density of the charged particles is often much smaller than the energy density of the field.  Initially (when $B_2 \approx B_1$), the ratio on the right side of Eq.\,(B7) is therefore very small.  When the field energy, $B_1 {\rm{,}}$ in the denominator on the left side of Eq.\,(B7)  decreases by a small amount to, $B_2 {\rm{,}}$  the fraction on the left side is very small; that is, a very small fraction of the field energy is transformed into kinetic energy, $3(mv_2^2  - mv_1^2)/2 {\rm{.}}~$

\indent We see thus that the magnetic field energy is rather stable.  It cannot transform easily into kinetic energy unless something else, such as the heating $\Delta Q$ from the plasma redshift contributes to the heating.  When the magnetic field's fluctuations reduce the field, only a small fraction of the field energy transforms into kinetic energy or heat due to the small ratio on the right side of Eq.\,(B7).  The field would then bounce back to nearly its initial value.  In solar atmosphere plasma redshift creates hot plasma ``bubbles'', which lead to increased plasma redshift and reduced cooling, which therefore leads to high temperatures in the ``bubble''.  This increase in temperature increases the ratio on the right side of Eq.\,(B7) and facilitates thereby transformation of the magnetic field energy to heat.  As the plasma temperature increases, the rate of transformation increases exponentially, and may lead to explosive phenomena such as large flares in the Sun.

\indent If the decrease in the magnetic field energy is to equal the increase in kinetic energy or heat, the left side of Eq.\,(B7) must be close to 1.  When by some means heat energy $\Delta Q$ is transferred to the plasma, the temperature will increase and a larger fraction of the magnetic energy is transformed into heat.  When during the field fluctuations the field energy decreases, less of the field energy will then bounce back.   We can write $\Delta Q = T\,\Delta S{\rm{,}}$ where $\Delta S$ is the entropy density of the plasma and field.  We will assume that initially the field energy density is high relative to the kinetic energy density.  When we then introduce heat $\Delta Q$ the particles entropy will increase and the field energy decrease.  But as we approach equilibrium, $
{{HB_2 } \mathord{\left/
 {\vphantom {{HB_2 } {\left( {8\pi } \right)}}} \right.
 \kern-\nulldelimiterspace} {\left( {8\pi } \right)}} = \sum {{{3mv_2^2} \mathord{\left/
 {\vphantom {{3mv_2^2 } {2{\rm{,}}}}} \right.
 \kern-\nulldelimiterspace} {2{\rm{,}}}}}$ the entropy reaches its maximum and the transformation ceases.  In a similar way, if the kinetic energy is initially greater than the field energy, small fluctuations induce the system either to increase the field energy or to reduce the kinetic energy of the system.

\indent The aggregate of a plasma and a magnetic field, thus, requires for an equilibrium that the magnetic field energy about equals the kinetic energy component perpendicular to the field, provided that chemical potentials are in balance; that is, when these chemical potentials neither release or absorb heat.

\indent The Fourier harmonics of the fields produced by moving and accelerating particles in a cavity of matter create a spectral energy density given by $
u_\omega   = {{\left( {E_\omega ^2  + H_\omega ^2 } \right)} \mathord{\left/
 {\vphantom {{\left( {E_\omega ^2  + H_\omega ^2 } \right)} {\left( {8\pi} \right)}}} \right.
 \kern-\nulldelimiterspace} {\left( {8\pi } \right)}}{\rm{,}}$ where
$E_\omega$ and $H_\omega$ are the electrical and magnetic field components with frequency $\omega {\rm{.}}$  The integrated electromagnetic energy density is $u = \int_0^\infty  {u_\omega  {\kern 1pt} d\omega } {\rm{.}}$  This energy density is created mainly by the accelerated charges and dipoles.  This electromagnetic field is the source of the blackbody radiation from blackbody cavity.  In plasmas, we may have magnetic fields with energy densities that far exceed the energy densities of the particles' thermal motions.

\indent The electrical field interacts more easily with the charged particles and the atoms through excitations and ionizations.  Therefore, when a plasma with the fields moves up through the photosphere from a position deep under the solar photosphere, the electrical field can quickly interact and adjust to the decrease in temperature, while the magnetic field, as shown by Eq.\,(B7), has difficulty transferring its energy to the particles when the plasma cools abruptly in the photosphere.  In the reversing layers of the Sun, we can therefore for a while have large magnetic fields, but rather normal electrical fields.  The electrical and magnetic fields are coupled; and the fields become equal, as we approach the equilibrium over huge dimensions of black body cavity.  The total pressure $p$ of the electromagnetic energy is equal to ${u \mathord{\left/
 {\vphantom {u {3{\rm{,}}}}} \right.
 \kern-\nulldelimiterspace} {3{\rm{,}}}}$ where $u$ is the total electromagnetic energy density.  In a quasi equilibrium, we must have that the electromagnetic pressure is about equal to the pressure in the plasma.  We get then that $p = NkT = {u \mathord{\left/
 {\vphantom {u {3{\rm{,}}}}} \right.
 \kern-\nulldelimiterspace} {3{\rm{,}}}}$ where $N$ is the total number density of particles and $k$ is Boltzmann constant.  The kinetic energy density is then $
\left( {{3 \mathord{\left/
 {\vphantom {3 2}} \right.
 \kern-\nulldelimiterspace} 2}} \right)NkT = {u \mathord{\left/
 {\vphantom {u {2{\rm{;}}}}} \right.
 \kern-\nulldelimiterspace} {2{\rm{;}}}}$ that is, {\it{we have equipartition between the kinetic energy, the electrical field energy, and the magnetic field energy}}.  This equipartition is principally valid as an average over the huge dimensions of a black body cavity, and not necessarily over smaller dimensions where the plasma may not be in equilibrium with the fields.

\indent Like the paramagnetic moments and the ferromagnetic moments, the diamagnetic moments are coupled.  This coupling seeks to direct all the diamagnetic moments in the same direction.  When the diamagnetic moments are small as they are in atoms, the thermal motion usually dominates the coupling and the direction of the magnetic moments is not aligned.  However, when the diamagnetic moment \textbf{M} is large, the energy ${{MH} \mathord{\left/
 {\vphantom {{MH} 2}} \right.
 \kern-\nulldelimiterspace} 2}$ in some external field, \textbf{H}, may become larger than the kinetic energy fluctuations.  Just outside the periphery of the domain containing the diamagnetic moments the induction field \textbf{B} becomes stronger.  The magnetic moments, both perpendicular and parallel to the field lines, can then align and gradually build up huge domains.  While the diamagnetic moments align along the field lines, they will reduce the \textbf{B}-field within the domain and thereby reduce the field energy within the domain.

\indent While these equations are deduced assuming a hot, sparse plasma, they are to some extent valid for diamagnetic moments of hot material with temperatures in excess of any paramagnetic and ferromagnetic Curie point, such as that of the hot interior of the Earth.  This possibly explains why the Earth has a magnetic field with large secular variation and even reversals of direction (see Brynjolfsson [126]) like the magnetic fields in the Sun.  The Moon and the planet Mars, which most likely have a relatively cold interior, have very small magnetic fields.


\section*{B2\, \,Repulsion of Diamagnetic Moments}

Sometimes, we also must consider that diamagnetic moments are pushed away in a direction of a decreasing field.  For example, the diamagnetic moments of protons and electrons are pushed away from the Sun, because of the outward decreasing magnetic field.  When we divide Eq.\,(B3) by ${{\partial R} \mathord{\left/
 {\vphantom {{\partial R} {\partial t{\rm{,}}}}} \right.
 \kern-\nulldelimiterspace} {\partial t{\rm{,}}}}$ the velocity along the field lines, we get the outward force, $F$, on the magnetic moment, which is
\be
F = \frac{{\partial A}}{{\partial R}} =  - \frac{{\partial B}}{{\partial R}}\frac{e}{c}\nu \pi r_ \bot ^2  =  - \frac{{\partial B}}{{\partial R}}M =  - \frac{{\partial B}}{{B\partial R}}\frac{1}{2}mv_ \bot ^2 {\rm{.}}
\ee
We have made use of that the diamagnetic moment is given by
\[ M = \frac{{e\nu \pi r_ \bot ^2 }}{c} = \frac{{e\nu \pi v_ \bot ^2 }}{{c\omega _c^2 }} = \frac{{ev_ \bot ^2 }}{{2c\omega _c }} = \frac{{mv_\bot ^2 }}{{2B}}{\rm{.}}
\]

\indent Close to a point P we use Eq.\,(B6) and set $v_ \bot ^2  = v_P^2 \left( {{{B_P } \mathord{\left/{\vphantom {{B_P } B}} \right.
 \kern-\nulldelimiterspace} B}} \right)$, where $B_P$ is the value of the \textbf{B}-field and $v_p$ the value of the particle velocity at right angle to the field at the point P.  From Eq.\,(B8) we get then that the outward force at P is
\be
F_P  = \frac{{\partial \left( {1/B} \right)}}{{\partial R}}\frac{1}{2}mv_P^2 B_P {\rm{ }}{\rm{.}}
\ee
If the field at P decreases outward as $B = B_P \left( {{{R_P } \mathord{\left/{\vphantom {{R_P } R}} \right.\kern-\nulldelimiterspace} R}} \right)^n {\rm{,}}$ we get
\be
F_P  = \frac{n}{2}\frac{{mv_P^2 }}{{R_P }}{\rm{ ,}}
\ee
We see from this equation that the force pushing the diamagnetic dipole outwards from the Sun is independent of the magnetic field strength at P (because when the field decreases the diamagnetic moment increases); however, the force depends on $n$; that is, the force depends on how fast the field decreases with $R$.  When $n\left( {{{mv_P^2 } \mathord{\left/{\vphantom {{mv_P^2 } 2}} \right. \kern-\nulldelimiterspace} 2}} \right)$ of the particle exceeds the numerical value of its gravitational potential, the particle will be pushed away from the Sun.  This uses energy and the magnetic field will decrease until it becomes amorphous.

\indent In thermal equilibrium the velocity distribution is often isotropic, and the average kinetic energy component ${{mv_P^2 } \mathord{\left/
 {\vphantom {{mv_P^2 } 2}} \right.
 \kern-\nulldelimiterspace} 2}$ in each direction is 1/3 of the average of the total kinetic energy of the particles.  However, as the protons accelerate outwards in sparse plasma of the outer corona, the scattering cross section may not be large enough to make the velocity distribution isotropic.  The velocity perpendicular to the field may be nearly constant as the velocity along the field lines increases.  The average value of ${mv_P^2 }$ perpendicular to the field at the point P may then be much smaller than 2/3 of the average total kinetic energy.

\indent As a rough guide, we can use Rutherford's formula for the angular scattering cross section $\sigma \left( \theta  \right)$ for elastic scattering in the center of mass system, which is valid in the classical limit for proton temperatures well below $2.8 \cdot 10^8$ K.  We have then that
\be
\frac{{d\sigma \left(\theta  \right)}}{{d\Omega }} = \frac{{Z_1^2 Z_2^2 e^4 }}{{4m^2 v^4{\rm{sin}}^{\rm{4}} \frac{\theta }{2}}}{\rm{, }}
\ee
where $d\Omega  = 2\pi \,{\rm{sin}}\theta \,d\theta$ and $\theta$ the scattering angle; the mass $m = {{m_1 m_2 } \mathord{\left/
 {\vphantom {{m_1 m_2 } {\left( {m_1  + m_2 } \right)}}} \right.
 \kern-\nulldelimiterspace} {\left( {m_1  + m_2 } \right)}}$ is the reduced mass of two particles with masses $m_1$ and $m_2$ and charges $Z_1 e$ and $Z_2 e{\rm{,}}$ respectively; and $v$ is their relative velocity.  This equation shows that the scattering is larger for multiply charged ions.  Eq.\,(B11) serves only to indicate the trend.  In the laboratory system the total cross section is the same, but variation with $\theta$ is slightly different, and good estimates require also that we take into account the shielding effects.  In a plasma with 5 to 10\% ${\rm{He}}^{{\rm{ +  + }}}$-ions and 95 to 90\% ${\rm{H}}^{\rm{ + }}$-ions, the velocities of the helium ions will become isotropic before the proton ion.  The average of the outward accelerating force on the helium ion can then be greater than that on the proton.  The average velocity of helium ions can even become equal to or slightly greater than that of the protons.


\renewcommand{\theequation}{C\arabic{equation}}
\setcounter{equation}{0}
\section*{Appendix C}


\section*{C1\, \,Emission and Absorption of CMB and X Rays from a Hot Intergalactic Plasma}


\subsection*{C1.1\, \,The basic equations for emission and absorption}

We define, see section 3 in Spitzer [127], the photon's energy flux, or the specific intensity $I_{\nu}(\vec{n},\,\vec{r},\,t)$ so that $I_{\nu}\,d\nu\,d\Omega\,dA\,dt$ is the energy of those photons, which during a time interval $dt$ pass through the area $d\vec{A}$, whose frequency lies within the interval $d\nu$ about $\nu$, and whose direction is within the solid angle $d\Omega$ about the photon's direction $\vec{n}$; the area $d\vec{A}$ is located at the position $ \vec{r} $ and is perpendicular to the photon's direction $\vec{n}{\rm{.}}~$

\indent  Similarly, we define the emission coefficient $j_{\nu}(\vec{n},\, \vec{r},\,t)$ so that $j_{\nu}\,d\nu\,d\Omega\,dV\,dt$ is the energy emitted by the volume $dV = ds\, dA$ in the intervals $d\nu {\rm{,}}$ $d\Omega {\rm{,}}$ and $dt {\rm{,}}$ whereas $\kappa_{\nu}\,I_{\nu}\,d\nu\,d\Omega\,dV\,dt$ is the corresponding energy absorbed from a beam of specific intensity $I_{\nu} {\rm{.}}~$

\indent  The average electron density in the intergalactic plasma is $N_e = 1.95\cdot 10^{-4}\cdot(H_0/60){\rm{,}}$ where $H_0$ is the Hubble constant.  The plasma frequency is then $\nu _p  \approx 8,979 \sqrt {N_e }\approx 125.4 ~ {\rm{Hz}}{\rm{.}}~ $  It is far below the CMB frequencies, which are in the range of $10^9 ~{\rm{to}}~ 10^{12} ~ {\rm{Hz}}{\rm{.}}~$ Usually, we can assume therefore that the dielectric constant is equal to one. 

\indent  If the photons travel in straight lines, the change in $I_{\nu}$ along a distance $ds $ is given by (see section 3 by Spitzer [127]), 
\be
\frac{{dI_{\nu}}}{{ds}} = -\kappa_{\nu}\,I_{\nu}~+~j_{\nu} ~~{\rm{erg \, cm}}^{-3} \, {\rm{s}}^{-1}\,{\rm{sr}}^{-1}\,{\rm{Hz}}^{-1}\,{\rm{,}} 
\ee
\noindent where $\kappa_{\nu} $ is the absorption coefficient and $I_{\nu}{\rm{,}}$ the photon's energy flux.

\indent  In thermodynamic equilibrium, the emission density, $j_{\nu}{\rm{,}}$ equals the absorption of $I_{\nu}$ per cubic cm.  We have then that $dI_{\nu}/ds = 0 $ and $I_{\nu} = $ constant in Eq.\,(C1).  Assuming Kirchhoff's law, we have then in extended uniform plasma in thermodynamic equilibrium that (see Spitzer [127] and Rybicki and Lightmann [128]), 
\be
\frac{{j_{\nu} }}{{\kappa_{\nu} }}=  I_{\nu} =  B_{\nu}(T) = \frac{{2 h {\nu}^3}}{{c^2}} \, \frac{{1}}{{\,\,e^{h\nu/kT}\,-\,1 \, }} ~~{\rm{erg \, cm}}^{-2} \, {\rm{s}}^{-1}\,{\rm{sr}}^{-1}\,{\rm{Hz}}^{-1} \, {\rm{,}}
\ee
\noindent  where $B_{\nu}(T)$ is the Planck's function, $h$ is Planck's constant, $c$ the velocity of light, $k$ the Boltzmann constant, and $T {\rm{~ in~  K}}$ units is the temperature of the blackbody radiation in the plasma.


\subsection*{C1.2\, \,Blackbody emission}

A cavity in a solid block heated to a temperature $T$ emits the well-known blackbody spectrum, $B_{\nu}(T)$.  The electrons in the plasma emit a background radiation, $j_{\nu}~ {\rm{erg \, s}}^{-1}\,{\rm{cm}}^{-3}\,{\rm{sr}}^{-1} \, {\rm{Hz}}^{-1} \, {\rm{,}} $ when they interact with the other charged particles, mainly the protons.  This emission accumulates over the absorption distance $ R_{\nu} = \kappa_{\nu}^{-1}~{\rm{cm}} $ for each frequency in the frequency interval between $\nu ~ {\rm{and}}~ (\nu + d\nu) \, {\rm{.}}~ $ When Eq.\,(C2) is integrated over all frequencies and over the absorption distance for each frequency in a uniform plasma in a steady state universe, we get the conventional expression,
\be
\int \!\frac{{j_{\nu} }}{{\kappa_{\nu} }} \, d{\nu} = \!\!\!\int \!\! R_{\nu}\,j_{\nu}\,d\nu = \!\!\!\int \!\! I_{\nu} \, d{\nu} = \!\!\int \!\! B_{\nu} \, d{\nu}=\!\!\! \int \frac{{2 h {\nu}^3}}{{c^2}}  \frac{{ d{\nu} }}{{\,\,e^{h\nu/kT}\,-\,1 \, }} = \frac{{\sigma \, T^4}}{{\pi}} ~~{\rm{erg \, cm}}^{-2} \, {\rm{s}}^{-1}\,{\rm{sr}}^{-1} \, {\rm{,}}
\ee
\noindent  where $\sigma = 5.67\cdot 10^{-5} ~ {\rm{erg\,s}}^{-1}\,{\rm{cm}}^{-2} \,{\rm{K}}^{-4} $ is the Stefan-Boltzmann constant for emission to one side of a plane surface, corresponding to $\Omega = \pi {\rm{.}}~$ The integrand in this equation has the same form as the blackbody spectrum.  The conventional absorption distance for each frequency is $R_{\nu} = \kappa_{\nu}^{-1}\, {\rm{.}}~ $ We may multiply Eq.\,(C3) by $4\pi/c$ to get the energy density of the photon field.  We get
\[
\frac{{4 \pi}}{{c}}\int\limits_0^\infty  \!\! B_{\nu} \, d{\nu}= \int\limits_0^\infty  \frac{{8 \pi h {\nu}^3}}{{c^3}}  \frac{{ d{\nu} }}{{\,\,e^{h\nu/kT}\,-\,1 \, }} = \frac{{4\, \sigma \, T^4}}{{c}}=a \,T^4 ~~{\rm{erg \, cm}}^{-3} \, {\rm{,}}
\]
\noindent  where $a = 4\, \sigma /c = 7.566 \cdot 10^{-15}~{\rm{erg \, cm}}^{-3}{\rm{K}}^{-4}{\rm{.}}~$
\vspace{4mm}

\noindent  \textbf{Caution}

\indent  We usually should not equate the blackbody temperature, $T{\rm{,}}$ with the particle temperature, $T_e{\rm{,}}$ in the plasma.  These two temperatures are defined differently.  The blackbody temperature, $T{\rm{,}}$ depends on the average electromagnetic energy per volume unit in the blackbody cavity, or in the plasma.  The particle temperature, $T_e{\rm{,}}$ in the plasma is defined through the average kinetic energy of the particle, usually the electron.  The field energy of a particle (except the very low frequency Fourier harmonics) reaches usually only a very short distance; see Eqs.\,(D7) and (D8).  The blackbody temperature in the plasma depends, therefore, not only on $T_e{\rm{,}}$ but also on the density $N_e{\rm{.}}~$ We will come back to this question in subsection C1.7.  Unfortunately, the blackbody temperature, $T{\rm{,}}$ is often, incorrectly, surmised to be equal to the particle temperature, $T_e{\rm{,}}$ which has created some confusion.


\subsection*{C1.3\, \,Free-free emission and absorption in a plasma}

The free-free emission, $j_{\nu} {\rm{,}}$ is obtained from the acceleration of mainly the electrons in the field of the nuclei.  It is given by (see for example Spitzer [127] and Rybicki and Lightmann [128])

\be
j'_{\nu} = \frac{{8}}{{3}}\,\left(\frac{{2\pi}}{{3}}\right)^{\!1/2}\frac{{\,\,Z_i^2\,e^6 \, \langle g_{ff}\rangle \,N_e \, N_i \,}}{{\, m_e^{3/2}\,c^3}\,(kT_e)^{1/2 \,}}  \, e^{-h\nu /kT_e}~~{\rm{erg \, cm}}^{-3} \, {\rm{s}}^{-1}\,{\rm{sr}}^{-1}\,{\rm{Hz}}^{-1}{\rm{,}}
\ee
where $T_e $ is the particle temperature defined by $(3/2) kT_e =(1/2)m_e v^2 {\rm{,}} $ and $\langle g_{ff}\rangle $ is the Gaunt factor averaged over the velocities.  For $h\nu \ll kT_e {\rm{,}}$ the value of $\langle g_{ff}\rangle {\rm{}}$ and therefore the intensity increases slightly (see Eq.\,(C10)) with decreasing frequency.  But as $h\nu$ approaches and exceeds $kT_e {\rm{,}}$ the free-free emission in Eq.\,(C4) falls off exponentially

\indent  Other processes contribute to the emission, especially in the sparse plasmas of intergalactic space.  In some of these other processes, the emissions and absorption are proportional to the electron density, $N_e {\rm{,}}$ while free-free emission given by Eq.\,(C4) is proportional to $N_e^2 {\rm{.}}~$  Relative to the processes that are proportional to $N_e^2 {\rm{,}}$ such as the free-free emission and absorption, the processes that are proportional to $N_e$ increase in importance, some times by a factor of $10^{18}{\rm{,}}$ as the particle densities decrease from about $10^{14}~{\rm{cm}}^{-3}{\rm{}}$ in a laboratory plasmas to about $10^{-4}~{\rm{cm}}^{-3}{\rm{}}$ in intergalactic plasmas.

\indent  Eq.\,(C4) is equivalent to
\be
j'_{\nu} = 5.444 \cdot 10^{-39} \,\frac{{Z_i^2 \, \langle g_{ff}\rangle \, N_e \, N_i \,}}{{ T_e^{1/2} }}  \, e^{- h\nu /kT_e}~~{\rm{erg \, cm}}^{-3} \, {\rm{s}}^{-1}\,{\rm{sr}}^{-1}\,{\rm{Hz}}^{-1}\,{\rm{.}}
\ee
For comparison with experiments, we often use different units.  Eq.\,(C5) is equivalent to 
\be
j'_{\nu} = 8.217 \cdot 10^{-13} \,\frac{{Z_i^2 \, \langle g_{ff}\rangle \, N_e \, N_i \,}}{{ T_e^{1/2} }}  \, e^{- h\nu /kT_e}~~{\rm{keV \, cm}}^{-3} \, {\rm{s}}^{-1}\,{\rm{sr}}^{-1}\,{\rm{keV}}^{-1}\,{\rm{.}}
\ee
\noindent  For each frequency, we may multiply this by the absorption length $R_{\nu}=(\kappa_{\nu})^{-1}{\rm{,}}$ where $(\kappa_{\nu})$ is the total absorption coefficient, including that from plasma redshift, and photoelectric absorptions and excitations.  We get then that the intensity spectrum is
\be
I_{\nu}= R_{\nu}\,j'_{\nu} = R_{\nu}\cdot 8.217 \cdot 10^{-13} \,\frac{{Z_i^2 \, \langle g_{ff}\rangle \, N_e \, N_i \,}}{{ T_e^{1/2} }}  \, e^{- h\nu /kT_e}~~{\rm{keV \, cm}}^{-2} \, {\rm{s}}^{-1}\,{\rm{sr}}^{-1}\,{\rm{keV}}^{-1}\,{\rm{.}}
\ee
\indent  When we multiply Eq.\,(C4) by $4\pi\,d\nu $ and integrate over all frequencies, we get that the total emission density is
\be
\int\limits_0^\infty  4\pi \,j'_{\nu} \, d\nu =  1.426\cdot 10^{-27} \, Z_i^2 \, \, \langle g_{ff}\rangle \,\,N_e \, N_i \,T_e^{1/2} ~~ {\rm{erg \, s}}^{-1}\,{\rm{cm}}^{-3}\,{\rm{.}}
\ee
\indent  Integration of the total emissivity over the absorption distance is obtained by multiplying the integrand in Eq.\,(C8) by $R_{\nu}=(\kappa_{\nu})^{-1} ~{\rm{cm .}}~ $  If $R_{\nu} \approx R_{pl}=(\kappa_{pl})^{-1} {\rm{,}}$ which is independent of the frequency, this integrated intensity is equal to
\be
\int\limits_0^\infty  R_{\nu}\,4\pi \,j'_{\nu} \, d\nu \approx R_{pl}\cdot 1.426\cdot 10^{-27}  \cdot  Z_i^2 \, \, \langle g_{ff}\rangle \,\,N_e \, N_i \,T_e^{1/2} ~~ {\rm{erg \, s}}^{-1}\,{\rm{cm}}^{-2}\,{\rm{.}}
\ee
\noindent In Eqs.\,(C8) and (C9), the Gaunt factor is averaged over both the velocities and the frequencies (see Rybicki and Lightmann [128]).

\indent  When the intergalactic space is filled with average electron densities of $N_e = 1.95 \cdot 10^{-4} ~ {\rm{cm}}^{-3} {\rm{,}} $ at an average electron temperatures, $ T_e \approx 3 \cdot 10^6 ~ {\rm{K}} {\rm{,}} $ as required by the plasma-redshift cosmology, the free-free emission without the plasma-redshift absorption would be very large and inconsistent with observations.  But plasma-redshift absorption and other absorption processes reduce the free-free emission by many orders of magnitudes.  These absorptions cause the X-ray intensity to match the observations, as shown by Eq.\,(C21) and the subsequent paragraphs in section C3.
 
\indent  For the very high frequencies, the averaged Gaunt factor is close to one.  But for $h\nu \ll kT_e {\rm{,}}$ the Gaunt factor is given by (see Spitzer [127])
\be
 \langle g_{ff}\rangle  = \frac{{\sqrt{3} }}{{\pi}}\left\{ {\rm{ln}}\frac{{(2kT_e)^{3/2}}}{{\pi e^2 m^{1/2} \nu}} - 1.443 \right\}= 9.769 \, \left( 1 \, + \, 0.130 \,{\rm{log}} \frac{{\,T_e^{3/2}\,}}{{\nu}} \right) \, {\rm{.}}
\ee
\indent  The maximum intensity of a blackbody radiation is at $h\nu_{max}/kT = 2.821 {\rm{.}}~$  We have then that $\nu_{max} = 5.879 \cdot 10^{10} \, T ~ {\rm{Hz.}}~$  If the CMB temperature is assumed to be $T_{CMB} = 2.728 \pm 0.002 ~ {\rm{K\,,}} $ as estimated by Fixen et al.~[100], the frequency at maximum intensity of the CMB is $\nu_{max} = 5.879 \cdot 10^{10} \, T_{CMB} = 1.604 \cdot 10^{11} ~ {\rm{Hz}} \,{\rm{.}}~$ 

\indent  For $T_e = 3 \cdot 10^6 ~ {\rm{K,}} $ and $\nu = 1.604 \cdot 10^{12} ~ {\rm{Hz}} \,{\rm{,}}$ the Gaunt factor given by Eq.\,(C10) is $ \langle g_{ff}\rangle  = 6.61 \, {\rm{;}}~ $  

\noindent  for $T_e = 3 \cdot 10^6 ~ {\rm{K,}} $ and $\nu_{max} = 1.604 \cdot 10^{11} ~ {\rm{Hz}} {\rm{}} $ at the maximum intensity, we get $\langle g_{ff}\rangle = 7.88\,{\rm{;}}~$  

\noindent  for $\nu = 1.604 \cdot 10^{10} ~ {\rm{Hz}} \,{\rm{,}}$  we get $ \langle g_{ff}\rangle  = 9.15 \, {\rm{;}} $

\noindent  for $\nu = 1.604 \cdot 10^{9} \,\,~ {\rm{Hz}} \,{\rm{,}}$  we get $ \langle g_{ff}\rangle   = 10.42 \, {\rm{;}}~ $ and

\noindent  for  $\nu = 1.604 \cdot 10^{8} \,\,~ {\rm{Hz}} \,{\rm{,}}$  we get $ \langle g_{ff}\rangle   = 11.69 \, {\rm{.}}~ $

\noindent   At the high temperature of $T_e = 3 \cdot 10^6 ~ {\rm{K\,,}} $ and at the frequencies of the CMB, the free-free emission, $j'_\nu {\rm{,}}$ per volume element increases thus with decreasing frequency.  This increase in the Gaunt factor (and therefore in the intensity) with decreasing frequency is related to the fact that the Fourier harmonics of the incident particle's field reach a distance $b = \gamma v/\omega \approx v/\omega $ from the path of the incident charged particle, as shown by Eqs.\,(D7) and (D8) in Appendix D, and where $\omega = 2 \pi \nu {\rm{.}}~$  In this equation $\gamma = (1-v^2/c^2)^{-1/2} $ is a relativistic factor, which in the plasma of main interest in this article is approximately equal to 1.

\indent  The increase in the emission density, $j'_{\nu} {\rm{,}}$ with the increase in the impact parameter, $b {\rm{,}}$ corresponds to a similar increase in the absorption coefficient with the Gaunt factor.  The ratio $j_{\nu}/\kappa_{\nu}$ in Eq.\,(C2) and (C3) is therefore independent of the Gaunt factor.


\subsection*{C1.4\, \,Absorption coefficient $\kappa'_{\nu}$ corresponding to free-free emission}

From Eq.\,(C2), we have in thermodynamic equilibrium, and disregarding the plasma-redshift absorption and other absorptions except the free-free absorption, that $\kappa'_{\nu} = j'_{\nu} / B_{\nu} {\rm{.}}~$  If the emission density is given by Eq.\,(C4), we have for the corresponding absorption coefficient that
\be
\kappa'_{\nu} = \frac{{4}}{{3}} \, \left( \frac{{2\pi}}{{3}} \right)^{1/2} \frac{{ Z_i^2 \, e^6 \, N_i \, N_e \, \, \langle g_{ff}\rangle \left(1 - e^{-h\nu/kT}\right) }}{{\,\, c \, m_e^{3/2} \,(kT_e)^{1/2} \, h\nu^3 \,}} = 3.692\cdot 10^8\, \frac{{ Z_i^2 \, N_i \, N_e  \, \langle g_{ff}\rangle \left(1 - e^{-h\nu/kT}\right) }}{{ \,T_e^{1/2} \, \nu^3 \,}} ~{\rm{cm}}^{-1} {\rm{.}}
\ee 
For $h \nu \ll kT_e  {\rm{,}}$ we have that (see Eq.\,(3-57) of Spitzer [127], or Eq.\,(5.14a) of Rybicki and Lightmann [128]) 
\be
\kappa'_{\nu} = \frac{{4}}{{3}} \, \left( \frac{{2\pi}}{{3}} \right)^{1/2} \frac{{ Z_i^2 \, e^6 \, N_e \, N_i \, \, \langle g_{ff}\rangle \, }}{{\,\, c \, m_e^{3/2} \,(kT_e)^{3/2} \, \nu^2 \,}} = 0.173 \left( 1 \, + \, 0.130 \, {\rm{log}}\frac{{ T_e^{3/2} }}{{ \nu }} \, \right) \frac{{ Z_i^2 \, N_e \, N_i }}{{ T_e^{3/2} \, \nu^2 }}~{\rm{cm}}^{-1} {\rm{.}}
\ee
\indent  We set $Z_i^2 N_i = 1.17\, N_e ~{\rm{cm}}^{-3} {\rm{,}} $ $N_e = 1.95 \cdot 10^{-4} \,(H_0/60) ~ {\rm{cm}}^{-3} {\rm{,}} $ and $ T_e = 3 \cdot 10^6 ~ {\rm{K}} {\rm{.}}~ $ The maximum intensity of the CMB is at $\nu_{max} = 1.604\cdot 10^{11}~ {\rm{Hz}} {\rm{.}} $  

\indent  For $\nu = 1.604\cdot 10^{12}~ {\rm{Hz}} {\rm{,}} $  we get $\kappa'_{\nu} = 5.30\cdot 10^{-43} ~ {\rm{cm}}^{-1} {\rm{,}} $ and that the absorption length $R'_{\nu} = (\kappa'_{\nu})^{-1} = R'_{12}\approx 1.85\cdot 10^{42}\,\,{\rm{cm ;}}~$ 

\noindent  for $\nu = 1.604\cdot 10^{11}~ {\rm{Hz}} {\rm{,}} $ we get $\kappa'_{\nu} = 4.64\cdot 10^{-41} ~ {\rm{cm}}^{-1} {\rm{,}} $ and $R'_{\nu} = R_{11} \approx 2.15\cdot 10^{40}~{\rm{cm ;}}~$  

\noindent  for $\nu = 1.604\cdot 10^{10}~ {\rm{Hz}} {\rm{,}} $  we get $\kappa'_{\nu} = 3.90\cdot 10^{-39} ~ {\rm{cm}}^{-1} {\rm{,}} $ and $R'_{\nu} = R'_{10} \approx 2.57\cdot 10^{38}~{\rm{cm ;}}~$ 

\noindent  for $\nu = 1.604\cdot 10^{9}\,\,~ {\rm{Hz}} {\rm{,}} $  we get $\kappa'_{\nu} = 3.15\cdot 10^{-37} ~ {\rm{cm}}^{-1} {\rm{,}} $ and $R'_{\nu} = R'_{9}~ \approx 3.18\cdot 10^{36}~{\rm{cm ;}}~$

\noindent  for $\nu = 1.604\cdot 10^{8}\,\,~ {\rm{Hz}} {\rm{,}} $  we get $\kappa'_{\nu} = 2.40\cdot 10^{-35} ~ {\rm{cm}}^{-1} {\rm{,}} $ and $R'_{\nu} = R'_{8}~ \approx 4.17\cdot 10^{34}~{\rm{cm ;}}~$

\indent  In the literature, the absorption length is often given by $R'_{\nu}=(\kappa'_{\nu})^{-1}{\rm{,}}$ where the free-free absorption, $\kappa'_{\nu}{\rm{,}}$ is given by Eq.\,(C11).  This $R'_{\nu}$ is, as shown above, many orders of magnitude larger than the plasma-redshift absorption length, $R_{pl}=\kappa_{pl}^{-1}= 1.54 \cdot 10^{28}\cdot (60/H_0) ~{\rm{cm .}}~$ We have also that the absorption length $R_{pl}$ is equal to the Hubble length $c/H_0\,{\rm{.}}~$ In the hot sparse plasmas of intergalactic space, the plasma redshift absorbs most of the free-free emission and reduces it to a tiny fraction, $R_{\nu}/R'_{\nu} {\rm{,}}$ of the intensity that would be measured if there was no plasma-redshift absorption.  For X-ray frequencies, we get further reduction because $R_{\nu} < R_{pl}{\rm{,}}$ as we can see from Table C1.


\subsection*{C1.5\, \,Compton scattering absorption}

\indent  For the small absorptions, $\kappa'_{\nu} {\rm{,}}$ given by Eq.\,(C11) and (C12), (small because of the low densities in intergalactic plasma) or the corresponding large absorption distances, $R'_{\nu} = (\kappa'_{\nu})^{-1} {\rm{,}} $ we cannot ignore the absorption (=energy transfer to the plasma) in Compton scattering with average energy transfer to the plasma of $h\nu/m_e c^2 $ per interaction for $h\nu \ll m_e c^2 {\rm{.}}~$ In this case, the absorption coefficient in Compton scattering is given by
\be
\kappa_{C} = \frac{{dh\nu}}{{h\nu\,dx}} = \frac{{ h\nu }}{{\, m_e c^2 \,}} 6.65 \cdot 10^{-25}\, N_e = 5.38 \cdot 10^{-45 } \, \nu  N_e \,~{\rm{cm}}^{-1}{\rm{,}} 
\ee
\noindent  This average absorption is due only to the average recoil energy transferred from the photon to the electron.  Like the plasma-redshift absorption, this process is linear in the electron density, $N_e{\rm{.}}~$ In the sparse plasmas of intergalactic space these processes are important when compared with the free-free emission and absorption.  In case of the densities of about $10^9~ {\rm{cm}}^{-3}$ in the transition zone to the solar corona, the Compton scattering is still not important.  The corresponding plasma-redshift heating can compensate the cooling from free-free emission and the larger cooling from the trace elements.  In laboratory plasmas, on the other hand, the free-free absorption may often dominate.

\indent  For $\nu_{max} = 1.604\cdot 10^{11}\,\, {\rm{Hz}} {\rm{}} $ for the CMB radiation, and for $N_e = 1.95 \cdot 10^{-4} ~ {\rm{cm}}^{-3} {\rm{,}} $ this gives $\kappa_{C} = 1.684 \cdot 10^{-35} ~ {\rm{cm}}^{-1} {\rm{,}}$ which corresponds to absorption distance of $R_{C} = \kappa_{C}^{-1} = 5.94 \cdot 10^{34} ~ {\rm{cm}} {\rm{.}}~ $  At this frequency, the Compton absorption is about 363,000 times the corresponding absorption coefficient, $\kappa'_{\nu} = 4.64\cdot 10^{-41} \,\, {\rm{cm}}^{-1} {\rm{,}} $ as calculated by the conventionally used Eq.\,(C12), but it is much smaller than the plasma-redshift absorption, $\kappa_{pl} = 6.49\cdot 10^{-29}\,(H_0/60)~ \,{\rm{cm}}^{-1} {\rm{.}}~$

\indent  The absorption distance for Compton scattering at the CMB frequency, $\nu  {\rm{,}}$ is equal to
\be
 R_{C} = \frac{{1.857 \cdot 10^{44}}}{{\nu  N_e}} \, ~\,{\rm{cm}} \, {\rm{.}} 
\ee
\noindent  For $N_e = 1.95 \cdot 10^{-4} ~ {\rm{cm}}^{-3} {\rm{,}} $ and $\nu = 1.604\cdot 10^{11}\,\,~ {\rm{Hz}} {\rm{,}} $ we get $R_{C} = \kappa_{C}^{-1} = 5.94 \cdot 10^{34} ~ {\rm{cm;}} $ for $\nu = 1.604\cdot 10^{8}\,\,~ {\rm{Hz}} {\rm{,}} $ we get $R_{C} = \kappa_{C}^{-1} = 5.94 \cdot 10^{37} ~ {\rm{cm;}} $ for  $\nu = 1.604\cdot 10^{12}\,\, {\rm{Hz}} {\rm{,}} $  we get $R_{C} = \kappa_{C}^{-1} = 5.94 \cdot 10^{33} ~ {\rm{cm}} {\rm{.}}~ $ For the CMB frequencies, $\nu \geq 2 \cdot 10^9~ {\rm{Hz}} {\rm{,}}$ the absorption lengths in Compton scatterings are shorter than the absorption lengths for the free-free absorptions, but usually much longer (by several orders of magnitude) than the absorption lengths for plasma redshift.

\indent  In Eq.\,(C12), the variations with frequency and electron density are different.  The conventional estimates, which usually rely on Eq.\,(C12), are therefore often misleading.


\subsection*{C1.6\, \,Absorptions and emissions from a plasma}

In addition to the free-free absorption, the Compton absorption, and the plasma-redshift absorption, we have many other absorptions and excitation processes.  Excitation and ionization of trace elements are important; see for example Sutherland and Dopita [11].  For solar abundance, the excitations and ionizations of trace elements, which like the free-free emissions are proportional to $N_e^2{\rm{,}}$ exceed the free-free emission by a factor of about 32 at $T_e = 10^6~{\rm{K}}{\rm{.}}~$  Besides that, when a ``free'' electron is pushed out of its position in the plasma, the plasma will adjust, and another ``free'' electron will fall into the empty position and emit microwave radiation.  These absorption and emission processes may far exceed the free-free emission; see Appendix D. 

\indent  We can evaluate the excitations given by Eqs.\,(D13) and (D19) of the Appendix D by replacing the frequency $\omega_0$ by the frequencies of the many states in the plasma that were deduced in (see section 3) and then weigh the transitions by the thermal distribution.  This is a cumbersome method, because it entails also stimulated absorptions and emissions.

\indent  For the purpose of deducing the CMB, it is adequate to use the fact that the plasma, like any other body, must emit a blackbody radiation based on the average temperature for the electromagnetic volume energy in the plasma.  In the next section, we will estimate this blackbody radiation.


\subsection*{C1.7\, \,Plasma redshift absorption and blackbody emission from a plasma}

We can use Eq.\,(C2) and (C3) to derive the emission density in the plasma.  From Eq.\,(C2), we get
\be
\j_{\nu}  =  \kappa_{pl} I_{\nu}=   \kappa_{pl} B_{\nu}(T) =  \kappa_{pl} \frac{{2 h {\nu}^3}}{{c^2}} \, \frac{{1}}{{\,\,e^{h\nu/kT}\,-\,1 \, }} ~~{\rm{erg \, cm}}^{-3} \, {\rm{s}}^{-1}\,{\rm{sr}}^{-1}\,{\rm{Hz}}^{-1} \, {\rm{.}}
\ee
\noindent  The plasma-redshift absorption, $\kappa_{pl} = 3.326\cdot 10^{-25}\cdot (N_e)_{av}= 6.486 \cdot 10^{-29}(H_0/60)~ {\rm{cm}}^{-1} {\rm{,}}$ dominates the other absorptions, as shown in sections C1.4 and C1.5.  $R_{pl} = (\kappa_{pl})^{-1} {\rm{}}$ is a constant independent of the frequency, and $T=T_{CMB}$ is the temperature of a ``blackbody cavity''.  We have that 
\be
 R_{pl}\, j_{\nu} = I_{\nu}=    B_{\nu}(T) =   \frac{{2 h {\nu}^3}}{{c^2}} \, \frac{{1}}{{\,\,e^{h\nu/kT}\,-\,1 \, }} ~~{\rm{erg \, cm}}^{-2} \, {\rm{s}}^{-1}\,{\rm{sr}}^{-1}\,{\rm{Hz}}^{-1} \, {\rm{.}}
\ee
\noindent In the CMB range of frequencies, the photon intensity emitted by the intergalactic plasma is thus a perfect blackbody radiation, because it dominates all other contributions by several orders of magnitude in the CMB range of frequencies.  

\indent  The temperature, $T {\rm{,}}$ of the blackbody spectrum could be derived from the equations in Appendix D.  A simpler way is to note that the absorption distance $R_{pl} = (\kappa_{pl})^{-1} = 1.542 \cdot 10^{28} \,\,{\rm{cm \,,}}$ is the radius of a blackbody cavity.  The walls of this blackbody cavity consist of the particle density, $N(x) {\rm{,}}$ along the line of the radius, $R_{pl} {\rm{,}}$ from the observer.  The column density, $\int N(x)\,dx = N_{av}\, R_{pl} {\rm{,}}$ forms the wall.  The radiation pressure, $p{\rm{,}}$ inside this cavity is given by $p = u/3 {\rm{,}}$ where $u$ is the photon's energy density.  We have that
\be
 p = \frac{{u}}{{3}}=  \frac{{4\pi}}{{3\,c}} \int\limits_0^\infty  B_{\nu}(T)\,d\nu =  \frac{{4\pi}}{{3\,c}} \int\limits_0^\infty  \frac{{2 h {\nu}^3}}{{c^2}} \, \frac{{d\nu}}{{\,\,e^{h\nu/k T }\,-\,1 \, }} = \frac{{4\,\sigma}}{{3\,c}}\, T^4 ~~{\rm{erg \, cm}}^{-3} \, {\rm{.}}
\ee
\noindent    This photon pressure, $p{\rm{,}}$ in the usual way must be equal to particle pressure, $N k T_e {\rm{,}}$ in the ``walls'' of the cavity, because of the second law of thermodynamics.  We replace $T$ with $T_{CMB}$ and get
\be
 \frac{{4\,\sigma }}{{c}}\,T_{CMB}^4 = a\,T_{CMB}^4 = 3 N k T_e ~~{\rm{dyne \, cm}}^{-2}  {\rm{,}}
\ee
where in Eq.\,(C17) and (C18), the pressure is $p = u/3 = (a/3)\, T_{CMB}^4 ~ {\rm{dyne \, cm}}^{-2} {\rm{,}}$ and where Stefan-Boltzmann constant for energy density is $a = 7.566\cdot 10^{-15}~ {\rm{dyne \, cm}}^{-2}\, {\rm{K}}^{-4}  {\rm{.}}~$  For the particle density, we use the approximation: $N \approx N_p + N_{He} + N_e = (2.3/1.2)\, N_e \approx 1.917\, N_e~ {\rm{cm}}^{-3}{\rm{.}}~ $  The electron density predicted by the plasma-redshift cosmology (see Eq.\,(57)) is $N_e = 1.95\cdot 10^{-4}\cdot (H_0/60)~ {\rm{cm}}^{-3}{\rm{.}}~$  If the photons' energy density is mainly due to the CMB radiation with temperature $T_{CMB} = 2.728 \pm 0.002 {\rm{}}$ (see Fixen [100]), we get that the particle temperature, $T_e {\rm{,}}$ is
\be
T_e = \frac{{a\,T_{CMB}^4}}{{3 N k}} = \frac{{4.1902 \cdot 10^{-13} }}{{3\cdot 1.917\cdot (N_e)_{av}\, 1.3807\cdot 10^{-16} }} = 2.706\cdot 10^6 \cdot\frac{{60}}{{H_0}}~~{\rm{K}}\,{\rm{,}} 
\ee
\noindent  where $(N_e)_{av} = 1.95\cdot 10^{-4}\cdot (H_0/60)~ {\rm{cm}}^{-3}{\rm{,}}$ and $k = 1.3807\cdot 10^{-16}$ is the Boltzmann constant.

\indent  As we saw in section C1.4 and C1.5, the plasma redshift, which is independent of the photon frequency, dominates by far the other absorptions in the CMB range of frequencies.  This dominance accounts for the fact that the CMB has such a nice blackbody radiation spectrum.   The blackbody temperature, $T_{CMB} {\rm{,}}$ is well defined through the pressure, $p {\rm{,}}$ or the product of $N_e$ and $T_e{\rm{}}$ in Eq.\,(C18), which is averaged over the long distance $R_{pl}\approx 5000~{\rm{Mpc}}{\rm{.}}~$

\indent  Below the frequency range of the CMB, the plasma-redshift cut-off gradually sets in below about $\nu \approx 10^9~{\rm{Hz}}{\rm{.}}~$  The cut-off is gradual, mainly because the cut-off begins in the coldest filaments of space and then in the colder regions.  Just below the cut-off the intensity accumulates and increases therefore slightly over the intensity that would be calculated from Eq.\,(C16).  At $T_e \approx 3\cdot 10^5{\rm{K,}}$ and density of about $N_e = 1.95\cdot 10^{-3}~{\rm{cm}}^{-3}{\rm{,}}$ the cut-off frequency is at about $\nu = 139 ~{\rm{MHz}}{\rm{.}}~$  For $T_e \approx 3\cdot 10^6{\rm{,}}$ and a frequency of $\nu = 44 ~{\rm{MHz}}{\rm{,}}$ the plasma redshift is about 50\,\% of its nominal value.  A hot region may have about 10 times lower density than the average and 10 times higher temperature. The 50\,\%cut-off is then at 
the frequency of $\nu = 14~{\rm{MHz}}{\rm{.}}~$   Below $\nu = 0.48 ~{\rm{MHz}}{\rm{,}}$ or at $\lambda \approx 520 ~{\rm{m,}}$ the plasma redshift is insignificant.  This is consistent with observations, see Keshet et al.~[101].

\indent  Above the CMB frequencies, we observe the continuum of redshifted photons from the high-frequency photons emitted by stars, active galactic nuclei (AGN), supernovae, and the high-pressure coronal plasma surrounding these objects.  These objects are mostly characterized by point sources.  The photons are gradually plasma redshifted in the coronas of the stars, the galaxies, the galaxy clusters, and in the intergalactic plasma.  Some of these radiations are absorbed in condensations of colder plasma and in cloud formations usually in the coronas of galaxies.  We should observe, therefore, a continuum of photon frequencies above about $7\cdot 10^{11} ~ {\rm{Hz}} $ to about $10^{16} ~ {\rm{Hz}} $ and beyond.  The sources of the high-energy radiations are usually found in and close to the galaxies.  The energy content is sometimes estimated to be on the order of 10\,\% of the CMB radiation.  If these other radiations contribute about 10\,\% of the CMB radiation, the average temperature is about $ T_e \approx 3\cdot 10^6\cdot (60/H_0)~{\rm{K,}}$ with a high-energy tail skewing the thermal distribution.  This is all consistent with observations.


\subsection*{C1.8\, \,Bubble formation in plasma}

``Bubble'' formation in a plasma affects the temperature and density variations in the plasma; and it affects therefore the absorption and emission processes in the plasma.  As mentioned in sections 5.1 and 5.5, the imbalance caused by plasma-redshift heating, which is a first order process with respect to the densities while the cooling processes are usually second order in the densities, results in formation of bubbles within the plasma.  ``Walls'' of cooler plasma surround the bubbles containing the hot plasma.  The pressure, $p{\rm{,}}$ which is proportional to the product $N_e \cdot T_e {\rm{,}} $ varies much less than the density, $N_e {\rm{,}}$ and the temperature, $T_e {\rm{,}}$ separately.

\indent  This is important, because as Eqs.\,(C17) and (C18) show, the electromagnetic field temperature,  $T_{CMB}{\rm{,}}$ of the CMB, is proportional to $p^{0.25}{\rm{.}}~$ Instead of the average values $( N_e {\rm{,}} ~T_e ) = ~(1.95 \cdot 10^{-4}, ~ 2.7 \cdot 10^6 ) {\rm{,}} $  we might have in a cold region at the surface of a ``bubble'' that the density and the temperature are about $(N_e,\,T_e) = (1.95\cdot 10^{-3},~2.7\cdot 10^5){\rm{,}}$ while close to the center of the ``bubble'', we might have about $(N_e,\,T_e) = (1.95\cdot 10^{-5},~2.7\cdot 10^7){\rm{.}}~$ The hotter volumes, the bubbles, are more extensive.  The above averages, $( N_e~{\rm{and}} ~T_e ) {\rm{,}} $ are averages per particle and not per volume unit, and the division is such that about 50\,\% of the particles are in the cold regions and 50\,\% in the hot region.  In practice, we have a continuum of intermediary groups.

\indent  The maximum temperatures in the bubbles are limited mostly by the steep increase in the heat conductivity coefficient, $\kappa_{th} \approx 10^{-6} \cdot T_e^{5/2} {\rm{,}}$ with temperature.  In the colder regions at the bubbles' surfaces, the temperatures are limited by the heat conductivity coefficient and by the X-ray heating.  The X-ray heating prevents usually condensation to atomic and molecular hydrogen.  The volumes of the hot regions, the bubbles, are much larger than those of the cold regions.  The hot and cold regions contain equal number of atoms.  The cold regions are often close to galaxies and galaxy clusters.  The softer X rays are absorbed heavily in the colder filaments at the surfaces of the ``bubbles''.

\indent  The pressure variations usually equilibrate through movement of material from the high pressure to a low-pressure region.  The pressure can also equilibrate by means of radiation, because high-pressure regions emit slightly more radiation than they absorb, while in a cold regions it is reversed.  The imbalance caused by plasma redshift is initially more powerful than gravitation in forming clouds.  Subsequently, the gravitation may dominate the formation of stars inside the clouds.  Where the intergalactic plasma meets the rotating Milky Way's corona, the cloud formation is sometimes increased.  These clouds may then sink as high-velocity clouds into the Milky Way.  At the inner periphery of the corona, clouds may form and with the rotational speed at the periphery drift into space.  In this way matter moves in to and out of the Galaxy.  This explains why the metallicity in intergalactic plasma is so similar to that in the galaxies.  In the big-bang cosmology it is assumed that the material flows on the average only into the galaxies, which is often difficult to reconcile with the clouds movements and with the observed trace elements in WHIM.


\section*{C2\, \,X rays from intergalactic plasma}

In section 5.7, we found that X rays from intergalactic space are needed to explain the high densities of the trace elements in the Milky Way's corona.  The high densities were derived, for example, from the observations of the 6374.51 {\AA} Fe\,X-line towards the bright supernova 1987A in the Large Magellanic Cloud (LMC) by Pettini et al.~[68].  Assuming solar abundance and collisional ionization of the iron Pettini et al. derived a column density of $N_H \approx 3.2\cdot 10^{21}~{\rm{cm}}^{-2} {\rm{,}}$ which corresponds to an average density of $N_e \approx 0.02~{\rm{cm}}^{-3}{\rm{}}$ along the 50 kpc distance to the SN 1987A and a temperature of about $ T\approx 1.25 \cdot 10^6~ {\rm{K}} {\rm{.}}$

\indent  Other effects such as a relatively high average density along the line towards LMC, and excessive solar abundance could possibly lower the average slightly.  We must conclude, however, that the low coronal densities, $N_e \approx 0.0005~{\rm{cm}}^{-3}{\rm{,}}$ conventionally surmised in the Galactic corona, are incorrect.  These low-density estimates are dictated by the conventional big-bang cosmology, which has no means of heating the intergalactic plasma or a denser coronal plasma.  Usually, it is surmised that somehow the debris from supernovae drifted into the corona to supply the heating for a low-density corona; but not enough supernovae could be envisaged to explain the higher densities.

\indent   As shown in section 5.7, the UV light, the supernovae, and the plasma redshift of direct light are inadequate for ionizing the dense corona of our Milky Way.  However, most of the light energy escaping from the galaxies is plasma-redshifted in the intergalactic plasma.  With the help of the hot intergalactic sparse plasma, the light energy from the galaxies is transformed into X rays, which return the energy to galaxies and to the clusters of galaxies and ionizes their coronae and the intracluster plasma.  This can explain the observed high coronal densities.

\indent  We saw in section C1.4 and C1.5 that for CMB frequencies the absorption length for the plasma redshift is many orders of magnitude shorter than the absorption length for the free-free absorption and Compton scattering.  But for X-ray frequencies, as seen from Table C1 (see the last row), the absorption length in the intergalactic plasma is, due to trace elements, shorter than the absorption length for plasma redshift.

\indent  The absorptions by the plasma redshifts and the trace elements reduce the intensity in the free-free emission.  The low-energy X rays are absorbed mainly in the cold filaments at the surface of the hot ``bubbles'', and in the transition zone and low in the corona of the Milky Way.

\indent  The Milky Way's corona is usually thinner in the polar region.  Towards the Lockman Hole, with $(l,\,b) = (150^{\circ},\,53^{\circ}){\rm{,}}$ the column density of hydrogen is only about $5 \cdot 10^{19}~ {\rm{cm}}^{-2} {\rm{.}}~$ However, the column density of the partially ionized elements is much greater, and X-ray absorption significant, as discussed in section C3.


\subsection*{C2.1\, \,The absorption distance for X rays}

Table C1 gives a rough estimate of the X-ray absorption.  We have used the new solar abundance values by Asplund, Grevesse, and Sauval [129], for the trace elements, which are usually significantly smaller than the widely used old values by Anders \& Grevesse (1989).  We have used the absorption coefficients from Saloman, Hubble, and Scofield [130], which are valid for the neutral atomic elements.  For X rays, the stripping of the outer electrons at high temperatures will reduce the absorptions coefficient for low-energy X-ray photons.  The low-energy X rays, however, are preferentially absorbed in the low temperature regions of intergalactic space and in the Galactic corona and its transition zone.  For X rays above about 250 eV, the removal of the outer electrons from the atoms increases slightly the absorption coefficients for the remaining electrons due to their increased binding energies.  Also the increased Stark broadening at higher temperature increases the cross sections.  For the frequencies of main interest, this will partially compensate for the lost electrons from the outer shells.  A detailed analysis would have to take these effects into account, including the changes with temperature and density.  For the present, we will use Table C1 as a rough guide.

\indent  The absorption distances $R_{\nu} $ like $R_{pl} $ is proportional to $60/H_0$, where $H_0$ is the Hubble constant.  In Table C1, the column density (in the penultimate row) is the number of hydrogen atoms per ${\rm{cm}}^{2}$ in a cold plasma.  It is independent of the Hubble constant.

\begin{table}[h]
\centering
{\bf{Table C1}} \, \, The absorption distance for hydrogen density of 
$ N_{e}=1.95 \cdot 10^{-4}\cdot (H_0/60)~{\rm{cm}}^{-3}$

\vspace{2mm}

\begin{tabular}{lllllllll}
	\hline
Atom-& New  &Absorp-&Absorp-&Absorp-&Absorp-&Absorp-&Absorp-&Absorp-\\
ic  &solar & tion& tion& tion&tion &tion&tion&tion\\
num- &abun- &in barn&in barn&in barn&in barn&in barn&in barn&in barn\\
ber  &dance\,\,* &per H at&per H at&per H at&per H at&per H at&per H at&per H at\\
      &      &0.25 keV&0.5 keV&0.75 keV&1 keV&2 keV&3 keV&5 keV\\
	 \hline \hline
1 H   & 1.00E-0 &1200.0&116.00 & 34.00& 12.00 & 1.80 &0.93& 0.71\\
2 He  & 8.51E-2 &2600.0&340.00 & 85.10& 33.61 & 4.25 &1.36& 0.33\\
6 C   & 2.46E-4 & 16.7& 68.90 & 24.60& 10.07 & 1.51 &0.48& 0.10\\
7 N   & 6.03E-5 &  8.19& 27.14 & 10.55&  4.82 & 0.68 &0.21& 0.05\\
8 O   & 4.57E-4 &100.5& 17.82 &114.25& 52.56 & 8.80 &2.74& 0.59\\
10 Ne & 6.92E-5 & 38.8&  6.92 &  2.35& 18.60 & 2.90 &0.80& 0.23\\
11 Na & 1.48E-6 &  1.2&  0.24 &  0.08&  0.04 & 0.09 &0.03& 0.01\\
12 Mg & 3.39E-5 & 39.0&  7.00 &  2.70&  1.33 & 2.60 &0.90& 0.22\\
13 Al & 2.34E-6 & 4.0&  0.73 &  0.29&  0.12 & 0.24 &0.08& 0.02\\
14 Si & 3.24E-5 & 66.4& 13.93 &  5.18&  2.33 & 4.50 &1.50& 0.39\\
16 S  & 1.38E-5 & 42.8& 10.10 &  3.86&  1.97 & 0.29 &0.98& 0.26\\
18 Ar & 1.51E-6 & 4.7&  1.82 &  0.68&  0.33 & 0.05 &0.02& 0.04\\
20 Ca & 2.04E-6 & 1.2&  3.50 &  1.37&  0.65 & 0.11 &0.04& 0.08\\
26 Fe & 2.82E-5 & 45.12& 12.10 &  39.4& 24.40 & 4.51 &1.49& 0.36\\
28 Ni & 1.70E-6 & 3.57&  0.98 &  0.43&  1.80 & 0.34 &0.12& 0.03\\
\hline
\multicolumn{2}{l}{$\Sigma$ barn/H-atom}&4172.2 &627.18&324.84&164.63&32.67&11.68&3.42\\
\hline
\multicolumn{2}{l}{$N_H$ in $ {\rm{cm}}^{-2}$\,\,**}&$2.4\cdot 10^{20}$&$1.6\cdot 10^{21}$&$3.1\cdot 10^{21}$&$6.1\cdot 10^{21}$&$3.1\cdot 10^{22}$&$8.6\cdot 10^{22}$&$3.0\cdot 10^{23}$\\
\hline
\multicolumn{2}{l}{$R_{\nu}$ in cm\,\,***}&$1.4\cdot 10^{24}$&$9.6\cdot 10^{24}$&$1.8\cdot 10^{25}$&$3.6\cdot 10^{25}$&$1.8\cdot 10^{26}$&$5.1\cdot 10^{26}$&$1.7\cdot 10^{27}$\\
\hline

\multicolumn{9}{l}{*\,The old abundance values are significantly greater.  For example, old abundances for oxygen}\\
\multicolumn{9}{l}{and nitrogen are 86\,\% higher than the new values and for iron 65\,\% higher}\\
\multicolumn{9}{l}{**\,The column density, $N_H {\rm{,}}$ of hydrogen atoms corresponding to one relaxation length, $R_{\nu}$.}\\
\multicolumn{9}{l}{***\,We assume that $N_e\! \approx \!1.17\! \cdot\! N_H = 1.95\cdot\! 10^{-4} (H_0/60)~{\rm{cm}}^{-3}{\rm{,}}$ where $H_0$ is the Hubble constant.} \\
\end{tabular}
\end{table}

\vspace{1mm}

\indent  From Table C1, we see that the absorption lengths for X rays are shorter than the absorption length for the plasma redshift, $ R_{pl} = 1.542 \cdot 10^{28}\cdot (60/H_0)~{\rm{cm.}}~ $  In case of 750 eV X rays, a few of the relevant electrons would be stripped from the atoms, but the remaining electrons would usually have a slightly stronger absorption.  A rough estimate from the Table C1 is an absorption length of $R_{\nu} = 1.8 \cdot 10^{25}~{\rm{cm}}{\rm{.}}~$ This is smaller than $R_{pl}$ by a factor of about 0.0012.  The corresponding column density is about $N_H = 3.1 \cdot 10^{21}~{\rm{cm}}^{-2}{\rm{.}}~$ This column density is comparable to column density $N_e = 2.78 \cdot 10^{21}~{\rm{cm}}^{-2}{\rm{}}$ for the Galactic corona at high latitudes.  This last mentioned column density corresponds to $\Delta z = 0.000925 {\rm{,}}$ which was determined to be the average for the Galactic corona at high latitudes from coronal densities (discussed in section 5.7) and from the supernovae SN\,Ia experiments (see the penultimate paragraph in section 5.9).  This means that the 750 eV X rays that we observe are produced mainly in the Milky Way's corona and in the transition zone to the corona.

\indent  The absorption of low and high-energy X rays is significant in intergalactic space, where the temperature in the cold regions can be about $(T_{e})_{cold} \approx 3\cdot 10^5 $ K.  The absorption is also significant in the transition zone to the Milky Way's corona, where the temperature may even be $(T_{e})_{cold}< 1\cdot 10^5$ K, as discussed in section 5.7.  The heating and ionization of the corona by X rays from intergalactic plasma is necessary for explaining the observed high densities in the Milky Way's corona.

\indent  We may divide the plasma into several groups with different temperatures and densities.  We could in the first approximation divide the plasma into two groups.  50\,\% of the particles are the colder group, with temperature and density approximated by $(T_e,\,\,N_e) \approx (6\cdot10^5~ {\rm{K}}, \,\,9.75\cdot 10^{-4})~ {\rm{cm}}^{-3} {\rm{.}}~$ The remaining 50\,\% have $(T_e,\,\,N_e) \approx (1.5 \cdot 10^7, \,\,3.90\cdot 10^{-5}) {\rm{.}}~$  For constant pressure the, the volume of the hot plasma is 25 times that of the cold plasma.  The hard X rays are produced mainly by the hot plasma, but the energy density will be much lower $(5^{-2.5}=0.0179 $ times lower), while the colder plasma emits X rays with lower energy and higher intensity.  The average absorption length for photons does not change much and is assumed to be nearly linear in density.


\section*{C3\, \,Comparison with observations}

\indent  We have shown in section C1.7 that temperature, $T_{CMB}{\rm{,}}$ of the microwave background is given by (see Eqs.\,(C17) and (C18))
\be
a\,T_{CMB}^4 = 3\, N \, k\, T_e {\rm{,}}\quad \quad {\rm{or}} \quad \quad  T_{CMB} = \left(\frac{{3\, N \, k \,T_e }}{{a}} \right)^{0.25}\,{\rm{.}}
\ee
We often can assume thermal equilibrium and use the approximations: $N \approx  1.917\, N_e~ {\rm{cm}}^{-3}{\rm{;}}$ and $ T_e = 2.706 \cdot 10^6~{\rm{K}}$ for the average kinetic temperature of the plasma particles.  For the electrons, we have: $(3/2) k T_e = (1/2)m_e (v^2)_{av} {\rm{,}}$ where $k$ is Boltzmann constant, $m_e$ the electron's mass, and $v$ its velocity.  $a = 7.566 \cdot 10^{-15}~ {\rm{erg\, cm}}^{-3}\,{\rm{K}}^{-4}$ is the Stefan-Boltzmann constant for the electromagnetic energy density.

\indent  From comparison with many experiments, such as the magnitude-redshift relation for SNe\,Ia, we know the average electron density of intergalactic space.  This magnitude-redshift relation predicts $N_e = 1.95\cdot 10^{-4} \cdot (H_0/60) ~{\rm{cm}}^{-3}{\rm{,}}$ which predicts also correctly the dimming that is caused by Compton scattering.  The Compton scattering thus independently predicts the same average electron density in intergalactic space.

\indent  From Eq.(C20), we can from $N$ then determine the average kinetic temperature $ T_e = 2.706~{\rm{K}} $ of the electrons, because $T_{CMB}$ is known accurately.  We can also independently determine $T_e$ from the X-ray emission measurements, although this last determination is less accurate.

\indent  Kuntz et al.~[102] estimated that the temperature of the warm-hot intergalactic medium (WHIM) can be approximated by $T_e = 2.63\cdot 10^6~{\rm{K .}}~$ They believe, however, that the true spectrum is more likely to come from a range of temperatures.  The temperature in the Local Hot Bubble (LHB) is believed to be about $T_e=1.3 \cdot 10^6 {\rm{;}}$ and the Galactic Halo with  $T_e=1.2 \cdot 10^6 {\rm{,}}$ and maybe $T_e=2.8 \cdot 10^6 {\rm{,}}$ as found by Snowden et al.~[131], and by Kuntz and Snowden [103].  These determinations are also consistent with the estimates in section 5.7.1 that were derived from the good observations by Pettini et al.~[68], who used the bright supernova 1987 in the Large Magellanic Cloud (LMC) to obtain good absorption lines. 

\indent  In the transition zone and low in the corona, the temperatures are lower.  The high temperatures of about 2.8 million K are likely to be high in the Galactic corona.  It follows that the coronal plasma must diffuse into the intergalactic space, because the gravitational potential can not hold it back.  These determinations thus contradict the assumptions often made by big-bang cosmologists that the intergalactic space is practically empty.

\indent  These densities and temperatures are consistent with those predicted by Eq.\,(C20), as well as, the temperatures in the Galactic corona predicted in section 5.7.  They are also consistent with the magnitude-redshift relation in the SN\,Ia experiments.  In section 5.9, we estimated that the column density in the Galactic corona at high latitudes corresponds is $\Delta z = 9.25\cdot 10^{-4} $ or $N_e  \approx  2.78 \cdot 10^{21} ~ {\rm{cm}}^{-2}{\rm{.}}~ $ This column density high in the corona nearly equals the column density, $3.1\cdot 10^{21}$ of one relaxation length for 0.75 keV photons (see Table C1).  It is therefore difficult to discern the temperatures in the WHIM beyond the Galactic corona, which decreases the accuracy in the temperatures of WHIM derived from the X-ray observations.  Besides Eq.\,(C19) many determination of the $T_e $ lead to similar values and besides Eq.\,(C20) could be used to determine the $T_{CMB}{\rm{.}}~ $
 
\indent  However, only plasma redshift gives a reasonable explanation of these million degree temperatures both in the LHB, the Galactic corona, and in intergalactic space.  The conventional excitations and ionizations by UV photons that produce a Str\"{o}mgren radius of a plasma do not result in million degree temperatures.  Although some invisible supernovae with no other supporting indicators have sometimes been suggested as possible explanations for these high temperatures, they can not explain the high densities in the Galactic corona and not the high temperatures in the solar corona, which are nicely explained by the plasma redshift, as shown in sections 5.1 and 5.2.
\vspace{3mm}

\noindent  \textbf{Spectrum and isotropy of CMB}
\vspace{1mm}

\noindent  It has been amazing to see how well the CMB fits the blackbody spectrum, and how isotropic it is.  In the frequency range of the CMB, the plasma redshift dominates all other absorption processes by several orders of magnitude.  Only at frequencies below the CMB frequency of about $10^9~{\rm{Hz ,}}$ does the cut-off for plasma redshift start to take effect gradually; first in the coldest filaments of space.  This is all consistent with observations. 

\indent  On the high frequency end, $\nu \geq 7\cdot 10^{11}~{\rm{Hz ,}}$ the limit is set by the gradual redshift of the light from stars and other objects and including infrared emission from grains.  The CMB frequencies, which exceed the cut-off frequencies, are also continually being redshifted.  The absorption (redshift) per cm of the CMB is continually replaced by equivalent emission of CMB.  The spectrum is therefore constant in space and time.  The isotropy follows from:

\begin{enumerate}   
\item The $T_{CMB} $ is given by Eq.\,(C20) and the average pressure is well defined.  The large dimensions, about $R_{pl}= 1.542\cdot 10^{28}~{\rm{cm}} \approx 5,000~{\rm{Mpc ,}} $ of the blackbody cavity lead to relatively good averages for the pressure $p{\rm{.}}$
\item The column density, $R_{pl}\,N_e = 3.006\cdot 10^{24}~{\rm{cm}}^{-2}{\rm{,}}$ of electrons, which determines the integration length for the blackbody radiation is a constant independent of the direction, the electron-density variations, and the spectral frequency.
\end{enumerate}
\vspace{3mm}

\noindent  \textbf{X-ray intensities}
\vspace{1mm}

\indent  For $T_e = 3 \cdot 10^6 ~ {\rm{K}}$ and X-ray frequencies, the velocity averaged Gaunt factor is about $\langle g_{ff}\rangle \approx 1$ (see Fig.\,\,5.3 of Rybicki and Lightmann [128]).   The maximum intensity in the free-free emission (a blackbody radiation) is at the frequency $\nu_{max} = 5.8790 \cdot 10^{10} \, T_e = 1.7637 \cdot 10^{17} ~ {\rm{Hz.}} ~$  This frequency corresponds to $h\nu_{max} = 729.4 $ eV, and to $h\nu_{max}/kT_e = 2.821{\rm{.}}~$ At this frequency the exponential factor in Eq.\,(C7) is $ {\rm{exp}}(- h\nu /kT_e) = 0.05952 {\rm{.}}~$

\indent  From Table C1, we get for 729.4 keV X rays that the absorption distance is $R_{\nu}=1.73 \cdot 10^{25}~{\rm{cm}}{\rm{.}}~$ At these frequencies it is thus much shorter than the absorption distance, about $R_{pl}=1.54 \cdot 10^{28}~{\rm{cm}}{\rm{,}}$ caused by the plasma redshift in intergalactic space.  From the SN\,Ia experiments, we estimated (see the penultimate paragraph in section 5.9) that the intrinsic redshift at high latitudes in the Galactic corona is about $\Delta z = 0.000925 {\rm{.}}~$  This corresponds to a column density of about $N_e  \approx  2.78 \cdot 10^{21} ~ {\rm{cm}}^{-2}{\rm{.}}~ $ The column densities at high latitudes in the Galactic corona would then reduce the intensity from intergalactic space by a factor of about ${\rm{exp}}(-2.78/3.05) = 0.4019 {\rm{,}}$ where $N_e  \approx  3.05\cdot 10^{21} ~ {\rm{cm}}^{-2}{\rm{}} $ is is obtained by interpolation from Table C1 in Appendix C.
  
\indent Using Eq.\,(C7), and the absorption in the corona, we get that the observed intensity at $h\,\nu = 729$ eV would be
\be
I_{\nu}= R_{\nu}\, 8.2166 \cdot 10^{-13} \,\frac{{Z_i^2 \, \langle g_{ff}\rangle \, N_e \, N_i \,}}{{ T_e^{1/2} }}  \, e^{- h\nu /kT_e}\,  \cdot {\rm{exp}}(-2.78\cdot 10^{21}/3.05\cdot 10^{21}) , \quad \quad \quad
\ee
or
\[ = 1.7\cdot 10^{25}\cdot 8.2 \cdot 10^{-13} \, \frac{{1.17\cdot 1\cdot (1.95\cdot 10^{-4})^2}}{{(3\cdot 10^6)^{1/2}}}\, e^{- 2.821}\,\cdot {\rm{exp}}(-2.78/3.05)
\]
\[
=8.7 ~~{\rm{keV \, cm}}^{-2} \, {\rm{s}}^{-1}\,{\rm{sr}}^{-1}\,{\rm{keV}}^{-1}\,{\rm{.}}\quad \quad \quad \quad \quad \quad \quad \quad \quad \quad \quad \quad \quad \quad \quad \quad \quad ~
\]
\noindent   At $h\nu = 750 $ eV, we could then expect slightly lower intensity, or about $I_{\nu}\approx 8~{\rm{keV \, cm}}^{-2} \, {\rm{s}}^{-1}\,{\rm{sr}}^{-1}\,{\rm{keV}}^{-1}\,{\rm{.}}~$  The intensity values given by Eq.\,(C21) would also be reduced by the inhomogeneties in intergalactic space.  Close to a center of a ``bubble'' the temperature would be higher, and the density lower, which reduces the ratio $N_e^2/T_e^{1/2}$ but increases the value of $ {\rm{exp}}(- h\nu /kT_e){\rm{.}}~$ For example, if the temperature $T_e$ increases from $3 \cdot 10^6$ to $4.5 \cdot 10^6 {\rm{,}}$ the ratio $N_e^2/T_e^{1/2}$ changes by a factor of 0.3629, while the factor $ {\rm{exp}}(- h\nu /kT_e){\rm{}}$ changes by a factor of 2.5608.  The total reduction is then only $0.3629\cdot 2.5608 = 0.9293 {\rm{.}}~$  This higher temperature of $T_e = 4.5 \cdot 10^6 {\rm{,}}$ in the bubble results also in a slower decrease in the intensity with increasing energy of the X rays.  In the cold regions the intensity will increase slightly, but the absorption in the Galactic corona would increase for the low-energy X-ray photons more than for the high-energy photons.  These modifications of Eq.\,(C21) are consistent with observations.  By analyzing the attenuation spectrum, it has been possible with the fine X-ray instrumentation to analyze the spectrum and separate the point sources from the diffuse X-ray background.
\vspace{1mm}

\indent  Kuntz et al.~[102] estimated that the upper limit for the emission strength for the cosmic X-ray background (CXB) is $7.5\pm 1.0~~{\rm{keV\,cm}}^{-2}\,{\rm{s}}^{-1} \,{\rm{sr}}^{-1}\,{\rm{keV}}^{-1}\,{\rm{}} $ in the 0.75 keV band, and that an unknown portion of this value may be due to the Galactic halo.  They found that the shape of the warm-hot intergalactic medium (WHIM) can be described as thermal emission with $T_e = 2.63 \cdot 10^6~{\rm{K ,}}$ although a true spectrum is more likely to come from a range of temperatures.  They concluded that the extra galactic power law component of the soft X-ray background spectrum must be $9.5\pm 0.9~~{\rm{keV \, cm}}^{-2}\, {\rm{s}}^{-1}\,{\rm{sr}}^{-1}\,{\rm{keV}}^{-1} \,{\rm{.}}~$ This small difference, 9.5 versus 7.5 between the intensity before and after the galactic absorption, indicates that they believed the Galactic absorption reduced the intensity only by a factor of about 7.5/8.5 = 0.79, while Eq.\,(C21) indicates a factor of 0.402.  This small Galactic absorption reflects the common opinion among big-bang cosmologists that the Galactic corona has much lower density than that derived from the plasma-redshift theory.  Many observations, such as those by Pettini et al.~[68] have confirmed the higher densities in the Galactic corona.  Kuntz et al.~[102] show in their Fig.\,1 that the observed intensity especially at high energies is dominated by unresolved extragalactic sources.  Beyond about 0.5 keV, even the X-ray intensity from unresolved Galactic stars after absorption in the Galactic disk is significant.

\indent  At 3 keV, De Luca and Molendi [104] estimated the CXB to be $7.4 \pm 0.27~{\rm{keV \, cm}}^{-2}\, {\rm{s}}^{-1}\,{\rm{sr}}^{-1}\,{\rm{keV}}^{-1} {\rm{,}}$ and in the 1 keV range, they estimated $11.6\pm 0.9~{\rm{keV \, cm}}^{-2}\, {\rm{s}}^{-1}\,{\rm{sr}}^{-1}\,{\rm{keV}}^{-1} {\rm{.}}~$   

\indent  In the 1 keV range, Vecchi et al.~[105] find that the CXB is $11.7 \pm 0.5~{\rm{keV \, cm}}^{-2}\, {\rm{s}}^{-1}\,{\rm{sr}}^{-1}\,{\rm{keV}}^{-1} \,{\rm{;}}$  and in the same range, Barcons et al.~[106] find the CXB to be: $9.8_{-1.0}^{+0.7}\,{\rm{,}}\,~ 11.1_{-2.3}^{+1.0}\,{\rm{,}}~\,{\rm{and}}~\, 12.7_{-1.9}^{+0.9}\,{\rm{,}}$ ${\rm{keV \, cm}}^{-2}\, {\rm{s}}^{-1}\,{\rm{sr}}^{-1}\,{\rm{keV}}^{-1} \,{\rm{,}}$ for ASCA, BeppoSAX, and ROSAT, respectively.

\indent  The values by De Luca and Molendi [104], Vechi et al. [105], and Barcons et al. [106] appear slightly higher than that predicted and may indicate contamination from point sources.  But all these values derived from observations are consistent with those predicted by Eq.\,(C21).


\section*{C4\, \,Conclusions}

The plasma-redshift cosmology gives a good explanation of the CMB, its well-defined temperature, spectrum, and isotropy.  The X-ray intensities from the diffuse WHIM are shown to be consistent with observations.

\indent  The fact that the plasma-redshift absorption is greater than the free-free absorption and the Compton scattering absorptions by many orders of magnitude in the CMB frequency range has the effect that the blackbody radiation is well defined.  The plasma-redshift cosmology, which uses only basic laws of physics, is incompatible with the big-bang cosmology.  Not only do the big-bang cosmologists overlook the plasma redshift, but they also incorrectly equate $T_{CMB}$ with $T_e{\rm{}}$ in the cosmic plasma at the time of decoupling.


\renewcommand{\theequation}{D\arabic{equation}}
\setcounter{equation}{0}
\section*{Appendix D}


\section*{D1\, \,Charged Particle's Field in a Plasma}

The electrical field of the incident particle that is moving with velocity $v~{\rm{cm\,s}^{-1}}$ in the direction of the x-axis can be Fourier analyzed.  In a rectangular coordinate system the particle has the coordinates $(x,\,y,\,z)=(vt,\,0,\,0)$  and charge $Ze{\rm{.}}~$  At a position $b$ on the y-axis the the field's Fourier harmonics with cyclic frequency $\omega $ are then given by 
\be
A_x(b,\,\omega)= \frac{{1}}{{2\pi}}\int\limits_{-\infty}^{~~\infty} \!\! F_x(b,\,t) \cdot e^{-i\omega t} \cdot dt = i\,\frac{{Ze}}{{\pi \gamma v b}}\left[ \frac{{\omega b}}{{\gamma v}} \cdot K_0\left(\frac{{\omega b}}{{\gamma v}}\right) \right]{\rm{,}}
\ee
and
\be
A_y(b,\,\omega)= \frac{{1}}{{2\pi}}\int\limits_{-\infty}^{~~\infty} \!\!  F_y(b,\,t) \cdot e^{-i\omega t} \cdot dt = \,\frac{{Ze}}{{\pi v b}}\left[ \frac{{\omega b}}{{\gamma v}} \cdot K_1\left(\frac{{\omega b}}{{\gamma v}}\right) \right]{\rm{,}}
\ee
\noindent where
\be
F_x(b,\,t) = F_x(0,\,b,\,0,\,t) = \frac{{ -Ze \,\gamma \, v \, t }}{{ (b^2 ~+~ \gamma^2 \, v^2 \, t)^{3/2} }}, 
\ee
\be
F_y(b,\,t) = F_x(0,\,b,\,0,\,t)  = \frac{{ Ze \,\gamma \, b  }}{{ (b^2 ~+~ \gamma^2 \, v^2 \, t)^{3/2} }}, 
\ee
\be
F_z (b,\,t)= H_x (b,\,t)= H_y(b,\,t) = 0 ,
\ee
\be
H_z(b,\,t) = H_z(0,\,b,\,0,\,t)  = \frac{{v}}{{c}}\frac{{ Ze \,\gamma \, b  }}{{ (b^2 ~+~ \gamma^2 \, v^2 \, t)^{3/2} }}, 
\ee
where $i = \sqrt{-1}$ and where $K_0(\omega b / \gamma v)$ and $K_1(\omega b / \gamma v)$ in Eqs.\,(D1) and (D2) are the modified Bessel functions of zero and first order, and where  $\gamma = (1-v^2/c^2)^{-1/2} {\rm{,}}$ is the relativistic factor, which is close to one in the plasmas of our main interest.  We can set
\be
A(b,\,\omega) = A_y + A_x = \frac{{Ze}}{{\pi v b}}\left\{ \frac{{\omega b}}{{\gamma v}} \cdot K_1\left(\frac{{\omega b}}{{\gamma v}}\right) + i\,\frac{{1}}{{\gamma}}\, \frac{{\omega b}}{{\gamma v}} \cdot K_0\left(\frac{{\omega b}}{{\gamma v}}\right) \right\}{\rm{.}}
\ee 
\noindent The component $A_x$ is in the direction of the particle's velocity $\vec{v}{\rm{,}}$ while $A_y$ is perpendicular to $\vec{v}{\rm{.}}~$ We have that
\be
A(b,\,\omega)\,A^*(b,\,\omega)=\frac{{Z^2 e^2}}{{\pi^2 v^2 b^2}}\left\{ \left[\frac{{\omega b}}{{\gamma v}} \cdot K_1\left(\frac{{\omega b}}{{\gamma v}}\right)\right]^2 + ~\left[ \frac{{1}}{{\gamma}}\, \frac{{\omega b}}{{\gamma v}} \cdot K_0\left(\frac{{\omega b}}{{\gamma v}}\right)\right]^2 \right\}{\rm{,}}
\ee
\noindent  where $A^*$ is the complex conjugate of $A{\rm{.}}$  This form, Eq.\,(D8), has the nice characteristic that the quantity within the braces is about equal to 1 for $\zeta = (\omega b / \gamma v)\leq 1 $ and falls of exponentially for $\zeta > 1{\rm{.}}$ 

\indent  We have also as Niels Bohr has shown in 1913 and 1915 [3] that any electron oscillator with charge $e{\rm{,}}$ mass $m$ and eigenfrequency $\omega_0$ when exposed to this the field will absorb energy $q$ equal to
\be
q = \frac{{2 \pi^2 e^2}}{{m}}\cdot A(b,\,\omega_0)A^*(b,\,\omega_0)= \frac{{2\,Z^2 e^4}}{{m v^2 b^2}}\left\{ \left[\frac{{\omega_0 b}}{{\gamma v}} \cdot K_1\left(\frac{{\omega_0 b}}{{\gamma v}}\right)\right]^2 +~ \left[ \frac{{1}}{{\gamma}}\, \frac{{\omega_0 b}}{{\gamma v}} \cdot K_0\left(\frac{{\omega_0 b}}{{\gamma v}}\right)\right]^2 \right\}{\rm{,}}
\ee
We assume that there are $N_{e,\,\omega_0}$ oscillators per ${\rm{cm}}^3$ with frequency $\omega_0$, and the density is so low that we can assume that the dielectric constant is 1.  The energy absorbed per infinitesimal distance $dx$ is obtained by multiplying Eq.\,(D9) by $N_{e,\,\omega_0} \, 2\, \pi \, b \,db$ and integrating over b.  We get
\be
Q = -\frac{{dE}}{{dx}} =  \frac{{4 \pi\,Z^2 e^4 \, N_{e,\,\omega_0}}}{{m v^2}}\!\!\!\!\int\limits_{~b_{min}}^{~~~b_{max}}\left\{ \left[\frac{{\omega_0 b}}{{\gamma v}} \cdot K_1\left(\frac{{\omega_0 b}}{{\gamma v}}\right)\right]^2 +~ \left[ \frac{{1}}{{\gamma}}\, \frac{{\omega_0 b}}{{\gamma v}} \cdot K_0\left(\frac{{\omega_0 b}}{{\gamma v}}\right)\right]^2 \right\} \frac{{db}}{{b}}{\rm{,}}
\ee
\noindent  where $ b_{min} = \hbar/\sqrt{ 2mE_{max} } $ and $ b_{max} = \gamma v / \omega_0 {\rm{.}}~$ For proton and heavier incident particles, we have that $E_{max} = 2m\gamma^2 v^2{\rm{,}}$ for incident electrons $E_{max} = (1/2)m c^2 (\gamma - 1) = (1/2)E_{kin} {\rm{,}}$ and for incident positrons $E_{max} = m c^2 (\gamma - 1) = E_{kin} {\rm{.}}~$ This integral can be integrated exactly and gives
\be
Q = -\frac{{ dE }}{{ dx }} =  \frac{{ 2 \pi\,Z^2 e^4 \, N_{e,\,\omega_0} }}{{ m v^2 }}\left[\frac{{v^2}}{{c^2}}\left\{ x^2 K_1^2(x) \,-\,x^2 K_0^2(x) \right \} \,-\,2 x K_0(x) K_1 (x)    \right]_{x_{min}}^{x_{max}} {\rm{,}}
\ee
where the minimum and maximum impact parameter correspond to  
\[
x_{min} = \frac{{\hbar \omega_0}}{{\gamma v \sqrt{2 m E_{max}}}} {\rm{,}}\quad {\rm{and}} \quad x_{max}=\frac{{b_{max}\, \omega_0}}{{\gamma v}}= 1.
\]
\noindent  We use the quantum mechanical limits (and not the classical limits), because the temperatures in the intergalactic plasma are very high.  When we insert these integration limits, we can bring Eq.\,(D11) on a simpler and more familiar form
\be
Q = -\frac{{ dE }}{{ dx }} =  \frac{{ 2 \pi\,Z^2 e^4 \, N_{e,\,\omega_0} }}{{ m v^2 }}\left[ {\rm{ln}} \frac{{ 2 m \gamma^2\,v^2 \, E_{max} }}{{ \delta_1^2 \hbar^2 \omega_0^2}} ~ - ~ \delta_2 \frac{{ v^2 }}{{ c^2 }} \right] {\rm{,}}
\ee
\noindent  where $\hbar \omega_0 $ is the binding energy of the oscillator, and $E_{max}$ is the maximum energy that can be transferred to the oscillator.  The constants $\delta_1 = 1.1474  {\rm{,}}$ and $\delta_2 = 0.8150$ are factors that take into account the quantum mechanical limits.  Hans A. Bethe used approximations for the modified Bessel functions, $K_0$ and $K_1$ and derived similar values for these constants.

\indent  We can write Eq.\,(D12) on a slightly different format 
\be
Q = -\frac{{ dE }}{{ dx }} =  \frac{{ 2 \pi\,Z^2 e^4 \, N_{e,\,\omega_0} }}{{ m v^2 }}\left[ {\rm{ln}} \frac{{ 2 m \gamma^2\,v^2 }}{{ \delta_1^2 \hbar \omega_0}} ~ - ~ \delta_2 \frac{{ v^2 }}{{ c^2 }}~+~{\rm{ln}} \frac{{E_{max} }}{{\hbar \omega_0}} \right] {\rm{,}}
\ee
\noindent  The two first terms inside the brackets are the resonance terms, and the third term is the hard collision term.   In this last term, the value of $E_{max}$ changes with the particles involved in the collision.


\section*{D2\, \,Quantum mechanical excitation in the plasma}

The states in the plasma are numerous.  In a hot plasma a few of the electrons are momentarily in a highly excited states of the hydrogen atom.  Farther away from the protons, the electrons can be considered as oscillating with the high multiples of the plasma frequency as shown in section 3.  The kinetic energies of the electrons exceed the ionization potentials for most of the atomic shells.  These electrons will oscillate in the high-energy plasma states as described in section 3.  The probability $\phi_{mn} $ for transition from state $n$ to a state $m$ due to action of a perturbing energy disturbance $U(r,\,t)$ acting for a short time $t\ll 2\pi/\omega_{mn}$ is known from quantum theory to be
\be
\phi_{mn} = \frac{{4 \pi^2}}{{\hbar^2}} |U(r,\,\omega_{mn})|^2 {\rm{,}}
\ee
where $U(r,\,\omega_{mn})$ is the Fourier component with frequency $\omega_{mn}$ of the matrix element of $U(r,\,t){\rm{,}}$ and $\hbar \omega_{mn}$ the energy difference between the two states.

\indent  We write the potential energy of an electron with charge $e$ in the electrical field $\vec{A}(t)$ as 
\be
U(r,\,t) = \vec{A}(t)\, e \, \vec{r} = A(t) \, \vec{u}\,e\,\vec{r} {\rm{,}}
\ee
where $\vec{r}$ is the space vector to the position of the charge $e{\rm{,}}$ and $\vec{u}$ is a unit vector in the direction of the acting electrical field $\vec{A}(t){\rm{.}}~$ The matrix element $U_{mn}(t)$ of $U(r,\,t)$ is
\be
U_{mn}(t) = \int \psi_m^*\, U(r,\,t)\,\psi_n \, d\tau {\rm{,}}
\ee
where the integration is performed over the entire volume of the state function $\psi_m^*$ and $\psi_n{\rm{.}}~$ If the function $U(r.\,t)$ is on the form of Eq.\,(D15) over integration volume of Eq.\,(D16), we get
\be
U_{mn}(t) = A(t) \,e\, \int \psi_m^*\, \vec{u}\,\vec{r}\,\psi_n \, d\tau = A(t)\, \vec{u}\,e\, |\vec{X}_{mn}| = A(t)\, \vec{u} |\vec{D}_{mn}|{\rm{,}}
\ee
\noindent  where $D_{mn} = e\, |X_{mn}|$ is the matrix element of any dipole moment in the plasma.
The Fourier harmonic of $A(t)$ is given by Eq.\,(D7).  When we insert these Fourier harmonics in to Eq.\,(D14), we get
\be
\phi_{mn}=\frac{{4 \pi^2}}{{\hbar^2}} |A(t)\, \vec{u}\,e\, \vec{X}_{mn}|^2 =\frac{{4 Z^2 e^4}}{{\hbar^2 v^2 b^2}}\left\{ \left[\frac{{\omega b}}{{\gamma v}} \cdot K_1\left(\frac{{\omega b}}{{\gamma v}}\right)\right]^2 + \left[ \frac{{1}}{{\gamma}}\, \frac{{\omega b}}{{\gamma v}} \cdot K_0\left(\frac{{\omega b}}{{\gamma v}}\right)\right]^2 \right\} \cdot |{X}_{mn}|^2{\rm{,}} 
\ee 
which is analogous to Eq.\,(D9).  When we multiply by $N_e \, 2\, \pi \, b \,db$ and integrate over b, we get that the cross section is given by
\be
\sigma_{mn} = \Phi_{mn} =  \frac{{ 4 \pi\,Z^2 e^4 \, N_e }}{{ \hbar^2 v^2 }}\cdot |{X}_{mn}|^2\cdot\left[ {\rm{ln}} \frac{{ 2 m \gamma^2\,v^2 \,  }}{{ \delta_1^2 \hbar \omega_0}} ~ - ~ \delta_2 \frac{{ v^2 }}{{ c^2 }} \right] {\rm{,}}
\ee
\noindent  which is analogous to the first two terms, the resonance terms in Eq.\,(D13).  When we integrate over all the levels $n$ above the level corresponding to $\omega_0 {\rm{,}}$ we should get the total absorption in transition to the resonance terms, the first two terms in Eq.\,(D13).  The values of $|{X}_{mn}|^2$ are usually very small, but in hot sparse plasma the collision broadening enhances the transition probabilities from one state to another in the plasma, as shown in section 3.



\begin{thebibliography}{10}

\bibitem{1}  W.~Heitler. {\it{The Quantum Theory of Radiation}}, 3rd ed.~Oxford Clarendon Press, 1954
\bibitem{2}  R.~J.~Gould. ApJ.~{\textbf{285}} (1984) 275
\bibitem{3}  N.~Bohr, Phil.~Mag. {\textbf{25}} (1913) 10, {\textbf{26}} (1913) 1, ibid. {\textbf{30}} (1915) 581.
\bibitem{4}  M.~Abramowitz, I.~A.~Stegun, editors.  {\it{Handbook of Mathematical Functions with Formulas, Graphs, and Mathematical Tables }}, National Bureau of Standards (now NIST), Applied Mathematics Series 55, June 1964, Fourth printing, December 1965, with corrections.  U.~S.~Government Printing Office, Washington, D.~C.
\bibitem{5}  P.~A.~Sturrock, {\it{Plasma Physics}}, Cambridge University Press 1994.  ISBN 0 521 44350 4 
\bibitem{6}  J.~E.~Vernazza, E.~H.~Averett, and R.~Loeser, ApJ.~Supplem.~Series {\textbf{45}} (1981) 635
\bibitem{7}  U.~Feldman, I.~E.~Dammasch, K.~Wilhelm, ApJ.~{\textbf{558}} (2001) 423
\bibitem{8}  H.~Friedman.  {\it{The Astronomers Universe}}, (Balatines Books, New York, 1990)
\bibitem{9}  J.~V.~Hollweg, ApJ.~\textbf{257} (1982) 345 
\bibitem{10}  M.~L.~Goodman,  ApJ.~\textbf{503} (1998) 938
\bibitem{11}  R.~S.~Sutherland and M.~A.~Dopita, ApJ., Supplement Series \textbf{88} (1993) 253
\bibitem{12}  H.~Holweger, Astron.~Astrophys.~\textbf{10} (1971) 128
\bibitem{13}  R.~W.~P.~McWhirter, P.~C.~Thonemann, R.~Wilson, Astron.~\& Astrophys \textbf{40} (1975) 63
\bibitem{14}  G.~L.~Withbroe, ApJ.~\textbf{325} (1988) 442
\bibitem{15}  G.~L.~Withbroe, ApJ.~\textbf{337} (1989) L49
\bibitem{16}   J.~T.~Gosling, Annual Review of Astronomy and Astrophysics, \textbf{34} (1996) 35
\bibitem{17}  J.~B.~Zirker, Solar Physics \textbf{148} (1993) 43
\bibitem{18}  H.~Zirin, Solar Physics \textbf{169} (1996) 313
\bibitem{19}  R.~A.~Frazin, and  P.~Janzen, ApJ.~ \textbf{570} (2002) 408
\bibitem{20}   E.~N.~Parker, ApJ.~\textbf{372} (1991) 719
\bibitem{21}   G.~Newkirk, Jr., Annual Review of Astronomy and Astrophysics \textbf{5} (1967) 213
\bibitem{22}   P.~A.~Sturrock, M.~S.~Wheatland,  L.W.~Acton, ApJ.~\textbf{461} (1996) L115 
\bibitem{23}   M.~S.~Wheatland,  P.~A.~Sturrock, L.~W.~Acton, ApJ.~\textbf{482} (1997) 510 
\bibitem{24}   N.~R.~Sheeley, Jr.~et al. (18 coauthors from USA, France, Germany, UK), ~ApJ.~\textbf{484} (1997) 472
\bibitem{25}  Y.-M.~Wang et al.~ApJ.~\textbf{508} (1998) 899
\bibitem{24}   D.~A.~Falconer, R.~L.~Moore, J.~G.~Porter, G.~A.~Gary, T.~Shimizu, ApJ.~\textbf{482} (1997) 519 
\bibitem{27}   R.~L.~Moore, D.~A.~Falconer, J.~G.~Porter, S.~T.~Suess, ApJ.~\textbf{526} (1999) 505 
\bibitem{28}  L.~Spitzer, {\it{Physics of fully ionized gases}}, New York Interscience, 1962 
\bibitem{29}   S.~R.~Spangler, S.~Mancuso,  ApJ.~\textbf{530} (2000) 491
\bibitem{30}   J.~T.~Steinberg, A.~J.~Lazarus, K.~W.~Ogilvie, R.~Lepping, J.~Byrnes, Geophys.~Res.~Lett.~\textbf{23} (1996) 1183
\bibitem{31}  J.-F.~De La Beaujardire, R.~C.~Canfield, H.~S.~Hudson, J.-P.~W\"{u}lser, L.~Acton, T.~Kosugi, S.~Masuda, ApJ.~\textbf{440} (1995) 386 
\bibitem{32}  R.~J.~Murphy, G.~H.~Share, K.~W.~DelSignore, X.-M.~Hua, ApJ.~\textbf{510} (1999) 1011
\bibitem{33}  M.~G.~Adam, Mon.~Not.~R.~astr.~Soc.~\textbf{119} (1959) 460 
\bibitem{34}  L.~A.~Higgs, Mon.~Not.~R.~astr.~Soc.~\textbf{121} (1960) 421 
\bibitem{35}   P.~N.~Brandt, E.~H.~Schr\"{o}ter, Solar Physics \textbf{79} (1982) 3   
\bibitem{36}   F.~Cavallini, G.~Ceppatelli, A.~Righini, Astrn.~Astrophys.~143 (1985) 116 
\bibitem{37}  R.~V.~Pound, G.~A.~Rebka, Jr., Phys.~Rev.~Lett.~\textbf{3} (1959) 439; ibid.~\textbf{3} (1959) 554 
\bibitem{38}  R.~V.~Pound, G.~A.~Rebka, Jr., Phys.~Rev.~Lett.~\textbf{4} (1960) 337 
\bibitem{39}  R.~V.~Pound, J.~L.~Snider, Phys.~Rev.~Lett.~\textbf{13} (1964) 539 
\bibitem{40}  R.~F.~C.~Vessot et al.,  Phys.~Rev.~Lett.~\textbf{45} (1980) 2081 
\bibitem{41}  T.~P.~Krisher, D.~D.~Morabito, J.~Anderson, Phys.~Rev.~Lett.~\textbf{70} (1993) 2213 
\bibitem{42} C.~Riveros, H.~Vucetich, Phys.~Rev.~D.~\textbf{34} (1986) 321 
\bibitem{43}  I.~I.~Shapiro, M.~E.~Ash, R.~P.~Ingalls, W.~B.~Smith, D.~B.~Campbell, R.~B.~Dyce, R.~F.~Jurgens, G.~H.~Pettengill, Phys.~Rev.~Lett.~\textbf{26} (1971) 1132 
\bibitem{44}  D.~Dravins, L.~Lindgren, \AA.~Norlund, Astron.~Astrophys.~\textbf{96} (1981) 345  
\bibitem{45} D.~Dravins, Ann.~Rev.~Astrophys.~\textbf{20} (1982) 61
\bibitem{46}  J.~E.~Vernazza, E.~H.~Averett, and R.~Loeser, ApJ.~Supplem.~Series \textbf{30} (1976) 1
\bibitem{47}  A.~K.~Pierce, J.~C.~LoPresto, Solar Phys.~\textbf{196} (2000) 41 
\bibitem{48}  H.~R.~Griem, {\it{Spectral line broadening by plasmas}}, Academic Press, New York, and London, ISBN 0-12-302850-7, 1974
\bibitem{49}   C.~E.~St.~John, ApJ.~ \textbf{67} (1929) 195
\bibitem{50}   L.~Herzberg, Can.~J.~Phys.~ \textbf{38} (1960) 853
\bibitem{51}  P.~Miller, P.~Foukal, P.~Keil, Solar Phys.~\textbf{149} (1984) 33 
\bibitem{52}  H.~Arp, Mon.~Not.~R.~astr.~Soc.~\textbf{258} (1992) 800.  See also, {\it{Seeing Red: Redshifts, Cosmology and Academic Science}}, published by Apeiron Montreal, Quebec, 1998, ISBN 0-9683689-0-5 
\bibitem{53}   A.~Brynjolfsson, {\it{Weightlessness of photons: A quantum effect}}, arXiv:astro-ph/0408312 v2 26 Aug 2004
\bibitem{54}   W.~S.~Adams, Proc.~Nat.~Acad.~Sci., \textbf{11} (1925) 382-387 
\bibitem{55}  G.~Str\"{o}mberg, Publs.~astr.~Soc.~Pacif.., \textbf{38} (1926) 44
\bibitem{56}  A.~S.~Eddington, in Mon.~Not.~R.~astr.~Soc., \textbf{848} (1924) 308
\bibitem{57}  J.~L.~Greenstein, J.~B.~Oke,  \& H.~L.~Shipman, Ap.J. \textbf{169} (1971) 563
\bibitem{58}   M.~A.~Barstow, H.~E.~Bond, J.~B.~Holberg, M.~R.~Burleigh, I.~Hubeny, \& D.~Koester, in  Mon.~Not.~R.~astr.~Soc., in press \textbf{848} (2005) 1-11 
\bibitem{59}   A.~Brynjolfsson, {\it{Hubble constant from lensing in plasma-redshift cosmology, and intrinsic redshift of quasars.}}  arXiv:astro-ph/04011666 v3 2Dec 2004
\bibitem{60}   L.~Spitzer Jr., J.~P.~Ostriker,  {\it{Dreams, Stars, and Electrons}}: Selected writings of Lyman Spitzer Jr.~ Princeton University Press, Princeton, New Jersey, 1997
\bibitem{61}   R.~J.~Reynolds, ApJ.~ \textbf{372} (1991) L17
\bibitem{62}   R.~J.~Reynolds, D.~P.~Cox, ApJ.~ \textbf{400} (1992) L33
\bibitem{63}  J.~M.~Cordes, J.~M.~Weisberg, D.~A.~Frail, S.~R.~Spangler, M.~Ryan, Nature \textbf{354} (1991) 121
\bibitem{64}  G.~C.~G\'{o}mez, R.~A.~Benjamin, D.~P.~Cox, Astron.~J.~\textbf{122} (2001) 908 
\bibitem{65}  B.~D.~Savage and D.~Massa, ApJ.~\textbf{314} (1987) 380
\bibitem{66}   D.~P.~Cox, R.~J.~Reynolds, Ann.~Rev.~Astron.~Astrophys.~\textbf{25} (1987) 303 
\bibitem{67}   R.~A.~Chevalier, W.~R.~Oegerle, ApJ.~ \textbf{227} (1979) 398 
\bibitem{68}   M.~Pettini, R.~Stathakis, S.~D'Odorico, P.~Molaro, G.~Vladilo, ApJ.~\textbf{340} (1989) 256
\bibitem{69}   P.~J.~E.~Peebles,  {\it{Principlees of Physical Cosmology}}, Princeton University Press, ISBN 0-691-01933-9, 1993
\bibitem{70}   T.~S.~van Albada, R.~Sancisi, Phil.~Trans.~R.~Soc.~London
\textbf{A320} (1986) 447 
\bibitem{71}   M.~Milgrom, ApJ.~ \textbf{270} (1983) 365 
\bibitem{72}   R.~Bottema, J.~L.~G.~Pesta\~{n}a, B.~Rothberg, R.~H.~Sanders.~ Astron.~Astrophys.~ \textbf{393} (2002)453
\bibitem{73}  T.~E.~Clarke, P.~P.~Kronberg, and H.~B\"{o}hringer, ApJ.~ \textbf{547} (2001) L111 
\bibitem{74}   M.~E.~Putman, B.~K.~Gibson, Publ.~Astron.~Soc.~Aust.~ \textbf{16} (1999) 70 
\bibitem{75}   M.~E.~Putman, Publ.~Astron.~Soc.~Aust.~ \textbf{17} (2000) 1 
\bibitem{76}  I.~Kaz\`{e}s, T.~H.~Troland, R.~M.~Crutcher, Astron.~Astrophys.~\textbf{245} (1991) L17
\bibitem{77}   A.~Aharonian, and A.~M.~Atoyan, Astron.~Astrophys.~\textbf{362} (2000) 937 
\bibitem{78}   R.~J.~Gould, ApJ.~\textbf{344} (1989) 232
\bibitem{79}  R.~J.~Gould, Phys.~Rev.~A~ \textbf{23} (1981) 2851
\bibitem{80}   W.~R.~Purcell, D.~A.~Grabelsky, M.~P.~Ulmer, W.~N.~Johnson, R.~L.~Kinzer, J.~D.~Kurfess, M.~S.~Strickman,  ApJ.~ \textbf{413} (1993) L85 
\bibitem{81}   R.~L.~Kinzer, W.~R.~Purcell, W.~N.~Johnson, J.~D.~Kurfess, G.~Jung, J.~Skibo,  Astron.~\& Astrophys, Supplem.~\textbf{120} (1996) 317
\bibitem{82}   C.~D.~Dermer, and J.~G.~Skibo, ApJ.~\textbf{487} (1997) L57 
\bibitem{83}   J.~I.~Trombka, C.~S.~Dyer, L.~G.~Evans, M.~J.~Bielefeld, S.~M.~Seltzer, A.~E.~Metzger, ApJ.~ \textbf{212} (1977) 925
\bibitem{84}  D.~N.~Borrows, J.A.~Mendenhall, Nature \textbf{351} (1991) 629 
\bibitem{85}  U.~Herbstmeier, U.~Mebold, S.L.~Snowden, D.~Hartmann, W.~B.~Burton, P.~Moritz, P.~M.~W.~Kalberla, R.~Egger, Astron.~Astrophys.~\textbf{298} (1995) 606 
\bibitem{86}  B.~P.~Wakker, H.~Van Woerden, Ann.~Rev.~Astron.~Astrophys.~\textbf{35} (1997) 217 
\bibitem{87}  W.~H.~Press, Understanding data better with Bayesian and global statistical methods.~ In {\it{Unsolved Problems in Astrophysics}}, Ed.~J.~N.~Bahcall and J.~P.~Ostriker, Princeton University Press, Princeton, NJ, ISBN 0-691-01607-0
\bibitem{88}   A.~Sandage, ApJ.~{\textbf{133}} (1961) 355 
\bibitem{89}   S.~Perlmutter, et al.~ApJ.~{\textbf{483}} (1997) 565 
\bibitem{90}   S.~Perlmutter et al., Nature {\textbf{391}} (1998) 51
\bibitem{91}   A.~G.~Riess, et al., Astron.~Journal \textbf{116} (1998) 1009
\bibitem{92}   A.~G.~Riess, et al., ApJ.~\textbf{504} (1998) 935
\bibitem{93}   A.~G.~Riess, L.-G.~Strolger, J.~Tonry, S.~Casertano, H.~C.~Ferguson, B.~Mobasher, P.~Challis, A.~V.~Filippenko, S.~Jha, W.~Li, R.~Chornock, R.~Kirshner, B.~Leibundgut, M.~Dickinson, M.~Livio, M.~Giavalisco, C.~C.~Steidel, T.~Ben\'{i}tez, \& Z.~Tsvetanov, 2004, ApJ, {\textbf{607}}, 665
\bibitem{94}   G.~Goldhaber, D.~E.~Groom, A.~Kim, G.~Aldering, P.~Astier, A.~Conley, S.~E.~Deustua, R.~Ellis, S.~Fabbro, A.~S.~Fruchter, A.~Goobar, I.~Hook, M.~Irwin, M.~Kim, R.~A.~Knop, C.~Lidman, R.~McMahon, P.~E.~Nugent, R.~Pain, N.~Panagia, C.~R.~Pennypacker, S.~Perlmutter, P.~Ruiz-Lapuente, B.~Schaefer, N.~A.~Walton, T.~York,  2001 ApJ, {\textbf{558}}, 359   
\bibitem{95}   A.~Brynjolfsson, {\it{Plasma redshift, time dilation, and supernovas Ia}} arXiv:astro-ph/0406437 v2 20 Jul 2004.
\bibitem{96}    R.~J.~Foley, A.~V.~Filippenko, D.~C.~Leonard, A.~G.~Riess, P.~Nugent, S.~Perlmutter, ApJ \textbf{626}\,(2005)\,L11
\bibitem{97}    A.~V.~Filippenko and A.~G.~Riess, {\it{Results from the high-z supernova search team.}} arXiv:astro-ph/9807008 v1 1 Jul 1998.
\bibitem{98}    M.~R.~S.~Hawkins, ApJ., \textbf{553}\,(2001)\,L97
\bibitem{99}    E.~J.~Lerner, {\it{Evidence for a non-expanding universe: Surface brightness data from HUDF}}astro-ph/0509611 (To be published in the Proceedings of the first Crisis in Cosmology Conference, AIP)
\bibitem{100}    D.~J.~Fixen, et al., ApJ. \textbf{473} (1996) 576
\bibitem{101}    U.~Keshet, E.~Waxman, A.~Loeb,  ApJ. {\textbf{617}} (2004) 281
\bibitem{102}   K.~D.~Kuntz, S.~L.~Snowden, R.~F.~Mushotzky, ApJ. {\textbf{548}}\,(2001)\,L119  
\bibitem{103}   K.~D.~Kuntz, S.~L.~Snowden, ApJ. {\textbf{543}} (2000) 195  
\bibitem{104}   A.~De Luca, S.~Molendi, to be published in A.\,\&\,A., {\textbf{419}}\,(2004)\,837; (arXiv:astro-ph/0311538).  See also Mem. S. A. It. Vol. 73,1\,(2002) (arXiv:astro-ph/0402233) 
\bibitem{105}   A.~Vecchi, S.~Molendi, M.~Guainazzi, F.~Fiore, A.~N. Parmar; A.\,\&\,A., {\textbf{548}}\,(1999)\,L11 
\bibitem{106}   X.~Barcons, S.~Mateos, M.~T.~Ceballos, Mon.\,Not.\,R.\,Astron.\,Soc.,\,{\textbf{316}}\,(2000)\,L13
\bibitem{107}   A.~Brynjolfsson, {\it{Weightlessness of photons: A quantum effect}}; arXiv: astro-ph/0408312 v2 26 Aug 2004  
\bibitem{108}   E.~G.~Adelberger, C.~W.~Stubbs, B.~R.~Heckel, Y.~Su, H.~E.~Swanson, G.~Smith, J.~H.~Gundlach, W.~F.~Rogers;  Phys. Rev. D {\textbf{42}}, (1990) 3267 
\bibitem{109}   Y.~Su, B.~R.~Heckel, E.~G.~Adelberger, J.~H.~Gundlach, M.~Harris, G.~L.~Smith, H.~E.~Swanson; Phys. Rev. D {\textbf{50}}, (1994) 3614  
\bibitem{110}   A.~Einstein, Sitzungsberichte der Preussuschen Akad. Wissenschaften, (1917) p.\,142
\bibitem{111}   C.~M{\o}ller, {\it{The Theory of Relativity}}, 2nd ed., Oxford University Press 1972, Delhi, Bombay, Calcutta, Madras, SBN 19 560539 
\bibitem{112}   R.~Narayan, Accreation flows around black holes.~ In {\it{Unsolved Problems in Astrophysics}}, Ed.~J.~N.~Bahcall and J.~P.~Ostriker, Princeton University Press, Princeton, NJ, ISBN 0-691-01607-0 
\bibitem{113}   M.~Ruderman, In and around neutron stars. In {\it{Unsolved Problems in Astrophysics}}, Ed.~J.~N.~Bahcall and J.~P.~Ostriker, Princeton University Press, Princeton, NJ, ISBN 0-691-01607-0
\bibitem{114}   K.~A.~Olive, Science \textbf{251} (1991) 1194 
\bibitem{115}   W.~B.~Burton, The large-scale distribution of neutral hydrogen in the galaxy.~ In {\it{Galactic and Extragalactic Radio Astronomy}}; pp.~82-117.~ Ed.~by G.~L.~Verschuur, and K.~I.~Kellermann, ISBN 0-387-06504-0, Springer Verlag, New York, Heidelberg, Berlin, 1974
\bibitem{116}   J.~H.~Oort, {\it{Nonstable phenomena in ``Galaxies.''}}  Proc.~IAU Symp.~No 29.~Yerevan: Izd-vo Akademiia Nauk Armianskoi SSR, (1966) 41
\bibitem{117}   J.H.~Oort, {\it{Nuclei of galaxies}}; ed.~D.J.K.~O'Conell, North Holland Publishing Co., Amsterdam, (1971) 321
\bibitem{118}   P.~C.~van der Kruit, Astron.~\& Astrophys.~\textbf{13} (1970) 405 
\bibitem{119}    W.~K.~H.~Panofsky and M.~Phillips, {\it{Classical Electricity and Magnetism}}, Addison-Wesley Publ.~Co., Inc., Reading, MA 1956
\bibitem{120}    R.~Becker, {\it{Theorie der Elektrizität, Band 2}}, B.~G.~Teubner, Verlagsgesellschaft, Leipzig, 1949 
\bibitem{121}    J.~D.~Jackson, {\it{Classical Electrodynamics}}, John \& and Sons, Inc., New York, London, 1972; see in particular Chapter 17
\bibitem{122}    F.~Rohrlich, {\it{Classical Charged Particles}}, Addison-Wesley, Reading, MA, 1965 (see in particular Chapters.~6 and 9)
\bibitem{123}    F.~Rohrlich, {\it{The Theory of the Electron}}, The-first Joseph Henry lecture, read before the Society May 11, 1962.~ http://philsoc.org/1962Sprin/1526transcript.html
\bibitem{124}    P.~A.~M.~Dirac, {\it{Classical theory of radiating electrons}}, Proc.~of the Roy.~Soc.~London, Ser.~A, Math.~and Phys.~Sci.~ \textbf{167} (1938) 148
\bibitem{125}    F.~V.~Hartemann, A.K.~Kerman, Phys.~Rev.~Lett.~\textbf{76} (1996) 624 
\bibitem{126}    A.~Brynjolfsson, Phil.~Mag., Supplem.~ \textbf{6} (1957) 247  
\bibitem{127}    L.~Spitzer, {\it{Physical Processes in the Interstellar Medium}}, John Willey\&Sons, N.Y. Chichester, Brisbane, Toronto, (ISBN 0-471-02232-2) 1978
\bibitem{128}    G.~B.~Rybicki, A.~P.~Lightmann, {\it{Radiative Processes in Astrophysics}}, A Wiley-Interscience Publication, John Wiley\,\&\,Sons, New York, Chichester, Brisbane, Toronto, 1979
\bibitem{129}    M.~Asplund, N.~Grevesse, A.~J.~Sauval, {\it{Cosmic Abundances as Records of Stellar Evolution and Nucleosynthesis}}, ASP Conference Series, Vol.\,{\textbf{XXX}},\,2005, F.\,N.\,Bash\,\&\,T.\,G.\,Barnes\, (ed.); arXiv:astro-ph/0410214, v2 10 Oct 2004
\bibitem{130}    E.~B.~Saloman, J.~H.~Hubble, J.~H.~Scofield, {\it{Atomic Data and Nuclear Tables}} {\textbf{38}}\,(1988)\,1-197  
\bibitem{131}    S.~L.~Snowden, R.~Egger, D.~Finkbeiner, M.~J.~Freyberg, P.~P.~Plucinsky, ApJ.\,{\textbf{493}}\,(1998)\,715




\end{thebibliography}
\end{document}